\documentclass[11pt,a4paper]{article}
\pdfoutput=1
\usepackage[nosort]{cite}

\usepackage[british]{babel} %dates in british format

\usepackage{epsfig,rotating}
\usepackage{amsmath,amssymb,amsthm}
\usepackage{amsfonts}
\usepackage{mathrsfs}
\usepackage{bbm}
\usepackage{bm}
\usepackage[normalem]{ulem}
\usepackage[all,knot,cmtip]{xy}
\usepackage{tabu}
\usepackage{slashed}
\usepackage{mathabx}
\usepackage{enumerate}
\usepackage{adjustbox}

\xyoption{arc}
\usepackage{tikz}
\usetikzlibrary{arrows}
\usetikzlibrary{patterns}
\usetikzlibrary{decorations.markings}

\usepackage{xcolor}
\usepackage[textwidth = 430 pt, textheight = 630 pt]{geometry}
\definecolor{MyDarkBlue}{rgb}{0.15,0.25,0.45}
 
\usepackage[linktocpage=true,hypertexnames=false]{hyperref}
\hypersetup{
colorlinks=true,
citecolor=MyDarkBlue,
linkcolor=MyDarkBlue,
urlcolor=MyDarkBlue,
pdfauthor={Branislav Jurco, Lorenzo Raspollini, Christian S\"amann, Martin Wolf},
pdftitle={},
pdfsubject={hep-th},
breaklinks=true
} 

\let\fn\footnote
\renewcommand{\footnote}[1]{\linespread{1.1}\fn{#1}\linespread{1.29}}

\makeatletter
\newcommand{\xRightarrow}[2][]{\ext@arrow 0359\Rightarrowfill@{#1}{#2}}

\renewcommand{\section}{\@startsection
{section}{1}{\z@}{-3.5ex plus -1ex minus
    -.2ex}{2.3ex plus .2ex}{\bf\mathversion{bold} }}
\renewcommand{\subsection}{\@startsection{subsection}{2}{\z@}{-3.25ex
plus -1ex minus
   -.2ex}{1.5ex plus .2ex}{\bf\mathversion{bold} }}
\renewcommand{\subsubsection}{\@startsection{subsubsection}{3}{\z@}{-3.25ex
plus -1ex minus
   -.2ex}{1.5ex plus .2ex}{\bf\mathversion{bold} }}
\renewcommand{\thesection}{\arabic{section}}
\renewcommand{\thesubsection}{\arabic{section}.\arabic{subsection}}
\renewcommand{\thesubsubsection}{\arabic{section}.\arabic{subsection}.\arabic{subsubsection}}
\renewcommand{\@seccntformat}[1]{\@nameuse{the#1}.~~}
\renewcommand{\theequation}{\thesection.\arabic{equation}}
\makeatletter \@addtoreset{equation}{section}

\renewcommand*\l@section{\@dottedtocline{1}{0em}{2em}}
\renewcommand*\l@subsection{\@dottedtocline{2}{2em}{2.4em}}
\renewcommand*\l@subsubsection{\@dottedtocline{4}{3.8em}{3.7em}}

\renewcommand\tableofcontents{%
    \section*{\large\contentsname
        \@mkboth{%
          \MakeUppercase\contentsname}{\MakeUppercase\contentsname}}%
       {\baselineskip=15pt plus 2pt minus 1pt
    \@starttoc{toc}}%
}

\newcommand{\acknowledgements}{\section*{Acknowledgements}
\addcontentsline{toc}{section}{Acknowledgements}}

\newcommand{\datamanagement}{\section*{Data Management}
\addcontentsline{toc}{section}{Data Management}}

\newcommand{\appendices}{
\section*{Appendices}\label{appendices}\setcounter{subsection}{0}
\addcontentsline{toc}{section}{Appendices}
\setcounter{equation}{0}
\setcounter{thm}{0}
\makeatletter
\renewcommand{\theequation}{\Alph{subsection}.\arabic{equation}}
\renewcommand{\thesubsection}{\Alph{subsection}}
\renewcommand{\thethm}{\Alph{subsection}.\arabic{thm}}
\renewcommand{\thesubsubsection}{\Alph{subsection}.\arabic{subsubsection}}
\@addtoreset{equation}{subsection}
\@addtoreset{thm}{subsection}
\makeatother
}

\providecommand*{\xmapstofill@}{%
  \arrowfill@{\mapstochar\relbar}\relbar\rightarrow
}
\providecommand*{\xmapsto}[2][]{%
  \ext@arrow 0395\xmapstofill@{#1}{#2}%
}

\providecommand*{\xhookrightfill@}{%
  \arrowfill@{\lhook\joinrel\relbar}\relbar\rightarrow
}
\providecommand*{\xhookrightarrow}[2][]{%
  \ext@arrow 0395\xhookrightfill@{#1}{#2}%
}

\makeatother

\renewcommand{\thethm}{\thesection.\arabic{thm}}

\def\periodb#1{\setbox0=\hbox{$#1$}#1\hskip-\wd0\hbox to\wd0{-}}

\newcommand{\unit}{\mathbbm{1}}   			% identity map/matrix
\newcommand{\im}{\mathrm{im}}   			% identity map/matrix
\newcommand{\id}{\mathrm{id}}   			% identity map/matrix
\newcommand{\CCC}{\mathscr{C}}
\newcommand{\CCE}{\mathscr{E}}

\newcommand{\CD}{\mathcal{D}}

\newcommand{\CF}{\mathcal{F}}

\newcommand{\CCG}{\mathscr{G}}

\newcommand{\CCS}{\mathscr{S}}
\newcommand{\CL}{\mathcal{L}}
\newcommand{\CM}{\mathcal{M}}

\newcommand{\CN}{\mathcal{N}}

\newcommand{\CO}{\mathcal{O}}

\newcommand{\CCV}{\mathscr{V}}

\newcommand{\frd}{\mathfrak{d}}				% frak-letters
\newcommand{\frg}{\mathfrak{g}}				% frak-letters
\newcommand{\frF}{\mathfrak{F}}
\newcommand{\frG}{\mathfrak{G}}

\newcommand{\frI}{\mathfrak{I}}
\newcommand{\frL}{\mathfrak{L}}

\newcommand{\FR}{\mathbbm{R}}     			% field of real numbers
\newcommand{\FC}{\mathbbm{C}}     			% field of complex numbers
\newcommand{\NN}{\mathbbm{N}}     			% set of natural numbers
\newcommand{\RZ}{\mathbbm{Z}}     			% ring of integers
\newcommand{\dd}{\mathrm{d}}     			% total differential
\newcommand{\dpar}{\partial}     			% partial differential
\newcommand{\dparb}{{\bar{\partial}}}     		% partial differential with bar
\newcommand{\embd}{{\hookrightarrow}}     		% embedded
\newcommand{\de}{\mathrm{e}}     			% Euler's number
\newcommand{\di}{\mathrm{i}}     			% imaginary unit
\newcommand{\eps}{{\varepsilon}}			% antisymmetric tensors
\newcommand{\upmu}{\hat \mu}

\newcommand{\ald}{{\dot{\alpha}}}     			% dotted letters
\newcommand{\bed}{{\dot{\beta}}}

\newcommand{\eand}{{~~~\mbox{and}~~~}}     		% and etc. in equations
\newcommand{\ewith}{{~~~\mbox{with}~~~}}
\newcommand{\efor}{{~~~\mbox{for}~~~}}

\newcommand{\der}[1]{\frac{\dpar}{\dpar #1}}   		% partielle ableitung, 1 argument
\newcommand{\Sh}{\overline{\mathrm{Sh}}}     			% trace
\newcommand{\asu}{\mathfrak{su}}
\newcommand{\aso}{\mathfrak{so}}

\newcommand{\sV}{\mathsf{V}}

\newcommand{\sSO}{\mathsf{SO}}
\newcommand{\sLie}{\mathsf{Lie}}
\newcommand{\sHom}{\mathsf{Hom}}

\newcommand{\sG}{\mathsf{G}}

\newcommand{\sL}{\sfL}

\newcommand{\sA}{\mathsf{A}}

\newcommand{\sSpin}{\mathsf{Spin}}
\newcommand{\sEnd}{\mathsf{End}\,}

\newcommand{\acton}{\vartriangleright}     			% span
\newcommand{\remark}[1]{}     				% remark
\newcommand{\myxymatrix}[1]{\vcenter{\vbox{\xymatrix{#1}}}}
\def\tyng(#1){\hbox{\tiny$\yng(#1)$}}			% small Young diagram
\def\tyoung(#1){\hbox{\tiny$\young(#1)$}}			% small Young diagram
\newcommand{\sff}{\mathsf{f}}

\newcommand{\sfa}{\mathsf{a}}
\newcommand{\sfb}{\mathsf{b}}
\newcommand{\sfc}{\mathsf{c}}

\newcommand{\sfl}{\mathsf{l}}

\newcommand{\pr}{{\rm pr}}

\newcommand{\sfL}{\mathsf{L}}

\newcommand{\intprod}{\mathbin{\raisebox{\depth}{\scalebox{1}[-1]{$\lnot$}}}}

\begin{document}
\begin{titlepage}

\setcounter{page}{0}
\renewcommand{\thefootnote}{\fnsymbol{footnote}}

\begin{flushright}
 EMPG--18--19\\ DMUS--MP--18/05
\end{flushright}

\begin{center}

{\LARGE\textbf{\mathversion{bold}$L_\infty$-Algebras of Classical Field Theories\\ and the Batalin--Vilkovisky Formalism}\par}

\vspace{1cm}

{\large
Branislav Jur\v co$^{a}$, Lorenzo Raspollini$^{b}$, Christian S\"amann$^{c}$, and Martin Wolf$^{\,b}$
\footnote{{\it E-mail addresses:\/}
\href{mailto:branislav.jurco@gmail.com}{\ttfamily branislav.jurco@gmail.com},
\href{mailto:l.raspollini@surrey.ac.uk}{\ttfamily l.raspollini@surrey.ac.uk},
\href{mailto:c.saemann@hw.ac.uk}{\ttfamily c.saemann@hw.ac.uk}, 
\href{mailto:m.wolf@surrey.ac.uk}{\ttfamily m.wolf@surrey.ac.uk}
}}

\vspace{.5cm}

{\it
$^a$ 
Charles University\\
Faculty of Mathematics and Physics, Mathematical Institute\\
Prague 186 75, Czech Republic\\[.3cm]

$^b$
Department of Mathematics,
University of Surrey\\
Guildford GU2 7XH, United Kingdom\\[.3cm]

$^c$ Maxwell Institute for Mathematical Sciences\\
Department of Mathematics,
Heriot--Watt University\\
Edinburgh EH14 4AS, United Kingdom

}

\vspace{1cm}

{\bf Abstract}
\end{center}
\vspace{-.5cm}
\begin{quote}
We review in detail the Batalin--Vilkovisky formalism for Lagrangian field theories and its mathematical foundations with an emphasis on higher algebraic structures and classical field theories. In particular, we show how a field theory gives rise to an $L_\infty$-algebra and how quasi-isomorphisms between $L_\infty$-algebras correspond to classical equivalences of field theories. A few experts may be familiar with parts of our discussion, however, the material is presented from the perspective of a very general notion of a gauge theory. We also make a number of new observations and present some new results. Most importantly, we discuss in great detail higher (categorified) Chern--Simons theories and give some useful shortcuts in usually rather involved computations.

%\vfill\noindent\today
\vfill\noindent 11th July 2019

\end{quote}

\setcounter{footnote}{0}\renewcommand{\thefootnote}{\arabic{thefootnote}}

\end{titlepage}

\tableofcontents
\bigskip
\bigskip
\hrule
\bigskip
\bigskip

\section{Introduction}

Categorified or higher mathematical structures appear very naturally within string theory. In particular, the Kalb--Ramond $B$-field is part of the connective structure on a gerbe,  which is the categorification of the notion of a connection on a circle bundle. Moreover, string field theory~\cite{Zwiebach:1992ie} is fundamentally based on homotopy Maurer--Cartan theory, the vastly generalised analogue of Chern--Simons theory to strong homotopy Lie algebras or $L_\infty$-algebras, which are $\infty$-categorifications of the notion of a Lie algebra.

If one believes in the fundamental nature of string theory, it is then not too surprising that remnants of these categorified structures are also found in ordinary classical field theories. In particular, the Batalin--Vilkovisky (BV) formalism~\cite{Batalin:1984jr,Batalin:1985qj,Batalin:1984ss,Batalin:1981jr,Batalin:1977pb} associates to each classical field theory an $L_\infty$-algebra and for interacting field theories, this $L_\infty$-algebra is {\em not} merely a differential graded Lie algebra. This fact is well-known to experts on BV quantisation, see for example~\cite{Fisch:1989rp,Barnich:1997ij}, in particular~\cite{Fulp:2002kk}, which is based on the earlier work~\cite{Berends:1984rq}, or the later works~\cite{Movshev:2003ib,Movshev:2004aw,Zeitlin:2007vv,Zeitlin:2007vd,Zeitlin:2007yf,Zeitlin:2007fp,Zeitlin:2008cc,Rocek:2017xsj}, but it seems to be much less known in general. The recent paper~\cite{Hohm:2017pnh} revived interest in the $L_\infty$-algebras of classical field theories, but only a very partial picture of the categorified structures and their origin was given.

The purpose of this paper is threefold. First of all, we wish to present an accessible, self-contained review\footnote{We do not claim that our review is historically accurate or complete regarding references.} of the complete picture, providing links to the standard mathematical terminology and collecting relevant references for further reading. Explicitly, we explain how $L_\infty$-algebras necessarily arise from the classical master equation of the BV formalism and why quasi-isomorphisms constitute the correct and very useful notion of equivalence which corresponds to the classical equivalence of field theories.  

Secondly, this paper is meant to be a starting point of a much deeper investigation of higher structures in both classical and quantum field theories which we plan to conduct in the near future. Here, we shall lay the ground work by fixing our notation and conventions and by reviewing the necessary basics of the BV formalism for future reference.

Thirdly, we would like to present some new results and observations we made. Most importantly, our perspective is adapted to an application towards higher gauge theory\footnote{We note that higher form fields have been treated in the BV formalism before, see e.g.~\cite{Baulieu:1996ep,Alfaro:1997ku} and, more recently,~\cite{Soncini:2014ara,Zucchini:2015ohw}, but not in the generality we discuss here. A rather general account from a somewhat different perspective than ours is found in~\cite{Zucchini:2017nax}.} from the outset. Nevertheless, we stress again that most of our language should be readily accessible to any theoretical physicist.

To be reasonably self-contained, we start with a detailed review on $L_\infty$-algebras, which are also known as strong homotopy Lie algebras. These arise as a particular categorification of the notion of a Lie algebra that is described in terms of graded vector spaces and higher brackets. Much more directly, they arise by generalising the equivalent definition of a Lie algebra via its differential graded algebra (dga) known as the Chevalley--Eilenberg algebra. Both, the higher bracket description as well as the dga-picture, have their individual advantages in computations, and it is occasionally very useful to switch\footnote{When performing this switch, one essentially converts a vector into a grade-shifted coordinate function. This transition is the reason for the fermionic character of the ghosts corresponding to gauge parameters and it can lead to confusion. We shall therefore be always very explicit regarding this point.} between those. This justifies our discussion of the technical details involved in this switch. They mostly boil down to carefully keeping track of signs arising from grade-shifting and permuting graded operators as well as working with the somewhat less familiar concept of a coalgebra.

Moreover, the classical BV complex provides precisely such a differential graded algebra corresponding to the classical $L_\infty$-algebra $\sL=\bigoplus_{k\in\RZ} \sL_k$ of a field theory. This $L_\infty$-algebra governs the field content, the (higher) gauge structure, the equations of motion and the (higher) Noether identities of a field theory and they have an underlying complex of the form
\begin{equation}
 \underbrace{\parbox{2.3cm}{\centering gauge\\symmetries\\[-.3cm]{}\phantom{*}}}_{\dots,~\sL_{-1},~ \sL_0}\ \longrightarrow \ \underbrace{\parbox{2.3cm}{\centering classical\\fields\\[-.3cm]{}\phantom{*}}}_{\sL_1}\ \longrightarrow \ \underbrace{\parbox{2.3cm}{\centering equations\\of motion\\[-.3cm]{}\phantom{*}}}_{\sL_2}\ \longrightarrow \ \underbrace{\parbox{2.3cm}{\centering Noether identities\\[-.3cm]{}\phantom{*}}}_{\sL_3,~\sL_4,~\dots}
\end{equation}
It is important to stress that the BV procedure does not only apply to gauge field theories and gauge symmetries. The usual discussion of the BV formalism involves two operations: the Chevalley--Eilenberg resolution of the gauge algebra, leading to the Becchi--Rouet--Stora--Tyutin (BRST) complex, as well as the Koszul--Tate resolution of the classical equations of motion, which introduces antifields. The first operation can also be applied to global symmetries of a field theory, even though the usual motivation for the BRST formalism, namely gauge fixing for subsequent quantisation, does not apply here. The second operation can be applied to any classical field theory. We shall discuss scalar field theory as an explicit example.

Besides the obvious isomorphisms, $L_\infty$-algebras come with a more general notion of isomorphism, called quasi-isomorphism. From a mathematical, $\infty$-categorical point of view, quasi-isomorphisms are the most appropriate ones to consider in most cases.\footnote{For example, all definitions of the gerbes governing the higher form fields work up to quasi-isomorphisms.} From a physical point of view, they will allow us to identify equivalent field theories with non-isomorphic field spaces (e.g.~field theories equivalent after integrating out fields, etc.). As an explicit example, we shall demonstrate that the $L_\infty$-algebras of the first-order and second-order formulations of classical Yang--Mills theory are quasi-isomorphic. We shall also review their descriptions in both the higher bracket and the dga-picture. 

These quasi-isomorphisms allow us to reduce the $L_\infty$-algebra, and, correspondingly, the field space, to the so-called {\em minimal model} in which the non-triviality of the action is fully absorbed in the higher products. For example, quasi-isomorphisms allow us to reduce the field space of Chern--Simons theory on a three-dimensional compact oriented manifold $M$ with gauge connection restricted to a trivial principal fibre bundle to the first de Rham cohomology $H^1_{\rm dR}(M)$ of $M$, which is to be seen as the dual to (topological) Wilson loops. In this paper, we merely make the observation that this quasi-isomorphism exists. We plan to study various aspects of this reduction to minimal models in future work.

The most relevant or natural example of a field theory for the BV formalism is the above-mentioned homotopy Maurer--Cartan theory defined for any $L_\infty$-algebra with an inner product of degree~$-3$, which is induced by a particular symplectic form of degree~$-1$. In the special case of an $L_\infty$-algebra arising from tensoring a Lie algebra with the differential forms $\Omega^\bullet(M)$ on a 3-dimensional compact oriented manifold $M$, we recover ordinary Chern--Simons theory. Generally, we construct higher Chern--Simons theory for a {\em Lie $n$-algebra} or {\em $n$-term $L_\infty$-algebra} in dimension $2+n$. 

One remarkable point about homotopy Maurer--Cartan theory for an arbitrary $L_\infty$-algebra $\sL$ is that the BV complex of fields, ghost, and their antifields is just the $L_\infty$-algebra $\sL$ itself. Moreover, the differential encoding $\sL$ in the differential graded algebra picture originates from a vector field with Hamiltonian, the latter of which is essentially the homotopy Maurer--Cartan action. Finally, this homotopy Maurer--Cartan action, at least formally, satisfies the quantum master equation of the BV formalism, which can simplify the computation of the quantum master equation. We have not found these and other observations we make in connection with the BV formalism in the literature.

A useful property we observe for the first time is the fact that homotopy Maurer--Cartan theory always allows for a supersymmetric extension by auxiliary fields, just as ordinary Chern--Simons theory. This is important if one wishes to compute path integrals via supersymmetric localisation techniques.

As stated above, this paper is also intended as the ground work for future research on higher structures in the context of classical and quantum BV formulations of field theories. A particularly interesting topic here is the relation between quasi-isomorphism and renormalisation group flow\footnote{For some recent work on the renormalisation in the BV context, see~\cite{Zucchini:2017ilk,Zucchini:2017ydo} and in particular~\cite{Costello:2007ei}.}, which has been established in some special cases, cf.~\cite{Sharpe:2003dr}.

\section{Mathematical tools}

\paragraph{Motivation.} The local description of gauge theories is based on Lie algebra-valued differential $k$-forms. Let $M$ be a manifold and $\frg$ a Lie algebra, then a corresponding gauge theory has a gauge potential $A\in\Omega^1(M,\frg):=\Omega^1(M)\otimes \frg$ and a curvature $F:=\dd A+\tfrac12[A,A]\in \Omega^2(M,\frg)$. Gauge parameters, Bianchi identities, Noether currents, and equations of motion also involve elements of $\Omega^\bullet(M,\frg)$. 

To obtain a natural (i.e.~category theoretical) description of gauge theory, we should therefore bring together differential forms and Lie algebras in a common framework. This framework is provided by differential graded algebras, and the differential graded algebras in which we are interested arise as functions on particular graded manifolds. On these manifolds we have a vector field $Q$ that induces a differential on the algebra of functions. Henceforth, they are referred to as {\it $Q$-manifolds}, cf.~\cite{Alexandrov:1995kv}.

Formulating gauge theories using differential graded algebras arising from $Q$-manifolds has several advantages. Firstly, one can define vast generalisations of ordinary gauge theory~\cite{Sati:2008eg} which appear naturally in string and M-theory in various contexts. Secondly, because of its mathematical naturality, it is not surprising that a powerful framework such as the Batalin--Vilkovisky (BV) formalism is best formulated in this language~\cite{Schwarz:1992nx}. Thirdly, this is how they appear directly in the Alexandrov--Kontsevich--Schwarz--Zaboronsky\linebreak (AKSZ) construction~\cite{Alexandrov:1995kv}.

In this section, we therefore review differential graded algebras with a focus on those arising as function algebras on $Q$-manifolds.

\subsection{Differential graded algebras}\label{ssec:DGAs}

We shall assume that the reader is familiar with $\RZ$-graded vector spaces. For more details on this topic, see e.g.~\cite{Cattaneo:2010re,Fairon:1512.02810}.

\paragraph{\mathversion{bold}$\RZ$-graded vector spaces and shifts.} For a $\RZ$-graded vector space $\sV=\bigoplus_{k\in\RZ}\sV_k$, we introduce the following notation for a degree shift by $l$:
\begin{equation}\label{eq:grade_shift}
 \sV[l]\ =\ \bigoplus_{k\in\RZ} (\sV[l])_k\ewith(\sV[l])_k\ :=\ \sV_{k+l}\efor l\ \in\ \RZ~.
\end{equation}
This convention (which is one of two commonly used ones) indicates the shift of the {\em coordinate functions} and the {\em opposite direction} of the shift of the vectors themselves. For example, given an ordinary vector space $V$, the degree-shifted vector space $V[1]$ consists of vectors $v$ of degree~$-1$, since only $V[1]_{-1}=V$ is non-trivial.  Note that the signs of the shifts are flipped when taking the dual vector space: the dual $\sV^*$ of a $\RZ$-graded vector space $\sV$ has the homogeneous subspaces $(\sV^*)_k=(\sV_{-k})^*$ for all $k\in\RZ$. 

Let us already now stress an important point for our whole discussion. With respect to a basis $\tau_\alpha$ of degree~$-1$ of the above example $V[1]$ of a grade-shifted ordinary vector space $V$, the coordinate functions $\xi^\alpha\in V[1]^*$ are of degree~1,
\begin{equation}
 \xi^\alpha\,:\, V[1]\ \rightarrow\  \FR\ewith\xi^\alpha(v)\ =\ \xi^\alpha(v^\beta \tau_\beta)\ :=\  v^\beta \underbrace{\xi^\alpha(\tau_\beta)}_{=:\,\delta^\alpha_\beta}\ =\ v^\alpha~.
\end{equation}
There is now much room for confusion between the coordinates $v^\alpha$ and the coordinate functions $\xi^\alpha$; Nick Woodhouse~\cite{Woodhouse:2009aa} coined the term {\em first fundamental confusion of calculus} for this phenomenon. While in the ungraded case, this confusion is usually reasonably controlled, it can get out of hands in the graded case, since the degree of the object and the coordinate functions acting on it will be inverse to each other.    

In the context of both the Becchi--Rouet--Stora--Tyutin (BRST) and the BV formalisms, this problem with degrees is exacerbated by the fact that there is an additional, implicit shift in degree by $-1$. For example, consider a gauge parameter $c\in\Omega^0(M,\frg)$ and a gauge potential $A\in \Omega^1(M,\frg)$, where, as before, $M$ is a manifold and $\frg$ a Lie algebra. These fields belong to the graded vector space 
\begin{equation}
 \sV\ =\ \sV_0\oplus \sV_1 \ :=\  \Omega^0(M,\frg)\oplus \Omega^1(M,\frg)~.
\end{equation}
The corresponding BRST complex, however, is that of $\sV[1]=\sV[1]_{-1}\oplus \sV[1]_0$. Consequently, we obtain coordinate functions of degrees~1~and~0,
\begin{equation}
 c^\alpha(x)\,:\, \Omega^0(M,\frg)[1] \ \rightarrow\  \FR\eand A^\alpha_\mu(x)\,:\,\Omega^1(M,\frg)\ \rightarrow\  \FR~,
\end{equation}
with $\tau_\alpha$ and $\dd x^\mu$ are bases\footnote{Note that  $\dd x^\mu$ is a basis of $\Omega^1(M)$ regarded as a module over $\CCC^\infty(M)$. More appropriately, we should be using the infinite-dimensional basis of $\Omega^1(M)$ regarded as a vector space over $\FR$. To avoid the related, potentially distracting technicalities, we are slightly sloppy here.} of $\frg$ and $\Omega^1(M)$, respectively. For convenience, we shall often contract the coordinate functions with the basis of $\sV$ to arrive at the {\em contracted coordinate functions}, which we denote by the same letters $c$ and $A$ as customary in the discussion of BRST/BV quantisation. These contracted coordinate functions are always of total degree~1. Whether we mean vectors or their coordinate functions should always be clear from the context, and we hope that no confusion will arise. The degree of the vectors $c$ and $A$ in $\sV$ will be called the {\em $L_\infty$-degree}, while the degree of the (uncontracted) coordinate functions $c^a(x)$ and $A_\mu^\alpha(x)$ will be called the {\em ghost degree}. The latter agrees with the general nomenclature.

\paragraph{Commutative dg-algebras.}
A {\it differential graded commutative\footnote{We shall only be concerned with commutative algebras and hence drop the adjective `commutative' in the following.} algebra} (or a dg(c)-algebra for short) is an associative unital commutative algebra $\sA$ which is simultaneously a $\RZ$-graded algebra and a differential algebra in a way that all structures are compatible. Specifically, the $\RZ$-grading means that we have the decomposition $\sA=\bigoplus_{k\in\RZ} \sA_k$ and non-zero elements of $\sA_k$ will be called {\em homogeneous} and {\em of degree~$k\in\RZ$}. In addition, the product $\sA\times\sA\to\sA$ is graded commutative,
\begin{equation}
 a_1a_2\ =\ (-1)^{|a_1||a_2|} a_2a_1
\end{equation}
for $a_{1,2}\in\sA$ of homogeneous degrees $|a_{1,2}|\in\RZ$. The graded commutative algebra $\sA$ becomes a differential algebra if it is equipped with differential derivations $\dd_k:\sA_k\to\sA_{k+1}$ of homogeneous degree~$1$, which we collectively denote by $\dd$. Specifically, $\dd$ satisfies $\dd^2=\dd\circ \dd=0$ $\Leftrightarrow$ $\dd_{k+1}\circ \dd_k=0$ and obeys the graded Leibniz rule 
\begin{equation}
 \dd (a_1a_2)\ =\ (\dd a_1)a_2+(-1)^{|a_1|} a_1 (\dd a_2)
\end{equation}
for $a_{1,2}\in\sA$ and $a_1$ of homogeneous degree~$|a_1|\in\RZ$. We shall write $(\sA,\dd)$ in the following to indicate a dg-algebra. 

\paragraph{Examples.}
The prime example in view of applications to gauge theory is the differential graded algebra given by the de Rham complex on a $d$-dimensional manifold $M$,
\begin{equation}
 \Omega^0(M)\ \xrightarrow{~\dd~}\ \Omega^1(M)\ \xrightarrow{~\dd~}\ \cdots \ \xrightarrow{~\dd~}\ \Omega^d(M)~.
\end{equation}
Another important example is that of the Chevalley--Eilenberg algebra described in Section~\ref{ssec:L_infty_algebras_and_algebroids}.

\paragraph{Cochain complexes.}
Notice that $\dd$ gives $\sA$ the structure of a cochain complex, 
\begin{equation}
 \cdots\ \xrightarrow{~\dd~}\ \sA_{-1}\ \xrightarrow{~\dd~}\ \sA_0\ \xrightarrow{~\dd~}\ \sA_1\ \xrightarrow{~\dd~}\ \cdots~,
\end{equation}
and its cohomology $H^\bullet(\sA,\dd)$ is a graded algebra. For convenience, we shall use the cochain convention and use the terms {\em cochain} and {\em cohomology} versus the slightly more common {\em chain} and {\em homology}, see the remark at the beginning of Appendix~\ref{app:Hodge}. More abstractly, we can define a dg-algebra also as a monoid object in the monoidal category of (co)chain complexes.

\paragraph{Morphisms of dg-algebras.} 
A {\it morphism} $f\,:\,(\sA,\dd)\to (\sA',\dd')$ between two dg-algebras $(\sA,\dd)$ and $(\sA',\dd')$ is a collection $f$ of maps $f_k:\sA_k\to\sA'_k$ for all $k\in\RZ$ of degree~0 which respects the differential in the sense that $f\circ\dd=\dd'\circ f$ $\Leftrightarrow$ $f_{k+1}\circ\dd_k=\dd'_k\circ f_k$.  An {\em isomorphism of dg-algebras} is an invertible morphism. This notion of isomorphism will turn out to be too strict for our purposes, mainly due to our interpretation of dg-algebras as categorified Lie algebras. More appropriately, we should use {\em quasi-isomorphisms of dg-algebras}. We shall return to this point and explain it in detail in Section~\ref{ssec:quasiisomorphism}.

\paragraph{Tensor algebras and shift isomorphism.} Recall that the (real) tensor algebra of $\sV$,
\begin{subequations}
\begin{equation}
 \bigotimes\nolimits^\bullet \sV\ :=\ \FR\oplus \sV \oplus (\sV\otimes \sV)\oplus~\cdots\ =\ \bigoplus_{k\geq0}\bigotimes\nolimits^k \sV~,
\end{equation}
has two totally graded symmetric and graded antisymmetric subalgebras,
\begin{equation}\label{eq:GradedSubTensorAlgebras}
\begin{gathered}
 \bigodot\nolimits^\bullet \sV\ :=\ \FR\oplus \sV \oplus (\sV\odot \sV) \oplus~\cdots\ =\ \bigoplus_{k\geq0}\bigodot\nolimits^k \sV~,\\
 \bigwedge\nolimits^\bullet \sV\ :=\ \FR\oplus \sV \oplus (\sV\wedge \sV) \oplus~\cdots\ =\ \bigoplus_{k\geq0}\bigwedge\nolimits^k \sV~,
\end{gathered}
\end{equation}
and their reduced counterparts\footnote{Other common notations are $\bar\bigodot^\bullet \sV$ and $\bar \bigwedge^\bullet \sV$.}
\begin{equation}\label{eq:reduced_algebras}
\begin{gathered}
\bigodot\nolimits^\bullet_0 \sV\ :=\ \sV \oplus (\sV\odot \sV) \oplus ~\cdots\ =\ \bigoplus_{k\geq1}\bigodot\nolimits^k \sV~,\\
 \bigwedge\nolimits^\bullet_0 \sV\ :=\ \sV \oplus (\sV\wedge \sV) \oplus~\cdots\ =\ \bigoplus_{k\geq1}\bigwedge\nolimits^k \sV~.
\end{gathered}
\end{equation}
\end{subequations}
The shift isomorphism $s:\sV\rightarrow \sV[1]$ induces an isomorphism of graded algebras,
\begin{equation}\label{eq:isomorphism_asym_sym}
\begin{gathered}
s^\bullet\,:\,\bigwedge\nolimits^\bullet \sV\ \rightarrow\  \bigodot\nolimits^\bullet \sV[1]~,\\
 s^{\otimes i}\,:\ v_1\wedge \ldots \wedge v_i\ \mapsto\ (-1)^{\sum_{j=1}^{i-1}(i-j)|v_j|} s v_1\odot \dots \odot s v_i
\end{gathered}
\end{equation}
for $v_1,\ldots,v_i\in\sV$. The sign arises from the usual {\it Koszul sign rule},
\begin{equation}
 (s\otimes s)(v_1\otimes v_2)\ :=\ (-1)^{|v_1|}sv_1\otimes s v_2
 \end{equation}
 for $v_1,v_2\in\sV$. The inverse map is given by
\begin{equation}
 (s^{\otimes i})^{-1}\ =\ (-1)^{\frac12i(i-1)} (s^{-1})^{\otimes i}~.
\end{equation}
This shift isomorphism will be crucial in treating Lie and higher Lie algebras as differential graded algebras.

\subsection{\texorpdfstring{$Q$}{Q}-manifolds}\label{ssec:QManifolds}

\paragraph{Motivation.} It is known that the differential forms $\Omega^\bullet(M)$ on a manifold $M$ can be regarded as the smooth functions $\CCC^\infty(T[1]M)$ on the degree-shifted tangent bundle $T[1]M$. Indeed, using local coordinates $x^\mu$, $\mu=1,\ldots,\dim(M)$, on $M$ and local coordinates $\xi^\mu$ on the fibres of $T[1]M$, functions on $T[1]M$ are simply polynomials in $\xi^\mu$, that is, they are of the form $f(x,\xi)=f^\circ(x)+\xi^\mu f_\mu(x)+\frac12 \xi^\mu\xi^\nu f_{\mu\nu}(x)+\cdots\in \CCC^\infty(T[1]M)$. Identifying $\xi^\mu$ with $\dd x^\mu$ amounts to the identification $\CCC^\infty(T[1]M)\cong\Omega^\bullet(M)$. In addition, the de Rham differential $\dd$ becomes the vector field $Q=\xi^\mu\der{x^\mu}$ under this identification with $Q^2=0$. The manifold $T[1]M$, together with $Q$, forms an important example of a {\em $Q$-manifold}~\cite{Schwarz:1992nx,Schwarz:1992gs,Alexandrov:1995kv}. These $Q$-manifolds also provide a very efficient way of encoding a categorified Lie algebra in the form of an $L_\infty$-algebra or a categorified Lie algebroid in the form of an $L_\infty$-algebroid, as we shall see later. Let us therefore recall some basic notions.

\paragraph{\mathversion{bold}$\RZ$-graded manifolds with body $\FR^d$.}
Consider $\FR^d$ as a manifold. Furthermore, let $\sV$ be a $\RZ$-graded vector space. We may enlarge the ring of smooth functions $\CCC^\infty(\FR^d)$ on $\FR^d$ by considering the tensor product $\bigodot^\bullet \sV^*\otimes \CCC^\infty(\FR^d)$, where $\bigodot^\bullet$ denotes the graded symmetric tensor algebra~\eqref{eq:GradedSubTensorAlgebras}. We call the result the algebra of functions
\begin{equation}\label{eq:LocalFunctionZManifoldRd}
 \CCC^\infty(M)\ :=\ \bigodot\nolimits^\bullet \sV^*\otimes \CCC^\infty(\FR^d)
\end{equation}
on the {\em $\RZ$-graded manifold} $M$ and the underlying ordinary manifold $\FR^d$ is called the {\em body} $M^\circ$ of $M$. By {\em coordinates} on the $\RZ$-graded manifold $M$, we mean a set of ordinary coordinates on $M^\circ$ together with a set of generators of $\bigodot^\bullet V^*$, say $\xi^\alpha$ with $\alpha\in I$ for some index set $I$.

Elements of $\CCC^\infty(M)$ are clearly polynomials in the generators $\xi^\alpha$ whose coefficients are functions on $M^\circ$,
\begin{equation}
 f(x,\xi)\ =\ f^\circ(x)+\xi^\alpha f_\alpha(x)+\tfrac{1}{2!}\xi^\alpha\xi^\beta f_{\alpha\beta}(x)+\cdots
\end{equation}
with $f^\circ,f_\alpha,f_{\alpha\beta},\ldots \in \CCC^\infty(M^\circ)$. We shall make extensive use of the natural decomposition 
\begin{equation}
 \CCC^\infty(M)\ \cong\ \bigoplus_{k\in \RZ} \CCC^\infty_k(M)~,
\end{equation}
where $\CCC^\infty_k(M)$ are the {\em homogeneous functions of degree~$k$}. These are spanned by the monomials of $\RZ$-degree~$k$, i.e.~monomials $\xi^{\alpha_1}\cdots \xi^{\alpha_n} f_{\alpha_1\cdots \alpha_n}(x)$, where $f_{\alpha_1\cdots \alpha_n}(x)\in \CCC^\infty(M^\circ)$ and $|\xi^{\alpha_1}|+\dots+|\xi^{\alpha_n}|=k$.

For most of our purposes, the above local picture~\eqref{eq:LocalFunctionZManifoldRd} of the ring of functions is sufficient. For some aspects, as e.g.~the correct definition of morphisms between $\RZ$-graded manifolds, however, the full mathematical definition\footnote{This definition also resolves naive paradoxes concerning super and graded manifolds.} can be helpful. We therefore recall it in the following.

\paragraph{General $\RZ$-graded manifolds.} A {\it ringed space} $M$ is a pair $(|M|,\CCS_M)$ where $|M|$ is a topological space and $\CCS_M$ a sheaf of rings on $|M|$ called the {\it structure sheaf} of $M$. A {\it locally ringed space} is a ringed space $(|M|,\CCS_M)$ such that all stalks of $\CCS_M$ are local rings, that is, they have unique maximal ideals. 

We then define a {\it morphism} of $(|M|,\CCS_M)\to (|M'|,\CCS_{M'})$ of locally ringed spaces to be a pair $(f,f^\sharp)$ where $f:|M|\to |M'|$ is a morphism of topological spaces and $f^\sharp\,:\,\CCS_{M'}\to f_*\CCS_M$ a comorphism of local rings, that is, a map that respects the maximal ideals. Here, $f_*\CCS_M$ is the zeroth direct image of $\CCS_M$ under $f$, that is, for any open subset $U'$ of $|M'|$ there is a comorphism $f^\sharp_{U'}:\CCS_{M'}|_{U'}\to\CCS_M|_{f^{-1}({U'})}$. If the structure sheaves carry extra structure such as a $\RZ$-grading, then we require the morphism to respect this structure. 

A {\it $\RZ$-graded manifold} is then defined to be a locally ringed space $M=(|M|,\CCS_M)$ for $|M|$ a topological manifold such that for each $x\in |M|$ there is an open neighbourhood $U\ni x$ and an isomorphism of locally ringed spaces,
\begin{equation}\label{eq:LocalIsoRingedSpaces}
(U,\CCS_M|_U)\ \cong\ (U',\bigodot\nolimits^\bullet\CCV^*_{U'}\otimes \CCC^\infty_{U'})~,
\end{equation}
where $U'\subseteq\FR^d$ open, $\CCC^\infty_{U'}$ the sheaf of smooth functions on $U'$, and $\CCV_{U'}$ a locally free $\RZ$-graded sheaf of $\CCC^\infty_{U'}$-modules on $U'$. Hence, we require the sheaf of functions to look locally like~\eqref{eq:LocalFunctionZManifoldRd}. We shall call the locally ringed space $(|M|,\CCC^\infty_M)$, where $\CCC^\infty_M$ is the sheaf of smooth functions\footnote{Recall that any smooth manifold can be defined as a locally ringed space $(|M|,\CCS_M)$ for a topological manifold $|M|$ such that for each $x\in |M|$ there is an open neighbourhood $U\ni x$ and an isomorphism of locally ringed spaces $(U,\CCS_M|_U)\cong (U',\CCC^\infty_{U'})$ where $\CCC^\infty_{U'}$ is the sheaf of smooth functions on the open set $U'\subseteq\FR^d$. The stalk of $\CCS_{M}$ at a point $x\in|M|$ is the set of all germs of smooth functions at $x\in|M|$, and the maximal ideal of the stalk are the functions vanishing at $x\in|M|$. Furthermore, if $f:|M|\to|M'|$ is a continuous function between two topological manifolds $|M|$ and $|M'|$ for two smooth manifolds $(|M|,\CCS_M)$ and $(|M'|,\CCS_{M'})$ and if there is a comorphism $F\,:\,\CCS_{M'}\to f_*\CCS_M$ of local rings, then $f$ must also be smooth and $F=f^\sharp$.} on $|M|$ (which is a subsheaf of $\CCS_M$), the {\it body} of $M=(|M|,\CCS_M)$ and denote it by $M^\circ$. We shall also write $\CCC^\infty(M):=\Gamma(|M|,\CCS_M)$ for the global functions on $M$.

\paragraph{Examples.}
A convenient way of obtaining $\RZ$-graded manifolds is by degree shifting the fibres of a vector bundle. We already mentioned the simplest example of the degree-shifted tangent bundle $T[1]M$. We can also consider more general vector bundles $E\rightarrow M^\circ$ over a manifold $M^\circ$ and shift the degrees of their fibre coordinates such that each fibre becomes a $\RZ$-graded vector space. The result is a {\em split $\RZ$-graded manifold}. Just as all real supermanifolds are diffeomorphic to split ones~\cite{JSTOR:1998201}, also all real $\RZ$-graded manifolds are diffeomorphic to split ones~\cite{Bonavolonta:2012fh}. Note, however, that complex $\RZ$-graded manifolds are not split in general which is basically due to the non-existence of a holomorphic partition of unity.

\paragraph{Vector fields and differential forms on \mathversion{bold}$\RZ$-graded manifolds.} 
A {\em vector field} $V$ on a $\RZ$-graded manifold $M$ is a graded derivation $V:\CCC^\infty(M)\rightarrow \CCC^\infty(M)$. That is, for homogeneous $V$ of degree~$|V|\in\RZ$ and homogeneous $f,g\in\CCC^\infty(M)$, we have
\begin{equation}
V(fg)\ =\ V(f)g+(-1)^{|V|\,|f|}f\,V(g)~.
\end{equation}
As in the ordinary case, we define the tangent bundle $TM$ of a $\RZ$-graded manifold $M$ to be the disjoint union of the tangent spaces which in turn are the vector space of derivations at particular points of $M$.

A particular example of a vector field is the {\em Euler vector field} $\Upsilon$ which is defined by its action $\Upsilon f:=|f| f$ on any homogeneous $f\in\CCC^\infty(M)$ of degree~$|f|\in\RZ$. The Euler vector field itself is homogeneous and of degree~0.

Note that the definition of differential forms on an ordinary manifold $M$ as functions on the degree-shifted tangent bundle, $\Omega^\bullet(M)\cong\CCC^\infty(T[1]M)$, as discussed above, generalises straightforwardly to $\RZ$-graded manifolds. That is, we may define 
\begin{equation}
\Omega^\bullet(M)\ :=\ \CCC^\infty(T[1]M)
\end{equation} 
also for $M$ a $\RZ$-graded manifold. The shift in the degree-shifted tangent bundle may now be regarded as an additional grading, extending that of the $\RZ$-graded manifold to a bi-grading. Since we are not interested in any further generalisation of this grading, we simply use the ordinary notation for differential forms on manifolds to $\RZ$-graded manifolds and write again $\dd$ and $V\intprod$ for de Rham differential and interior product on a $\RZ$-graded manifold,
\begin{equation}
 \dd\ \Longleftrightarrow\ \xi^\mu\der{x^\mu}\eand V^\mu\der{x^\mu}\intprod\ \Longleftrightarrow\ V^\mu\der{\xi^\mu}~,
\end{equation}
where the $x^\mu$ are now local $\RZ$-graded coordinates on $M$ and $\xi^\mu$ local $\RZ$-graded fibre coordinates on $T[1]M$.

In the following, we shall also make use of the {\em Lie derivative} which is now defined by the graded version of Cartan's formula
\begin{equation}\label{eq:CartanMagicFormula}
 \CL_V\omega \ :=\  V\intprod \dd \omega +(-1)^{|V|}\,\dd(V\intprod\omega)
\end{equation}
for $\omega\in\Omega^\bullet(M)$ and $V$ a homogeneous vector field of degree~$|V|\in\RZ$. Note that $[\CL_V,\dd]=0$, which implies $\CL_\Upsilon\dd f=\dd\CL_\Upsilon f = |f|\,\dd f$. Consequently, $\CL_\Upsilon$ extracts the $\RZ$-degree of a differential form while ignoring its form degree. We say that $\omega\in\Omega^\bullet(M)$ is of {\em degree} $k\in\RZ$ if and only if $\CL_\Upsilon\omega = k\omega$.

\paragraph{\mathversion{bold}$\RZ$-graded vector bundles.}
The tangent bundle of a $\RZ$-graded manifold as introduced above is an example of a $\RZ$-graded vector bundle. Generally, a {\it $\RZ$-graded vector bundle} over a $\RZ$-graded manifold $M=(|M|,\CCS_M)$ is defined to be a locally free sheaf $\CCE_M$ of $\RZ$-graded $\CCS_M$-modules over $M$. In addition, for a morphism $(f,f^\sharp):(|M|,\CCS_M)\to (|M'|,\CCS_{M'})$ of locally ringed spaces, the {\it pull-back} of a $\RZ$-graded vector bundle $\CCE_{M'}$ over $M'$ to $M$ is the locally free sheaf $f^*\CCE_{M'}:=\CCS_M\otimes_{f^{-1}\CCS_{M'}} f^{-1}\CCE_{M'}$ over $M$.\footnote{Here,  $f^{-1}\CCS_{M'}$ denotes the topological inverse of $\CCS_{M'}$ (and likewise for $\CCE_{M'}$) defined by the pre-sheaf $U\mapsto\Gamma(f(U),\CCS_{M'})$ for $U\subseteq|M|$ open.}

\paragraph{\mathversion{bold}$Q$-manifolds.}
A {\it $Q$-manifold} is a $\RZ$-graded manifold $M$ endowed with a {\it homological vector field} $Q$, that is, a homogeneous vector field $Q$ of degree~$1$ which satisfies $Q^2=0$. Note that the homological vector field induces a differential on the algebra of functions and the pair $(\CCC^\infty(M),Q)$ is a dg-algebra. If the $\RZ$-grading reduces to a non-negative or $\NN$-grading, then we also speak of an {\em N$Q$-manifold}.

The simplest example of a $Q$-manifold is any ordinary manifold $M$ together with $Q=0$. Another simple but more interesting example is the degree-shifted tangent bundle $T[1]M$ with the canonical vector field turning into the de Rham differential on the algebra of functions given in the motivational paragraph. We shall encounter many more examples in Section~\ref{ssec:L_infty_algebras_and_algebroids}.

\paragraph{Symplectic \mathversion{bold}$Q$-manifolds.}
A {\it graded symplectic structure of degree~$k$} on a $\RZ$-graded manifold $M$ is a closed non-degenerate differential two-form $\omega\in\Omega^2(M)$ of degree~$k$. The non-degeneracy means that $V\intprod \omega=0$ is equivalent to $V=0$ for vector fields $V$. 

A {\it symplectic $Q$-manifold of degree~$k$} is a $Q$-manifold $(M,Q)$ equipped with a graded symplectic structure $\omega$ of degree~$k$ for which $Q$ is symplectic, that is, $\CL_Q\omega=0$. It is rather straightforward to see that $\omega$ must be exact for $k\neq0$ and $Q$ Hamiltonian for  $k\neq-1$, respectively~\cite{Roytenberg:0203110}. Indeed, 
\begin{equation}\label{eq:omexact}
 k\omega\ =\ \CL_\Upsilon\omega\ =\ \Upsilon\intprod\dd\omega+\dd(\Upsilon\intprod\omega)\ =\ \dd(\Upsilon\intprod\omega)\quad\Longrightarrow\quad\omega\ =\ \dd(\tfrac1k\Upsilon\intprod\omega)~.
\end{equation}
Likewise, to verify that $Q$ is Hamiltonian, we first note that $\dd(Q\intprod\omega)=0$ since $Q$ is symplectic. Then,\footnote{Cartan's formula~\eqref{eq:CartanMagicFormula} together with the fact that the Lie derivative is a graded derivation that commutes with the contraction, $\CL_V(W\intprod\omega)=[V,W]\intprod\omega+(-1)^{|V||W|}W\intprod\CL_V\omega$, imply that $[V,W]\intprod\omega=V\intprod W\intprod\dd\omega+V\intprod\dd(W\intprod \omega) -(-1)^{|V|(|W|+1)}W\intprod\dd(V\intprod \omega)+(-1)^{|V|}\dd(V\intprod W\intprod\omega)$.} 
\begin{subequations}
\begin{equation}
\begin{aligned}
 -Q\intprod\omega\ =\ [Q,\Upsilon]\intprod\omega\ &=\ Q\intprod\dd(\Upsilon\intprod\omega)-\dd(Q\intprod\Upsilon\intprod\omega)\\
 \ &=\ k Q\intprod\omega+\dd(\Upsilon\intprod Q\intprod\omega)~,
 \end{aligned}
\end{equation}
where in the second step we have used~\eqref{eq:omexact}. Consequently,
\begin{equation}\label{eq:HamiltonianQ}
 Q\intprod\omega\ =\ \dd S\ewith S\ :=\ \tfrac{1}{k+1}Q\intprod\Upsilon\intprod\omega~.
\end{equation}
\end{subequations}

\paragraph{Poisson structure.}
As on ordinary manifolds, a symplectic structure on a $Q$-manifold induces a Poisson structure. Concretely, let $(M,Q,\omega)$ be a smooth symplectic degree~$k$ manifold. For any $f\in\CCC^\infty(M)$ let $V_f$ be the corresponding Hamilton vector field given by 
\begin{equation}
 V_f\intprod\omega\ =\ \dd f~.
\end{equation}
For homogeneous $f\in\CCC^\infty(M)$ of degree~$|f|\in\RZ$, this equation implies $|V_f|=|f|-k$ because $|V_f\intprod \omega|=|V_f|+|\omega|$ and $|\dd f|=|f|$. We then define the {\it graded Poisson structure}
\begin{equation}
 \{f,g\}\ :=\ V_f\intprod V_g\intprod\omega\ =\ V_fg~.
\end{equation}
For homogeneous $f,g\in\CCC^\infty(M)$ of degrees $|f|,|g|\in\RZ$, we have that $|\{f,g\}|=|f|+|g|-k$. Using Cartan's formula~\eqref{eq:CartanMagicFormula}, we immediately find the standard result $V_{\{f,g\}}=(-1)^{|f|-k}[V_f,V_g]$. Furthermore, the Poisson structure is graded antisymmetric,
\begin{subequations}
\begin{equation}
 \{f,g\}\ =\ -(-1)^{(|f|-k)(|g|-k)}\{g,f\}~,
\end{equation}
satisfies a graded Jacobi identity,
\begin{equation}\label{eq:GradedJacobi}
\begin{aligned}
&\{f,\{g,h\}\}+(-1)^{(|f|-k)(|g|+|h|)+|f|-k}\{g,\{h,f\}\}\,+\\
&\kern2cm+(-1)^{(|h|-k)(|f|+|g|)+|h|-k}\{h,\{f,g\}\}\ =\ 0~,
\end{aligned}
\end{equation}
as well as a graded Leibniz rule,
\begin{equation}
\{f,gh\}\ =\ \{f,g\}h+(-1)^{(|f|-k)|g|}g\{f,h\}~.
\end{equation}
\end{subequations}

Using~\eqref{eq:HamiltonianQ} and the fact that $|S|=k+1$, we find
\begin{equation}
 Q\ =\ \{S,-\}~.
\end{equation}
The Jacobi identity then implies that $Q^2=\frac12\{\{S,S\},-\}$. Consequently, $\{S,S\}=QS$ must be locally constant for $Q^2=0$. Since $|\{S,S\}|=2+k$, we may conclude that for $k\neq-2$ the condition $Q^2=0$ is equivalent to saying that
\begin{equation}\label{eq:ClassicalMasterSymplecticQManifold}
\{S,S\}\ =\ 0~.
\end{equation}
For $S$ the (classical) Batalin--Vilkovisky action, this is called the {\it classical master equation}. In this special case, the Poisson bracket is of degree~1. General Poisson algebras of degree~1 are known as {\em Gerstenhaber algebras}. We shall return to the Batalin--Vilkovisky action in Section~\ref{ssec:ClassicalMasterEquation}.

\paragraph{Examples.} The following examples were first given in~\cite{Severa:2001aa}, see also~\cite{Roytenberg:0203110} and~\cite{Cattaneo:2010re} for further details. A symplectic N$Q$-manifold of degree~0 is simply a symplectic manifold. A symplectic N$Q$-manifold of degree~1 is a Poisson manifold. Such a manifold can be shown to be symplectomorphic to $T^*[1]M$ with canonical symplectic structure. A compatible homological vector field $Q$ corresponds to a bi-vector field on $M$ and the condition $Q^2=0$ amounts to this bivector being a Poisson tensor. A symplectic N$Q$-manifold of degree~2 is a Courant algebroid.

\subsection{\texorpdfstring{$L_\infty$}{L-infty}-algebras and \texorpdfstring{$L_\infty$}{L-infty}-algebroids}\label{ssec:L_infty_algebras_and_algebroids}

\paragraph{Motivation.}
The previous example of a Courant algebroid as well as the $Q$-manifolds $T[1]M$ and $T^*[1]M$ connect $Q$-manifolds to Lie algebroids. This connection can be vastly generalised as we shall see in the following.

Firstly, let us see what happens for a $Q$-manifold $M$ with body $M^\circ$ a point. In that case, we simply have a $\RZ$-graded vector space $\sV$. Let us further simplify $\sV$ such that it is non-trivial only in degree~$-1$, i.e.~$\sV=\frg[1]$ for a vector space $\frg$. Let $\xi^\alpha$ be the coordinates on $\sV$ of degree~1. Then, the homological vector field $Q$ is necessarily of the form
\begin{equation}\label{eq:Q_for_Lie_algebra}
 Q\ =\ -\tfrac12{f_{\alpha\beta}}^\gamma \xi^\alpha\xi^\beta\der{\xi^\gamma}
\end{equation}
for some constants ${f_{\alpha\beta}}^\gamma=-{f_{\beta\alpha}}^\gamma$. The identity $Q^2=0$ amounts to the Jacobi identity
\begin{equation}
 {f_{\alpha\beta}}^\delta {f_{\gamma\delta}}^\epsilon+{f_{\beta\gamma}}^\delta {f_{\alpha\delta}}^\epsilon+{f_{\gamma\alpha}}^\delta {f_{\beta\delta}}^\epsilon\ =\ 0
\end{equation}
so that the constants ${f_{\alpha\beta}}^\gamma$ are, in fact, the structure constants of a Lie algebra structure on $\frg$. A $Q$-manifold concentrated in degree~$-1$ is therefore a Lie algebra $\frg$. 

The latter generalises to the following statements, which we shall explain in more detail in the remainder of this section. A $Q$-manifold with body a point is an {\em $L_\infty$-algebra}. If the only non-trivial coordinates are of degrees $1,\dots, n$ we shall speak of a {\em Lie $n$-algebra}\footnote{Strictly speaking, they are $n$-term $L_\infty$-algebras, but for all intents and purposes, they can be regarded as (categorically) equivalent to Lie $n$-algebras. The categorical equivalence has been proven for Lie 2-algebras and 2-term $L_\infty$-algebras~\cite{Baez:2003aa}; the extension to Lie $n$-algebras and $n$-term $L_\infty$-algebras should be very involved, but ultimately a mere technicality.}.  A $Q$-manifold with non-trivial body and coordinates of degrees $0,\dots,n$ is a {\em Lie $n$-algebroid} and a general $Q$-manifold is an {\em $L_\infty$-algebroid}.

\paragraph{Chevalley--Eilenberg complex and $Q$-manifolds.} Let us now link the above discussion to standard mathematical nomenclature, introducing the language that we shall be using later on. This language is well-known from Lie algebra cohomology and it extends to the case of $L_\infty$-algebras. The differential graded algebra\footnote{The isomorphism is the shift isomorphism $s^\bullet$ defined in~\eqref{eq:isomorphism_asym_sym}.} $(\bigwedge^\bullet \frg^*,\dd_{\rm CE})\cong(\CCC^\infty(\frg[1]),Q)$ for some finite dimensional\footnote{Special care has to be taken in the infinite-dimensional case.} Lie algebra $\frg$ is called the {\em Chevalley--Eilenberg algebra} ${\rm CE}(\frg)$ of $\frg$ and the differential $\dd_{\rm CE}$ induced by the homological vector field $Q$ is identified with the {\it Chevalley--Eilenberg differential}.

The Chevalley--Eilenberg algebra is a special case of the Chevalley--Eilenberg complex for a $\frg$-module $\CCE$, 
\begin{equation}
 0 \ \xrightarrow{~\phantom{\dd_{\rm CE}}~}\ \sHom(\bigwedge\nolimits^0 \frg,\CCE)\ \cong\ \CCE \ \xrightarrow{~\dd_{\rm CE}~}\ \sHom(\bigwedge\nolimits^1 \frg,\CCE)  \ \xrightarrow{~\dd_{\rm CE}~}\ \sHom(\bigwedge\nolimits^2 \frg,\CCE)  \ \xrightarrow{~\dd_{\rm CE}~}\ \cdots
\end{equation}
with Chevalley--Eilenberg differential
\begin{equation}
\begin{gathered}
 \dd_{\rm CE}\, :\, \sHom(\bigwedge\nolimits^p \frg,\CCE)\ \rightarrow\  \sHom(\bigwedge\nolimits^{p+1} \frg,\CCE)~,\\
 \kern-9cm(\dd_{\rm CE} F)(X_1,\dots, X_{p+1}) \ :=\\
  \kern-2cm:=\ \sum_{i=1}^{p+1} (-1)^{i+1}X_i\acton F(X_1,\dots,\hat X_i,\dots, X_{p+1})\,+\\
 \kern3cm+\sum_{1\leq i<j\leq p+1} (-1)^{i+j}F([X_i,X_j],X_1,\dots,\hat X_i,\dots,\hat X_j,\dots,X_{p+1})~.
\end{gathered}
\end{equation}

We now explain that in the case of the trivial $\frg$-module $\CCE=\FR$, we recover the Chevalley--Eilenberg algebra and the action of $\dd_{\rm CE}$ is essentially that of $Q$. First, note that a function $F\in\sHom(\bigwedge\nolimits^k \frg,\FR)$ corresponds to an element in $\CCC^\infty_k(\frg[1])$ according to
\begin{equation}
 F(\tau_{\alpha_1}\wedge \ldots \wedge \tau_{\alpha_k})\ =\ F_{\alpha_1\cdots \alpha_k}~~~\Longleftrightarrow~~~\frac{1}{k!}\xi^{\alpha_1}\cdots \xi^{\alpha_k}F_{\alpha_1\cdots \alpha_k}~.
\end{equation}
Under this isomorphism, the Chevalley--Eilenberg differential is indeed mapped to $Q$. Concretely, we have that
\begin{equation}
\begin{aligned}
 (\dd_{\rm CE}F)(\tau_{\alpha_1}\wedge \ldots \wedge \tau_{\alpha_{k+1}})\ &=\ \sum_{1\leq i<j\leq k+1} (-1)^{i+j}F([\tau_{\alpha_i},\tau_{\alpha_j}],\tau_{\alpha_1},\dots,\hat \tau_{\alpha_i},\dots,\hat \tau_{\alpha_j},\dots,\tau_{k+1})\\
&=\ -\frac{(k+1)!}{2(k-1)!}f_{[\alpha_1\alpha_2}{}^{\beta}F_{\beta\alpha_3\dots \alpha_{k+1}]}
\end{aligned}
\end{equation}
corresponds to
\begin{equation}
 Q\frac{1}{k!}\xi^{\alpha_1}\cdots \xi^{\alpha_k}F_{\alpha_1\cdots \alpha_k}\ =\ -\frac{1}{2(k-1)!}\xi^{\alpha_1}\cdots \xi^{\alpha_{k+1}}f_{\alpha_1\alpha_2}{}^{\beta}F_{\beta\alpha_3\cdots \alpha_{k+1}}~.
\end{equation}

\paragraph{Contracted coordinate functions I.} We now come to an important technicality which, however, will greatly simplify our notation. Relying on basis dependent equations as $Q\xi^\alpha=-\tfrac12 f_{\beta\gamma}{}^\alpha\xi^\beta\xi^\gamma$ is both inconvenient and inelegant. We can, however, contract both sides of the equation by the basis vector $\tau_\alpha$, obtaining the basis independent version
\begin{equation}\label{eq:Qxi}
 Q\xi \ =\  -\tfrac12 [\xi,\xi]\ewith\xi\ :=\ \xi^\alpha\tau_\alpha~.
\end{equation}
Note that $\xi$ is {\em not} an element of $\frg[1]$ but rather an element of $(\frg[1])^*\otimes\frg$ which, in turn, is a subset of $\frg_\CCC:=\CCC^\infty(\frg[1])\otimes \frg$. We thus continued the Lie bracket from $\frg$ to $\frg_\CCC$. Since $[-,-]$ does not carry any degree, this extension by linearity is unique. This is similar to the extension of the Lie bracket from $\frg$ to $\Omega^\bullet(M,\frg)$, the set of Lie algebra valued differential forms, which is often used in gauge theories. In the general case of $L_\infty$-algebras, however, we will have to be more careful.

\paragraph{\mathversion{bold}$L_\infty$-algebras.}
As stated above, the differential graded algebra $(\CCC^\infty(\sV) ,Q)$ for a $\RZ$-graded vector space $\sV=\bigoplus_{k\in\RZ}\sV_k$ corresponds to an $L_\infty$-algebra. It is, in fact, the {\em Chevalley--Eilenberg algebra} CE$(\sL)$ of an {\it $L_\infty$-algebra} or {\it strong homotopy Lie algebra} $\sL=\sV[-1]$. The homological vector field $Q$ dualises to the codifferential $D$ of homogeneous degree~$1$ on $\bigodot^\bullet_0 \sV\cong\bigwedge^\bullet_0\sL$, where the isomorphism is the shift isomorphism $s^\bullet$ defined in~\eqref{eq:isomorphism_asym_sym}. The details of the translation between differential graded algebra and coalgebra are found in Appendix~\ref{app:L-infty-morphisms}. Here, we merely note that the codifferential decomposes into a sum 
\begin{equation}
 D\ =\ \sum_{i\in\NN}\tilde D_i~,
\end{equation}
which, when considering the restriction $D_i:=\tilde D_i|_{\bigodot^i_0\sV}:\bigodot^i_0\sV\to \sV$, defines a set of {\em higher products}
\begin{equation}\label{eq:def_higher_products}
  \mu_i\ :=\ (-1)^{\frac12i(i-1)+1}s^{-1}\circ D_i\circ s^{\otimes i}~,
\end{equation}
where $s^{\otimes i}$ is again the shift isomorphism~\eqref{eq:isomorphism_asym_sym}. These higher products are graded totally anti-symmetric multilinear maps\footnote{Please note that our notation differs from another commonly used one, where higher products are denoted by $l_i$ or $\ell_i$. For us, the latter are often elements of the $L_\infty$-algebras.}
\begin{subequations}
\begin{equation}\label{eq:HigherProducts}
 \mu_i\,:\, \underbrace{\sL\times\cdots\times\sL}_{i~{\rm copies}}\ \rightarrow\ \sL~,
\end{equation}
of degree~$2-i$, which satisfy the {\it higher} or {\em homotopy Jacobi identities},
\begin{equation}\label{eq:homotopyJacobi}
 \sum_{j+k=i}\sum_{\sigma\in \Sh(j;i) }\chi(\sigma;\ell_1,\ldots,\ell_{i})(-1)^{k}\mu_{k+1}(\mu_j(\ell_{\sigma(1)},\ldots,\ell_{\sigma(j)}),\ell_{\sigma(j+1)},\ldots,\ell_{\sigma(i)})\ =\ 0
\end{equation}
for all $i\in\NN$ and $\ell_1,\ldots,\ell_{i}\in \sL$. Here, the sum is taken over all $(j;i)$ {\it unshuffles} $\sigma$ which consist of permutations $\sigma$ of $\{1,\ldots,i\}$ such that the first $j$ and the last $i-j$ images of $\sigma$ are ordered: $\sigma(1)<\cdots<\sigma(j)$ and $\sigma(j+1)<\cdots<\sigma(i)$. Moreover, $\chi(\sigma;\ell_1,\ldots,\ell_i)$ is the {\it graded Koszul sign} defined via the equation
\begin{equation}\label{eq:DefKoszulSign}
 \ell_1\wedge \ldots \wedge \ell_i\ =\ \chi(\sigma;\ell_1,\ldots,\ell_i)\,\ell_{\sigma(1)}\wedge \ldots \wedge \ell_{\sigma(i)}
\end{equation}
\end{subequations}
in the free graded algebra for homogeneous elements. The homotopy Jacobi identities~\eqref{eq:homotopyJacobi} for $i=1$ and $i=2$ state that $\mu_1$ is a differential which is compatible with the product $\mu_2$. The corresponding relation for $i=3$ describes the controlled violation of the graded Jacobi identity. Note that the homotopy Jacobi identities follow from $Q^2=0$ $\Leftrightarrow$ $D^2=0$. The full details of this point are found in Appendix~\ref{app:L-infty-morphisms}, where we also derive the following, alternative form of the homotopy Jacobi identity
\begin{equation}\label{eq:homotopy_Jacobi_circ}
 \sum_{j+k=i} (-1)^{k}\mu_{k+1}\circ (\mu_j\otimes \id^{\otimes k})\circ \sum_{\sigma\in {\rm Sh}(j;i)} \chi(\sigma,-) \sigma(-)\ =\ 0~.
\end{equation}
Here, $\sigma(-)$ is the application of the $(j;i)$-unshuffle.

Many interesting Lie algebras are obtained by antisymmetrising the product on an associative (e.g.~matrix) algebra to the commutator. The higher analogue of an associative algebra is an $A_\infty$-algebra as introduced in~\cite{Stasheff:1963aa,Stasheff:1963ab}. Such an $A_\infty$-algebra comes with higher homotopy associative products  and antisymmetrising these yields $L_\infty$-algebra products $\mu_i$ on the underlying graded vector space.

\paragraph{Special cases.}
We call the $L_\infty$-algebra 
\begin{equation}
 L\ =\ \cdots\ \longrightarrow\ *\ \longrightarrow\ *\ \longrightarrow\ \cdots~,
\end{equation}
where $*$ denotes the 0-dimensional vector space consisting of a point, the {\em trivial $L_\infty$-algebra}. An $L_\infty$-algebra is called {\em minimal} whenever $\mu_1=0$. For example, Lie algebras form minimal $L_\infty$-algebras, while differential Lie algebras are not minimal in general. Furthermore, an $L_\infty$-algebra is called {\em linearly contractible} whenever $\mu_i=0$ for $i\geq 2$ and the cohomology of the differential $\mu_1$ is trivial. For example, the $L_\infty$-algebra $V[1]\xrightarrow{~\id~} V$ with $\mu_1=\id$ and all higher products trivial is linearly contractible. Also the trivial pairs introduced later for gauge fixing in the BV formalism are linearly contractible. Finally, $L_\infty$-algebras concentrated in degrees $-n+1,\dots, 0$ are called {\em $n$-term $L_\infty$-algebras} or {\em Lie $n$-algebras}, as already mentioned above.

\paragraph{Cyclic \mathversion{bold}$L_\infty$-algebras.}
The appropriate notion of an an {\it inner product} on an $L_\infty$-algebra $\sL$ is a graded symmetric non-degenerate bilinear pairing
\begin{subequations}
\begin{equation}
 \langle -,-\rangle_\sL\,:\,\sL\times\sL\ \to\ \FR
\end{equation}
which is cyclic in the sense that
\begin{equation}\label{eq:cyclicity}
 \langle\ell_1,\mu_i(\ell_2,\ldots,\ell_{i+1})\rangle_\sL\ =\ (-1)^{i+i(|\ell_1|_\sL+|\ell_{i+1}|_\sL)+|\ell_{i+1}|_\sL\sum_{j=1}^{i}|\ell_j|_\sL}\langle\ell_{i+1},\mu_i(\ell_1,\ldots,\ell_{i})\rangle_\sL
\end{equation}
\end{subequations}
for all $i\in\NN$ for homogeneous $\ell_1,\ldots,\ell_{i+1}\in\sL$ with $|\ell_i|_\sL$ the $L_\infty$-degree of $\ell_i\in\sL$, cf.~\cite{Kontsevich:1992aa,Penkava:9512014} for the original introduction of cyclic structures.

Whenever an $L_\infty$-algebra is equipped with such an inner product, we shall call it a {\it cyclic $L_\infty$-algebra}. When a cyclic $L_\infty$-algebra is Hilbert (i.e.~complete), the non-degeneracy of the inner product induces the isomorphisms $\sL\cong(\sL[k])^*\cong\sL^*[-k]$ where $k:=|\langle-,-\rangle_\sL|_\sL$.

In the $Q$-manifold setting, a cyclic inner product corresponds to a symplectic form. Let us illustrate this using the example of a Lie algebra $\frg$. Let $\xi^\alpha$ be coordinates on $\frg[1]$ with respect to a basis $\tau_\alpha$ of $\frg$. As we have seen above, the homological vector field is then given by $Q=-\frac12{f_{\alpha\beta}}^\gamma \xi^\alpha\xi^\beta\der{\xi^\gamma}$. A symplectic structure on $\sL[1]$ is necessarily of degree~2 and thus takes the form $\omega=\frac12\omega_{\alpha\beta}\dd\xi^\alpha\wedge\dd\xi^\beta$ with $\omega_{\alpha\beta}=\omega_{\beta\alpha}\in\FR$. The fact that $Q$ is symplectic implies $\CL_Q\omega=-\dd (Q\intprod \omega)=0$ which, together with $Q\intprod \dd \xi^\alpha=Q\xi^\alpha$ leads to 
\begin{equation}\label{eq:CyclicityFromLQ}
  \tfrac12\omega_{\alpha\beta}\left(\dd (Q \xi^\alpha)\wedge \dd \xi^\beta-\dd \xi^\alpha\wedge \dd (Q \xi^\beta)\right)\ =\ 0 \ ~~\Longleftrightarrow~~\ \omega_{\delta(\alpha}{f_{\beta)\gamma}}^\delta\ =\ 0~.
\end{equation}
This is precisely the cyclicity condition for a symmetric inner product $\langle \tau_\alpha,\tau_\beta\rangle_\frg :=\omega_{\alpha\beta}$ on the Lie algebra $\frg$,
\begin{equation}
 \langle \tau_\alpha,[\tau_\beta,\tau_\gamma]\rangle_\frg\ =\ \langle \tau_\gamma,[\tau_\alpha,\tau_\beta]\rangle_\frg~.
\end{equation}

Finally, note that the Hamiltonian $\eqref{eq:HamiltonianQ}$ for $Q$ is given by $S=\frac{1}{3!} \xi^\alpha\xi^\beta\xi^\gamma\omega_{\delta[\alpha}{f_{\beta\gamma]}}^\delta$. It is not difficult to see that this treatment generalises to $\sL[1]$ with $\sL=\bigoplus_{k\in\RZ}\sL_k$. That is, a smooth symplectic $Q$-manifold of the form $(\sL[1],Q,\omega)$ encodes a cyclic $L_\infty$-algebra $\sL$.

\paragraph{\mathversion{bold}$L_\infty$-algebras from tensor products.}
An important observation which we shall heavily rely upon is that the graded vector space obtained from the tensor product of an $L_\infty$-algebra and a differential graded commutative algebra carries a natural $L_\infty$-algebra structure, see e.g.~\cite{Jurco:2014mva}. 

Let $(\sA,\dd)$ be a differential graded commutative algebra and $(\sL,\mu_i)$ be an $L_\infty$-algebra. Then we have a new $L_\infty$-algebra $\sL_\sA$ with underlying graded vector space
\begin{subequations}\label{eq:LinftyExtension}
\begin{equation}
 \sL_\sA\ :=\ \bigoplus_{k\in\RZ} (\sA\otimes \sL)_k\ewith (\sA\otimes \sL)_k\ :=\ \bigoplus_{i+j=k}~\sA_i\otimes \sL_j
\end{equation}
so that the homogeneous degree in $\sL_\sA$ is given by $|a\otimes \ell|_{\sL_\sA}=|a|_\sA+|\ell|_\sL$ for homogeneous $a\in \sA$ and $\ell\in \sL$. The higher products $\upmu_i$ on $\sL_\sA$ read as 
\begin{equation}
 \upmu_i\ :=\ \left\{
 \begin{array}{ll}
 \dd \otimes \id + \id \otimes \mu_1 & i= 1~,\\
 m_i\otimes \mu_i & \mbox{else}~,
 \end{array}\right.
\end{equation}
\end{subequations}
where $m_i(a_1,\ldots, a_i):=a_1\cdots a_i$ is the commutative, associative product on $\sA$. 

Applied to $(a_1\otimes \ell_1,\ldots,a_i\otimes \ell_i)$ with homogeneous $a_1,\ldots,a_i\in \sA$ and $\ell_1,\ldots,\ell_i\in \sL$, we obtain
\begin{subequations}\label{eq:LinftyExtensionHP}
\begin{equation}
\begin{aligned}
 \upmu_1(a_1 \otimes \ell_1)\ &:=\ \dd a_1 \otimes \ell_1+(-1)^{|a_1|_\sA}a_1\otimes \mu_1(\ell_1)~,\\[5pt]
 \upmu_i(a_1\otimes \ell_1,\ldots,a_i\otimes \ell_i)\ &:=\ (-1)^{i\sum_{j=1}^i|a_j|_\sA+\sum_{j=2}^{i} |a_j|_\sA\sum_{k=1}^{j-1} |\ell_k|_{\sL}}\,\times\\[-5pt]
  &\kern2cm\times(a_1 \cdots a_i)\otimes \mu_i(\ell_1,\ldots,\ell_i)
\end{aligned}
\end{equation}
for $i\geq 2$, and they extend to general elements by linearity.  It is shown in Appendix~\ref{app:lemmata} that these products satisfy the homotopy Jacobi identities~\eqref{eq:homotopyJacobi}. 

In addition, if both $\sA$ and $\sL$ come with inner products $\langle-,-\rangle_\sA$ and $\langle-,-\rangle_\sL$, then $\sL_\sA$ admits a natural inner product defined by
\begin{equation}\label{eq:LinftyExtensionIP}
 \langle a_1\otimes\ell_1,a_2\otimes\ell_2\rangle_{\sL_\sA}\ :=\ (-1)^{|a_2|_\sA |\ell_1|_\sL+(|a_1|_\sA+|a_2|_\sA)(\langle-,-\rangle_\sL)}\langle a_1,a_2\rangle_\sA\langle\ell_1,\ell_2\rangle_\sL
\end{equation}
\end{subequations}
for homogeneous $a_1,a_2\in \sA$ and $\ell_1,\ell_2\in \sL$. Clearly, this inner product is graded symmetric and its cyclicity is shown in Appendix~\ref{app:lemmata}. We shall come back to such extensions in Sections~\ref{ssec:HBVA} and~\ref{ssec:HCST}.

\paragraph{\mathversion{bold}$L_\infty$-algebra structures on graded modules.} Let us remark that our previous discussion of $L_\infty$-algebra structures on graded vector spaces translates to the case of graded modules. This is particularly important for the application to field theory, as here the field of real numbers is essentially always replaced by a ring of functions or, more generally, by a ring of sections of some vector bundle. In the following, a (cyclic) $L_\infty$-algebra will have an underlying graded module. The most important example for our purposes is discussed in the next paragraph.

\paragraph{Example: \mathversion{bold}$\Omega^\bullet(M,\sL)$.} The $L_\infty$-algebra $\Omega^\bullet(M,\sL)=\sL_{\Omega^\bullet(M)}$ is the tensor product of some $L_\infty$-algebra $\sL$ and $(\Omega^\bullet(M),\dd)$, the de Rham complex on some smooth manifold $M$ of dimension $d$. The tensor product yields the graded vector space 
\begin{subequations}\label{eq:HCSLInfinity}
\begin{equation}
 \Omega^\bullet(M,\sL)\ :=\ \bigoplus_{k\in \RZ} \Omega^\bullet_k(M,\sL)
\end{equation}
with
\begin{equation}
\begin{aligned}
 \Omega^\bullet_k(M,\sL)\ &:=\ \Omega^0(M)\otimes\sL_k~\oplus~\Omega^1(M)\otimes \sL_{k-1}~\oplus~\cdots~\oplus~ \Omega^d(M)\otimes \sL_{k-d}\\
 \ &\phantom{:}=\ \bigoplus_{\substack{i+j=k\\ 0\leq i\leq d\\j\in \RZ}}~\Omega^i(M)\otimes \sL_{j}~,
\end{aligned}
\end{equation}
and the total degree is the sum of the individual degrees, $|\alpha\otimes \ell|_{\Omega^\bullet(M,\sL)}=|\alpha|_{\Omega^\bullet(M)}+|\ell|_\sL$ for homogeneous $\alpha\in\Omega^\bullet(M)$ and $\ell\in \sL$. This graded vector space carries an $L_\infty$-algebra structure which is the linear extension of the higher products
\begin{equation}
\begin{aligned}
 \hat\mu_1(\alpha_1 \otimes \ell_1) \ &:=\ \dd \alpha_1 \otimes \ell_1+(-1)^{|\alpha_1|_{\Omega^\bullet(M)}}\alpha_1\otimes \mu_1(\ell_1)~,\\[5pt]
 \hat\mu_i(\alpha_1\otimes \ell_1,\ldots,\alpha_i\otimes \ell_i) \ &:=\ (-1)^{i\sum_{j=1}^i|\alpha_j|_{\Omega^\bullet(M)}+\sum_{j=0}^{i-2}|\alpha_{i-j}|_{\Omega^\bullet(M)}\sum_{k=1}^{i-j-1}|\ell_k|_\sL}\,\times\\[-5pt]
  &\kern2cm\times(\alpha_1\wedge \ldots\wedge \alpha_i)\otimes \mu_i(\ell_1,\ldots,\ell_i)
\end{aligned}
\end{equation}
\end{subequations}
for $i\geq 2$, $\alpha_1,\ldots,\alpha_i\in \Omega^\bullet(M)$ and $\ell_1,\ldots,\ell_i\in \sL$. 

If $\sL$ is cyclic and $M$ is compact, oriented, and without boundary, then we have a natural cyclic inner product on $\Omega^\bullet(M,\sL)$,
\begin{equation}
\langle \alpha_1\otimes \ell_1,\alpha_2\otimes \ell_2\rangle_{\Omega^\bullet(M,\sL)}\ :=\ (-1)^{|\alpha_2|_{\Omega^\bullet(M)} |\ell_1|_\sL}\int_M \alpha_1\wedge \alpha_2~\langle \ell_1,\ell_2\rangle_\sL~,
\end{equation}
where $\alpha_{1,2}\in \Omega^\bullet(M)$ and $\ell_{1,2}\in \sL$.

\paragraph{Contracted coordinate functions II.} Recall that to use the simplifying notation in Equation~\eqref{eq:Qxi},  we introduced the contracted coordinate functions $\xi:=\tau_\alpha \xi^\alpha$, extending the Lie algebra $\frg$ to the tensor product $\frg_\CCC:=\CCC^\infty(\frg[1])\otimes \frg$. 

In the case of an $L_\infty$-algebra, the analogue extension is 
\begin{subequations}\label{eq:def_contracted_coordinate_functions}
\begin{equation}
 \sL_{\CCC}\ :=\ \CCC^\infty(\sL[1])\otimes \sL~,
\end{equation}
where we regard $\CCC^\infty(\sL[1])$ as a differential graded algebra with trivial differential. This leads to higher products obtained by linearly extending
\begin{equation}\label{eq:GhostProd}
\begin{aligned}
 \upmu_1(\zeta_1 \otimes \ell_1) \ &:=\  (-1)^{|\zeta_1|_{\rm gh}}\zeta_1\otimes \mu_1(\ell_1)~,\\
  \upmu_i(\zeta_1 \otimes \ell_1,\ldots, \zeta_i \otimes \ell_i) \ &:=\ (-1)^{i\sum_{j=1}^i|\zeta_i|_{\rm gh}+\sum_{j=2}^{i} |\zeta_j|_{\rm gh}\sum_{k=1}^{j-1} |\ell_k|_{\sL}} (\zeta_1\cdots \zeta_i)\otimes\mu_i(\ell_1,\ldots, \ell_i)
\end{aligned}
\end{equation}
for homogeneous $\zeta_j\in \CCC^\infty(\sfL[1])$, of degree $|\zeta_j|_{\rm gh}\in\RZ$, and $\ell_j\in \sL$. For $\sL$ a cyclic $L_\infty$-algebra with cyclic inner product of degree~$k$, we also define a non-degenerate graded symmetric pairing $\sL_\CCC\times \sL_\CCC\rightarrow \CCC^\infty(\sfL[1])$ given by 
\begin{equation}\label{eq:BVInnerProduct}
\langle\zeta_1\otimes \ell_1,\zeta_2\otimes \ell_2\rangle_{\sL_\CCC}\ :=\ 
(-1)^{k(|\zeta_1|_{\rm gh}+|\zeta_2|_{\rm gh})+|\ell_1|_{\sL} |\zeta_2|_{\rm gh}}(\zeta_1\zeta_2)\langle\ell_1,\ell_2\rangle_{\sL} 
\end{equation}
\end{subequations}
for homogeneous $\zeta_{1,2}\in \CCC^\infty(\sfL[1])$ and $\ell_{1,2}\in \sL$.

This tensor product $L_\infty$-algebra now allows us to write the action of $Q$ in a very compact form, extending formula~\eqref{eq:Qxi}. We introduce again the contracted coordinate functions $\xi=\xi^\alpha\otimes \tau_\alpha\in \sL_\CCC$ with $|\xi|=1$, where $\tau_\alpha$ is a basis for $\sL$. As shown in~\eqref{eq:comp_contracted_ccf}, we then have
\begin{equation}\label{eq:Qxi_L_infty}
 Q\xi\ =\ -\sum_{i\geq 1} \frac{1}{i!} \hat \mu_i(\xi,\dots,\xi)~.
\end{equation}
This formula should be interpreted as acting on elements of $\bigodot^\bullet \sL[1]$. The $\mu_i(\xi,\dots,\xi)$ act non-trivially only on elements of $\bigodot^i \sL[1]$ and when moving $\xi$ past elements in $\sL[1]$, one should insert Koszul signs accordingly.

\paragraph{\mathversion{bold}$L_\infty$-algebroids.}
It remains to discuss the case of a $Q$-manifold $(M,Q)$ with non-trivial body $M^\circ$. As stated above, these correspond to $L_\infty$-algebroids. We call the dg-algebra $(\CCC^\infty(M),Q)$ the Chevalley--Eilenberg algebra of an {\it $L_\infty$-algebroid}.

The $Q$-manifolds underlying ordinary Lie algebroids are simply degree-shifted vector bundles. Let $E\rightarrow M^\circ$ be a vector bundle over an ordinary manifold $M^\circ$. On the degree-shifted vector bundle  $M=E[1]$, we introduce local coordinates $x^\mu,\xi^\alpha$ with $|x^\mu|=0$ and $|\xi^\alpha|=1$, the homological vector field $Q$ must be of the form 
\begin{equation}\label{eq:QLieAlgebroid}
Q\ =\ \rho^\mu_\alpha(x)\xi^\alpha\der{x^\mu} -\tfrac12{f_{\alpha\beta}}^\gamma(x) \xi^\alpha\xi^\beta\der{\xi^\gamma}~.
\end{equation}
The condition $Q^2=0$ then amounts to requiring that the ${f_{\alpha\beta}}^\gamma$ encode a Lie bracket on the space of sections of $E$ satisfying a Leibniz rule and the $\rho^\mu_\alpha$ encode a Lie algebra morphism $E\to TM^\circ$. 

\paragraph{Example: action Lie algebroid.} An important example of a Lie algebroid is the {\it action Lie algebroid}. To define it, let $M^\circ=(|M|,\CCC^\infty_M)$ be an ordinary manifold together with an action $\rho:\sG\times M^\circ\to M^\circ$ of a Lie group $\sG$. We are then often interested in the orbit space $M^\circ/\sG$. This space can be badly behaved and hard to get under control. For example, the action of $\sG$ may not be free (i.e.~it contains fixed points) which leads to singularities when trying to regard $M^\circ/\sG$ as a smooth manifold. But even if this is not the case, it may be hard to find an explicit and useful description of the quotient space $M^\circ/\sG$.

One way to circumvent this problem is to use the {\em derived quotient} instead.  This is just modern terminology for considering the {\em action Lie groupoid}. The morphisms of the latter are the maps transforming $x\in M^\circ$ by a group element $g\in \sG$ to $g\acton x\in M^\circ$,
\begin{equation}
 x\ \xrightarrow{~(g,x)~}\ g\acton x~.
\end{equation}
The action Lie groupoid is thus the category $\sG \ltimes M^\circ \rightrightarrows M^\circ$. The structure maps are rather obvious. The identity map $\id_x:M^\circ \rightarrow \sG \ltimes M^\circ $ is simply $x\mapsto (\unit_\sG,x)$ and composition of morphisms  $(g_2,g_1\acton x)$ and $(g_1,x)$ is given by $(g_2,g_1\acton x)\circ(g_1,x) =((g_2g_1)\acton x,x)$. The inverse of a morphism $(g,x)$ is $(g,x)^{-1}=(g^{-1},g\acton x)$.

Just as a Lie group differentiates to a Lie algebra, a Lie groupoid differentiates to a Lie algebroid and a very general prescription for the Lie differentiation of $L_\infty$-groupoids is found in~\cite{Severa:2006aa}, see also~\cite{Jurco:2016qwv} for all details.

The action Lie algebroid is now the trivial vector bundle $\frg\times M^\circ \rightarrow M^\circ$, where $\frg$ is the Lie algebra of $\sG$. The  corresponding $Q$-manifold is $M=\frg[1]\times M^\circ$ with local coordinates $x^\mu$ and $\xi^\alpha$ of degrees $0$ and $1$, respectively, and the homological vector field reads as~\eqref{eq:QLieAlgebroid}. Here, $\rho^\mu_\alpha$ is given by linearising the Lie group action $\rho:\sG\times M^\circ\to M^\circ$ and $f_{\alpha\beta}{}^\gamma$ are the structure constants of $\frg$.

As we shall see in Section~\ref{sec:GeneralBVFormalism}, the action Lie algebroid is the mathematical structure underlying the {\it Becchi--Rouet--Stora--Tyutin (BRST) complex}. Finally, also note that this construction generalises to actions of $L_\infty$-algebras on manifolds.

\paragraph{Comments on generalisations.} Our treatment of $L_\infty$-algebras and $L_\infty$-algebroids as particular $Q$-manifolds extends to cases where the $\RZ$-graded vector bundles become infi\-nite-dimensional. Here, however, care needs to be taken in the dualisation from the dg-algebra picture involving the Chevalley--Eilenberg differential $Q$ to the coalgebra picture with codifferential $D$.

Also, the cyclicity condition for $L_\infty$-algebras~\eqref{eq:cyclicity} can certainly be extended to bilinear maps on a module $A$ over a ring $R$ to that ring $R$, just as in the case of Lie algebras. Recall that such maps are used e.g.~when defining Lagrangians, where $A$ are representation space-valued differential forms and $R$ is the ring of functions (or even densities) on a manifold.

\subsection{Morphisms of \texorpdfstring{$L_\infty$}{L-infty}-algebras and quasi-isomorphisms}\label{ssec:quasiisomorphism}

The description of $L_\infty$-algebras in terms of differential graded algebras induces the natural notion of morphism of $L_\infty$-algebras. In the following, we shall translate this notion to the multilinear maps $\mu_i$ and clarify the appropriate form of equivalence, called {\em quasi-isomorphism}. A key references for this section is~\cite{Kajiura:0306332} where corresponding definitions and results are found in the more general case of $A_\infty$-algebras.

\paragraph{\mathversion{bold}$L_\infty$-morphisms.}
Morphisms of $L_\infty$-algebras can be defined via their descriptions in terms of $Q$-manifolds or via their interpretation as codifferential coalgebras, and both lead to the same result. The technical details of the derivation are found in Appendix~\ref{app:L-infty-morphisms} which we summarise here as follows. A {\em morphism} between two $L_\infty$-algebras $(\sL,\mu_i)$ and $(\sL',\mu'_i)$ is a collection of homogeneous maps $\phi_i:\sL\times\cdots\times\sL\rightarrow\sL'$ of degree~$1-i$ for $i\in\NN$ which are multilinear and totally graded anti-symmetric and obey
\begin{subequations}\label{eq:L_infty_morphism}
\begin{equation}
\begin{aligned}
   &\sum_{j+k=i}\sum_{\sigma\in \Sh(j;i)}~(-1)^{k}\chi(\sigma;\ell_1,\ldots,\ell_i)\phi_{k+1}(\mu_j(\ell_{\sigma(1)},\dots,\ell_{\sigma(j)}),\ell_{\sigma(j+1)},\dots ,\ell_{\sigma(i)})\\
   \ &=\ \sum_{j=1}^i\frac{1}{j!} \sum_{k_1+\cdots+k_j=i}\sum_{\sigma\in{\rm Sh}(k_1,\ldots,k_{j-1};i)}\chi(\sigma;\ell_1,\ldots,\ell_i)\zeta(\sigma;\ell_1,\ldots,\ell_i)\,\times\\
   &\kern1cm\times \mu'_j\Big(\phi_{k_1}\big(\ell_{\sigma(1)},\ldots,\ell_{\sigma(k_1)}\big),\ldots,\phi_{k_j}\big(\ell_{\sigma(k_1+\cdots+k_{j-1}+1)},\ldots,\ell_{\sigma(i)}\big)\Big)~,
\end{aligned}
\end{equation}
where $\chi(\sigma;\ell_1,\ldots,\ell_i)$ is the Koszul sign and $\zeta(\sigma;\ell_1,\ldots,\ell_i)$ for a
$(k_1,\dots, k_{j-1};i)$-unshuffle $\sigma$ is defined as
\begin{equation}
 \zeta(\sigma;\ell_1,\ldots,\ell_i)\ :=\ (-1)^{\sum_{1\leq m<n\leq j}k_mk_n+\sum_{m=1}^{j-1}k_m(j-m)+\sum_{m=2}^j(1-k_m)\sum_{k=1}^{k_1+\cdots+k_{m-1}}|\ell_{\sigma(k)}|_\sL}~.
\end{equation}
\end{subequations}
For the sake of brevity, we shall also call morphisms of $L_\infty$-algebras {\it $L_\infty$-morphisms}.

An $L_\infty$-morphism for which $\phi_i=0$ for $i\geq 2$ is called {\em strict}. Clearly, $L_\infty$-morphisms concentrated in degree~0 are of this type, and for those the relation~\eqref{eq:L_infty_morphism} reduces to
\begin{equation}
 \phi_1(\mu_2(\ell_1,\ell_2)))\ =\ \mu'_2(\phi_1(\ell_1),\phi_1(\ell_2))~,
\end{equation}
that is, the expected relation for a morphism of Lie algebras. The notion of a weak morphism between 2-term $L_\infty$-algebras was derived in~\cite{Baez:2003aa}, where also many more details on 2-term $L_\infty$-algebras can be found. Morphisms of $L_\infty$-algebras are composable, and the formulas for the composition map can be derived using the coalgebra picture in Appendix~\ref{app:L-infty-morphisms} in which composition is evident.

\paragraph{\mathversion{bold}$L_\infty$-isomorphisms.}
An $L_\infty$-morphism is (strictly) invertible if and only if the map $\phi_1$ is invertible. This is already suggested by~\eqref{eq:L_infty_morphism}, which shows that the higher products on either $(\sL,\mu_i)$ or $(\sL',\mu'_i)$ can be reconstructed from the respective others if $\phi_1^{-1}$ is known. In this case, the map $\phi_1:\sL\to\sL'$ is an isomorphism, and, correspondingly, we call such $L_\infty$-morphisms {\em isomorphisms of $L_\infty$-algebras} or {\it  $L_\infty$-isomorphisms}. They allow us to formulate the next theorem.

\paragraph{Decomposition theorem.} 
Any $L_\infty$-algebra $\sL$ is $L_\infty$-isomorphic to the direct sum of a minimal $L_\infty$-algebra (that is, an  $L_\infty$-algebra with $\mu_1=0$) and a linearly contractible $L_\infty$-algebra (that is, an  $L_\infty$-algebra with $\mu_i=0$ for $i>1$ and trivial cohomology), see~\cite{Kajiura:0306332} for the more general case of $A_\infty$-algebras. 

Because an $L_\infty$-isomorphism is in particular a cochain map, it follows that the minimal $L_\infty$-algebra in the decomposition should have the cohomology complex $(H^\bullet_{\mu_1}(\sL),0)$ of the complex $(\sL,\mu_1)$ as its differential graded vector space. Thus, we have in general
\begin{equation}
 (\sL,\mu_i)\ \cong\ (\sL',\mu_i') \ :=\ H^\bullet_{\mu_1}(\sL)~\oplus~ \sL/H^\bullet_{\mu_1}(\sL)~, 
\end{equation}
and $\sL/H^\bullet_{\mu_1}(\sL)$ is the linearly contractible part.

\paragraph{\mathversion{bold}$L_\infty$-quasi-isomorphisms.}
Often, $L_\infty$-isomorphisms do not yield the physically relevant equivalence classes; see e.g.~Section~\ref{ssec:HMCE}. Instead, one should consider the following weaker notion of an isomorphism: a {\em quasi-isomorphism of $L_\infty$-algebras} or simply a {\it  quasi-isomorphism}  is an $L_\infty$-morphism $(\sL,\mu_i)\rightarrow (\sL',\mu'_i)$ for which $\phi_1$ induces an isomorphism $H^\bullet_{\mu_1}(\sL)\cong H^\bullet_{\mu'_1}(\sL')$  of graded vector spaces. We will argue below that $L_\infty$-quasi-isomorphy indeed induces an equivalence on the space of $L_\infty$-algebras. Since a Lie algebra is isomorphic to its cohomology, the difference between quasi-isomorphisms and $L_\infty$-isomorphisms is a new feature of $L_\infty$-algebras, which is not present in the case of Lie algebras.

We also call two differential graded algebras $(\CCC^\infty(\sL[1]),Q)$ and $(\CCC^\infty(\sL'[1]),Q')$ {\em quasi-isomorphic}, if they form the Chevalley--Eilenberg algebras of two quasi-isomorphic $L_\infty$-algebras $\sL$ and $\sL'$.

\paragraph{Minimal model theorem.} 
The decomposition theorem~\cite{Kajiura:0306332} now guarantees that there are $L_\infty$-quasi-isomorphisms 
\begin{equation}\label{eq:QuasiIsosCohomL}
\begin{gathered}
 p~:~\sL\ \xrightarrow{~\cong~}\ H^\bullet_{\mu_1}(\sL)~\oplus~\sL/H^\bullet_{\mu_1}(\sL)\ \overset{\pi}{\twoheadrightarrow}\  H^\bullet_{\mu_1}(\sL)~,\\
 e~:~H^\bullet_{\mu_1}(\sL)\ \overset{\iota}{\hookrightarrow}\ H^\bullet_{\mu_1}(\sL)~\oplus~\sL/H^\bullet_{\mu_1}(\sL)\ \xrightarrow{~\cong~}\ \sL~,  
 \end{gathered}
\end{equation}
where $\pi:H^\bullet_{\mu_1}(\sL)~\oplus~\sL/H^\bullet_{\mu_1}(\sL)\twoheadrightarrow H^\bullet_{\mu_1}(\sL)$ is the projection and $\iota:H^\bullet_{\mu_1}(\sL)\hookrightarrow H^\bullet_{\mu_1}(\sL)\oplus \sL/H^\bullet_{\mu_1}(\sL)$ the inclusion. Both $\pi$ and $\iota$ are strict, but neither $p$ nor $e$ are, in general. The existence of the second $L_\infty$-quasi-isomorphism in~\eqref{eq:QuasiIsosCohomL} is known as the {\em minimal model theorem}~\cite{kadeishvili1982algebraic}, which historically predates the decomposition theorem. The $L_\infty$-structure on $H^\bullet_{\mu_1}(\sL)$ is also called a {\em minimal model}. Minimal models are unique up to $L_\infty$-isomorphisms. Roughly speaking, the restriction to cohomology $\sL\rightarrow H^\bullet_{\mu_1}(\sL)$ is the restriction from the kinematical data of a gauge theory to its physical states, and we shall come back to this in Section~\ref{sec:L_infty-structures}.

To construct a minimal model explicitly, let $(\sL,\mu_i)$ be an $L_\infty$-algebra and write $\dd_k:\sL_k\to\sL_{k+1}$ for $\mu_1$. Consider the complex $(\sL,\dd)$ and denote its cohomology complex by $(H^\bullet_\dd(\sL), 0)$. The minimal model theorem then tells us that we have cochain maps $p$ and $e$
\begin{subequations}\label{eq:splitComplex}
\begin{equation}
 \myxymatrix{\ar@(dl,ul)[]^h \sL\ar@<+2pt>@{->>}[rr]^{\kern-20pt p} & & ~~H^\bullet_\dd(\sL) \ar@<+2pt>@{^(->}[ll]^{\kern-20pt e}},
\end{equation}
with $p\circ e=1$ and $h$ is a {\em contracting homotopy}. Specifically, $h$ is a collection of maps $h_k:\sL_k\rightarrow \sL_{k-1}$ such that 
\begin{equation}
 \dd_k\ =\ \dd_k\circ h_{k+1}\circ \dd_k~.
\end{equation}
\end{subequations}
It follows that we can construct the three projectors
\begin{subequations}
\begin{equation}
 P_k\ :=\ e_k\circ p_k~,~~~h_{k+1}\circ \dd_k~,\eand\dd_{k-1}\circ h_k
\end{equation}
with
\begin{equation}\label{eq:ProjectorDecomposition}
 1\ =\ P_k+h_{k+1}\circ \dd_k+\dd_{k-1}\circ h_k~,
\end{equation}
\end{subequations}
that is, they allow for the decomposition
\begin{equation}
  \sL\ \cong\ \im(P)\oplus \im(h\circ \dd)\oplus \im(\dd\circ h)\ewith \im(P)\ \cong\ H^\bullet_\dd(\sL)~.
\end{equation}
This decomposition is also known as the {\it abstract Hodge--Kodaira decomposition}, see e.g.~\cite{Kajiura:0306332} as well as Appendix~\ref{app:Hodge} for more details.

\paragraph{Explicit minimal model.}
To write down the $L_\infty$-structure on $H^\bullet_\dd(\sL)$, let us set $\sL':=H^\bullet_\dd(\sL)$ and $\mu'_1:=0$. Following~\cite{Kajiura:0306332}, we define totally graded anti-symmetric multilinear maps $\phi_i:\sL'\times\cdots\times\sL'\to\sL$ of homogeneous degree~$1-i$ recursively by setting
\begin{equation}\label{eq:minimalQuasiIsomorphism}
\begin{aligned}
 \phi_1(\ell'_1)\ &:=\ e(\ell')~,\\
  \phi_2(\ell'_1,\ell'_2)\ &:=\ - h(\mu_2(e(\ell'_1),e(\ell'_2)))~,\\
  &~~\vdots\\
  \phi_i(\ell'_1,\ldots,\ell'_i)\ &:=\ -\sum_{j=2}^i\frac{1}{j!} \sum_{k_1+\cdots+k_j=i}\sum_{\sigma\in{\rm Sh}(k_1,\ldots,k_{j-1};i)}\chi(\sigma;\ell'_1,\ldots,\ell'_i)\zeta(\sigma;\ell'_1,\ldots,\ell'_i)\,\times\\
   &\kern-2cm\times h\left\{\mu_j\Big(\phi_{k_1}\big(\ell'_{\sigma(1)},\ldots,\ell'_{\sigma(k_1)}\big),\ldots,\phi_{k_j}\big(\ell'_{\sigma(k_1+\cdots+k_{j-1}+1)},\ldots,\ell'_{\sigma(i)}\big)\Big)\right\},
\end{aligned}
\end{equation}
where $\ell'_1,\ldots,\ell'_i\in\sL'$. Here, $h$ and $e$ are again the maps from~\eqref{eq:splitComplex}, $\chi(\sigma;\ell'_1,\ldots,\ell'_i)$ is the Koszul sign defined in~\eqref{eq:DefKoszulSign}, and $\zeta(\sigma;\ell'_1,\ldots,\ell'_i)$ the sign factor introduced in~\eqref{eq:L_infty_morphism}. Recall that $e$ is a cochain map and thus so is $\phi_1$. The maps $\phi_i$ form an $L_\infty$-quasi-isomorphism from $\sL'$ to $\sL$ provided  the higher products $\mu'_i$ on $\sL'$ are given by
\begin{equation}\label{eq:minimalHigherProducts}
\begin{aligned}
 \mu'_1(\ell'_1)\ &:=\ 0~,\\
 \mu'_2(\ell'_1,\ell'_2)\ &:=\  p(\mu_2(e(\ell'_1),e(\ell'_2)))~,\\
  &~~\vdots\\
   \mu'_i(\ell'_1,\ldots,\ell'_i)\ &:=\ \sum_{j=2}^i\frac{1}{j!} \sum_{k_1+\cdots+k_j=i}\sum_{\sigma\in{\rm Sh}(k_1,\ldots,k_{j-1};i)}\chi(\sigma;\ell'_1,\ldots,\ell'_i)\zeta(\sigma;\ell'_1,\ldots,\ell'_i)\,\times\\
   &\kern-2cm\times p\left\{\mu_j\Big(\phi_{k_1}\big(\ell'_{\sigma(1)},\ldots,\ell'_{\sigma(k_1)}\big),\ldots,\phi_{k_j}\big(\ell'_{\sigma(k_1+\cdots+k_{j-1}+1)},\ldots,\ell'_{\sigma(i)}\big)\Big)\right\}.
\end{aligned}
\end{equation}
Using the identities $p\circ \phi_1=p\circ e=1$, $p\circ \mu_1=p\circ \dd=0$, and $\mu_1(e(\ell'))=\dd(e(\ell'))=0$ for all $\ell'\in\sL'=H^\bullet_\dd(\sL)$ together with the decomposition~\eqref{eq:ProjectorDecomposition} and the higher homotopy Jacobi identities~\eqref{eq:homotopyJacobi} for the products $\mu_i$ on $\sL$, it is rather straightforward to see that~\eqref{eq:minimalQuasiIsomorphism} together with~\eqref{eq:minimalHigherProducts} satisfy the definition~\eqref{eq:L_infty_morphism} of an $L_\infty$-morphism.\footnote{The calculations are much simplified if one instead works with the contracting homotopy $\tilde h:=h-h\circ h\circ\mu_1$ in~\eqref{eq:minimalQuasiIsomorphism} since then $\tilde h\circ\tilde h\circ \mu_1=\tilde h\circ\tilde h\circ \dd=0$.}

\paragraph{Weak inverses of \mathversion{bold}$L_\infty$-quasi-isomorphisms.} 
Using the decomposition theorem, any $L_\infty$-quasi-isomorphism $\phi:(\sL,\mu_i)\rightarrow (\sL',\mu'_i)$ can be {\it weakly inverted} by going through the corresponding minimal models. Specifically, using the $L_\infty$-quasi-isomorphisms~\eqref{eq:QuasiIsosCohomL}, we obtain an $L_\infty$-isomorphism 
\begin{equation}
 \phi_{\rm rd}\,:\,H^\bullet_{\mu_1}(\sL)\ \xrightarrow{~~~}\ \sL\ \xrightarrow{~\phi~}\ \sL'\ \xrightarrow{~~~}\ H^\bullet_{\mu'_1}(\sL')~,
\end{equation}
which can be inverted and composed to give the inverse $L_\infty$-quasi-isomorphism, 
\begin{equation}
 \phi^{-1}\,:\,\sL'\ \xrightarrow{~~~}\ H^\bullet_{\mu'_1}(\sL')\ \xrightarrow{~\phi^{-1}_{\rm rd}~}\ H^\bullet_{\mu_1}(\sL)\, \xrightarrow{~~~} \ \sL~.
\end{equation}
For instance, the quasi-isomorphisms~\eqref{eq:QuasiIsosCohomL} are trivially weakly inverses of each other. Since $L_\infty$-quasi-isomorphism can be weakly inverted, $L_\infty$-quasi-isomorphy induces an equivalence relation on the space of all $L_\infty$-algebras.

\paragraph{Examples.} Note that linearly contractible $L_\infty$-algebras, such as $V[1]\xrightarrow{~\id~} V$, have trivial cohomology and  therefore they are quasi-isomorphic to the trivial $L_\infty$-algebra. The decomposition theorem therefore implies the minimal model theorem.

It is always possible to extend an $L_\infty$-algebra with underlying graded vector space
\begin{equation}
 \sL\ =\ \cdots\ \rightarrow\ *\ \rightarrow\ *\ \rightarrow\ \sL_{i}\ \xrightarrow{~\mu_1~}\ \sL_{i+1}\ \xrightarrow{~\mu_1~}\ \sL_{i+2}\ \rightarrow\ \cdots~,
\end{equation}
to an $L_\infty$-algebra structure on 
\begin{equation}
 \sL'\ =\ \cdots\ \xrightarrow{~~~}\ *\ \xrightarrow{~~~}\ \ker(\mu_1)\ \embd\ \sL_{i}\ \xrightarrow{~\mu_1~}\ \sL_{i+1}\ \xrightarrow{~\mu_1~}\ \sL_{i+2}\ \xrightarrow{~~~}\ \cdots~.
\end{equation}
Note, however, that $\sL'$ is $L_\infty$-quasi-isomorphic to an $L_\infty$-algebra structure on 
\begin{equation}
 \sL''\ =\ \cdots\ \xrightarrow{~~~}\ *\ \xrightarrow{~~~}\ *\ \xrightarrow{~~~}\ {\rm coker}(\mu_1)\ \xrightarrow{~\mu_1~}\ \sL_{i+1}\ \xrightarrow{~\mu_1~}\ \sL_{i+2}\ \xrightarrow{~~~}\ \cdots~,
\end{equation}
effectively reducing the $L_\infty$-algebra $L$ to $\sL''$ by extending it to $\sL'$. It is therefore important to distinguish between the $L_\infty$-algebras $\sL$ and $\sL'$.

\paragraph{Morphisms of cyclic \mathversion{bold}$L_\infty$-algebras.} The definition of a morphism of cyclic $L_\infty$-algebras is induced from the description in terms of a differential graded algebra: it is simply a morphism of symplectic differential graded vector spaces. However, such a morphism $\Phi:(M,Q,\omega)\rightarrow (M',Q',\omega')$ would imply that $\omega=\Phi^*\omega'$ which, due to the non-degeneracy of $\omega$ implies that $\Phi$ is injective. This is often too restrictive, and one usually switches to {\em Lagrangian correspondences}, see e.g.~\cite{Weinstein:1977aa}. For our purposes, however, this is not necessary.

Moreover, we shall restrict our morphisms a bit further with an eye to homotopy Maurer--Cartan theory, which we shall discuss in Section~\eqref{sec:HMCT}. In this context, we are dealing with constant symplectic forms, and we require that morphisms of cyclic $L_\infty$-algebras preserve the homotopy Maurer--Cartan action. Following~\cite{Kajiura:0306332}, we thus define a {\em morphism of cyclic $L_\infty$-algebras} $\phi:(\sL,\mu_i,\langle-,-\rangle_\sL)\rightarrow (\sL',\mu'_i,\langle-,-\rangle_{\sL'})$ as an $L_\infty$-morphism $\phi:(\sL,\mu_i)\rightarrow (\sL',\mu'_i)$ such that in addition
\begin{subequations}
\begin{equation}
 \langle \phi_1(\ell_1),\phi_1(\ell_2)\rangle_{\sL'}\ =\ \langle \ell_1,\ell_2\rangle_\sL
\end{equation}
and for all $i\geq 3$ and $\ell_1,\dots, \ell_i\in \sL$,
\begin{equation}
 \sum_{\substack{j+k=i\\j,k\geq 1}}\langle \phi_j(\ell_1,\dots,\ell_j),\phi_k(\ell_{j+1},\dots,\ell_{j+k}))\rangle_{\sL'}\ =\ 0~.
\end{equation}
\end{subequations}
As before, for the sake of brevity, we shall also refer to such morphisms as  {\em cyclic $L_\infty$-morphisms}.

\paragraph{Decomposition theorem for cyclic \mathversion{bold}$L_\infty$-algebras.}
As shown in~\cite{Kajiura:0306332} for cyclic $A_\infty$-algebras, the decomposition theorem extends to cyclic $L_\infty$-algebras. That is, any cyclic $L_\infty$-algebra is isomorphic to the direct sum of a minimal cyclic $L_\infty$-algebra and a linearly contractible cyclic $L_\infty$-algebra.

\paragraph{Quasi-isomorphisms of cyclic \mathversion{bold}$L_\infty$-algebras.} 
We indicated above that quasi-isomor\-phisms allow us to describe an equivalence between data that is the same up to some gauge symmetry. To extend this notion to action principles, we need a preservation of the cyclic inner product on the relevant parts, which are the cohomology. We therefore define that a {\em quasi-isomorphism of cyclic $L_\infty$-algebras} is a morphism of cyclic $L_\infty$-algebras $\phi:\sL\rightarrow \sL'$, which descends to an isomorphism of cyclic $L_\infty$-algebras between $H^\bullet_{\mu_1}(\sL)$ and $H^\bullet_{\mu_1}(\sL')$.

\subsection{Representations of \texorpdfstring{$L_\infty$}{L-infty}-algebras}\label{ssec:representations}

To define (higher) supersymmetric field theories with matter content, we need to specify what we mean by a representation of an $L_\infty$-algebra. The first ingredient is a higher analogue of a vector space carrying the representation. There is a variety of definitions in the literature already for the simplest case of a 2-vector space. Fortunately, supersymmetry requires us to use the same type of categorified vector space that underlies our $L_\infty$-algebras. We can thus restrict ourselves to dg-vector spaces. Note that these can be regarded as Abelian $L_\infty$-algebras with all higher brackets $\mu_i$ trivial for $i\geq 2$.

\paragraph{\mathversion{bold}$L_\infty$-representations.}
There are now (at least) three evident ways of defining a representation of an $L_\infty$-algebra $(\sL,\mu_i)$ on a dg-vector space $(\sV,\dd)$:
\begin{enumerate}[(i)]
 \item Via an action of elements of $\sL$ on $V$ with compatibility relations as done in~\cite[Definition~5.1]{Lada:1994mn}, see also~\cite{Lada2004aa};
 \item As an $L_\infty$-morphism of $L_\infty$-algebras from $\sL$ to $\sEnd(\sV) $, cf.~e.g.~\cite[Definition~4.3]{Mehta:2012ppa};
 \item As a semidirect product of $L_\infty$-algebras $\sL\ltimes \sV$, which can be regarded as a short exact sequence of $L_\infty$-algebras $\sV \hookrightarrow \sL\ltimes \sV\rightarrow \sL$, cf.~e.g.~\cite[Definition~11.1.1.1]{costello2016factorization}.
\end{enumerate}

Theorem~5.4 of~\cite{Lada:1994mn} shows that (i) and (ii) are equivalent, and we choose to work with the latter. Recall that any dg-vector space $(\sV,\dd)$ comes with a dg-algebra $(\sEnd(\sV),\dd_{\sEnd(\sV) })$, which is defined by
\begin{subequations}
\begin{equation}
 \sEnd(\sV) \ :=\ \bigoplus_{i\in \RZ} \sEnd_i(\sV)  \ewith \sEnd_i(\sV) \ :=\ \bigoplus_{j\in\RZ}\sHom(\sV_j,\sV_{j+i})
 \end{equation}
together with
\begin{equation}
 \dd_{\sEnd(\sV) }T\ :=\ \dd \circ T-(-1)^{|T|}T \circ \dd
\end{equation}
\end{subequations}
for $T\in\sEnd(\sV) $. Together with the commutator $[S,T]:=S\circ T-T\circ S$ for $S,T\in\sEnd(\sV) $, $\sEnd(\sV) $ becomes a dg-Lie algebra.

An {\em $L_\infty$-representation} of an $L_\infty$-algebra $(\sL,\mu_i)$ on a differential graded vector space $(\sV,\dd)$ is an $L_\infty$-morphism, as defined in~\eqref{eq:L_infty_morphism}, from $(\sL,\mu_i)$ to $(\sEnd(\sV) ,\dd_{\sEnd(\sV) })$. 

\paragraph{Example.}
As an example, let us consider the case of a representation of a Lie 2-algebra, that is, an $L_\infty$-algebra $(\sL,\mu_i)=(\sL_{-1}\oplus \sL_0,\mu_i)$ concentrated in degrees~$-1$ and~$0$ on the differential graded vector space $(\sV,\dd)=(\sV_{-1}\oplus \sV_0,\dd)$. We note that
\begin{equation}
 \sEnd(\sV) \ =\ \big(\sHom(\sV_0,\sV_{-1})\xrightarrow{~ \dd_{\sEnd(\sV) }~}\sEnd(\sV_0)\oplus \sEnd(\sV_{-1})\xrightarrow{~\dd_{\sEnd(\sV) }~}\big(\sHom(\sV_{-1},\sV_0)\big)
\end{equation}
and therefore a representation of $(\sL,\mu_i)$ on $(\sV,\dd)$ consists of a cochain map $\phi_1$
\begin{subequations}
\begin{equation}
\myxymatrix{
    \sL_{-1} \ar@{->}[rr]^{\mu_1} \ar@{->}[d]^{\phi_1} && \sL_0 \ar@{->}[rr]^{0} \ar@{->}[d]^{\phi_1}&&  {*} \ar@{->}[d]^{\phi_1}\\
    \sHom(\sV_0,\sV_{-1}) \ar@{->}[rr]^-{\dd_{\sEnd(\sV) }}& & \sEnd(\sV_0)\oplus \sEnd(\sV_{-1}) \ar@{->}[rr]^-{\dd_{\sEnd(\sV) }} & &\sHom(\sV_{-1},\sV_0)
    }
\end{equation}
together with a map
\begin{equation}
 \phi_2\,:\,\sL_0\times \sL_0\ \rightarrow\ \sHom(\sV_0,\sV_{-1})
\end{equation}
such that 
\begin{equation}
 \begin{aligned}
  \phi_1(\mu_1(y))\ &=\ \dd_{\sEnd(\sV) }\phi_1(y)\ =\ \dd\circ \phi_1(y)+\phi_1(y)\circ \dd~,\\
  \phi_1(\mu_2(x_1,x_2))\ &=\ [\phi_1(x_1),\phi_1(x_2)]+\dd_{\sEnd(\sV) } \phi_2(x_1,x_2)\\
  \ &=\ [\phi_1(x_1),\phi_1(x_2)]+\dd\circ\phi_2(x_1,x_2)+\phi_2(x_1,x_2)\circ \dd~,\\
  \phi_1(\mu_2(x,y))\ &=\ [\phi_1(x),\phi_1(y)]+\phi_2(x,\mu_1(y))~,\\
  \phi_1(\mu_3(x_1,x_2,x_3))\ &=\ \phi_2(\mu_2(x_1,x_2),x_3)-[\phi_1(x_1),\phi_2(x_2,x_3)]+\mbox{cyclic}
 \end{aligned}
\end{equation}
\end{subequations}
for al $x_{1,2,3}\in \sL_0$ and $y\in \sL_{-1}$, see~\eqref{eq:L_infty_morphism}.

As examples of concrete applications, let us specialise to the case $\phi_2=0$. For $(\sV,\dd)=(\sL_{-1}\oplus \sL_0,\mu_1)$ and $\phi_1:\sL_{-1}\to \sHom(\sL_0,\sL_{-1})$ trivial, we recover the representations underlying the models of~\cite{Samtleben:2011fj}. For a suitable choice of $\sL$, $\sV_{-1}=*$ and $\sV_0$ the tensor fields on a manifold, we obtain the representations relevant in generalised geometry and double field theory, see the discussion in~\cite{Deser:2016qkw}.

\section{Batalin--Vilkovisky formalism}\label{sec:GeneralBVFormalism}

This section consists of an outline of the BV formalism as developed and explained in a series of papers, see~\cite{Batalin:1984jr,Batalin:1985qj,Batalin:1984ss,Batalin:1981jr,Batalin:1977pb,ZinnJustin:1974mc} as well as~\cite{Schwarz:1992nx} and the famous~\cite{Alexandrov:1995kv} in the context of topological field theories. This formalism is also known as the BV/BRST formalism, the antifield formalism or it is included in the term BRST formalism. For more detailed reviews, we refer the reader to Section~8 of~\cite{Henneaux:1992} as well as the papers~\cite{Henneaux:1990:47-105,Gomis:1994he,Barnich:2000zw,Fiorenza:0402057,Mnev:2017oko} and~\cite{Stasheff:1997fe,Stasheff:1997iz,Felder:2012kn}.

In this section, we shall illustrate most constructions in the context of ordinary gauge theory. We try to highlight the conceptual origins of the BV formalism and its formulation in terms of the language of symplectic $Q$-manifolds introduced in the previous section. In a following section we explain the much more general example of homotopy Maurer--Cartan theory, which directly contains Chern--Simons and higher Chern--Simons theory and, via a more indirect construction, any BV quantisable gauge theory.

\subsection{Motivation and outline}

The BV formalism is certainly best known as a very general approach to the quantisation of classical field theories with complicating symmetries such as gauge theories. This was also the historical motivation for its development. The corresponding setup of a classical field theory for quantisation, however, exposes much of the theory's internal structure in a way that is mathematically very precise and useful at the same time. The underlying language is mostly that of homological algebra, i.e.~that of cochain complexes and differential graded algebras.

\paragraph{Classical structure.}
The classical part of the BV formalism consists essentially of two important steps. The first one is the usual starting point of BRST quantisation\footnote{We shall always distinguish between the BRST formalism, which involves ghosts and where antifields only enter when gauge fixing, and the BV formalism, which involves antifields from the outset.}~\cite{Becchi:1975nq,Tyutin:1975qk}, which expresses the symmetries\footnote{Usually, one is only concerned with the local symmetries, as these are the ones complicating the quantisation.} of a field theory as the action Lie algebroid introduced in Section~\ref{ssec:L_infty_algebras_and_algebroids}. This interpretation may seem at first unnecessarily abstract, but it clarifies the mathematical origin of the fermionic ghosts as well as the generalisation to arbitrarily complicated gauge theories. Concretely, this yields a complex $(\frF_{\rm BRST},Q_{\rm BRST})$ also known as the {\em minimal set of fields}, which can be extended to the set of fields $\frF_{\rm eBRST}$ needed for gauge fixing in the BRST formalism.

In many cases, the latter complex exists only on-shell, and therefore needs to be lifted before quantisation. This is true in particular in higher gauge theories if one regards the vanishing of fake curvatures as a dynamical instead of a kinematical condition. The off-shell lift is provided by a Koszul--Tate resolution, cf.~\cite{Stasheff:1997fe,Stasheff:1997iz}, which amounts to introducing antifields and, if necessary, anti-ghosts and higher anti-ghosts. Explicitly, we extend the set of fields $\frF_{\rm BRST}$ further to $\frF_{\rm BV}:=T^*[-1]\frF_{\rm BRST}$, which is endowed with the differential $Q_{\rm BV}$ given by a derived bracket $Q_{\rm BV}:=\{S_{\rm BV},-\}$. Here, $S_{\rm BV}$ is the classical BV action satisfying the classical master equation $\{S_{\rm BV},S_{\rm BV}\}=0$. The fact that, under some reasonable conditions, $S_{\rm BV}$ exists and is unique for a general $k$-th stage reducible theory (that is, $k$-th level gauge invariance) was proved in~\cite{Fisch:1989rp} using homological perturbation theory.

The result is then an $L_\infty$-algebra\footnote{or $L_\infty$-algebroid in the most general setting} that encodes the fields, the gauge structure, gauge invariant observables, field equations, Noether identities and consistent deformations of the theory; in short, everything one needs to know about a classical theory. The minimal model of this $L_\infty$-algebra yields a minimal representation of the classical dynamical data and an equivalence of classical theories is a quasi-isomorphism of $L_\infty$-algebras.

\paragraph{Quantisation.}
To quantise a classical field theory means to make sense of the path integral
\begin{equation}
 Z(S)\ =\ \int_{\frF} \mu_{\frF}(\Phi)~\de^{\frac{\di}{\hbar} S[\Phi]}~,
\end{equation}
where $\frF$ is the space of fields, $\Phi\in \frF$, $\mu_{\frF}(\Phi)$ is a measure on this space and $S$ is the action functional. Path integrals of quantum field theories on spaces $M$ with Minkowski signature are oscillatory functional integrals. Such functional integrals can be computed perturbatively by the stationary phase formula (see~\cite[Section~1.2.4]{Mnev:2017oko} for details), but this requires the stationary points to be isolated or, equivalently, the Hessian to be non-degenerate at the stationary points.

The problem is that the stationary phase formula requires critical points of f to be isolated (more precisely, we need the Hessian of f at critical points to be non-degenerate). However, diffeomorphism invariant classical field theories are gauge theories, i.e. there is a tangential distribution E on FM which preserves the action SM (for instance, E corresponds to an action of a group G, the gauge group, on the space of fields FM ). Thus, critical points of SM come in E-orbits and therefore are not isolated (the Hessian of SM is degenerate in the direction of E).
So, the stationary phase formula cannot be applied to the path integral in the case of a gauge theory.

This is not the case in gauge theories due to the large degeneracy arising from gauge orbits. Let $\sG$ be a Lie group inducing a group $\frG$ of gauge transformations acting on the space $\frF$. A gauge theory action $S$ is invariant under the action of $\frG$ and therefore we could in principle restrict to gauge orbits,
\begin{equation}\label{eq:PI1}
 \int_{\frF} \mu_{\frF}(\Phi)~\de^{\frac{\di}{\hbar} S[\Phi]}~~~\ \rightarrow\ ~~~\int_{\frF/\frG} \mu_{\frF/\frG}(\Phi)~\de^{\frac{\di}{\hbar} S[\Phi]|_{\frF/\frG}}~,
\end{equation}
where $\mu_{\frF/\CCG}(\Phi)$ is the measure induced on $\frF/\frG$ by $\mu_{\frF}(\Phi)$. There are various reasons why a restriction to the orbit space is not feasible in practise, chief of all the fact that the orbit space is not well-behaved in general.

The classical BV formalism, however, provides a starting point for a very convenient gauge fixing procedure. As observed above, introducing antifields in the Koszul--Tate resolution corresponds to extending the space of fields and ghosts $\frF_{\rm BRST}$ to its cotangent bundle $\frF_{\rm BV}:=T^*[-1]\frF_{\rm BRST}$. The original action functional corresponds to evaluating the BV action for the zero section, but we can choose a different Lagrangian submanifold. This can be done such that the resulting restricted BV action functional has isolated stationary points and that corresponding functional integral equals\footnote{after applying heuristics generalised from ordinary integration to functional integration} the original functional integral. The choice of section is encoded in a functional known as the gauge fermion. Very roughly speaking, this procedure is analogous to the computation of a real integral by going to the complex plane: one doubles the number of variables and extends the original integrand to the new variables. The final integral is still taken along a half-dimensional contour.

Clearly, expectation values should be independent of the choice of gauge fermion, which is tantamount to the {\em quantum master equation}, a deformation of the classical master equation in $\CO(\hbar)$. This requires, in general, to deform the classical BV action to a formal power series in $\hbar$, the quantum BV action. The latter action is then the starting point for all further, e.g.~perturbative, computations.

\subsection{Gauge Lie algebroid}\label{ssec:GaugeStructureGeneral}

As stated above, we shall focus on the example of ordinary gauge theories, but we shall present the steps in a way that allows for a straightforward extension to higher gauge theory. For a detailed discussion along traditional lines, see also~\cite{Barnich:2010xq} and in particular~\cite{Henneaux:1990:47-105} for the case of {\it open algebras}, i.e.~gauge algebroids where the gauge symmetries close only on-shell.

\paragraph{Action Lie algebroid.}
Quotient spaces as e.g.~$\frF/\frG$ appearing in~\eqref{eq:PI1} are often badly behaved and a useful way to circumvent this issue is to consider the {\em derived quotient}\footnote{or model for the homotopy quotient} $[\frF/\frG]$, which amounts to considering the corresponding {\em action Lie groupoid} as discussed in Section~\ref{ssec:L_infty_algebras_and_algebroids}. 

Let $\frF$ and $\frG$ be again the space of fields and the group of gauge transformations, respectively. Then the action Lie groupoid has objects $\frF$ and morphisms $\frG\ltimes \frF$. A morphism $(g,\Phi)$ is of the form 
\begin{equation}
 \Phi\ \xrightarrow{~(g,\Phi)~}\ g\acton \Phi
\end{equation}
with the obvious concatenation and identity morphisms, cf.~Section~\ref{ssec:L_infty_algebras_and_algebroids}.

For many purposes, and in particular for the BV formalism over contractible manifolds or with trivial principal (gauge) bundle, the infinitesimal picture in terms of Lie algebra actions is sufficient. The corresponding action Lie algebroid is most readily described in terms of $Q$-manifolds, as seen in Section~\ref{ssec:L_infty_algebras_and_algebroids}, and looks like
\begin{equation}
\frF_{\rm BRST} \ :=\ \sLie(\frG)[1]\ltimes\frF~,
\end{equation}
where $\sLie(\frG)[1]$ is the Lie algebra $\sLie(\frG)$ of the group of gauge transformations $\frG$, whose underlying vector space is degree-shifted by $-1$, cf.~Equation~\eqref{eq:grade_shift}. The group product and the action of the gauge transformation are now encoded in the homological vector field $Q_{\rm BRST}$, whose form is determined by actions on the contracted coordinate functions on $\frF_{\rm BRST}$,\footnote{Recall our remark about the first fundamental confusion of calculus in Section~\ref{ssec:DGAs}.}
\begin{equation}\label{eq:first_BRST}
 Q_{\rm BRST}\Phi\ :=\ \delta_c \Phi\eand Q_{\rm BRST} c\ :=\ -\tfrac12[c,c]~.
\end{equation}
Here, $\delta_c\Phi$ denotes an infinitesimal gauge transformation of $\Phi$ parametrised by $c\in\sLie(\frG)[1]$ and $[-,-]$ is the Lie bracket on $\sLie(\frG)$. Hence, together with $Q_{\rm BRST}$, the algebra of functions on $\frF_{\rm BRST}$ forms a dg-algebra.

Let us stress here the important distinction between elements $c$ of $\sLie(\frG)$, which parametrise gauge transformations via 
\begin{equation}
 A\ \mapsto\ A'\ :=\ A+\delta A\ewith \delta A\ :=\ \dd c+[A,c]
\end{equation}
and the corresponding coordinate functions $c\in \CCC^\infty(\frF_{\rm BRST})$ appearing in~\eqref{eq:first_BRST}, which is of degree~1. Nevertheless, we follow the common convention of using the same letter for a vector and its (contracted) coordinate functions. The degree shift is due to the $Q$-manifold description of the gauge algebroid $\frF_{\rm BRST}$ in which the degree of $\sLie(\frG)$ is shifted by $-1$ and therefore the coordinate function on $\sLie(\frG)[1]$ has degree~$+1$. The coordinate functions $c$ are known as {\em ghosts} and the degree shift is the origin of their fermionic character. In general, we call the natural degree of functions on $\frF_{\rm BRST}$ the {\em ghost number}.

If we are dealing with a higher action Lie algebroid encoding gauge symmetries between gauge symmetries, as will be the case for higher gauge theory, then we also have {\it ghosts for ghosts}, which are functions of homogeneous degree greater than one in $\frF_{\rm BRST}$. In this case, $Q_{\rm BRST}^2=0$ only on the proper kinematical data which we shall discuss later.

\paragraph{\mathversion{bold}$Q_{\rm BRST}$-cohomology.}
Note that the functions on $\frF_{\rm BRST}$ form a cochain complex with differential $Q_{\rm BRST}$:
\begin{equation}\label{eq:CE-complex}
 0\ \xrightarrow{~\phantom{Q_{\rm BRST}}~}\ \CCC^\infty_0(\frF_{\rm BRST})\ \xrightarrow{~Q_{\rm BRST}~}\ \CCC^\infty_1 (\frF_{\rm BRST})\ \xrightarrow{~Q_{\rm BRST}~}\ \cdots~,
\end{equation}
where $\CCC^\infty_i(\frF_{\rm BRST})$ are functions of ghost degree~$i$. Gauge invariant functionals $F[\Phi]\in\CCC^\infty(\frF)$, such as the action $S[\Phi]$, satisfy
\begin{equation}
 Q_{\rm BRST} F[\Phi]\ =\ 0
\end{equation}
because $Q_{\rm BRST}$ encodes the action of gauge transformations on $\frF$. The analogue of restricting to the isomorphism classes in the derived quotient $[\frF/\frG]$ therefore corresponds to restricting to the $Q_{\rm BRST}$-cohomology $H^0(\frF_{\rm BRST})$,
\begin{equation}\label{eq:QResolution}
\begin{gathered}
 \CCC^\infty(\frF/\frG)\ \cong\ H^0(\frF_{\rm BRST})~.
\end{gathered}
\end{equation}
We shall return to this point below.

Note that the idea that cohomological considerations should play a key role in functional integration is motivated by the following heuristics: let $M$ be a smooth compact manifold with volume form $\mu$. Then the expectation value of an observable $F\in \CCC^\infty(M)$ is computed as 
\begin{equation}
 \langle F\rangle_\mu\ :=\ \frac{\int_M \mu F}{\int_M \mu}\ =\ \frac{[F\mu]}{[\mu]}~,
\end{equation}
where $[-]$ denotes the cohomology class of a differential form and we used Stokes theorem and the fact that $\dd \mu=0$. Thus, the computation of the expectation value of an observable can be reduced from an integral to a comparison of cohomology classes. The resolution~\eqref{eq:QResolution} is also a first step into this direction for the case of path integrals.

\paragraph{The \mathversion{bold}$Q_{\rm BRST}$-complex as a resolution.}
Let us briefly consider the above from a mathematical perspective. Recall that we replaced the naive quotient $\frF/\frG$ by the derived quotient $[\frF/\frG]$, where $\frF/\frG$ equals the isomorphism classes of objects in the action Lie groupoid $[\frF/\frG]$. At the infinitesimal level, this corresponds to considering the cohomology of the {\em Chevalley--Eilenberg (cochain) complex} for the $\sLie(\frG)$-module $\CCC^\infty(\frF)$, which is given in~\eqref{eq:CE-complex}.

Recall from Section~\ref{ssec:L_infty_algebras_and_algebroids} that the $p$-cochains of the Chevalley--Eilenberg complex for a Lie algebra $\frg$ and a $\frg$-module $\CCE$ are given by $\sHom(\bigwedge^p \frg,\CCE)$, and the differential arises from the action of $\frg$ on $\CCE$ as well as the Lie algebra structure on $\frg$. In our case $\CCE=\CCC^\infty(\frF)$, and we have
\begin{equation}
 \sHom(\mbox{$\bigwedge^p$} \sLie(\frG),\CCC^\infty(\frF))\ \cong\ \CCC^\infty_p(\frF_{\rm BRST})~,
\end{equation}
which indeed reproduces the complex~\eqref{eq:CE-complex} in the case of ordinary gauge theory. Note also that the Chevalley--Eilenberg complex of a Lie algebra has a straightforward generalisation for modules of $L_\infty$-algebras. 

This complex provides a resolution to the gauge invariant functionals, since we can extended the Chevalley--Eilenberg complex on the left to
\begin{equation}\label{eq:CE-complex2}
 0 \ \xrightarrow{~\phantom{Q_{\rm BRST}}~}\ \CCC^\infty(\frF/\frG)\ \cong\ H^0(\frF/\frG)\ \xhookrightarrow{~~~~~~~}\ \CCC^\infty_0(\frF_{\rm BRST})\ \xrightarrow{~Q_{\rm BRST}~}\ \cdots~.
\end{equation}
Note that two other low cohomology groups have an interesting interpretation. Firstly, $H^1(\frF_{\rm BRST})$ is the set of derivations modulo inner derivations. Here, a {\em derivation} is a map 
\begin{subequations}
\begin{equation}
 \delta\,:\,\sLie(\frG)\ \rightarrow\ \CCC^\infty(\frF_{\rm BRST})
\end{equation}
so that
\begin{equation}
\delta([ c, c'])\ =\  c\acton \delta( c')- c'\acton \delta( c)\efor c, c' \in \sLie(\frG)
\end{equation}
\end{subequations}
and {\em inner derivations} are derivations of the form $\delta_f( c)= c\acton f$ for some $f\in \CCC^\infty(\frF)$.

Secondly, $H^2(\frF_{\rm BRST})$ is isomorphic to the equivalence classes of Lie algebra extensions $\widehat{\sLie(\frG)}$ by $\CCC^\infty(\frF)$, i.e.~short exact sequences
\begin{equation}
 0\ \longrightarrow\ \CCC^\infty(\frF) \ \longrightarrow\ \widehat{\sLie(\frG)} \ \longrightarrow\ \sLie(\frG) \ \longrightarrow\ 0~.
\end{equation}
Further details are found, e.g., in~\cite{0387948236}.

Below, we shall encounter a second type of resolution which reduces $H^0(\frF_{\rm BRST})$ to the classical observables, which are obtained after taking the quotient by the ideal of functionals vanishing on classical solutions.

\subsection{Becchi--Rouet--Stora--Tyutin quantisation}\label{ssec:BRST_quant}

If the symmetries of a classical theory close off-shell, which amounts to $Q^2_{\rm BRST}=0$ without any further restriction on the fields $\frF_{\rm BRST}$, then the BRST formalism is sufficient for quantisation. We briefly outline this approach in the following.

\paragraph{Gauge fixing and Faddeev--Popov determinant.}
The gauge fixing itself is encoded in the {\em gauge fixing function}, which is a map $F:\frF\rightarrow \sLie(\frG)$ such that each point in $F^{-1}(0)$ represents a different orbit of $\frG$. We can restrict to $F^{-1}(0)$ by inserting a factor of $\delta(F(\Phi))$ into the functional integral~\eqref{eq:PI2} with $\delta$ the functional analogue of the $\delta$-distribution. This also requires the insertion of the {\em Faddeev--Popov (FP) determinant} ${\rm det}(M_{\rm FP}(\Phi))$ to render the construction invariant under deformations of $F$. Schematically, we obtain
\begin{equation}\label{eq:PI2}
 \int_{\frF} \mu_{\frF}(\Phi)~\de^{\frac{\di}{\hbar} S[\Phi]}~~~\ \rightarrow\ ~~~ \int_{\frF} \mu_{\frF}(\Phi)~{\rm det}(M_{\rm FP}(\Phi))~\delta(F(\Phi))~\de^{\frac{\di}{\hbar} S[\Phi]}~.
\end{equation}

\paragraph{BRST quantisation.}
Instead of dealing with these two insertions as they are, we can encode them in an extended action functional on an extension of $\frF_{\rm BRST}$ by fields of negative degrees.\footnote{More appropriately, one should speak of the graded ring of functions on the action Lie algebroid and extending it by generators of negative degree.} It turns out that an appropriate choice is $\frF_{\rm eBRST}$ which contains the fields $\Phi$ and ghost $c$ as well as the Lagrange multipliers $b\in\sLie(\frG)[0]$ and the antighosts $\bar c\in\sLie(\frG)[-1]$, that is,
\begin{equation}
 \frF_{\rm eBRST} \ :=\ (\sLie(\frG)[1]\oplus\sLie(\frG)[0]\oplus\sLie(\frG)[-1])\ltimes \frF~.
\end{equation}
The homological vector field $Q_{\rm BRST}$ is extended to the homological vector field
\begin{equation}
 Q_{\rm eBRST}\Phi\ :=\ \delta_c \Phi~,~~~Q_{\rm eBRST} c\ :=\ -\tfrac12[c,c]~,~~~Q_{\rm eBRST} \bar c\ :=\ b~,~~~Q_{\rm eBRST}b\ :=\ 0~.
\end{equation}
Note that the antighosts and Lagrange multipliers  form the dg-subalgebra $\sLie(\frG)[-1]\xrightarrow{~\id~}\sLie(\frG)[0]$ of $\frF_{\rm eBRST}$. The corresponding $L_\infty$-algebra is linearly contractible, and therefore the action algebroid $\frF_{\rm eBRST}$ is quasi-isomorphic to $\frF_{\rm BRST}$. In this sense, we have not extended the data of the theory.

Using elements of $\frF_{\rm eBRST}$, we now rewrite~\eqref{eq:PI2} as
\begin{equation}
 \int_{\frF} \mu_{\frF}(\Phi)~\de^{\frac{\di}{\hbar} S[\Phi]}~~~\ \rightarrow\ ~~~ \int_{\frF_{\rm eBRST}} \mu_{\rm eBRST}(\Phi,b,c,\bar c)~\de^{\frac{\di}{\hbar}S[\Phi]+\langle b,F(\Phi)\rangle+\langle\bar c, M_{\rm FP}(\Phi) c\rangle}~.
\end{equation}
Here $\mu_{\rm eBRST}(\Phi,b,c,\bar c)$ is a natural extension of $\mu_{\frF}(\Phi)$ to $\frF_{\rm eBRST}$ and $\langle-,-\rangle$ is an appropriate pairing between Lie algebra valued fields, including the integral over spacetime. We thus achieved our goal of replacing a functional integral with degenerate Hessian at stationary points by a technically equivalent, non-degenerate functional integral over a larger function space. 

Moreover, there is a function\footnote{We shall describe this function in more detail in Section~\ref{ssec:BV_gauge_fixing}.} $\Psi\in \CCC^\infty(\frF_{\rm eBRST})$ of homogeneous degree~$-1$, called the {\it gauge fixing fermion}, such that
\begin{equation}\label{eq:quantum_BRST}
\begin{aligned}
 &\int_{\frF_{\rm eBRST}} \mu_{\rm eBRST}(\Phi,b,c,\bar c)~\de^{\frac{\di}{\hbar} S[\Phi]+\langle b,F(\Phi)\rangle+\langle\bar c, M_{\rm FP}(\Phi) c\rangle}\ =\ \\
 &\kern4cm\ =\ \int_{\frF_{\rm BRST}} \mu_{\rm BRST}(\Phi,b,c,\bar c)~\de^{\frac{\di}{\hbar}(S[\Phi]+Q_{\rm eBRST}\Psi)}~,
 \end{aligned}
\end{equation}
and the measure $\mu_{\rm eBRST}(\Phi,b,c,\bar c)$ is compatible with $Q_{\rm eBRST}$ in the sense that 
\begin{equation}\label{eq:quantum_compatibility_Q_mu}
 \int_{\frF_{\rm eBRST}} \mu_{\rm eBRST}(\Phi,b,c,\bar c)~Q_{\rm eBRST}f\ =\ 0
\end{equation}
for all reasonable test functions $f\in \CCC^\infty(\frF_{\rm eBRST})$. 

Just as before gauge fixing, we have again a cochain complex $(\CCC^\infty(\frF_{\rm eBRST}),Q_{\rm eBRST})$. Its cohomology encodes potential observables: because of~\eqref{eq:quantum_compatibility_Q_mu} and $Q_{\rm eBRST}S=0$, $Q_{\rm eBRST}$-exact terms vanish under the functional integral. Also, only the expectation values of $Q_{\rm eBRST}$-closed functions are independent of the gauge fixing. Thus, equation~\eqref{eq:quantum_BRST} shows that the BRST approach to quantisation renders gauge invariance manifest.

\subsection{Batalin--Vilkovisky complex and classical master equation}\label{ssec:ClassicalMasterEquation}

The BRST formalism is not suitable for the general treatment of gauge theories. In the case of {\it open symmetries}, which are symmetries that are only satisfied on-shell, the BRST complex is only a complex up to equations of motion. For many purposes including quantisation, however, we require an off-shell description. This can be obtained by a further extension of the BRST complex, and this extension is known as the BV formalism. The idea is to double the field content and to construct the $Q$-manifold
\begin{equation}
\frF_{\rm BV}\ :=\ T^*[-1]\frF_{\rm BRST}~,
\end{equation}
which allows for a homological vector field $Q_{\rm BV}$ for which $Q_{\rm BV}^2=0$ off-shell. The functional integral is then performed over a Lagrangian submanifold of $\frF_{\rm BV}$ which extends $\frF_{\rm BRST}$. 

\paragraph{Symplectic structure.}
Since $\frF_{\rm BV}$ is a cotangent bundle, we have a natural symplectic structure $\omega_{\rm BV}$ of degree~$-1$. Let $\Phi^A$ be local coordinates on $\frF_{\rm BRST}$ (i.e.~the fields) and let $\Phi^+_A$ (i.e.~the antifields) be fibre coordinates on $\frF_{\rm BV}\rightarrow \frF_{\rm BRST}$, where $A,B,\ldots$ are multi-indices constituting Lorentz indices, gauge algebra indices, etc. In terms of these Darboux coordinates, the canonical symplectic form reads as 
\begin{equation}\label{eq:BVSymplecticStructureGeneral}
 \omega_{\rm BV}\ :=\ (-1)^{|\Phi_A|}\delta\Phi^A\wedge\delta\Phi_A^+~,
\end{equation}
where $\delta$ is the exterior differential on $\frF_{\rm BV}$. We shall denote the induced Poisson structure by $\{-,-\}_{\rm BV}$, and we have $|\{F,G\}_{\rm BV}|=|F|+|G|+1$ for homogeneous $F,G\in\CCC^\infty(\frF_{\rm BV})$. As seen in Section~\ref{ssec:QManifolds}, $\{-,-\}_{\rm BV}$ is graded antisymmetric and obeys a graded Leibniz rule and a graded Jacobi identity. This Poisson bracket is also known as the {\it antibracket}. It is of degree~1 and therefore $\CCC^\infty(\frF_{\rm BV})$ forms a {\em Gerstenhaber algebra}. 

\paragraph{Batalin--Vilkovisky complex.}
We now wish to extend the homological vector field $Q_{\rm BRST}$ to a homological vector field $Q_{\rm BV}$ such that
\begin{subequations}
\begin{equation}\label{eq:bc_Q_BV}
 Q_{\rm BV}|_{\frF_{\rm BRST}}\ =\ Q_{\rm BRST}
\end{equation}
and  $Q_{\rm BV}$ is Hamiltonian with respect to the symplectic structure  $\omega_{\rm BV}$, that is,
\begin{equation}
 Q_{\rm BV}\intprod\omega_{\rm BV}\ =\ \delta S_{\rm BV}\ewith S_{\rm BV}\ \in\ \CCC^\infty(\frF_{\rm BV})
\end{equation}
or, equivalently,
\begin{equation}\label{eq:derived-BV}
 Q_{\rm BV}\ =\ \{S_{\rm BV},-\}_{\rm BV}~.
\end{equation}
\end{subequations}
This makes $(\frF_{\rm BV},Q_{\rm BV},\omega_{\rm BV})$ a symplectic $Q$-manifold of degree~$-1$. Recall from Section~\ref{ssec:QManifolds} that the Hamiltonian condition is equivalent to requiring that $Q_{\rm BV}$ generates a symplectomorphism on symplectic $Q$-manifolds of degree~$k$ except for $k=-1$, which is the case at hand.

Equation~\eqref{eq:derived-BV}, together with the Jacobi identity of the Poisson bracket, provides the equivalence between $Q_{\rm BV}^2=0$ and the {\em classical master equation},
\begin{equation}\label{eq:ClassicalMasterGeneral}
\big\{S_{\rm BV},S_{\rm BV}\big\}_{\rm BV}\ =\ 0~.
\end{equation}

\paragraph{Solutions to the classical master equation.}
We have some freedom in choosing a solution $S_{\rm BV}$ to equation~\eqref{eq:ClassicalMasterGeneral}, and we use this to impose a boundary condition beyond~\eqref{eq:bc_Q_BV}. We require that 
\begin{equation}\label{eq:bc_S_BV}
 \left.S_{\rm BV}\right|_{\frF_{\rm BRST}}\ =\ S~,
\end{equation}
where $S\in \CCC^\infty(\frF)$ is the original action of our field theory. Thus, $S_{\rm BV}$ encodes simultaneously our action and the gauge structure of the fields. One important consequence of the choice~\eqref{eq:bc_S_BV} is that the classical equations of motion are now encoded in $Q_{\rm BV}$ via
\begin{equation}\label{eq:eom_from_Q+}
 \left.\{S_{\rm BV},\Phi_A^+\}\right|_{\frF_{\rm BRST}}\ =\ \delta_{\Phi^A} S~.
\end{equation}

A solution $S_{\rm BV}$ also defines a Lagrangian subspace $\frL_{S_{\rm BV}}$ of $\frF_{\rm BV}$ through its stationary points. It is called {\it proper} provided the rank of the Hessian of $S_{\rm BV}$ on $\frL_{S_{\rm BV}}$ equals the number of fields $\Phi^A$. For such a proper solution, one finds that $S_{\rm BV}$ has precisely the gauge invariance required to eliminate all auxiliary fields. It can be shown that a proper solution always exists, see~\cite{Gomis:1994he} and references therein for details.

A proper solution can be written as a power series in the antifields,
\begin{equation}
 S_{\rm BV}\ =\ S+\Phi^+_{A}\,{R^A}_B\,\Phi^B+\CO((\Phi^+_A)^2)~,
\end{equation}
where the coefficients ${R^A}_B$ vanish unless the ghost number of $\Phi^A$ is one less than that of $\Phi^B$ so that the total ghost number of $S_{\rm BV}$ indeed vanishes. From the power series expansion, we can iteratively determine the relevant proper solution for a given action $S$ and gauge symmetries $Q_{\rm BRST}$.

\paragraph{Koszul--Tate resolution.}
Let us briefly look at the BV complex induced by $Q_{\rm BV}$ on $\CCC^\infty(\frF_{\rm BV})$ from a more mathematical perspective. Given a field theory with a set of classical fields $\frF$, the classical states are given by the subset that solves the equations of motion of the theory. The functionals which vanish on solutions to the equations of motion form an ideal $\frI$ of $\CCC^\infty(\frF)$, and the classical observables are given by $\CCC^\infty(\frF)/\frI$. Just as we replaced the gauge quotient $\frF/\frG$ by its Chevalley--Eilenberg resolution, we should also replace this quotient by a resolution encoded in a suitable differential graded algebra. This is precisely what the BV formalism does. 

Consider first the case of a general field theory with action $S$, ignoring potential gauge symmetries. Let $\CCC^\infty(\frF)$ be the functionals on the fields $\frF$ and $\frI$ the ideal induced by the critical locus of $S$. We then have the cochain complex of functions on $T^*[-1]\frF$, 
\begin{equation}
\cdots\ \xrightarrow{~Q_{\rm BV}~}\ \CCC^\infty_{-1}(T^*[-1]\frF)\ \xrightarrow{~Q_{\rm BV}~}\ \CCC^\infty_{0}(T^*[-1]\frF)\ \cong\ \CCC^\infty(\frF)\ \xrightarrow{~\phantom{Q_{\rm BV}}~}\ 0~,
\end{equation}
where $Q_{\rm BV}=\{S_{\rm BV},-\}$ with a BV action $S_{\rm BV}$ satisfying~\eqref{eq:bc_S_BV}. Because $\CCC^\infty_{-1}(T^*[-1]\frF)$ consists of functionals linear in the antifields, equation~\eqref{eq:eom_from_Q+} implies that the ideal $\frI$ is simply the image of $Q_{\rm BV}$
\begin{equation}
 Q_{\rm BV}(\CCC^\infty_{-1}(T^*[-1]\frF))\ =\ \frI~. 
\end{equation}
Thus, the cohomology group $H^0(T^*[-1]\frF)$ consists of the desired quotient $\CCC^\infty(\frF)/\frI$. We can extend the above cochain complex by the projection onto the latter, which leads to the resolution
\begin{equation}
\cdots\ \xrightarrow{~Q_{\rm BV}~}\ \CCC^\infty_{-1}(T^*[-1]\frF)\ \xrightarrow{~Q_{\rm BV}~}\ \CCC^\infty_{0}(T^*[-1]\frF)\ \xrightarrow{~\eps~} \ H^0(T^*[-1]\frF)\ \xrightarrow{~\phantom{Q_{\rm BV}}~}\ 0~,
\end{equation}
and this is the {\em Koszul--Tate resolution} of $\CCC^\infty(\frF)/\frI$.

To incorporate gauge symmetry, we replace $\frF$ by $\frF_{\rm BRST}$ and construct a new homological vector field $Q_{\rm BV}$ satisfying the boundary condition~\eqref{eq:bc_Q_BV}. The result is the complex
\begin{equation}
\cdots\ \xrightarrow{~Q_{\rm BV}~}\ \CCC^\infty_{-1}(\frF_{\rm BV})\ \xrightarrow{~Q_{\rm BV}~}\ \CCC^\infty_{0}(\frF_{\rm BV})\ \xrightarrow{~Q_{\rm BV}~}\  \CCC^\infty_{1}(\frF_{\rm BV})\ \xrightarrow{~Q_{\rm BV}~}\ \cdots~.
\end{equation}
The image of $Q_{\rm BV}$ in $\CCC^\infty_{0}(\frF_{\rm BV})$ are now the functionals vanishing on the equations of motions (for fields, ghosts, etc.) and the kernel of $Q_{\rm BV}$ in $\CCC^\infty_{0}(\frF_{\rm BV})$ are the gauge invariant functionals. The cohomology therefore contains in particular the classical observables: classical states which are gauge invariant. Moreover, the cohomology groups in negative degree encode Noether and higher Noether identities, and we shall come back to this point in Section~\ref{ssec:Classical_FT_and_L_infty}.

\paragraph{Classical \mathversion{bold}$L_\infty$-algebra structure.} We note already here that $Q_{\rm BV}$ induces an $L_\infty$-algebra structure on the graded vector space $\frF_{\rm BV}[-1]$. This structure captures the essence of a classical (gauge) field theory and we shall return to a detailed discussion in Section~\ref{sec:L_infty-structures}.

\subsection{Quantum master equation}\label{ssec:BV_gauge_fixing}

\paragraph{Gauge fixing.}
Having constructed the BV action, we now need to implement gauge fixing in the BV formalism before we can quantise the theory. To this end, we return to the gauge fixing fermion $\Psi\in\CCC^\infty(\frF_{\rm BV})$, a field of ghost number~$-1$ which we have already encountered in Section~\ref{ssec:BRST_quant}. The idea is then to eliminate all antifields by imposing the equation~\cite{Batalin:1981jr}
\begin{equation}\label{eq:elimination_antifields}
 \Phi^+\ =\ \frac{\delta}{\delta \Phi} \Psi~,
\end{equation}
which defines a Lagrangian submanifold $\frL_{\Psi}$ in $\frF_{\rm BV}$. The gauge fixed action is then $S_{\rm BV}|_{\frL_{\Psi}}$.

In a functional integral, gauge fixing would be implemented by a delta functional, and we would define expectation values of an observable $F\in\CCC^\infty(\frF_{\rm BV})$ as
\begin{equation}
 \langle F\rangle_\Psi\ :=\ \int_{\frF_{\rm BV}} \mu_{\rm BV}(\Phi,\Phi^+)~\delta\left(\Phi^+-\frac{\delta}{\delta \Phi} \Psi\right)F[\Phi,\Phi^+]~\de^{\frac{\di}{\hbar} S^\hbar_{\rm BV}[\Phi,\Phi^+]}~.
\end{equation}
Here, $S^\hbar_{\rm BV}$ is the quantum generalisation of $S_{\rm BV}$ with the boundary condition
\begin{equation}\label{eq:bounday_S_hbar}
 S^\hbar_{\rm BV}|_{\hbar=0}\ =\ S_{\rm BV}~.
\end{equation}
Moreover, $\mu_{\rm BV}(\Phi,\Phi^+)$ is the functional measure on $\frF_{\rm BV}$ compatible with the symplectic structure $\omega_{\rm BV}$, i.e.~coordinate transformations between Darboux charts are measure preserving.

\paragraph{Quantum master equation.}
Clearly, for physically meaningful statements, we would want $\langle F\rangle_\Psi$ to be independent of the gauge fixing fermion $\Psi$. One may check that the expectation value $\langle F\rangle_\Psi$ is invariant under deformations of $\Psi$ if and only if
\begin{subequations}
\begin{equation}\label{eq:BVQA1}
\Delta_{\rm BV}\left(F[\Phi,\Phi^+]~\de^{\frac{\di}{\hbar} S^\hbar_{\rm BV}[\Phi,\Phi^+]}\right)\ =\ 0~,
\end{equation}
where
\begin{equation}\label{eq:BVLaplacianGeneral}
 \Delta_{\rm BV}\ :=\ (-1)^{|\Phi^A|+1}\overset{\leftarrow}{\frac{\delta}{\delta \Phi^A}}\overset{\leftarrow}{\frac{\delta}{\delta \Phi^+_A}}~.
\end{equation}
\end{subequations}
is the {\it Batalin--Vilkovisky Laplacian}; see e.g.~\cite{Gomis:1994he} for details. The BV Laplacian satisfies 
\begin{equation}
\begin{gathered}
 \Delta^2_{\rm BV}\ =\ 0~,~~~\Delta_{\rm BV}(FG)\ =\ F\Delta_{\rm BV} G+(-1)^{|G|}(\Delta_{\rm BV} F)G+(-1)^{|G|}\big\{F,G\big\}_{\rm BV}~,\\
 \Delta_{\rm BV}\big\{F,G\big\}_{\rm BV}\ =\ \big\{F,\Delta_{\rm BV} G\big\}_{\rm BV}-(-1)^{|G|}\big\{\Delta_{\rm BV} F,G\big\}_{\rm BV}
\end{gathered}
\end{equation}
for $F,G\in\CCC^\infty(\frF_{\rm BV})$.

For $F=1$, the condition~\eqref{eq:BVQA1} reduces to 
\begin{equation}
 \Delta_{\rm BV} \de^{\frac{\di}{\hbar} S^\hbar_{\rm BV}[\Phi,\Phi^+]}\ =\ 0\quad\Longleftrightarrow\quad\{S^\hbar_{\rm BV},S^\hbar_{\rm BV}\}_{\rm BV}-2\di\hbar\Delta_{\rm BV} S^\hbar_{\rm BV}\ =\ 0~,
\end{equation}
which is known as the {\it quantum master equation}. Using this equation and the boundary condition~\eqref{eq:bounday_S_hbar}, one can compute constraints on the coefficients of the power series expansion of $S^\hbar_{\rm BV}$ in $\hbar$. A solution to the quantum master equation can then be found iteratively.

In many cases, and in particular in Chern--Simons theory, it turns out that a solution $S_{\rm BV}$ of the classical master equation satisfies $\Delta_{\rm BV}S_{\rm BV}=0$ (at least formally, before regularisation) and therefore also solves the quantum master equation.

\paragraph{Quantum \mathversion{bold}$L_\infty$-algebra structure.} 
It is now an interesting question to ask what survives of the classical $L_\infty$-algebra structure noticed in Section~\ref{ssec:ClassicalMasterEquation} and discussed further in Section~\ref{sec:L_infty-structures}. We may define the nilquadratic differential operator
\begin{equation}
 \frd_{\rm BV}\ =\ -\di \hbar \Delta_{\rm BV}+\{S_{\rm BV},-\}_{\rm BV}\ewith \frd^2_{\rm BV}\ =\ 0~.
\end{equation}
Just as the homological vector field $Q_{\rm BV}=\{S_{\rm BV},-\}_{\rm BV}$ induces a classical $L_\infty$-algebra structure on $\frF_{\rm BV}$, so $\frd_{\rm BV}$ induces a {\em quantum $L_\infty$-algebra} or {\em loop homotopy Lie algebra} on the same graded vector space. For more details, see~\cite{Zwiebach:1992ie} and~\cite{Markl:9711045,Mnev:2006ch,LOSEV:2007:3-30,Mnev:2008sa,Granaker:0803.1763,Cattaneo:0811.2045,Doubek:2017naz}. In particular, the decomposition theorem can be proved and the minimal model constructed explicitly, e.g.~by a direct application of the homological perturbation lemma. This approach also leads directly to a homotopy between a quantum $L_\infty$-algebra and its minimal model. 
Finally, let us mention some further highly relevant papers addressing closely related issues: the fibre BV integral and its relation to the homotopy transfer of quantum $L_\infty$-structures is subject of~\cite{Mnev:2006ch,LOSEV:2007:3-30,Mnev:2008sa,Cattaneo:0811.2045}, see also~\cite{Cattaneo:2015vsa} and references therein. BV canonical transformations as isomorphisms of homotopy algebraic structures are discussed in~\cite{Mnev:2006ch,LOSEV:2007:3-30,Mnev:2008sa,Cattaneo:0811.2045}.
The fact that the propagator is the chain homotopy is explained in~\cite{Mnev:2006ch,Mnev:2008sa,Cattaneo:0811.2045}, see also~\cite{Cattaneo:2015vsa} and  references therein.

\paragraph{Gauge fixing and trivial pairs.}
Since the gauge fixing fermion $\Psi$ is a function of degree~$-1$ and because we wish to use it to eliminate the antifields via~\eqref{eq:elimination_antifields}, we have to introduce additional fields of negative degree to construct such a $\Psi$, and these fields must be trivial in a certain sense. For this to be consistent, these fields have to have an interpretation as coordinate functions on a symplectic $Q$-manifold of degree~$-1$.

We note that one can always add {\it trivial pairs} $\bar c\in V[l]$ and $b\in V[l+1]$ for $V$ some vector space and $l\in\RZ$ and consider the $Q$-manifold $V[l]\oplus V[l+1]$ with $Q\bar c=b$. For example, $V=\sLie(\frG)$ and $l=-1$ are used in BRST quantization, cf.~Section~\ref{ssec:BRST_quant}. The corresponding $L_\infty$-algebra has trivial cohomology and it is therefore quasi-isomorphic to the trivial $L_\infty$-algebra, cf.~Section~\ref{ssec:quasiisomorphism}. Adding trivial pairs therefore does not affect the data of the classical theory.

To use a trivial pair in the BV formalism, we have to minimally extend it to a symplectic $Q$-manifold of degree~$-1$. This is simply done by adding corresponding antifields $\bar c^+$ and $b^+$, which yields
\begin{equation}
 \frF_{\rm tp}\ :=\ T^*[-1]\big( \sLie(\frG)[l]\oplus  \sLie(\frG)[l+1]\big)
\end{equation}
with symplectic structure
\begin{equation}
 \omega_{\rm tp}\ :=\  (-1)^l\langle \delta\bar c, \delta \bar c^+\rangle+(-1)^{l+1}\langle\delta b, \delta b^+\rangle
\end{equation}
of degree~$-1$ and Hamiltonian
\begin{equation}\label{eq:TrivialPairHamiltonian}
 S_{\rm tp}\ :=\ -\langle b, \bar c^+\rangle
\end{equation}
of homogeneous degree~$0$. The resulting homological vector field $Q_{\rm tp}=\{S_{\rm tp},-\}_{\rm tp}$ is still a shift isomorphism,
\begin{equation}
 Q_{\rm tp}\bar c\ =\ b~,~~~Q_{\rm tp}\bar c^+\ =\ 0~,~~~Q_{\rm tp}b\ =\ 0~,~~~Q_{\rm tp}b^+\ =\ (-1)^l\bar c^+~.
\end{equation}
Consequently, the resulting cyclic $L_\infty$-algebra is still trivial by quasi-isomorphism.

It is now rather straightforward to see that all structures add up properly when adding trivial pairs to the outcome of the BV formalism. We define $\omega_{\rm eBV}:=\omega_{\rm BV}+\omega_{\rm tb}$, $\{-,-\}_{\rm BV}:=\{-,-\}_{\rm BV}+\{-,-\}_{\rm tp}$ and $\Delta_{\rm eBV}:=\Delta_{\rm BV}+(-1)^{l+1}\big\langle\overset{\leftarrow}{\frac{\delta}{\delta\bar c}},\overset{\leftarrow}{\frac{\delta}{\delta\bar c^+}}\big\rangle+(-1)^{l}\big\langle\overset{\leftarrow}{\frac{\delta}{\delta b}},\overset{\leftarrow}{\frac{\delta}{\delta b^+}}\big\rangle$. If an action functional $S_{\rm BV}^\hbar$ satisfies the quantum master equation for $\{-,-\}_{\rm BV}$ and then 
$S_{\rm eBV}^\hbar:=S_{\rm BV}^\hbar + S_{\rm tp}$ satisfies the quantum master equation for $\{-,-\}_{\rm eBV}$ and $\Delta_{\rm eBV}$.

\section{Homotopy Maurer--Cartan theory}\label{sec:HMCT}

Next, let us introduce homotopy Maurer--Cartan theory and show how the BV formalism applies to it. Homotopy Maurer--Cartan theory was first developed in the context of string field theory in~\cite{Zwiebach:1992ie}, where also $L_\infty$-algebras were defined for the first time, taking inspiration from the definition of $A_\infty$-algebras~\cite{Stasheff:1963aa,Stasheff:1963ab}. This theory is a vast generalisation of Chern--Simons theory, which contains higher Chern--Simons theories as special cases. As we shall show in Section~\ref{sec:L_infty-structures}, essentially any BV quantisable theory can be rewritten as a homotopy Maurer--Cartan theory.

In the following, let $\sL$ be an $L_\infty$-algebra with higher products $\mu_i$ and define $|\ell|_\sL\in\RZ$ as the {\em $\sL$-degree} of a homogeneous element $\ell\in\sL$.

\subsection{Homotopy Maurer--Cartan equation}\label{ssec:HMCE}

\paragraph{Gauge potentials and curvatures.}
We call an element $a\in\sL_1$ a {\em gauge potential}, and we define its {\em curvature} $f\in\sL_2$ as 
\begin{equation}\label{eq:Curvature}
f\ :=\ \mu_1(a)+\tfrac12 \mu_2(a,a)+\cdots\ =\ \sum_{i\geq 1}\frac{1}{i!}\mu_i(a,\ldots,a)~.
\end{equation}
This terminology originates from particular choices of $\sL$ in which $a$ and $f$ indeed reduce to the gauge potential and the curvature of (higher) gauge theory. 

A gauge potential $a\in\sL_1$ is called a {\em Maurer--Cartan (MC) element} provided it satisfies the {\em homotopy Maurer--Cartan equation}
\begin{equation}\label{eq:MCEquation}
 f\ =\ \mu_1(a)+\tfrac12 \mu_2(a,a)+\cdots\ =\ 0~.
\end{equation}

Due to the higher homotopy Jacobi identities~\eqref{eq:homotopyJacobi}, the curvature satisfies the {\em Bianchi identity}
\begin{equation}\label{eq:BianchiIdentity}
 \sum_{i\geq0}\frac{(-1)^i}{i!}\mu_{i+1}(f,a,\ldots,a)\ =\ 0~.
\end{equation}
The proof of the Bianchi identity is found in Appendix~\ref{app:lemmata}, where we collect computational proofs like this one to avoid them cluttering our discussion. We give, in fact, two proofs: one by direct but tedious computation and another, shorter one using the contracted coordinate function $\xi$ and formula~\eqref{eq:Qxi_L_infty}. The existence of such shortcuts from using $\xi$ and/or the coalgebra picture is a recurring theme in proofs involving $L_\infty$-algebras.

\paragraph{Examples.}
Let $M$ be a manifold and $\sL$ a Lie $n$-algebra. For the $L_\infty$-algebra $(\Omega^\bullet(M,\sL),\hat\mu_i)$ as defined in~\eqref{eq:HCSLInfinity}, one obtains the potentials and curvatures of higher gauge theory on topologically trivial\footnote{For topologically non-trivial higher principal bundles, one would have to consider local gauge potentials on patches of the manifold and provide gluing prescriptions on overlaps of the patches.} higher principal bundles, cf.~also~\cite{Jurco:2014mva}. We shall use this case to illustrate our constructions throughout the remainder of this section.

As a concrete example, let $\sL$ be an ordinary Lie algebra $\frg$. Here, a gauge potential is a $\frg$-valued one-form $a=A\in \Omega^1(M,\frg)$ and its curvature is simply 
\begin{equation}\label{eq:CS3EoM}
 f\ =\ F\ewith F\ :=\ \dd A+\tfrac12\hat \mu_2(A,A)\ =\ \dd A+\tfrac12[A,A]~.
\end{equation}
The Bianchi identity reads as $\nabla F=0$ and the homotopy MC equation reduces to the ordinary MC equation $\dd A+\tfrac12 [A,A]=0$. 

As a second concrete example, let $\sL$ be a Lie $2$-algebra $\sL=\sL_{-1}\oplus \sL_0$. The gauge potential $a$ decomposes as 
\begin{subequations}\label{eq:CS4EoM}
\begin{equation}
 a\ =\ A+B\ewith A\in \Omega^1(M,\sL_0)\eand B\in \Omega^2(M,\sL_{-1})~,
\end{equation}
and the generalisation to Lie $n$-algebras is obvious. The curvatures read as 
\begin{equation}\label{eq:Lie-2-algebra_curvatures}
\begin{aligned}
 f\ &=\ \hat\mu_1(a)+\tfrac12\hat\mu_2(a,a)+\tfrac{1}{3!}\hat\mu_3(a,a,a)\\
 &=\ \underbrace{\dd A + \tfrac12\mu_2(A,A)+\mu_1(B)}_{=:\,\CF\,\in\,\Omega^2(M,\sL_0)}+\underbrace{\dd B +\mu_2(A,B)-\tfrac{1}{3!}\mu_3(A,A,A)}_{=:\,H\,\in\,\Omega^3(M,\sL_{-1})}~,
\end{aligned}
\end{equation}
\end{subequations}
where the higher products $\mu_i$ only see the gauge algebra, not the degree of the differential forms in the arguments, yielding sign factors e.g.~$\hat \mu_3(a,a,a)=\hat \mu_3(A,A,A)=-\mu_3(A,A,A)$. All components of $f$ except for the form of highest degree are usually called {\em fake curvatures}. In the case of a Lie 2-algebra, there is only one fake curvature, $\CF$. The MC equation is simply total flatness, $\CF=0$ and $H=0$.

\paragraph{Gauge transformations.}
Elements of $\sL_{-k}$ for $k\geq 0$ are the gauge parameters of infinitesimal gauge transformations (also called level~$0$ gauge transformations for $k=0$) and infinitesimal higher gauge transformations (also called level~$k$ gauge transformations for $k\geq1$). In particular, level~$0$ gauge transformations between two gauge potentials, the standard gauge transformations, are encoded in partially flat homotopies between them. These homotopies are captured by gauge potentials for the tensor product $L_\infty$-algebra $\Omega^\bullet(I,\sL):=\Omega^\bullet(I)\otimes\sL$ with $I:=[0,1]\subseteq\FR$, where the tensor product is as defined in Section~\ref{ssec:L_infty_algebras_and_algebroids}. We can decompose $\Omega^\bullet_1(I,\sL)\cong\CCC^\infty(I,\sL_1)\oplus \Omega^1(I,\sL_0)$ and $\sfa\in\Omega^\bullet_1(I,\sL)$ takes the form
\begin{equation}
 \sfa(t)\ =\ a(t)+\dd t\otimes  c(t)
\end{equation}
with $t\in I$, $a(t)\in \CCC^\infty(I,\sL_1)$ and $ c(t)\in \CCC^\infty(I,\sL_0)$. Since $\Omega^\bullet_2(I,\sL)\cong \CCC^\infty(I,\sL_2)\oplus \Omega^1(I,\sL_1)$, the curvature $\sff\in\Omega^\bullet_2(I,\sL)$ reads as
\begin{equation}
\begin{aligned}
 \sff(t)\ &=\ \sum_{i\geq 1}\frac{1}{i!} \hat\mu_{i}(\sfa(t),\dots,\sfa(t), \sfa(t))\\
 &=\ f(t)+\dd t\otimes \left\{\der{t} a(t)-\sum_{i\geq 0}\frac{1}{i!} \mu_{i+1}(a(t),\dots,a(t), c(t))\right\},
\end{aligned}
\end{equation}
where the higher products $\mu_i$ are agnostic about the form degree of their arguments. Partial flatness $\der{t}\intprod f=0$ or, equivalently, $\sff\in \CCC^\infty(I,\sL_2)$, leads to the differential equation
\begin{equation}
\der{t} a(t)-\sum_{i\geq 0}\frac{1}{i!} \mu_{i+1}(a(t),\dots,a(t), c(t))\ =\ 0~,
\end{equation}
which describes the changes of $a(t)$. At $t=0$, we can read off the gauge transformation $a\mapsto a+\delta_{ c_0} a$ of the gauge potential $a:=a(0)$ parametrised by the gauge parameter $c_0:= c(0)\in\sL_0$, 
\begin{equation}\label{eq:GaugeTrafo}
 \delta_{ c_0} a\ :=\ \left.\der{t}\right|_{t=0}  a(t)\ =\ \sum_{i\geq0} \frac{1}{i!}\mu_{i+1}(a,\ldots,a, c_0)~.
\end{equation}

As demonstrated in Appendix~\ref{app:lemmata}, the commutator of two successive gauge transformations with gauge parameters $ c_0, c'_0\in\sL_0$ is given by
\begin{subequations}\label{eq:CommutatorGT}
\begin{equation}
  [\delta_{ c_0},\delta_{ c'_0}]a\ =\ \delta_{ c''_0}a+\sum_{i\geq0}\frac{1}{i!}(-1)^{i}\mu_{i+3}(f,a,\ldots,a, c_0, c'_0)
\end{equation}
with
\begin{equation}
   c''_0\ :=\ \sum_{i\geq0}\frac{1}{i!}\mu_{i+2}(a,\ldots,a, c_0, c'_0)~,
\end{equation}
\end{subequations}
which shows that for general $L_\infty$-algebras, gauge transformations only close up to terms proportional to the curvature $f$. 

Furthermore, the curvature transforms under gauge transformations as
\begin{equation}\label{eq:GaugeTrafoCurvature}
 \delta_{ c_0} f\ :=\ \left.\der{t}\right|_{t=0}  f(t)\ =\ \sum_{i\geq 0}\frac{(-1)^i}{i!}\mu_{i+2}(f,a,\ldots,a, c_0)~,
\end{equation}
the proof of this equation is found in Appendix~\ref{app:lemmata}.

In the special case $\Omega^\bullet(M,\frg)$ with $\frg$ a Lie algebra, the formulas for the gauge transformations~\eqref{eq:GaugeTrafo} and~\eqref{eq:GaugeTrafoCurvature} reproduce the familiar transformations,
\begin{equation}
\begin{gathered}
 A\ \mapsto\  A+\delta_{ c_0} A\ =\ A+\dd  c_0+[A, c_0]~,\\
 F\ \mapsto\  F+\delta_{ c_0} F\ =\ F+[F, c_0]~.
\end{gathered} 
\end{equation}
Since $\mu_3=0$, the gauge algebra closes also for $F\neq 0$. The same is true in the case of higher gauge theories corresponding to $\Omega^\bullet (M,\sL)$ for $\sL$ an $L_\infty$-algebra with $\mu_i=0$ for $i\geq 3$, but here the higher gauge transformations, which we shall discuss next, generically fail to close. 

\paragraph{Higher gauge transformations.}
Higher gauge transformations are described by homotopies between homotopies. In particular, consider a level~$1$ gauge transformation between two level~$0$ gauge transformations. These are captured by gauge potentials on the tensor product $\Omega^\bullet(I^2,\sL)$ with $I^2:=I\times I$ and $I:=[0,1]\subseteq\FR$. This time, we have the decomposition $\Omega^\bullet_1(I^2,\sL)\cong\CCC^\infty(I^2,\sL_1)\oplus \Omega^1(I^2,\sL_0)\oplus \Omega^2(I^2,\sL_{-1})$ and hence, $\sfa\in \Omega^\bullet_1(I^2,\sL)$ takes the form
\begin{subequations}
\begin{equation}
 \sfa(t,s)\ =\ a(t,s)+\dd t\otimes  c^{(1)}(t,s)+\dd s\otimes c^{(2)}(t,s)+(\dd t\wedge \dd s)\otimes\sigma(t,s)~,
\end{equation}
where $(t,s)\in I^2$, $a(t,s)\in \CCC^\infty(I^2,\sL_1)$, $ c^{(1,2)}(t,s)\in \CCC^\infty(I^2,\sL_0)$, and $\sigma(t,s)\in\CCC^\infty(I^2,\sL_{-1})$. The fact that $\sfa$ is a homotopy between homotopies is reflected in the boundary conditions
\begin{equation}
 a(0,s)\ =\ a\eand   c^{(1)}(t,0)\ =\  c(t)~.
\end{equation}
The geometric shape underlying this homotopy between homotopies is not a square but a bigon with a coordinate degeneracy in $s$ at $t=0$ and $t=1$. Therefore, we have to supplement the above boundary conditions by
\begin{equation}
  c^{(2)}(0,s)\ =\  c^{(2)}(1,s)\ =\ 0~.
\end{equation}
\end{subequations}
Moreover, $\Omega^\bullet_2(I^2,\sL)$ decomposes as $\Omega^\bullet_2(I^2,\sL)\cong \CCC^\infty(I^2,\sL_2)\oplus \Omega^1(I^2,\sL_1)\oplus \Omega^2(I^2,\sL_0)$, and upon imposing the partial flatness condition $\sff\in \CCC^\infty(I^2,\sL_2)$, we obtain the level~$0$ gauge transformation~\eqref{eq:GaugeTrafo} with the gauge parameter $ c_0:= c^{(1)}(0,0)\in\sL_0$ together with the level~$1$ gauge transformation
\begin{equation}\label{eq:HigherGaugeTrafoLevel1}
  \delta_{ c_{-1}}  c_0\ :=\ \left. \der{s}\right|_{t=s=0}  c^{(1)}(t,s)\ =\ \sum_{i\geq0} \frac{1}{i!}\mu_{i+1}(a,\ldots,a, c_{-1})~,
\end{equation}
where $ c_{-1}:=\sigma(0,0)\in\sL_{-1}$. 

The derivation of level~$k$ gauge transformations from certain components of a partially flat curvature on $I^{k+1}$ makes it clear that this can be iterated further by considering $\Omega^\bullet(I^{k+1},\sL)$ for $k\geq0$. Ultimately, we obtain the level~$k+1$ gauge transformation,
\begin{equation}\label{eq:HigherGaugeTransformation}
  \delta_{ c_{-k-1}}  c_{-k}\ =\ \sum_{i\geq0} \frac{1}{i!}\mu_{i+1}(a,\ldots,a, c_{-k-1})~,
\end{equation}
for $ c_{-k}\in\sL_{-k}$. 

It is important to stress that as for the commutator of two successive gauge transformations~\eqref{eq:CommutatorGT}, also gauge-of-gauge transformations close only up to terms proportional to $f$. Indeed, we have
\begin{equation}\label{eq:ClosureGaugeTrafo}
\begin{aligned}
  \delta_{c_{-1}} (\delta_{ c_{0}}a) \ &:=\ \sum_{i\geq0} \frac{1}{i!} \mu_{i+1}(a,\ldots,a, \delta_{ c_{-1}} c_{0})\\
  & \phantom{:}=\ \sum_{i\geq0} \frac{1}{i!}\mu_{i+2}(f, a,\ldots,a, c_{-1})~,\\
  \delta_{ c_{-k-2}} (\delta_{ c_{-k-1}} c_{-k})\ &:=\ \sum_{i\geq0} \frac{1}{i!}\mu_{i+1}(a,\ldots,a, \delta_{ c_{-k-2}} c_{-k-1})\\
  & \phantom{:}=\ \sum_{i\geq0} \frac{1}{i!}\mu_{i+2}(f, a,\ldots,a, c_{-k-2})
  \end{aligned}
\end{equation}
for all $k\geq0$ as is demonstrated in Appendix~\ref{app:lemmata}. Hence, for MC elements~\eqref{eq:MCEquation}, this vanishes identically and gauge transformations of level~$k$ gauge parameters leave the outcome of level~$k$ gauge transformations unchanged (as expected).\footnote{Strictly speaking, it is only in this case that the geometric shape underlying the homotopy between homotopies becomes a bigon.}  

\paragraph{Vanishing curvature and kinematical data.} The observation that general gauge transformations only close for $f=0$ can now be interpreted in two possible ways. Firstly, we can regard $f=0$ as a {\em dynamical} equation and see the gauge structure as an {\it open algebra}, which only closes on-shell. As discussed before, this requires then using the BV formalism in the quantisation of the theory. Secondly, we can regard $f=0$ as a constraint on the {\em kinematical} data of the higher gauge theory and therefore as an equation that is also imposed off-shell. 

We should note that the approach to the kinematical data of higher gauge theory presented above fits the interpretation of higher connections as a way of capturing a higher-dimensional parallel transport. From this perspective, the second interpretation is favoured and the fake curvature should indeed be part of the kinematical data. Beyond equations~\eqref{eq:CommutatorGT} and~\eqref{eq:ClosureGaugeTrafo}, there are a number of crucial points observed in the literature. Firstly, a consistent, reparametrisation-invariant parallel transport for a Lie 2-algebra requires the fake curvature $\CF$ defined in~\eqref{eq:Lie-2-algebra_curvatures} to vanish~\cite{Baez:0511710}. Secondly, we observed in our previous work~\cite{Jurco:2014mva} that for semistrict Lie 2-algebras with non-trivial $\mu_3$, infinitesimal gauge transformations can only be concatenated if the fake curvature vanishes. This is simply a special case of equation~\eqref{eq:CommutatorGT}. Thirdly, equation~\eqref{eq:GaugeTrafoCurvature} shows that the curvature appears itself in gauge transformations of the curvature, which makes it essentially impossible to write down covariant equations of motions beyond $f=0$ in the general setting.

Let us stress, however, that there is an alternative approach to defining higher potentials, curvatures and their gauge transformations that has been worked out for the special case of (twisted) string structures in~\cite{Sati:2008eg,Sati:2009ic}. In this approach, things become significantly simpler, and the resulting structures have been applied in the context of self-dual strings and six-dimensional superconformal field theories e.g.~in~\cite{Saemann:2017zpd,Saemann:2017rjm}. The precise relation between both approaches has not been fully worked out yet.

\paragraph{Covariant derivative.}
Consider $\varphi\in\sL$ and require that $\varphi$ transforms under gauge transformations {\it adjointly}, that is,
\begin{equation}\label{eq:GaugeTrafoMatter}
\delta_{ c_0}\varphi\ :=\ -\sum_{i\geq0}\frac{1}{i!}\mu_{i+2}(a,\ldots,a, c_0,\varphi)
\end{equation}
for $ c_0\in\sL_0$. We define the {\it covariant derivative}  $\nabla\varphi$ of $\varphi$ by
\begin{equation}\label{eq:CovariantDerivative}
\nabla\varphi\ :=\ \sum_{i\geq0}\frac{1}{i!}\mu_{i+1}(a,\ldots,a,\varphi)~.
\end{equation}
Note that $\nabla:\sL_k\to\sL_{k+1}$. Under the gauge transformations~\eqref{eq:GaugeTrafo} and~\eqref{eq:GaugeTrafoMatter}, $\nabla\varphi$ behaves as
\begin{equation}\label{eq:GTNabla}
 \delta_{ c_0}(\nabla\varphi)\ =\  -\sum_{i\geq0}\frac{1}{i!}\mu_{i+2}(a,\ldots,a, c_0,\nabla\varphi)+\sum_{i\geq0}\frac{(-1)^i}{i!}\mu_{i+3}(f,a,\ldots,a, c_0,\varphi)
\end{equation}
as is demonstrated in Appendix~\ref{app:lemmata}. Hence, $\nabla\varphi$ transforms adjointly up to terms proportional to the curvature $f$. 
Furthermore, as is shown in Appendix~\ref{app:lemmata} as well, we have the standard result
\begin{equation}\label{eq:NableSquared}
\nabla^2\varphi\ =\ \sum_{i\geq0}\frac{(-1)^i}{i!}\mu_{i+2}(f,a,\ldots,a,\varphi)~.
\end{equation}

For the special case $\Omega^\bullet(M,\frg)$ with $\frg$ a Lie algebra, the covariant derivative is the usual one. In the case of a Lie 2-algebra $\sL=\sL_{-1}\oplus \sL_0$, we have
\begin{equation}
 \nabla \varphi\ =\ \dd \varphi+\mu_1(\varphi)+\mu_2(A,\varphi)+\mu_2(B,\varphi)+\tfrac12 \mu_3(A,A,\varphi)
\end{equation}
for $\varphi\in \Omega^\bullet(M,\sL)$.

\paragraph{Maurer--Cartan elements and \mathversion{bold}$L_\infty$-morphisms} Let us now come to the interplay between Maurer--Cartan elements and gauge transformations with an $L_\infty$-morphism $\phi:\sL\rightarrow \sL'$ as introduced in~\eqref{eq:L_infty_morphism}. The details of the computations we present in this paragraph are found in Appendix~\ref{app:lemmata}.\footnote{See also~\cite{Kajiura:2001ng} for a similar discussion in the case of $A_\infty$-algebras.}

Naively, one may be led to assume that $\phi$ should act on a gauge potential $a$ as $a'=\phi_1(a)$. This, however, does not give the desired compatibility with the $L_\infty$-algebra structures. Instead, one should either regard the shifted exponential\footnote{This expression is used in proofs in Appendix~\ref{app:lemmata}.} $\de^{sa}_0:=sa+\tfrac12 sa\odot sa+\tfrac{1}{3!}sa\odot sa \odot sa+\cdots$ in the coalgebra picture as the natural invariant object, or, equivalently in the $L_\infty$-picture, evaluate~\eqref{eq:L_infty_morphism} at $\ell_1=\cdots=\ell_i=a\in\sL_1$. Both approaches eventually lead to 
\begin{subequations}
\begin{equation}\label{eq:morphism_on_a}
 a\ \mapsto\ a'\ :=\ \sum_{i\geq 1} \frac{1}{i!}\phi_i(a,\dots,a)
\end{equation}
and
\begin{equation}\label{eq:morphism_on_f}
  f\ \mapsto\ f'\ =\ \sum_{i\geq 0}\frac{(-1)^i}{i!}\phi_{i+1}(f,a,\ldots,a)
\end{equation}
for the curvatures
\begin{equation}
  f\ =\ \sum_{i\geq1}\frac{1}{i!}\mu_i(a,\ldots,a)\eand  f'\ =\ \sum_{i\geq1}\frac{1}{i!}\mu'_i(a',\ldots,a')~.
\end{equation}
\end{subequations}
Hence, we may conclude MC elements are mapped to MC elements. 

At the level of gauge transformations $a\mapsto a+\delta_{ c_0}a$ and $a'\mapsto a'+\delta_{ c'_0}a'$ for two MC elements $a$ and $a'$ related by~\eqref{eq:morphism_on_a}, we obtain
\begin{equation}\label{eq:morphism_on_omega}
  c_0\ \mapsto\  c_0'\ :=\ \sum_{i\geq 0} \frac{1}{i!}\phi_{i+1}(a,\dots,a, c_0)~,
\end{equation}
so that the gauge orbits of MC elements $a$ are mapped to the gauge orbits of MC elements of $a'$ under $L_\infty$-morphisms. 

It also follows that in the case of quasi-isomorphic $L_\infty$-algebras $\sL$ and $\sL'$, we have an isomorphism 
\begin{equation}
 \CM_\sL\ \cong\ \CM_{\sL'}
\end{equation}
between the moduli spaces of MC elements (i.e.~the spaces of MC elements modulo gauge transformations) in $\sL$ and $\sL'$.

\subsection{Homotopy Maurer--Cartan action}\label{ssec:HMCAction}

Let $\sL$ now be a cyclic $L_\infty$-algebra with an inner product $\langle -,-\rangle_\sL$ of degree~$-3$. Then the MC equation~\eqref{eq:MCEquation} is variational.

\paragraph{Action.} 
The MC equation~\eqref{eq:MCEquation} describes the stationary locus of the action functional
\begin{equation}\label{eq:MCAction}
 S_{\rm MC}[a]\ :=\ \sum_{i\geq1} \frac{1}{(i+1)!}\langle a,\mu_i(a,\ldots,a)\rangle_\sL~.
\end{equation}
Using the cyclicity~\eqref{eq:cyclicity}, it is a straightforward exercise to show that the extrema of $ S_{\rm MC}$ are given by solutions to the MC equation~\eqref{eq:MCEquation}. We shall refer to the action~\eqref{eq:MCAction} as the {\em homotopy Maurer--Cartan action}.

The homotopy MC action~\eqref{eq:MCAction} is invariant under the gauge transformations~\eqref{eq:GaugeTrafo}. Indeed, we have
\begin{equation}
\delta_{ c_0} S_{\rm MC}[a]\ =\ \langle f,\delta_{ c_0} a\rangle_\sL\ =\ -\sum_{i\geq0} \frac{(-1)^i}{i!}\langle  c_0,\mu_{i+1}(f,a,\ldots,a)\rangle_\sL\ =\ 0~,
\end{equation}
where we have used graded symmetry and the cyclicity~\eqref{eq:cyclicity} of the inner product as well as the Bianchi identity~\eqref{eq:BianchiIdentity}. 

Note that the homotopy MC action is a vast generalisation of the Chern--Simons action functional. In particular, if $\sL=\Omega^\bullet(M,\frg)$ with $\frg$ a metric Lie algebra, $M$ a compact oriented three-dimensional manifold, and $\sL$ endowed with the obvious cyclic structure given in Section~\ref{ssec:L_infty_algebras_and_algebroids}, then $a=A\in\Omega^1(M,\frg)$ and 
\begin{equation}\label{eq:CS_action_from_MC}
 S_{\rm MC}[a]\ =\ \int_M \Big\{\tfrac12 \langle A,\dd A\rangle_\frg +\tfrac{1}{3!}\langle A,[A,A]\rangle_\frg\Big\}~.
\end{equation}
We shall return to higher Chern--Simons theory, for which $\frg$ is replaced by a general Lie $n$-algebra $\sL$, in Section~\ref{sec:L_infty-structures}.

\paragraph{Further bosonic symmetries.}
Whilst gauge transformations themselves do not close off-shell in general, their action on the action functional~\eqref{eq:MCAction} does. This is due to an additional invariance of $S_{\rm MC}[a]$ under transformations of the form
\begin{equation}\label{eq:ExtraSymmetries}
  \delta_{\ell_1,\ldots,\ell_i} a\ :=\ \sum_{j\geq0} \frac{\gamma_j}{j!}\mu_{i+j+1}(f,a,\ldots,a,\ell_1,\ldots,\ell_i)
  \end{equation}
for $\ell_1,\ldots,\ell_i\in\sL$ with $\sum_{j=1}^i|\ell_j|_{\sL}\ =\ i-2$ and $\gamma_j\in\FR$. The invariance follows directly from cyclicity of the inner product~\eqref{eq:cyclicity} and the fact that $\mu_{i+2}(f,f,\ldots )\ =\ 0$ for $i\ \geq\ 0$. Since these symmetries vanish on-shell and therefore do not affect classical observables, they are referred to as {\em trivial symmetries}. They are of no physical significance. In particular, they neither lead to conserved quantities nor do they pose any obstacle for perturbation theory. 

\paragraph{Supersymmetric extension.} A useful property of Chern--Simons theory in three dimensions is that it is trivially $\CN\ =\ 2$ supersymmetric, cf.~\cite{Ivanov:1991fn}. That is, it can be extended to an $\CN\ =\ 2$ supersymmetric action such that all superpartners of the gauge potential are auxiliary fields. The latter come with algebraic equations that can be integrated out and one is left with the usual Chern--Simons action. The supersymmetries can be linearly combined into the odd symmetry required for computing path integrals of Chern--Simons theory using localisation, see e.g.~\cite{Kapustin:2009kz}.

We now show that a similar supersymmetric extension exists in general homotopy MC theory. To avoid introducing the spinors involved in a true supersymmetric extension, we consider here an example of, roughly, an analogue of a topological twist with only one real supercharge (which would be sufficient for localisation). 

We introduce superpartners $(\varphi_k,\vartheta_k)\in\sL_k\oplus\Pi\sL_k$  for $k=0,\ldots,3$, where $\Pi$ is the Gra{\ss}mann-parity changing functor, which  transform under gauge transformations~\eqref{eq:GaugeTrafoMatter}. These fields are thus a generalisation of matter fields transforming in the adjoint representation of some gauge Lie algebra. A gauge invariant action functional is then
\begin{equation}\label{eq:fermionicMCAction}
 S_{\rm SMC}[a,\varphi,\vartheta]\ :=\ S_{\rm MC}[a]+\langle\vartheta_0,\vartheta_3\rangle_\sL+\langle\vartheta_1,\vartheta_2\rangle_\sL+\langle\varphi_0,\varphi_3\rangle_\sL+\langle\varphi_1,\varphi_2\rangle_\sL
\end{equation}
with $S_{\rm MC}$ the homotopy MC action~\eqref{eq:MCAction}. Note that gauge invariance follows directly from the cyclicity~\eqref{eq:cyclicity} and the graded symmetry of the inner product $\langle-,-\rangle_\sL$. Since the extra fields $\varphi_k$ and $\vartheta_k$ appear only algebraically and do not mix with the gauge potential, the two actions $S_{\rm SMC}$ and $S_{\rm MC}$ are clearly equivalent. We stress that the new fields are auxiliary, but physical. They are {\em not} to be regarded as ghosts or antifields for any value of $k$.

The action $S_{\rm SMC}$ is also invariant under the following fermionic transformations:
\begin{subequations}\label{eq:QFermionicGeneral}
\begin{equation}
\begin{aligned}
Qa\ &:=\ \vartheta_1~,\\
Q\varphi_1\ &:=\ \vartheta_1~,\\
Q\varphi_2\ &:=\ -\tfrac12\nabla\vartheta_1+\tfrac12\sum_{i\geq0}\frac{1-2\gamma_{i0}}{i!}\mu_{i+2}(\underbrace{a,\ldots,a}_{i~{\rm copies}},\vartheta_1,\varphi_1)\,+\\
&\kern1cm+\sum_{i,j\geq0}\frac{\gamma_{i+1j}+\gamma_{ij+1}}{i!(j+2)!}\mu_{i+j+3}(\underbrace{a,\ldots,a}_{i~{\rm copies}},\vartheta_1,\underbrace{\varphi_1,\ldots,\varphi_1}_{j+2~{\rm copies}})~,\\
Q\vartheta_2\ &:=\ f+\varphi_2-\tfrac12\nabla\varphi_1+\tfrac12\sum_{i\geq0}\frac{\gamma_{i0}}{i!}\mu_{i+2}(\underbrace{a,\ldots,a}_{i~{\rm copies}},\varphi_1,\varphi_1)\,+\\
&\kern1cm+\sum_{i,j\geq0}\frac{\gamma_{ij+1}}{i!(j+3)!}\mu_{i+j+3}(\underbrace{a,\ldots,a}_{i~{\rm copies}},\underbrace{\varphi_1,\ldots,\varphi_1}_{j+3~{\rm copies}})
\end{aligned}
\end{equation}
and 
\begin{equation}
 Q\varphi_0\ :=\ 
Q\varphi_3\ :=\ 
Q\vartheta_0\ :=\ 
Q\vartheta_1\ :=\ 
Q\vartheta_3\ :=\ 0~,
\end{equation}
for
\begin{equation}
 \gamma_{i0}\ =\ \frac{1}{3}\eand \gamma_{i+1j}+\gamma_{ij+1}\ =\ -\frac{1}{j+3}~.
\end{equation}
\end{subequations}
Here, $\nabla$ is the covariant derivative defined in~\eqref{eq:CovariantDerivative}, $f$ is the curvature~\eqref{eq:Curvature} and it is easy to see that  $Q^2=0$. We shall see an explicit example in Section~\ref{ssec:HCST}.

\subsection{Batalin--Vilkovisky complex of homotopy Maurer--Cartan theory}\label{ssec:HBVA}

We now explain how the BV formalism is applied to homotopy MC theory, starting with the BRST complex. As before, consider a cyclic $L_\infty$-algebra $\sL$ with inner product $\langle-,-\rangle_\sL$ of $\sL$-degree~$-3$ and ghost degree~0.

\paragraph{Becchi--Rouet--Stora--Tyutin complex.} The action of infinitesimal gauge symmetries on the gauge potential $a$ and the gauge parameters $c_{-k}$ and their symmetry structure are captured by the elements of the $L_\infty$-algebra $\sL$ of degree~$i\leq 1$ as displayed in Table~\ref{tab:BRSTfields}. The BRST complex therefore corresponds to a truncation of $\sL$,
\begin{equation}
  \sL_{\rm trunc}\ =\ \bigoplus_{i\leq 1}\sL_i~,
\end{equation}
which we endow with the higher products $\mu_i$ of $\sL$, but putting all $\mu_i$ with images outside of $\sL_{\rm trunc}$ to zero.

\begin{table}[h]
\begin{center}
\begin{tabu}{|l|c|c|c|c|c|c|}
\hline
 & $a$ & $c_0$ & $c_{-1}$ & $\cdots$ & $c_{-k}$ & $\cdots$ \\ \hline
$L_\infty$-degree & $1$ & $0$ & $-1$ & $\cdots$ & $-k$& $\cdots$ \\
\hline
\end{tabu}
\caption{\label{tab:BRSTfields} Becchi--Rouet--Stora--Tyutin fields}
\end{center}
\vspace{-.6cm}
\end{table}

In the case where $\sL$ is concentrated in degrees $0$ and $1$ (e.g.~in the case corresponding to ordinary Chern--Simons theory), as well as in many other special cases of $L_\infty$-algebras, the truncated higher products still satisfy the homotopy Jacobi identity~\eqref{eq:homotopyJacobi}. The truncated $L_\infty$-algebra $\sL_{\rm trunc}$ forms then an action $L_\infty$-algebroid and the BRST complex is the Cheval\-ley--Eilenberg algebra of this $L_\infty$-algebroid as discussed in~Sections~\ref{ssec:GaugeStructureGeneral} and~\ref{ssec:BRST_quant}. In general, however $\sL_{\rm trunc}$ is not an $L_\infty$-algebra, as we shall explain now.

We still can switch to a graded manifold description underlying $\sL_{\rm trunc}$, performing the usual shift by~1 to obtain
\begin{equation}
 \frF_{\rm BRST}\ :=\ \sL_{\rm trunc}[1]\ =\ \bigoplus_{i\leq 1}\sL_i[1]~,
\end{equation}
where the higher brackets $\mu_i$ yield a degree~1 vector field $Q_{\rm BRST}$. The fact that $\sL_{\rm trunc}$ is no longer an $L_\infty$-algebra amounts then to $Q_{\rm BRST}$ being no longer homological, that is, $Q_{\rm BRST}^2=0$ is no longer guaranteed.

To simplify our notation, we again combine the coordinate functions on $\frF_{\rm BRST}$ with the basis on $\sL_{\rm trunc}$ to form the graded vector space
\begin{equation}
\sL_{\rm trunc,\CCC}\ :=\ \CCC^\infty(\frF_{\rm BRST})\otimes\sL_{\rm trunc}
\end{equation}
with higher products $\hat\mu_i$ as defined in~\eqref{eq:GhostProd}. Objects in $\sL_{\rm trunc,\CCC}$ now have a bi-degree, and we refer to the one from $\sL_{\rm trunc}$ as the $L_\infty$-degree and the one from $\CCC^\infty(\frF_{\rm BRST})$ as the {\em ghost degree}. The latter indeed matches the usual nomenclature. We use here the same letter for both the elements of $\sL_{\rm trunc}$ and their contracted coordinate functions, hoping to avoid the first fundamental confusion of calculus.
\begin{table}[h]
\begin{center}
\begin{tabu}{|l|c|c|c|c|c|c|}
\hline
 & $a$ & $c_0$ & $c_{-1}$ & $\cdots$ & $c_{-k}$ & $\cdots$ \\ \hline
$L_\infty$-degree & $1$ & $0$ & $-1$ & $\dots$ & $-k$& $\cdots$ \\
ghost degree & 0 & 1 & 2 & $\cdots$ & $k+1$ & $\cdots$ \\
field type & b & f & b & $\cdots$ & f/b & $\cdots$ \\
\hline
\end{tabu}
\caption{Becchi--Rouet--Stora--Tyutin fields; a `b' stands for boson and an `f' for fermion.}
\end{center}
\vspace{-.6cm}
\end{table}

We see that the field type is determined by the parity of the ghost degree, i.e.~by the field regarded as a contracted coordinate function, as expected. 

The action of the BRST operator $Q_{\rm BRST}$ on elements of $\frF_{\rm BRST}$ is then described using formula~\eqref{eq:Qxi_L_infty}, combining all contracted coordinate functions into a single object,
\begin{equation}
  \sfa\ :=\ a+\sum_{k\geq0} c_{-k}~.
\end{equation}
We then have
\begin{equation}\label{eq:BRST}
 Q_{\rm BRST} \sfa\ =\ -\sum_{i\geq 1} \frac{1}{i!}\hat \mu_i(\sfa,\dots,\sfa)~.
\end{equation}
Let us briefly look at the individual components of $\sfa$. First of all, we have 
\begin{subequations}\label{eq:BRST2}
\begin{equation}
 Q_{\rm BRST} a\ :=\ -\sum_{i\geq0} \frac{1}{i!}\hat \mu_{i+1}(a,\ldots,a,c_0)\ =\ \sum_{i\geq0} \frac{1}{i!}\mu_{i+1}(a,\ldots,a,c_0)~,
\end{equation}
where the $\mu_i$ only respect the $L_\infty$-degree of the arguments and they are agnostic about the ghost degree. This explains the additional sign in going from $\hat \mu_{i+1}$ to $\mu_{i+1}$: we need to move the coordinate function in $c_0$, which is of degree~1, past the degree~$1$ basis vectors of the $i$ arguments $a$ and the bracket $\hat \mu_{i+1}$. Similarly, we have
\begin{equation}
\begin{aligned}
  &Q_{\rm BRST} c_{-k}\ :=\\
  &~~\ :=\ \sum_{\substack{i,n\geq0\\i+n\geq1}}\frac{1}{i!}\sum_{\pi_n} \frac{1}{n_0!\cdots n_{k+1}!}\hat \mu_{i+n}(\underbrace{a,\dots,a}_{i},\underbrace{c_0,\dots,c_0}_{n_0},\underbrace{c_{-1},\dots,c_{-1}}_{n_1},\dots,\underbrace{c_{-k-1},\dots,c_{-k-1}}_{n_{k+1}})\\
  &~~\ \phantom{!}=\ \sum_{\substack{i,n\geq0\\ i+n\geq1}}\frac{1}{i!}\sum_{\pi_n} \frac{\chi(\pi_n)}{n_0!\cdots n_{k+1}!}\mu_{i+n}(\underbrace{a,\dots,a}_{i},\underbrace{c_0,\dots,c_0}_{n_0},\underbrace{c_{-1},\dots,c_{-1}}_{n_1},\dots,\underbrace{c_{-k-1},\dots,c_{-k-1}}_{n_{k+1}})~,
  \end{aligned}
\end{equation}
with the sum over $\pi_n$ running over all weighted partitions $\pi_n$ of $n$ with $n=n_0+\dots+n_{k+1}$.  The sign
\begin{equation}
 \chi(\pi_n)\ =\ \sum_{j=0}^{k+1}\sum_{m=1}^{n_k+1} (j+1)\left(n_j-m+ \sum_{l=j+1}^{k+1} n_l\right)
\end{equation}
\end{subequations}
arises again by moving all coordinate functions past the basis vectors and the higher product $\hat \mu_{i+n}$.

We note that $Q_{\rm BRST}$ governs the gauge transformations of fields and ghosts,
\begin{equation}
 Q_{\rm BRST} a \ :=\ \delta_{c_0} a\eand
 Q_{\rm BRST} c_{-k} \ :=\ \delta_{c_{-k-1}} c_{-k}+\cdots~,
\end{equation}
and it also incorporates the symmetry structure of the ghosts themselves.

As shown in Appendix~\ref{app:lemmata}, we have
\begin{equation}\label{eq:BRSTSquared}
Q^2_{\rm BRST}a\ =\ \sum_{i\geq0}\frac{(-1)^i}{i!}\Big[-\mu_{i+2}(f, a,\ldots,a,c_{-1})+\frac{1}{2!}\mu_{i+3}(f,a,\ldots,a,c_0,c_0)\Big]
\end{equation}
and similar equations for $c_{-k}$. This reflects the fact that the truncation from $\sL$ to $\sL_{\rm trunc}$ breaks the homotopy Jacobi relation on the truncated higher products. We see that for ordinary Lie algebras and, consequently, ordinary gauge theory, we have neither $\mu_i$ for $i>2$ nor the higher ghosts $c_{-k}$ with $k>0$ and so, $Q^2_{\rm BRST}=0$. Therefore, the BRST formalism is sufficient for ordinary gauge theory. In the general case, however, we would have to impose $f=0$ to close the gauge algebra, which is usually described by saying that the gauge algebra only closes on-shell.

To obtain an off-shell formulation, e.g.~for a quantisation of the field theory, we need to extend the BRST formalism. In the case of homotopy MC theory it is very obvious what this extension should be. Instead of truncating the original $L_\infty$-algebra $\sL$ to $\sL_{\rm trunc}$ yielding the BRST complex, we should have simply kept all of $\sL$ and put $\frF_{\rm BV}=\sL[1]$. This is indeed what the BV formalism does.

\paragraph{Batalin--Vilkovisky fields.}
As discussed in Section~\ref{ssec:ClassicalMasterEquation}, we need an antifield for every field and ghost, so that
\begin{equation}\label{eq:DefFBV}
\frF_{\rm BV}\ :=\ T^*[-1]\frF_{\rm BRST}~.
\end{equation}
Note that in the case of homotopy MC theory, an inner product $\langle \tau_\alpha,\tau_\beta\rangle_\sL=\omega_{\alpha\beta}$ of degree~$-3$ with respect to some basis $\tau_\alpha$ of $\sL$ induces a symplectic form $\omega=\tfrac12 \omega_{\alpha\beta}\dd \xi^\alpha \wedge\dd \xi^\beta$ on $\sL[1]$ of degree~$-|\tau_\alpha|+1-|\tau_\beta|+1=-3+2=-1$. Non-degeneracy of $\langle -,-\rangle_\sL$ therefore implies that $T^*[-1]\frF_{\rm BRST}\cong \sL[1]$ as claimed above.

For clarity, let us summarise the $L_\infty$-degrees and ghost degrees again in Table~\ref{tab:BV_fields}.
\begin{table}[h]
\begin{center}
\begin{tabu}{|l|c|c|c|c|c|c|c|c|c|c|c|c|}
\hline
 & $\cdots$ & $c^+_{-k}$ & $\cdots$ & $c^+_{-1}$ & $c_0^+$ &  $a^+$ & $a$ & $c_0$ & $c_{-1}$ & $\cdots$ & $c_{-k}$ & $\cdots$ \\ \hline
$L_\infty$-degree & $\cdots$ & $3+k$ & $\cdots$ & $4$ & $3$ & $2$ & $1$ & $0$ & $-1$ & $\cdots$ & $-k$& $\dots$ \\
ghost degree & $\cdots$ & $-k-2$ & $\cdots$ & $-3$ & $-2$ & $-1$ & 0 & 1 & 2 & $\cdots$ & $k+1$ & $\dots$ \\
field type & $\cdots$ & f/b & $\cdots$ & f & b & f & b & f & b & $\cdots$ & f/b & $\cdots$ \\
\hline
\end{tabu}
\caption{Batalin--Vilkovisky fields; a `b' stands for boson and an `f' for fermion.}
\label{tab:BV_fields}
\end{center}
\vspace{-.6cm}
\end{table}

Note that the above does not yet include the additional trivial pairs needed for full gauge fixing; we shall come to these later.

Since $\frF_{\rm BV}=T^*[-1]\frF_{\rm BRST}$, it comes with the canonical symplectic structure~\eqref{eq:BVSymplecticStructureGeneral}
\begin{equation}
\omega_{\rm BV}\ =\ \langle\dd a,\dd a^+\rangle_{\sL}+\sum_{k\geq 0}(-1)^{k+1}\langle\dd c_{-k},\dd c^+_{-k}\rangle_{\sL}~.
\end{equation}
Note that $\omega_{\rm BV}$ is of degree~$-1$ precisely when $\langle -,-\rangle_{\sL}$ is of degree~$-3$ after exchanging coordinate functions for the actual fields. We can conveniently combine all fields, ghosts and all their antifields into the contracted coordinate function
\begin{equation}\label{eq:BVGaugePotential}
 \sfa\ :=\ a+a^++\sum_{k\geq 0}(c_{-k}+c_{-k}^+)~,
\end{equation}
in terms of which the symplectic form simply reads as 
\begin{equation}\label{eq:BVSympForm}
 \omega_{\rm BV}\ :=\ -\tfrac12 \langle \dd \sfa,\dd \sfa\rangle_{\sL_{\CCC}}~.
\end{equation}
In terms of $\sfa$, this symplectic form induces the Poisson structure 
\begin{subequations}\label{eq:InducedBVPoisson}
\begin{equation}
 \{F,G\}_{\rm BV}\ =\  F\left\langle \overset{\leftarrow}{\frac{\delta }{\delta \sfa}},\overset{\rightarrow}{\frac{\delta }{\delta \sfa}}\right\rangle_{\!\sL^*_{\CCC}}\kern-5pt G
 \end{equation}
 for  $F,G\in \CCC^\infty(\frF_{\rm BV})$ and
 \begin{equation}
   F \overset{\leftarrow}{\frac{\delta }{\delta \sfa}}\ =\ (-1)^{|F|_{\sL_\CCC}+1}\overset{\rightarrow}{\frac{\delta }{\delta \sfa}} F~.
 \end{equation}
 \end{subequations}
It remains to construct the BV action $S_{\rm BV}$ satisfying the classical master equation\linebreak $\{S_{\rm BV},S_{\rm BV}\}_{\rm BV}\ =\ 0$ and which induces the homological vector field $Q_{\rm BV}\ :=\ \{S_{\rm BV},-\}$ on $\frF_{\rm BV}$.

\paragraph{Batalin--Vilkovisky action.} We could follow the construction of $S_{\rm BV}$ discussed in Section~\ref{ssec:ClassicalMasterEquation}, but for homotopy MC theory, there exists a significant shortcut. Recall that we require $S_{\rm BV}$ to agree with $S_{\rm MC}$ after all ghosts and antifields are put to zero. Also, we require
\begin{equation}
 \left.\{S_{\rm BV},-\}_{\rm BV}\right|_{\frF_{\rm BRST}}\ =\ Q_{\rm BRST}~,
\end{equation}
where $Q_{\rm BRST}$ has the action~\eqref{eq:BRST}. An obvious ansatz is therefore
\begin{equation}\label{eq:BVaction} 
S_{\rm BV}[\sfa]\ :=\ \sum_{i\geq 1}\frac{1}{(i+1)!}\langle \sfa,\hat \mu_i(\sfa,\ldots,\sfa)\rangle_{\sL_{\CCC}}
\end{equation}
with $\sfa$ defined in~\eqref{eq:BVGaugePotential}. Note that $S_{\rm BV}[\sfa]$ is still a function on $\frF_{\rm BV}$ and we compute 
\begin{subequations}
\begin{equation}
\big\{S_{\rm BV},S_{\rm BV}\big\}_{\rm BV}\  = \ -\langle \sff,\sff\rangle_{\sL_{\CCC}}~,
\end{equation}
where $\sff$ is the curvature of $\sfa$,
\begin{equation}\label{eq:CurvatureSFA}
 \sff\ =\ \sum_{i\geq0}\frac{1}{i!}\hat \mu_i(\sfa,\ldots,\sfa)~.
\end{equation}
\end{subequations}
By virtue of the identity~\eqref{eq:CurvatureIdentity} proved in Appendix~\ref{app:lemmata}, the expression $\langle \sff,\sff\rangle_{\sL_\CCC}$ vanishes identically. Consequently, $S_{\rm BV}$ satisfies the classical master equation
\begin{equation}
\big\{S_{\rm BV},S_{\rm BV}\big\}_{\rm BV}\  \ =\ 0~.
\end{equation}

Together with the homological vector field
\begin{equation}
Q_{\rm BV} \ :=\ \big\{S_{\rm BV},-\big\}_{\rm BV}~,
\end{equation}
$(\frF_{\rm BV},Q_{\rm BV},\omega_{\rm BV})$ becomes a symplectic $Q$-manifold of degree~$-1$. We note that
\begin{equation}\label{eq:QBVonFields}
 Q_{\rm BV}\sfa\ =\ -\sff\eand Q_{\rm BV}\sff\ =\ 0~.
\end{equation}
From~\eqref{eq:QBVonFields}, we can derive the action of $Q_{\rm BV}$ on the individual contracted coordinate functions on $\frF_{\rm BV}$.

\paragraph{Example.} As an explicit example, consider the case $\sL=\sL_{-1}\oplus \sL_0\oplus \sL_1\oplus \sL_2$ and the only non-trivial higher brackets being $\mu_{1,2,3}$. We then have $\sfa\ =\ a+a^++c_0+c_0^++c_{-1}+c_{-1}^+$ with
\begin{subequations}
\begin{equation}\label{eq:Example2TermAction}
\begin{aligned}
 S_{\rm BV}[\sfa]\ &=\  \sum_{i\geq 1}\frac{1}{(i+1)!}\langle \sfa,\hat \mu_i(\sfa,\ldots,\sfa)\rangle_{\sL_{\CCC}}\\ 
 &=\  S_{\rm MC}[a]-\langle c_0,\mu_1(a^+)\rangle_\sL+\langle c_{-1},\mu_1(c^+_{0})\rangle_\sL\,+\\
  &\kern1cm+\langle a,\mu_2(c_{-1},c^+_{0})\rangle_\sL+\langle a,\mu_2(a^+,c_{0})\rangle_\sL+\tfrac12\langle a^+,\mu_2(a^+,c_{-1})\rangle_\sL\,+\\
  &\kern1cm+\tfrac12\langle c^+_{0},\mu_2(c_{0},c_{0})\rangle_\sL-\langle c^+_{-1},\mu_2(c_{-1},c_{0})\rangle_\sL\,+\\
  &\kern1cm+\tfrac12\langle a,\mu_3(a,a^+,c_0)\rangle_\sL+\tfrac12\langle a,\mu_3(a,c^+_0,c_{-1})\rangle_\sL+\tfrac12\langle a,\mu_3(a^+,a^+,c_{-1})\rangle_\sL\,+\\
  &\kern1cm+\tfrac12\langle a,\mu_3(c_0,c_0,c_0^+)\rangle_\sL-\langle a,\mu_3(c_0,c_{-1},c_{-1}^+)\rangle_\sL-\tfrac{1}{2\cdot 2}\langle a^+,\mu_3(a^+,c_0,c_0)\rangle_\sL\,+\\
   &\kern1cm-\langle a^+,\mu_3(c_0,c_0^+,c_{-1})\rangle_\sL-\tfrac12\langle a^+,\mu_3(c_{-1},c_{-1},c_{-1}^+)\rangle_\sL\,+\\
   &\kern1cm-\tfrac{1}{3!}\langle c_0,\mu_3(c_0,c_0,c_{-1}^+)\rangle_\sL+\tfrac{1}{2\cdot 2}\langle c_0^+,\mu_3(c_0^+,c_{-1},c_{-1})\rangle_\sL~,
 \end{aligned}
 \end{equation}
where the higher products $\mu_i$ are agnostic about the ghost degree of the enclosed fields, and the signs arise again from moving coordinate functions past graded basis vectors and the $\hat \mu_i$. Moreover, $S_{\rm MC}$ is the homotopy MC action  for $\sL$,
\begin{equation}
S_{\rm MC}[a]\ =\ \tfrac12\langle a,\mu_1(a)\rangle_\sL+\tfrac{1}{3!}\langle a,\mu_2(a,a)\rangle_\sL+\tfrac{1}{4!}\langle a,\mu_3(a,a,a)\rangle_\sL~.
\end{equation}
\end{subequations}

The homological vector field induced by $S_{\rm BV}$ acts as follows on the individual contracted coordinate functions on $\sL$:
\begin{equation}\label{eq:QBVonComponentFields2Term}
\begin{aligned}
 Q_{\rm BV}a\ &=\ \mu_1(c_0)+\mu_2(a,c_0)+\tfrac12\mu_3(a,a,c_0)+\mu_2(c_{-1},a^+)\,-\\
  &\kern1cm-\mu_3(a,a^+,c_{-1})-\mu_3(c_{-1},c_0,c_0^+)+\tfrac12\mu_3(c_0,c_0,a^+)+\tfrac12\mu_3(c_{-1}, c_{-1}, c^{+}_{-1})~,\\
  Q_{\rm BV}c_{0}\ &=\ -\mu_1(c_{-1})-\mu_2(a,c_{-1})-\tfrac12\mu_3(a,a,c_{-1})-\tfrac12\mu_2(c_0,c_0)\,-\\
  &\kern1cm-\tfrac12\mu_3(a,c_0,c_0)+\mu_3(c_0,c_{-1},a^+)-\tfrac12\mu_3(c_{-1},c_{-1},c_0^+)~,\\
 Q_{\rm BV}c_{-1}\ &=\ \mu_2(c_{-1},c_0)+\mu_3(a,c_{-1},c_0)+\tfrac{1}{3!}\mu_3(c_0,c_0,c_0)+\tfrac12\mu_3(c_{-1},c_{-1},a^+)~,\\
 Q_{\rm BV}a^+\ &=\ -\mu_1(a)-\tfrac12\mu_2(a,a)-\tfrac{1}{3!}\mu_3(a,a,a)-\mu_2(c_0,a^+)-\mu_2(c_{-1},c_0^+)\,+\\
 &\kern1cm+\mu_3(a,c_0,a^+)-\mu_3(a,c_{-1},c_0^+)-\tfrac12\mu_3(c_{-1},a^+,a^+)\,-\\
  &\kern1cm-\tfrac12\mu_3(c_0,c_0,c_0^+)-\mu_3(c_{-1},c_0,c_{-1}^+)~,\\
 Q_{\rm BV}c^+_{0}\ &=\ \mu_1(a^+)+\mu_2(a,a^+)+\tfrac12\mu_3(a,a,a^+)-\mu_2(c_0,c_0^+)+\mu_2(c_{-1},c_{-1}^+)\,-\\
 &\kern1cm-\mu_3(a,c_0,c_0^+)+\mu_3(a,c_{-1},c_{-1}^+)+\mu_3(a,c_{-1},c_{-1}^+)\,+\\
  &\kern1cm+\tfrac12\mu_3(a^+,a^+,c_0)-\mu_3(c_{-1},a^+,c_0^+)+\tfrac12(c_0,c_0,c_{-1}^+)~,\\
  Q_{\rm BV}c^+_{-1}\ &=\ -\mu_1(c_0^+)-\tfrac12\mu_2(a,c_0^+)-\tfrac12\mu_3(a,a,c_0^+)-\tfrac12\mu_2(a^+,a^+)-\mu_2(c_0,c_{-1}^+)\,-\\
   &\kern1cm-\tfrac12\mu_3(a,a^+,a^+)-\mu_3(a^+,c_0,c_0^+)+\mu_3(a^+,c_{-1},c_{-1}^+)\,-\\
   &\kern1cm-\tfrac12\mu_3(c_{-1},c_0^+,c_0^+)- \mu_3 (a, c_0, c^{+}_{-1})~.
\end{aligned}
\end{equation}

\paragraph{Quantum master equation.}
Following~\eqref{eq:BVLaplacianGeneral}, we introduce the BV Laplacian by its action on an $F\in\CCC^\infty(\frF_{\rm BV})$,
\begin{equation}
 \Delta_{\rm BV} F\ :=\ -\frac12 F \left\langle \overset{\leftarrow}{\frac{\delta}{\delta \sfa}},\overset{\leftarrow}{\frac{\delta}{\delta \sfa}}\right\rangle_{\!\sL^*_\CCC}~.
\end{equation}

Since the inner product $\langle-,-\rangle_{\sL_\CCC}$ is graded symmetric and since the higher products $\hat \mu_i$ for $i\geq2$ are graded anti-symmetric, it follows immediately from the cyclicity of the inner product that
\begin{subequations}
\begin{equation}
 \Delta_{\rm BV}\, \langle \sfa,\hat \mu_i(\sfa,\ldots,\sfa)\rangle_{\sL_\CCC} \ =\ 0
 \end{equation}
 for $i\ \geq\ 2$. We also have
 \begin{equation}
  \Delta_{\rm BV}\, \langle \sfa,\hat \mu_1(\sfa)\rangle_{\sL_\CCC}\ =\ 0
 \end{equation}
 \end{subequations}
 since both the $\sL$-degree and the ghost degree of $\langle \sfa,\hat \mu_1(\sfa)\rangle_{\sL_\CCC}$ are zero so that a field and its antifield cannot pair up in $\langle \sfa,\hat \mu_1(\sfa)\rangle_{\sL_\CCC}$.  Hence, the BV action~\eqref{eq:BVaction} obeys
\begin{equation}\label{eq:Delta_S}
 \Delta_{\rm BV} S_{\rm BV}\ =\ 0~.
 \end{equation}
Altogether, we conclude that $S_{\rm BV}$ satisfies the quantum master equation 
 \begin{equation}
   \big\{S_{\rm BV},S_{\rm BV}\big\}_{\rm BV} -2\di\hbar\,  \Delta_{\rm BV} S_{\rm BV}\ =\ 0\quad\Longleftrightarrow\quad  \Delta_{\rm BV} \de^{\frac{\di}{\hbar}S_{\rm BV}}\ =\ 0~.
 \end{equation}
It is important to stress, however, that this is only true formally since the BV Laplacian is singular and regularisation needs to be taken into account in general. Furthermore, the above observation, namely that the two terms in the quantum master equation vanish separately, was made earlier in the context of $BF$-theory~\cite{Cattaneo:0010172}.

\subsection{Gauge fixing}

While gauge fixing is mostly relevant to the quantisation of our theory which is beyond the scope of this paper, let us briefly summarise the classical part and add an outlook on the quantum master equation.

\paragraph{Additional fields.}
To gauge-fix the BV action~\eqref{eq:BVaction}, we will need to introduce trivial pairs, that is, {\em antighosts} $\bar c_{i,j}$ and {\em Lagrange multipliers} $b_{i,j}$ together with their antifields, the {\em antifield antighost} $\bar c^+_{i,j}$ and the {\em antifield Lagrange multiplier} $b^+_{i,j}$ as done in Sections~\ref{ssec:BRST_quant} and~\ref{ssec:BV_gauge_fixing}.

In the $L_\infty$-algebra picture, the necessary extension is given in Table~\ref{tab:BVLAlgPic}. In general, we have additional quadruples of fields for all $i\leq 0$ and $i-1\leq j\leq -i-1$, with $j$ increased in steps of 2, as displayed in Table~\ref{tab:LGDegs}.
\begin{table}[h]
\begin{center}
\begin{tabu}{|l|c|c|c|c|}
\hline
 & $\bar c_{i,j}$ & $b_{i,j}$ & $\bar c^+_{i,j}$ & $b^+_{i,j}$ \\ \hline
takes values in a copy of & $\sL_i$ & $\sL_i$ & $\sL_{3-i}$ & $\sL_{3-i}$ \\
which is added to $\sL$ in ghost degree & $j$ & $j+1$ & $-j-1$ & $-j-2$\\
or, equivalently, in $L_\infty$-degree & $1-j$ & $-j$ & $2+j$ & $j+3$ \\
\hline
\end{tabu}
\caption{\label{tab:LGDegs} $L_\infty$-degrees and ghost degrees of the trivial pairs and their antifields.}
\end{center}
\vspace{-.6cm}
\end{table}

\begin{table}
\begin{center}
\begin{adjustbox}{angle=90,width=410pt}
\tabulinesep = 10pt
\tabcolsep = 2pt
\begin{tabu}{|c|ccccccccccccccccc|}
\hline
$L_\infty$-degree &&& $-2$ && $-1$ && $0$ && $1$ && $2$ && $3$ && $4$ &&\\
ghost degree &&& $3$ && $2$ && $1$ && $0$ && $-1$ && $-2$ && $-3$ &&\\
\hline
BV fields & $\cdots$ & $\xrightarrow{~\mu_1~}$ & $(\sL_{-2},c_{-2})$ & $\xrightarrow{~\mu_1~}$ & $(\sL_{-1},c_{-1})$ & $\xrightarrow{~\mu_1~}$ & $(\sL_{0},c_0)$ & $\xrightarrow{~\mu_1~}$ & $(\sL_1,a)$ & $\xrightarrow{~\mu_1~}$ & $(\sL_2,a^+)$ & $\xrightarrow{~\mu_1~}$ & $(\sL_3,c_0^+)$ & $\xrightarrow{~\mu_1~}$ & $(\sL_4,c_{-1}^+)$ & $\xrightarrow{~\mu_1~}$ & $\cdots$\\
&&&&&&&&& $\oplus$ && $\oplus$ &&&&&&\\
trivial pair &&&&&&&&& $(\sL_0,b_{0,-1})$ & $\xrightarrow{~\id~}$ & $(\sL_0,\bar c_{0,-1})$ &&&&&&\\
&&&&&&&&& $\oplus$ && $\oplus$ &&&&&&\\
antifields &&&&&&&&& $(\sL_3,\bar c_{0,-1}^+)$ & $\xrightarrow{~-\id~}$ & $(\sL_3,b_{0,-1}^+)$ &&&&&&\\
&&&&&&& $\oplus$ && $\oplus$ && $\oplus$ && $\oplus$ &&&&\\
trivial pairs &&&&&&& $(\sL_{-1},b_{-1,0})$ &$ \xrightarrow{~\id~}$ & $(\sL_{-1},\bar c_{-1,0})$ && $(\sL_{-1},b_{-1,-2})$ & $\xrightarrow{~\id~}$ & $(\sL_{-1},\bar c_{-1,-2})$ &&&&\\
&&&&&&& $\oplus$ && $\oplus$ && $\oplus$ && $\oplus$ &&&&\\
antifields &&&&&&& $(\sL_{4},\bar c_{-1,-2}^+)$ &$\xrightarrow{~\id~}$ & $(\sL_{4},b_{-1,-2}^+)$ && $(\sL_{4},\bar c_{-1,0}^+)$ & $\xrightarrow{~\id~}$ & $(\sL_{4}, b_{-1,0}^+)$ &&&&\\
&&&&& $\oplus$ && $\oplus$ && $\oplus$ && $\oplus$ && $\oplus$ && $\oplus$ &&\\
trivial pairs &&&&& $(\sL_{-2},b_{-2,1})$ & $\xrightarrow{~\id~}$ & $(\sL_{-2},\bar c_{-2,1})$ & & $(\sL_{-2},b_{-2,-1} )$ & $\xrightarrow{~\id~}$ & $(\sL_{-2},\bar c_{-2,-1})$ &  & $(\sL_{-2},b_{-2,-3})$ & $\xrightarrow{~\id~}$ & $(\sL_{-2},\bar c_{-2,-3})$ &&\\
&&&&& $\oplus$ && $\oplus$ && $\oplus$ && $\oplus$ && $\oplus$ && $\oplus$&&\\
antifields &&&&& $(\sL_{5},\bar c_{-2,-3}^+)$ & $\xrightarrow{~-\id~}$ & $(\sL_{5},b_{-2,-3}^+)$ & & $(\sL_{5},\bar c_{-2,-1}^+)$ & $\xrightarrow{~-\id~}$ & $(\sL_{5},b_{-2,-1}^+)$ & & $(\sL_{5},\bar c_{-2,1}^+)$ & $\xrightarrow{~-\id~}$ & $(\sL_{5},b_{-2,1}^+)$ &&\\
$\vdots$ &&&&&&&&& $\vdots$ &&&&&&&&\\
\hline
\end{tabu}
\end{adjustbox}
\caption{\label{tab:BVLAlgPic} $L_\infty$-algebra picture of the Batalin--Vilkovisky fields including trivial pairs}
\label{tab:full_aux_fields}
\end{center}
\end{table}

\noindent
Put differently, we extend $\sL$ to
\begin{equation}
 \sL_{\rm e}\ :=\ \sL \oplus \bigoplus_{\substack{i\leq 0\\ 0\leq k\leq i-1}}\big(\,\sL_i[i-2-2k]\oplus \sL_i[i-1-2k]\oplus \sL_{3-i}[-i-1+2k]\oplus \sL_{3-i}[-i-2+2k]\,\big)~,
\end{equation}
and the BV complex $\frF_{\rm BV}$ correspondingly reads as 
\begin{equation}
\begin{aligned}
 \frF_{\rm eBV}\ &:=\ \sL_{\rm e}[1] \\
 &\phantom{:}\cong \ T^*[-1]\left(\frF_{\rm BRST}\bigoplus_{\substack{i\leq 0\\ 0\leq k\leq i-1}}\big(\,\sL_i[i-1-2k]\oplus \sL_i[i-2k]\,\big)\right)\\
 &\phantom{:}\cong \ \frF_{\rm BV}\bigoplus_{\substack{i\leq 0\\ 0\leq k\leq i-1}}\big(\,\sL_i[i-1-2k]\oplus \sL_i[i-2k]\oplus \sL_{3-i}[-i+2k]\oplus \sL_{3-i}[-i-1+2k]\,\big)~.
\end{aligned} 
\end{equation}
A diagram of the additional fields is found in Table~\ref{tab:full_aux_fields}.

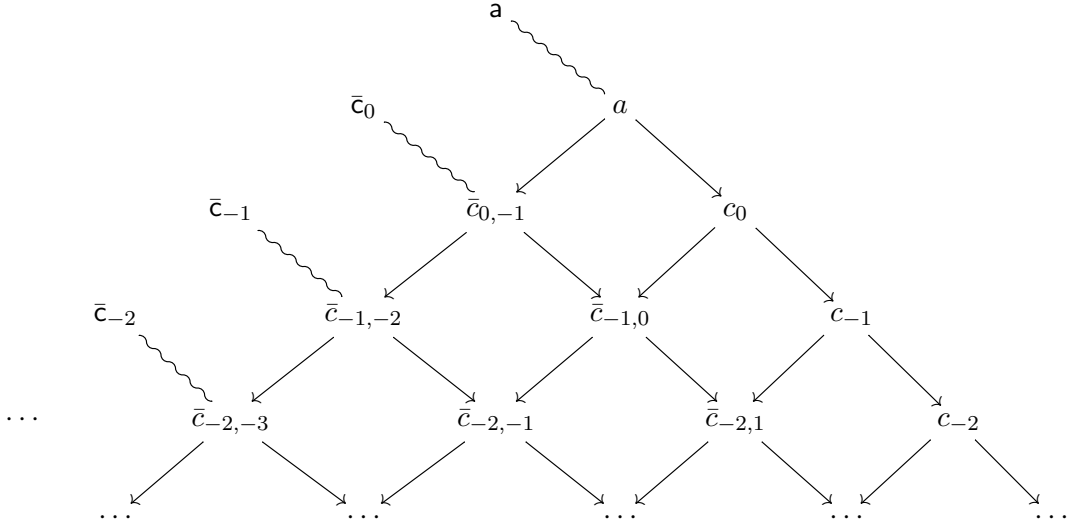
The reason for introducing families of antighosts and Lagrange multipliers for each level~$k$ can be understood as follows: the lowest level antighosts and Lagrange multipliers are needed to fix the gauge symmetries of the fields and ghosts, the next-to-lowest level antighosts and Lagrange multipliers are needed to fix the gauge symmetries of the lowest level antighosts, and so on~\cite{Batalin:1984jr}. Reducing to merely the antighost of each quadruple of new fields, one obtains the so-called {\em Batalin--Vilkovisky triangle} displayed in Table~\ref{tab:BVTriangle}.

\begin{table}[h]
$$
\xymatrixcolsep{1.2pc}
\myxymatrix{
    & & &   & \sfa \ar@{~}[dr] &  & & & &\\
    & & &  \bar \sfc_0  \ar@{~}[dr] &  & \ar[dl] a  \ar[dr]  & & & &\\
    & & \bar \sfc_{-1}  \ar@{~}[dr] & & \ar[dl]  \bar c_{0,-1}  \ar[dr]  & &  \ar[dl] c_0  \ar[dr] & & \\
    & \bar \sfc_{-2}  \ar@{~}[dr] & & \ar[dl] \bar c_{-1,-2}  \ar[dr] &  &   \ar[dl] \bar c_{-1,0}  \ar[dr] & & \ar[dl] \ c_{-1}  \ar[dr] & &\\
   \cdots & &  \ar[dl] \bar c_{-2,-3}  \ar[dr]   & &   \ar[dl] \bar c_{-2,-1}  \ar[dr]  & &  \ar[dl] \bar c_{-2,1} \ar[dr]  & &  \ar[dl] \ c_{-2}  \ar[dr] &\\
     & \cdots  &   & \cdots &   &  \cdots&& \cdots& & \cdots 
    }
$$
\caption{\label{tab:BVTriangle} Batalin--Vilkovisky triangle.}
\end{table}

\paragraph{Symplectic \mathversion{bold}\texorpdfstring{$Q$}{Q}-manifold structure.} The graded vector space $\frF_{\rm eBV}$ comes with the canonical symplectic structure
\begin{equation}\label{eq:eBVsymplectic_form}
\begin{gathered}
\kern-4cm \omega_{\rm eBV}\ =\ \langle\dd a,\dd a^+\rangle_{\sL}+\sum_{i\leq 0}(-1)^{i+1}\langle\dd c_{i},\dd c^+_{i}\rangle_{\sL}\,+\\
\kern4cm+\sum_{i\leq 0}\sum_{j=i-1}^{-i-1} \Big[(-1)^j\langle \dd \bar c_{i,j}, \dd\bar c_{i,j}^+\rangle_\sL+(-1)^{j+1}\langle \dd b_{i,j},\dd b_{i,j}^+\rangle_\sL\Big]~.
\end{gathered}
\end{equation}
Also, the extension of the BV action to a solution of the classical master equation is given by
\begin{equation}\label{eq:eBVaction}
S_{\rm eBV}[a,c_i,\dots,\bar c_{i,j},\dots] \ =\  S_{\rm BV}[a,c_i,\dots]-\sum_{i\leq 0}\sum_{j=i-1}^{-i-1} \langle b_{i,j},\bar c_{i,j}^+\rangle_\sL~,
\end{equation}
as discussed in Section~\ref{ssec:BV_gauge_fixing}. We set again $Q_{\rm eBV}\ :=\ \{S_{\rm eBV},-\}_{\rm eBV}$ and its action on all the fields is~\eqref{eq:QBVonFields} together with
\begin{equation}
 Q_{\rm eBV}\bar c_{i,j}\ =\ b_{i,j}~,~~
 Q_{\rm eBV} b_{i,j}\ =\ 0~,~~
 Q_{\rm eBV}\bar c_{i,j}^+\ =\ 0~,~~
 Q_{\rm eBV} b_{i,j}^+\ =\ (-1)^j\bar c^+_{i,j}~.
\end{equation}
Note that the new fields $\bar c_{i,j}$, etc., denote contracted coordinate functions here. 

As before, it is convenient to combine the additional fields arising from trivial pairs into superfields $\bar \sfc_i$ and $\sfb_i$ such that 
\begin{equation}
 Q_{\rm eBV}\bar \sfc_i\ =\ \sfb_i\eand Q_{\rm eBV}\sfb_i\ =\ 0~.
\end{equation}
We can put
\begin{subequations}
\begin{equation}
 \bar \sfc_{i}\ :=\ \begin{cases}
  \sum_{j\geq 0} (-1)^j\bar c_{i-j,i+j-1} &\efor i\ \leq \ 0~,\\
   \sum_{j\geq 0} (j) b^+_{1-i-j,j-i} &\efor i\ >\  1
\end{cases}
\end{equation}
and
\begin{equation}
 \sfb_{i}\ :=\ \begin{cases}
  \sum_{j\geq 0} (-1)^j b_{i-j,i+j-1} &\efor i\ \leq \ 0~,\\
 \sum_{j\geq 0} (-1)^{j-i}\bar c^+_{1-i-j,j-i}&\efor i\ >\ 1~.
 \end{cases}
\end{equation}
\end{subequations}
To obtain the component fields, as before, one simply projects onto the corresponding ghost degree. 

The symplectic form~\eqref{eq:eBVsymplectic_form} and the extended BV action~\eqref{eq:eBVaction} then read as 
\begin{equation}
\begin{gathered}
\omega_{\rm eBV}\ :=\ -\tfrac12\langle\dd\sfa,\dd\sfa\rangle_{\sL_\CCC}+\sum_{i+j=1}\langle\dd \sfb_{i},\dd\bar \sfc_{j}\rangle_{\sL_{{\rm e}\CCC}}~,\\
S_{\rm eBV}[\sfa,\sfb] \ :=\  S_{\rm BV}[\sfa]-\sum_{\substack{i+j=1\\ i\leq j}}(-1)^{i+1}\langle \sfb_i,\sfb_j\rangle_{\sL_{{\rm e}\CCC}}~.
\end{gathered}
\end{equation}
 
\paragraph{Quantum master equation.} The fact that the BV action satisfies the quantum master equation is preserved after the extension by trivial pairs. In particular, the BV Laplacian $\Delta_{\rm eBV}$ on $\sL_{\rm e}$ annihilates $S_{\rm eBV}$, and we have altogether 
\begin{equation}
 \{S_{\rm eBV},S_{\rm eBV}\}_{\rm eBV}\ =\ 0\eand  \Delta_{\rm eBV}S_{\rm eBV}\ =\ 0
 \end{equation}
so that 
\begin{equation}
   \big\{S_{\rm eBV},S_{\rm eBV}\big\}_{\rm eBV} -2\di\hbar\,  \Delta_{\rm eBV} S_{\rm eBV}\ =\ 0\quad\Longleftrightarrow\quad  \Delta_{\rm eBV} \de^{\frac{\di}{\hbar}S_{\rm eBV}}\ =\ 0~.
\end{equation}
As before, this is only true formally as regularisation needs to be taken into account.

\paragraph{Gauge fixing.} To gauge-fix the extended BV action~\eqref{eq:eBVaction}, we introduce a gauge fixing fermion 
\begin{equation}
\Psi \ :=\ \sum_{i\leq0} \langle \bar \sfc_{i}, \sG_{i}(\sfa,\bar \sfc_0, \bar \sfc_{-1},\ldots)\rangle_{\sL_{\rm e}}~,
\end{equation}
such that $\frac{\dpar}{\dpar \phi}\Psi$ for any field $\phi\in \sL_e$ takes values in the same homogeneously graded vector subspace of $\sL_e$ as its antifield, $\phi^+$. The gauge fixed quantum BV action is then given by 
\begin{equation}
S_{\rm qBV}[a,c,\bar c,b]\ :=\ \kern-10pt\left.\phantom{\frac12}S_{\rm eBV}[\sfa,\sfb]\right|_{\phi^+=\overset{\rightarrow}{\frac{\dpar}{\dpar \phi}}\Psi}~.
\end{equation}

\section{Classical \texorpdfstring{$L_\infty$}{L-infty}-structure of field theories}\label{sec:L_infty-structures}

\subsection{Classical field theories and \texorpdfstring{$L_\infty$}{L-infty}-algebras}\label{ssec:Classical_FT_and_L_infty}

\paragraph{Outline.} The BV formalism maps classical fields $\frF$ and a classical action $S$ to a set of BV fields $\frF_{\rm BV}$ together with a BV action $S_{\rm BV}$. As explained in Section~\ref{ssec:L_infty_algebras_and_algebroids}, via the BV bracket, the BV action defines a homological vector field $Q_{\rm BV}$, which in turn encodes an $L_\infty$-algebra structure on the graded vector space $\frF_{\rm BV}$. This $L_\infty$-algebra encodes all relevant classical information on the field theory. It captures the field content and its gauge symmetry structure, the equations of motion, as well as the Noether identities. At the classical level, Lagrangian field theories\footnote{This relation should extend to non-Lagrangian field theories, and we plan to study these in future work.} are thus equivalently described by cyclic  $L_\infty$-algebras. 

Equivalent field theories should have $L_\infty$-algebras which are isomorphic in some sense. The only plausible candidates for such an isomorphism are isomorphisms and quasi-iso\-morphisms of $L_\infty$-algebras. Since classical field theories can be extended to equivalent field theories by adding auxiliary fields with algebraic equations of motions, we are left with the quasi-isomorphisms of $L_\infty$-algebras. Below, it will become clear that this is indeed the correct choice and that quasi-isomorphic $L_\infty$-algebras belong to field theories with the same observables.

Before discussing explicit examples, let us make a few more observations about the $L_\infty$-algebra structure which we expect.

\paragraph{\mathversion{bold}$L_\infty$-algebra structure.} The vector space $\frF_{\rm BV}$ is graded in particular with respect to the ghost degree, $\frF_{\rm BV}:=\bigoplus_{i\in \RZ} \frF_{\rm BV}^i$. The usual correspondence between $Q$-manifolds and $L_\infty$-algebras suggests that we need to shift the degree by one and invert it, for an $L_\infty$-algebra with higher products $\mu_i$ of degree~$2-i$. We thus arrive at the $L_\infty$-algebra
\begin{equation}
\cdots \ \xrightarrow{~\mu_1~}\  \underbrace{\frF^{-1}_{\rm BV}}_{=:\,\sL_0}\ \xrightarrow{~\mu_1~}\ \underbrace{\frF^{0}_{\rm BV}}_{=:\,\sL_1}\ \xrightarrow{~\mu_1~}\ \underbrace{\frF^{1}_{\rm BV}}_{=:\,\sL_2}\ \xrightarrow{~\mu_1~}\ \underbrace{\frF^{2}_{\rm BV}}_{=:\,\sL_3}\ \xrightarrow{~\mu_1~}\ \cdots
\end{equation}
That is, $\sL_0$ is given by the ghosts, $\sL_1$ by the fields, $\sL_2$ by the antifields and $\sL_3$ by the antighosts. This extends in an obvious manner to cases with ghosts-for-ghosts and trivial pairs. 

The map $\mu_1$ is encoded in the linear part of the action $Q_{\rm BV}:=\{S_{\rm BV},-\}_{\rm BV}$ on the field corresponding to the {\em image} of $\mu_1$. Explicitly, $\mu_1:\sL_0\rightarrow \sL_1$ is encoded in the linear part of the explicit expression for $Q_{\rm BV} A$ and therefore encodes the linearised gauge transformations. The map $\mu_1:\sL_1\rightarrow \sL_2$ is obtained by linearising $Q_{\rm BV}A^+$, which yields the linearised variation of the classical action with respect to the field and therefore the linearised classical equations of motion. The map $\mu_1:\sL_2\rightarrow \sL_3$ is the linearised part of the action of $Q_{\rm BV}$ on $c^+$ which precisely encodes the Noether identities as we shall explain below. This is all the structure necessary to describe a classical (gauge) field theory; for higher gauge theories, one obtains an extension beyond the homogeneously graded vector subspaces $L_j$ with $0\leq j\leq 3$.

The higher brackets then fulfil the task of making the linearised expressions covariant and to allow for higher interaction terms. In general, an interaction term of $n$th order in the fields will be encoded in a higher product $\mu_i$ with $i=n-1$. 

From our discussion in Section~\ref{sec:HMCT}, it is also clear that the homotopy Maurer--Cartan action for $\sL$ reproduces the original action $S$ and the homotopy Maurer--Cartan action for $\sL_{\CCC}$ reproduces the BV action $S_{\rm BV}$. 

\begin{table}[h]
\begin{center}
 \begin{tabu}{|c|c|c|c|c|c|c|c|}
\hline
 $\cdots$ & $\sL_{-1}$ & $\sL_0$ & $\sL_1$ & $\sL_2 $ & $\sL_3$ & $\sL_4$ & $\cdots$ \\ \hline
 $\cdots$ & gauge-of-gauge  & gauge & physical & equations of & Noether & higher & $\cdots$\\
 & transf. & transf. & fields & motion & identities & Noether &\\
\hline
\end{tabu}
\end{center}
\caption{\label{tab:Meaning_L_infty} Summary of the structure of the $L_\infty$-algebra of a classical field theory. While the labels under $\sL_i$ for $i\leq 1$ describe the spaces, the meaning of the labels changes for $i\geq 2$: $\sL_2$, for instance, is {\em not} the space of the equations of motion, but the element $\ell$ of $\sL$ that is forced to zero by the equations of motion $\ell=0$ takes values in $\sL_2$. }
\end{table}

\paragraph{Noether identities.} Next, let us discuss the Noether identities in somewhat more detail. Here, we are concerned with Noether's second theorem, generalising the more familiar first one. In this picture, also gauge symmetries give rise to Noether symmetries. 

Our motivation for considering Noether identities is twofold. First of all, they are at the heart of the BV formalism: they correspond precisely to the degeneracies of the Hessian which make the application of the stationary phase formula in the interpretation of the path integral impossible. Secondly, they are an important part of the classical structure of a field theory and also contained in its $L_\infty$-algebra. 

Let $M$ be a manifold with local coordinates $x^\mu$. Consider an infinitesimal group action on a set of fields $\Phi^A$ on $M$ parametrised by infinitesimal parameters $\eps=(\eps^I)$ as
\begin{equation}\label{eq:Noether_gauge_symmetry}
 \delta \Phi^A(x)\ :=\ R^A_I(\Phi)\eps^I(x)~,
\end{equation}
where $R^A_I(\Phi)$ are field-dependent differential operators, possibly containing terms of order~0. Alternatively, we can write
\begin{equation}
 \delta \Phi^A(x)\ =\ \int_M \mu(y)R^A_I(x,y,\Phi)\eps^I(y)~,
\end{equation}
where $\mu$ is a suitable measure on $M$. If this action is a symmetry of an action $S[\Phi]$, then we have the Noether identity
\begin{equation}
\int_M \mu(x)\frac{\delta S[\Phi]}{\delta \Phi^A(x)}\frac{\delta \Phi^A(x)}{\delta \eps^I(y)}\ =\ 0
\end{equation}
or
\begin{equation}
 \int_M \mu(x)\frac{\delta S[\Phi]}{\delta \Phi^A(x)}R^A_I(x,y,\Phi)\ =\ 0~.
\end{equation}

If we vary this equation with respect to $\delta \Phi^B(z)$ and restrict $\Phi$ to the stationary surface, we have
\begin{equation}
 \frac{\delta^2 S[\Phi]}{\delta \Phi^B(z)\delta \Phi^A(x)} R_I^A(x,y,\Phi)\ =\ 0~,
\end{equation}
which implies that the $R_I^A$ encode degeneracies of the Hessian, that is, they are the eigenvectors of the Hessian with eigenvalue zero.

The Noether identities also imply that the vector fields $Q_{\rm BV}$ decompose as
\begin{equation}
 Q_{\rm BV}\ =\ Q_{\rm KT}+\cdots~,
\end{equation}
where $Q_{\rm KT}$ is the part purely responsible for the Koszul--Tate resolution, acting non-trivially only on the antifields of fields and ghosts, with $Q_{\rm KT}^2=0$. In the BV picture, the symmetry transformation~\eqref{eq:Noether_gauge_symmetry} is encoded in $Q_{\rm BV}\Phi^A$, which contains the operators $R^A_I$. Since $Q_{\rm BV}\Phi^A$ is related to the variation of $S_{\rm BV}$ with respect to $\Phi^+_A$, we have a term in the BV action of the form $\langle \Phi^+_A,R^A_I c^I\rangle$, where the inner product is usually given by an integral over some spacetime. This implies that the adjoint of $R^A_I$ appears in the $Q_{\rm BV} c^+_I=\pm(R^\dagger)^A_I \Phi^+_A$, which is the variation of $S_{\rm BV}$ with respect to $c^I$. Here, $R^\dagger$ denotes the adjoint of $R$ with respect to $\langle-,-\rangle$. The Noether identity then implies that $Q_{\rm KT}^2=0$, at least when acting on the antifields of ghosts:
\begin{equation}
 Q_{\rm KT}^2 c^+_I\ =\ \pm Q_{\rm KT} \Phi^+_A (R^\dagger)^A_I \ =\ \pm\frac{\delta S[\Phi]}{\delta \Phi^A}(R^\dagger)^A_I\ =\ 0~.
\end{equation}
For more details on Noether identities, see e.g.~\cite{Gomis:1994he,Stasheff:1997iz,Fulp:2002fm,Henneaux:1992}.

\paragraph{Example: scalar field theory.} To stress the point that the BV formalism also makes sense for theories without gauge symmetries, let us consider scalar field theory on Minkow\-ski space $\FR^{1,d}$ as a simple example. Further examples will follow below. As another unusual point, let us include global symmetries into the BRST formalism. This is clearly not necessary for the quantisation of the path integral, and it is usually not even desirable, as it reduces the space of solutions to globally symmetric ones. It will, however, allow us to obtain the usual Noether identities in the $L_\infty$-algebra picture.

Let $\varphi\in \CCC^\infty(\FR^{1,d})$ be a real scalar field with action functional
\begin{equation}
 S\ :=\ \int_{\FR^{1,d}} \dd^{d+1} x~\Big\{\tfrac12 (\dpar_\mu \varphi)^2-\tfrac12 m^2 \varphi^2-\tfrac{\lambda}{4!}\varphi^4\Big\}~.
\end{equation}
We extend the field space $\frF=\CCC^\infty(\FR^{1,d})$ to the action groupoid for the Poincar\'e group,
\begin{equation}
 (\sSO(1,d)\ltimes \FR^{1,d})\ltimes \CCC^\infty(\FR^{1,d})\ \rightrightarrows\  \CCC^\infty(\FR^{1,d})~,
\end{equation}
which differentiates to the action algebroid
\begin{equation}
 \frF_{\rm BRST}\ =\ (\aso(1,d)\ltimes \FR^{1,d})\ltimes \CCC^\infty(\FR^{1,d})~.
\end{equation}
In addition to $\varphi$, we also have ghosts $c=c^I=c^{\mu}{}_{\nu}+c^\mu \in \aso(1,d)\ltimes \FR^{1,d}$. Those are not fields but rather constants on Minkowski space $\FR^{1,d}$. The actions of the BRST operator $Q_{\rm BRST}$ on $c$ and $\varphi$ capture the Poincar\'e Lie algebra as well as its action on the field $\varphi$ and read as
\begin{equation}
\begin{aligned}
 Q_{\rm BRST}(c^\mu{}_\nu+c^\mu)\ &:=\ c^\mu{}_\kappa c^\kappa{}_\nu +c^\mu{}_\kappa c^\kappa ~,\\
 Q_{\rm BRST} \varphi\ &:=\ c\acton\varphi\ :=\ c^\mu \dpar_\mu \varphi+c^\mu{}_\nu x^\nu \dpar_\mu \varphi~.
\end{aligned}
\end{equation}

We now perform the Koszul--Tate resolution by including antifields $\varphi^+$ and $c^+$. The BV bracket is induced by the canonical symplectic form,
\begin{equation}
 \omega_{\rm BV}\ :=\ -\dd c^I\wedge \dd c^+_I+\int_{\FR^{1,d}} \dd^{d+1} x\left\{\delta \varphi(x)\wedge \delta \varphi^+(x)\right\}
\end{equation}
and the BV action functional reads as
\begin{equation}\label{eq:phi4_BV_action}
 S_{\rm BV}\ :=\ c^+_I [c,c]^I+\int_{\FR^{1,d}} \dd^{d+1} x~\Big\{\tfrac12 (\dpar_\mu \varphi)^2-\tfrac12 m^2 \varphi^2-\tfrac{\lambda}{4!}\varphi^4+\varphi^+(c\acton \varphi)\Big\}~,
\end{equation}
from which the action of $Q_{\rm BV}$ is read off as $Q_{\rm BV}=\{S_{\rm BV},-\}_{\rm BV}$.

The differential graded vector space underlying the $L_\infty$-algebra is
\begin{subequations}
\begin{equation}
 \underbrace{\aso(1,d)\ltimes \FR^{1,d}}_{=:\,\sL_0}\ \xrightarrow{~0~}\ \underbrace{\CCC^\infty(\FR^{1,d})}_{=:\,\sL_1}\ \xrightarrow{~-\dpar_\mu\dpar^\mu-m^2~}\ \underbrace{\CCC^\infty(\FR^{1,d})}_{=:\,\sL_2}\ \xrightarrow{~0~} \underbrace{(\aso(1,d)\ltimes \FR^{1,d})^*}_{=:\, \sL_3}
\end{equation}
and the non-trivial higher brackets take the form
\begin{equation}
\begin{gathered}
 \mu_2(c_1,c_2)\ :=\ [c_1,c_2]~,~~~
 \mu_2(c_1,\varphi_1)\ :=\ c_1\acton \varphi_1~,\\
 \mu_2(c_1,\varphi^+_1)\ :=\ c_1\acton\varphi^+~,~~~\mu_2(c_1,c^+_1)\ :=\ c_1\acton c^+_1~,\\
 \mu_3(\varphi_1,\varphi_2,\varphi_3)\ :=\ -\lambda \varphi_1\varphi_2\varphi_3
\end{gathered}
\end{equation}
\end{subequations}
for $c_1\in \sL_0$, $\varphi_{1,2,3}\in \sL_1$, $\varphi^+_1\in \sL_2$ and $c^+_1\in \sL_3$.

One sees that the homotopy MC action~\eqref{eq:BVaction}  of this $L_\infty$-algebra is indeed the BV action~\eqref{eq:phi4_BV_action}. In addition, we note that the Noether identities follow. For example, we have
\begin{equation}
 Q^2_{\rm KT}c^+_\mu\ =\ Q_{\rm KT}(\varphi^+ \dpar_\mu \varphi+c^+_\nu c_\mu{}^\nu)\ =\ \frac{\delta S}{\delta \varphi}\dpar_\mu \varphi+\cdots
 \ =\ \dpar_\nu \left(\frac{\dpar \CL}{\dpar (\dpar_\nu \varphi)}\right) \dpar_\mu \varphi+\cdots~,
\end{equation}
where the ellipsis denote ghost terms. We thus see indeed the emergence of the usual Noether identities.

\paragraph{Why quasi-isomorphisms?} 
Finally, let us comment a bit more on the role of quasi-isomorphisms. Besides the many mathematical reasons for using them rather than ordinary isomorphisms, they also allow us to identify field theories that are related by integrating out auxiliary fields as we shall explain now using a simple example. Consider two classical field theories with actions
\begin{equation}
 \begin{aligned}
   S\ &:= \ \int_{\FR^{1,d}} \dd^{d+1} x~\Big\{\tfrac12 \varphi(-\dpar_\mu\dpar^\mu-m^2)\varphi-\tfrac{\lambda}{4!} \varphi^4\Big\}~,\\
  \tilde S\ &:= \ \int_{\FR^{1,d}} \dd^{d+1} x~\Big\{\tfrac12 \varphi(-\dpar_\mu\dpar^\mu-m^2)\varphi+\tfrac12 X^2+\tfrac12\sqrt{\tfrac{\lambda}{3}}X\varphi^2\Big\}~,
 \end{aligned}
\end{equation}
where $\varphi$ and $X$ are real scalar fields on Minkowski space $\FR^{1,d}$. The equations of motion read as
\begin{equation}
\begin{aligned}
 S~&:~(-\dpar_\mu\dpar^\mu-m^2)\varphi-\tfrac{\lambda}{3!}\varphi^3\ =\ 0~,\\
 \tilde S~&:~(-\dpar_\mu\dpar^\mu-m^2)\varphi+\sqrt{\tfrac{\lambda}{3}} X\varphi\ =\ 0 \eand X+\tfrac12\sqrt{\tfrac{\lambda}{3}}\varphi^2\ =\ 0~
\end{aligned}
\end{equation}
and $S$ clearly arises from $\tilde S$ by integrating out the auxiliary field $X$.

Following the BV formalism and introducing antifields (which amounts to the Koszul--Tate resolution), leads to two $L_\infty$-algebras $\sL$ and $\tilde \sL$ given by 
\begin{subequations}
\begin{equation}
\begin{gathered}
\underbrace{*}_{=:\,\sL_0}\ \longrightarrow\ \underbrace{\CCC^\infty(\FR^{1,d})}_{=:\,\sL_1} \ \longrightarrow\  \underbrace{\CCC^\infty(\FR^{1,d})}_{=:\, \sL_2} \ \longrightarrow\  \underbrace{*}_{=:\,\sL_3}~,\\
\mu_1(\varphi_1)\ :=\ (-\dpar_\mu\dpar^\mu-m^2)\varphi_1~,\\
  \mu_3(\varphi_1,\varphi_2,\varphi_3)\ :=\ -\lambda \varphi_1\varphi_2\varphi_3
\end{gathered}
\end{equation}
and
\begin{equation}
\begin{gathered}
 \underbrace{*}_{=:\,\tilde \sL_0}\ \longrightarrow\ \underbrace{\CCC^\infty(\FR^{1,d})\oplus\CCC^\infty(\FR^{1,d})}_{=:\,\tilde\sL_1} \ \longrightarrow\  \underbrace{\CCC^\infty(\FR^{1,d})\oplus \CCC^\infty(\FR^{1,d})}_{=:\,\tilde\sL_2} \ \longrightarrow\  \underbrace{*}_{=:\,\tilde\sL_3}~,\\
 \tilde \mu_1(\varphi_1+ X_1)\ :=\ (-\dpar_\mu\dpar^\mu-m^2)\varphi_1 + X_1~,\\
 \tilde \mu_2(\varphi_1+ X_1,\varphi_2+ X_2)\ :=\ \sqrt{\tfrac{\lambda}{3}} \Big((X_1\varphi_2+X_2\varphi_1)+\varphi_1\varphi_2\Big)~.
\end{gathered}
\end{equation}
\end{subequations}
The identity map contained in $\tilde \mu_1$ makes it obvious that the graded vector spaces underlying $\sL$ and $\tilde \sL$ have the same cohomology and we define a cochain map $\phi_1:\tilde \sL\to\sL$ by setting
\begin{equation}
\begin{gathered}
 \phi_1\,:\, \tilde\sL_1\ \rightarrow\  \sL_1\ewith \phi_1(\varphi+ X)\ :=\ \varphi~,\\
 \phi_1\, :\, \tilde \sL_2\ \rightarrow\  \sL_2\ewith \phi_1(\zeta+ Y)\ :=\ \zeta~,
\end{gathered}
\end{equation}
that is, we obtain a quasi-isomorphism of cochain complexes. To extend this to a quasi-isomorphism between $\tilde \sL$ and $\sL$, we note that all higher products are of the form $\tilde \mu_i:\tilde \sL_1\times\cdots\times \tilde\sL_1\rightarrow \tilde\sL_2$ and $\mu_i:\sL_1\times\cdots\times \sL_1\rightarrow \sL_2$. Thus, we are only interested in the defining equation of $L_\infty$-morphisms,~\eqref{eq:L_infty_morphism}, for all arguments in $\tilde \sL_1$. Upon reducing to the non-trivial higher products and truncating the morphism at the level 2, we obtain
\begin{equation}
 \begin{aligned}
  i\ =\ 1~&:~\mu_1(\phi_1(\varphi_1+X_1))\ =\ \phi_1(\tilde \mu_1(\varphi_1+X_1))~,\\
  i\ =\ 2~&:~\mu_1(\phi_2(\varphi_1+X_1,\varphi_2+X_2))\ =\\
   &\kern1cm =\ \phi_1(\tilde\mu_2(\varphi_1+X_1,\varphi_2+X_2))\, -\\
  &\kern2cm -\phi_2(\tilde\mu_1(\varphi_1+X_1),\varphi_2+X_2)-\phi_2(\tilde\mu_1(\varphi_2+X_2),\varphi_1+X_1)~,\\
  i\ =\ 3~&:~\mu_3(\phi_1(\varphi_1+X_1),\phi_1(\varphi_2+X_2),\phi_1(\varphi_3+X_3))\ = \\
  &\kern1cm=\ -[\phi_2(\tilde \mu_2(\varphi_1+X_1,\varphi_2+X_2),\varphi_3+X_3)+{\rm cyclic}]~.
 \end{aligned}
\end{equation}
These equations are fulfilled by setting
\begin{equation}
\begin{gathered}
\phi_2\, :\, \tilde\sL_1\times \tilde\sL_1\ \rightarrow\  \sL_1\ewith\phi_2(\varphi_1+ X_1,\varphi_2+ X_2)\ :=\ 0~,\\
\phi_2\,:\,\tilde\sL_1\times \tilde\sL_2\ \rightarrow\  \sL_2\ewith \phi_2(\varphi_1+ X_1,\zeta_1+ Y_1)\ :=\ \sqrt{\tfrac{\lambda}{3}}\varphi_1Y_1~. 
\end{gathered}
\end{equation}
Hence, we may conclude that the $L_\infty$-algebras $\sL$ and $\tilde \sL$ corresponding to classically equivalent field theories are quasi-isomorphic.

This observation can be generalised: two classical field theories are classically equivalent if their $L_\infty$-algebras are quasi-isomorphic. We note that a related notion of equivalence was discussed in~\cite{Barnich:2004cr}.

\subsection{Chern--Simons theory and its higher analogues}\label{ssec:HCST}

As a first detailed example, let us specialise from the general homotopy MC action~\eqref{eq:MCAction} to (higher) Chern--Simons theory.

 Let $M$ be a $d$-dimensional smooth compact oriented manifold with $d\geq 3$. Let $\sL:=\bigoplus_{k=-d+3}^0\sL_k$ be a cyclic $L_\infty$-algebra to which we shall refer as the {\em gauge $L_\infty$-algebra} in the following. The tensor product $L_\infty$-algebra $\Omega^\bullet(M,\sL)$ as defined in~\eqref{eq:HCSLInfinity} then has a cyclic structure of degree~$-d-(-d+3)=-3$ and we can write down the corresponding homotopy MC action~\eqref{eq:MCAction}. This action defines higher Chern--Simons theory with trivial underlying principal $\infty$-bundles. For instance, in the case when $d=3$, the gauge $L_\infty$-algebra $\sL$ is an ordinary Lie algebra and we recover ordinary Chern--Simons theory, see~\eqref{eq:CS3EoM} and~\eqref{eq:CS_action_from_MC}. In the case when $d=4$, $\sL$ is a 2-term $L_\infty$-algebra or, equivalently, a Lie 2-algebra, see~\eqref{eq:CS4EoM}. To obtain the classical $L_\infty$-structure, we simply enlarge the space of fields consisting of the gauge potentials by all ghosts and higher ghosts and then once more by all corresponding antifields.

\paragraph{\mathversion{bold}Case $d=3$.} In the case of ordinary Chern--Simons theory in three dimensions, we have additional ghosts $c\in\Omega^0(M,\frg)[1]$ and antifields $A^+\in\Omega^2(M,\frg)[-1]$ and $c^+\in\Omega^3(M,\frg)[-2]$. Hence, with $\sfa=c+A+A^++c^+$, the symplectic form $\omega_{\rm BV}$ on $\frF_{\rm BV}$ defined in~\eqref{eq:BVSympForm} becomes
\begin{equation}
 \omega_{\rm BV}\ =\ \int_M\left\{\langle \delta A, \delta A^+\rangle_\frg-\langle \delta c, \delta c^+\rangle_\frg\right\}~.
\end{equation}
Thus, the induced the Poisson bracket~\eqref{eq:InducedBVPoisson} reads explicitly as
\begin{equation}
\begin{aligned}
 \{F,G\}_{\rm BV}\ &=\ \int^*_M \left\{F\left\langle\frac{\overset{\leftarrow}{\delta}}{\delta A^+},\frac{\overset{\rightarrow}{\delta}}{\delta A}\right\rangle_{\frg^*} G-F\left\langle\frac{\overset{\leftarrow}{\delta}}{\delta A},\frac{\overset{\rightarrow}{\delta}}{\delta A^+}\right\rangle_{\frg^*} G\,-\right.\\
 &\kern2cm-\left.F\left\langle\frac{\overset{\leftarrow}{\delta}}{\delta c^+},\frac{\overset{\rightarrow}{\delta}}{\delta c}\right\rangle_{\frg^*} G+F\left\langle\frac{\overset{\leftarrow}{\delta}}{\delta c},\frac{\overset{\rightarrow}{\delta}}{\delta c^+}\right\rangle_{\frg^*} G\right\}
\end{aligned}
\end{equation}
for $F,G\in\CCC^\infty(\frF_{\rm BV})$ and $\int^*_M \langle -, -\rangle_{\frg^*}$ is the dual inner product induced by $\int_M \langle - ,-\rangle_\frg$. The BV action~\eqref{eq:BVaction} reads as 
\begin{equation}
\begin{aligned}
 S_{\rm BV} \ &= \ \int_M\Big\{\tfrac12\langle A,\dd A\rangle_\frg+\tfrac{1}{3!}\langle A,[A,A]\rangle_\frg\,-\\
  &\kern2cm-\langle c,\dd A^+\rangle_\frg-\langle c,[A,A^+]\rangle_\frg+\tfrac12\langle c^+,[c,c]\rangle_\frg\Big\}~,
 \end{aligned}
\end{equation}
which is the Hamiltonian for the vector field $Q_{\rm BV}$, which acts on an element $F\in\CCC^\infty(\frF_{\rm BV})$ according to
\begin{equation}
\begin{aligned}
 Q_{\rm BV} F\ &:=\ \{S_{\rm BV},F\}_{\rm BV}\\
 & \phantom{:}=\ \int^*_M\left\{-
  \left\langle\frac{\delta S_{\rm BV}}{\delta A^+},\frac{\delta F}{\delta A}\right\rangle_{\frg^*}-
 \left\langle \frac{\delta S_{\rm BV}}{\delta A},\frac{\delta F}{\delta A^+}\right\rangle_{\frg^*}\,-\right.\\
 &\hspace{3cm}\left.
 -\left\langle\frac{\delta S_{\rm BV}}{\delta c^+},\frac{\delta F}{\delta c}\right\rangle_{\frg^*}-
 \left\langle\frac{\delta S_{\rm BV}}{\delta c},\frac{\delta F}{\delta c^+}\right\rangle_{\frg^*}\right\}.
\end{aligned}
\end{equation}
For the coordinate functions, we obtain explicitly
\begin{equation}
\begin{gathered}
  Q_{\rm BV}c\ =\ -\tfrac12[c,c]~,~~~Q_{\rm BV}A\ =\ \dd c+[A,c]~,\\
  Q_{\rm BV}A^+\ =\ -\dd A-\tfrac12[A,A]-[c,A^+]~,~~~Q_{\rm BV}c^+\ =\ \dd A^++[A,A^+]-[c,c^+]~.
\end{gathered}
\end{equation}

Note that we can also specialise the extended BV action~\eqref{eq:eBVaction}, which is suitable for gauge fixing. Here, the above field content is further extended by the trivial pair $(\bar c,b)\in\Omega^1(M,\frg)[-1]\oplus \Omega^1(M,\frg)[0]$ together with a trivial pair of corresponding antifields $(\bar c^+,b^+)\in \Omega^2(M,\frg)[0]\oplus \Omega^2(M,\frg)[-1]$. The extended BV action~\eqref{eq:eBVaction} then reads as 
\begin{equation}
\begin{aligned}
 S_{\rm eBV}\ &=\ \int_M\Big\{\tfrac12\langle A,\dd A\rangle_\frg+\tfrac{1}{3!}\langle A,[A,A]\rangle_\frg\,-\\
 &\kern2cm-\langle c,\dd A^+\rangle_\frg+\langle A,[A^+,c]\rangle_\frg+\tfrac12\langle c^+,[c,c]\rangle_\frg-\langle b,\bar c^+\rangle_\frg\Big\}~,
 \end{aligned}
\end{equation}
resulting in 
\begin{equation}
\begin{gathered}
 Q_{\rm eBV}c\ =\ -\tfrac12[c,c]~,~~~Q_{\rm eBV}A\ =\ \dd c+[A,c]~,\\
  Q_{\rm eBV}A^+\ =\ -\dd A-\tfrac12[A,A]-[c,A^+]~,~~~Q_{\rm eBV}c^+\ =\ \dd A^++[A,A^+]-[c,c^+]~,\\
 Q_{\rm eBV}\bar c\ =\ b~,~~~Q_{\rm eBV}\bar c^+\ =\ 0~,~~~Q_{\rm eBV}b^+\ =\ -\bar c^+~,~~~Q_{\rm eBV} b\ =\ 0~.\\ 
\end{gathered}
\end{equation}

\paragraph{\mathversion{bold}Case $d=4$.} 
Next, let us discuss the simplest higher case $d=4$ in detail. Here, the gauge algebra is of the form $\sL=\sL_{-1}\oplus\sL_0$. The inner product identifies $(\sL_{-1})^*\cong\sL_0$ so that $\sL_{-1}$ and $\sL_0$ must be of the same dimension. We have the decomposition $\Omega_1^\bullet(M,\sL)\cong \Omega^1(M,\sL_0)\oplus\Omega^2(M,\sL_{-1})$ of homogeneous degree~1 elements in $\Omega^\bullet(M,\sL)$. Consequently, we obtain a $1$-form gauge potential $A\in\Omega^1(M,\sL_0)[0]$ and a $2$-form gauge potential $B\in\Omega^2(M,\sL_{-1})[0]$, respectively.  The homotopy MC action~\eqref{eq:MCAction} then becomes with $a=A+B$
\begin{equation}\label{eq:4dHCSAction}
 S_{\rm MC}\ =\ \int_M\Big\{\langle B,\dd A+\tfrac12\mu_2(A,A)+\tfrac12\mu_1(B)\rangle_\sL+\tfrac{1}{4!}\langle\mu_3(A,A,A), A\rangle_\sL\Big\}~,
\end{equation}
and the curvature~\eqref{eq:Curvature} reduces to
\begin{equation}\label{eq:4DCurvatures}
\begin{gathered}
 \CF\ =\ \dd A+\tfrac12\mu_2(A,A)+\mu_1(B)\ \in\ \Omega^2(M,\sL_0)~,\\
 H\ =\ \dd B+\mu_2(A,B)-\tfrac{1}{3!}\mu_3(A,A,A)\ \in\ \Omega^3(M,\sL_{-1})~,
\end{gathered}
\end{equation}
cf.~\eqref{eq:Lie-2-algebra_curvatures}. As mentioned before, in these and the following formulas, the higher products $\mu_i$ will not see the form degree of the various fields. Since any homogeneous degree~0 element of $\Omega^\bullet(M,\sL)$ decomposes into $c\in\Omega^0(M,\sL_0)[0]$ and $\Lambda\in\Omega^1(M,\sL_{-1})[0]$, the gauge transformations~\eqref{eq:GaugeTrafo} and~\eqref{eq:GaugeTrafoCurvature}  reduce to
\begin{subequations}
\begin{equation}
\begin{gathered}
 \delta_{c,\Lambda}A\ =\ \dd c+\mu_2(A,c)+\mu_1(\Lambda)~,\\
 \delta_{c,\Lambda}B\ =\ -\mu_2(c,B)+\dd\Lambda+\mu_2(A,\Lambda)+\tfrac12\mu_3(c,A,A)
\end{gathered}
\end{equation}
and
\begin{equation}
\begin{gathered}
 \delta_{c,\Lambda}\CF\ =\ -\mu_2(c,\CF)~,\\
 \delta_{c,\Lambda}H\ =\ -\mu_2(c,H)+\mu_2(\CF,\Lambda)-\mu_3(\CF,A,c)~.
\end{gathered}
\end{equation}
\end{subequations}
Recall also the discussion of the meaning of the fake curvature $\CF$ in Section~\ref{ssec:HMCE} and in the previous paragraph. 

To write down the BV action~\eqref{eq:BVaction}, see also~\eqref{eq:Example2TermAction}, we first note that the fields decompose as
\begin{subequations}
\begin{equation}
\begin{gathered}
 a\ =\ A+B\ \in\ \Omega^1(M,\sL_0)[0]\oplus\Omega^2(M,\sL_{-1})[0]~,\\
 a^+\ =\ A^++B^+\ \in\  \Omega^3(M,\sL_{-1})[-1]\oplus\Omega^2(M,\sL_0)[-1]~,
 \end{gathered}
\end{equation}
while for the ghosts we obtain
 \begin{equation}
\begin{gathered}
 c_0\ =\ c_0^0+c_0^1\ \in\ \Omega^0(M,\sL_0)[1]\oplus\Omega^1(M,\sL_{-1})[1]~,\\
 c_0^+\ =\ c_0^{0+}+c_0^{1+}\ \in\ \Omega^4(M,\sL_{-1})[-2]\oplus\Omega^3(M,\sL_0)[-2]~,\\
 c_{-1} \ \in\ \Omega^0(M,\sL_{-1})[2]~,~~~ c_{-1}^+\ \in\ \Omega^4(M,\sL_0)[-3]~.
  \end{gathered}
\end{equation}
\end{subequations}
The full BV action~\eqref{eq:BVaction} is then $S_{\rm BV}\ =\ S_{\rm MC}+S_{\rm gh}$ with $S_{\rm MC}$ given by~\eqref{eq:4dHCSAction} and
\begin{equation}
\begin{aligned}
 S_{\rm gh} \ &:=\ \int_M\Big\{\langle \dd A^+, c_0^0\rangle_\sL+\langle \dd B^+, c_0^1\rangle_\sL+\langle c_0^1,\mu_1(A^+)\rangle_\sL\,+\langle c_{-1},\dd c_0^{1+}+\mu_1(c_0^{0+})\rangle_\sL\,-\\
 &\kern1.5cm-\langle \mu_2(c_0^{1+},c_{-1}) + \mu_2(B^+,c_0^1) + \mu_2(A^+, c_0^0),A\rangle_\sL\,+\langle B,\mu_2(B^+, c_0^0)\rangle_\sL-\\
  &\kern1.5cm-\tfrac12\langle\mu_2(B^+,c_{-1}), B^+\rangle_\sL\,+\tfrac12\langle c_0^{0+},\mu_2( c_0^0, c_0^0)\rangle_\sL-\langle\mu_2( c_0^0,c_0^1),c_0^{1+}\rangle_\sL\,-\\
  &\kern1.5cm-\langle \mu_2( c_0^0,c_{-1}), c_{-1}^+\rangle_\sL+\tfrac12\langle \mu_3(A,B^+, c_0^0),A\rangle_\sL-\tfrac12\langle\mu_3(A,c_0^{1+}, c_0^0),A\rangle_\sL+\\
  &\kern1.5cm+\tfrac12\langle \mu_3( c_0^0, c_0^0,c_0^{1+}),A\rangle_\sL\,+\tfrac{1}{2\cdot 2}\langle\mu_3(B^+, c_0^0, c_0^0),B^+\rangle_\sL+\\
  &\kern1.5cm+\tfrac{1}{3!}\langle \mu_3( c_0^0, c_0^0,c_{-1}^+),c_0^0\rangle_\sL\Big\}~.
  \end{aligned}
 \end{equation}
It induces a homological vector field $Q_{\rm BV}$ on $\frF_{\rm BV}$ acting on $F\in \CCC^\infty(\frF_{\rm BV})$ as
\begin{equation}
\begin{aligned}
 Q_{\rm BV} F\ &:=\ 
 \int^*_M\left\{- \left\langle\frac{\delta S_{\rm BV}}{\delta A^+},\frac{\delta F}{\delta A}\right\rangle_{\sL^*}-
  \left\langle\frac{\delta S_{\rm BV}}{\delta A},\frac{\delta F}{\delta A^+}\right\rangle_{\sL^*}-
   \left\langle\frac{\delta S_{\rm BV}}{\delta B^+},\frac{\delta F}{\delta B}\right\rangle_{\sL^*}-\right.\\
     &\kern2cm-\left\langle\frac{\delta S_{\rm BV}}{\delta B},\frac{\delta F}{\delta B^+}\right\rangle_{\sL^*}-
 \left\langle\frac{\delta S_{\rm BV}}{\delta c_0^{0+}},\frac{\delta F}{\delta c_0^0}\right\rangle_{\sL^*}-
  \left\langle\frac{\delta S_{\rm BV}}{\delta c_0^0},\frac{\delta F}{\delta c_0^{0+}}\right\rangle_{\sL^*}-\\
   &\kern2cm- \left\langle\frac{\delta S_{\rm BV}}{\delta c_0^{1+}},\frac{\delta F}{\delta c_0^1}\right\rangle_{\sL^*}-
    \left\langle\frac{\delta S_{\rm BV}}{\delta c_0^1},\frac{\delta F}{\delta c_0^{1+}}\right\rangle_{\sL^*}+
 \left\langle\frac{\delta S_{\rm BV}}{\delta c_{-1}^{+}},\frac{\delta F}{\delta c_{-1}}\right\rangle_{\sL^*}+\\
  &\kern2cm+ \left.\left\langle\frac{\delta S_{\rm BV}}{\delta c_{-1}},\frac{\delta F}{\delta c_{-1}^{+}}\right\rangle_{\sL^*}\right\}.
\end{aligned}
\end{equation}
Explicitly, we have 
 \begin{equation}
 \begin{aligned}
   Q_{\rm BV}  A\ &=\ \mathrm{d} c_0^0 + \mu_2 (A, c_0^0)  - \mu_1( c_0^1)  ~,\\
 Q_{\rm BV}  B\ &=\ -\mu_2(c_0^0,B) + \mathrm{d} c_0^1 + \mu_2 (A, c_0^1) + \tfrac{1}{2} \mu_3(c_0^0, A, A)\,-   \\ 
  &\kern1cm -\mu_2( B^+, c_{-1})+ \tfrac{1}{2} \mu_3( B^+ ,c_0^0 , c_0^0) ~,\\
 Q_{\rm BV}  A^+\ &=\  -\dd B-\mu_2(A,B)  + \tfrac{1}{3!} \mu_3( A, A, A) - \mu_2(A^+, c_0^0) -  \mu_2(B^+, c_0^1)\, +\\
 &\kern1cm+ \mu_2(c_{-1}, c_0^{1+}) +  \mu_3( A, B^+, c_0^0) 
 + \tfrac{1}{2} \mu_3( c_0^0, c_0^0, c_0^{1+})  ~,\\
 Q_{\rm BV}  B^+\ &=\ - \dd A -\tfrac{1}{2} \mu_2(A,A) - \mu_1(B) - \mu_2(B^+, c_0^0) ~,\\
Q_{\rm BV}  c_0^{0+} \ &=\ - \mathrm{d}A^+ + \mu_2(A, A^+) + \mu_2(B, B^+) - \mu_2(c_0^0, c_0^{0+})  
   + \mu_2(c_0^1, c_0^{1+}) \\
   &\kern1cm +  \mu_2(c_{-1}, c_{-1}^+) + \tfrac{1}{2} \mu_3(A, A,  B^+) + \tfrac{1}{2} \mu_3( B^+, B^+, c_0^0) \\
     &\kern1cm  -   \mu_3( A ,c_0^0 , c_0^{1+}) + \tfrac{1}{2}   \mu_3(c_0^0 ,c_0^0 , c_{-1}^+)   ~,\\
 Q_{\rm BV} \ c_0^1\ &=\ -\mathrm{d} c_{-1}  - \mu_2(A , c_{-1}) - \mu_2(c_0^0 , c_0^1) 
 + \tfrac{1}{2} \mu_3(A, c_0^0 , c_0^0)  ~, \\
 Q_{\rm BV}  c_0^{1+}\ &=\ - \mathrm{d}B^+  -\mu_1(A^+) +\mu_2(A,B^+) -\mu_2(c_0^0, c_0^{1+}) ~, \\
 Q_{\rm BV}  c_{-1}\ &=\ -\mu_2(c_0^0, c_{-1}) + \tfrac{1}{3!} \mu_3(c_0^0, c_0^0, c_0^0)  ~, \\
 Q_{\rm BV}  c_{-1}^+\ &=\  - \mathrm{d} c_0^{1+}  - \mu_1(c_0^{0+}) + \mu_2(A, c_0^{1+})  -\tfrac{1}{2}\mu_2(B^+, B^+)  - \mu_2(c_{-1}^+, c_0^0)~,
\end{aligned}
 \end{equation}
 cf.~\eqref{eq:QBVonComponentFields2Term}. 
 
The extension~\eqref{eq:eBVaction} of $S_{\rm BV}$ by trivial pairs requires the introduction of the additional quadruples
\begin{subequations}
\begin{equation}
\begin{gathered}
 \bar c_{0,-1}\ =\ \bar c_{0,-1}^0+\bar c_{0,-1}^1\ \in \ \big(\Omega^0(M,\sL_0)\oplus \Omega^1(M,\sL_{-1})\big)[-1]~,\\
 b_{0,-1}\ =\ b_{0,-1}^0+b_{0,-1}^1\ \in \ \big(\Omega^0(M,\sL_0)\oplus \Omega^1(M,\sL_{-1})\big)[0]~,\\
 \bar c^+_{0,-1}\ =\ \bar c^{3+}_{0,-1}+\bar c^{2+}_{0,-1}\ \in \ \big(\Omega^3(M,\sL_0)\oplus \Omega^2(M,\sL_{-1})\big)[0]~,\\
 b^+_{0,-1}\ =\ b^{3+}_{0,-1}+b^{2+}_{0,-1}\ \in \ \big(\Omega^3(M,\sL_0)\oplus \Omega^2(M,\sL_{-1})\big)[-1]~,\\
\end{gathered}
\end{equation}
and
\begin{equation}
\begin{gathered}
 \bar c_{-1,-2}\ \in \ \Omega^0(M,\sL_{-1})[-2]~,~~~b_{-1,-2}\ \in \ \Omega^0(M,\sL_{-1})[-1]~,\\
 \bar c^+_{-1,-2}\ \in \ \Omega^3(M,\sL_{-1})[1]~,~~~b^+_{-1,-2}\ \in \ \Omega^3(M,\sL_{-1})[0]~,\\
 \bar c_{-1,0}\ \in \ \Omega^0(M,\sL_{-1})[0]~,~~~b_{-1,0}\ \in \ \Omega^0(M,\sL_{-1})[1]~,\\
 \bar c^+_{-1,0}\ \in \ \Omega^3(M,\sL_{-1})[-1]~,~~~b^+_{-1,0}\ \in \ \Omega^3(M,\sL_{-1})[-2]~,
\end{gathered}
\end{equation}
\end{subequations}
where all $\Omega^i(M,\sL_j)$ are regarded as ungraded vector spaces with elements of degree~0. The additional contribution to the extended BV action~\eqref{eq:eBVaction} is
\begin{equation}
\begin{aligned}  
     S_{\rm tp}\ &:=\ \int_M\Big\{ -\langle b^0_{0,-1},\bar c^{3+}_{0,-1}\rangle_\sL-\langle b^1_{0,-1},\bar c^{2+}_{0,-1}\rangle_\sL+\langle b_{-1,-2},\bar c^{+}_{-1,-2}\rangle_\sL+\langle b_{-1,0},\bar c^{+}_{-1,0}\rangle_\sL\Big\}~.
\end{aligned}
\end{equation}

\paragraph{\mathversion{bold}Minimal model and $L_\infty$-quasi-isomorphism.} 
We now construct the minimal model of the gauge $L_\infty$-algebra $\Omega^\bullet(M,\sL)$ underlying higher Chern--Simons theory following our discussion in Section~\ref{ssec:quasiisomorphism}. 

We start by noting that the cochain complex underlying the $L_\infty$-algebra~\eqref{eq:HCSLInfinity} is the tensor product of two cochain complexes: the de Rham complex and the complex arising from the gauge $L_\infty$-algebra $\sL$. Since the cohomology $H^\bullet_{\mu_1}(\sL)$ of $\sL$ is evidently free, K{\"u}nneth's theorem, see e.g.~\cite{Weibel:1994aa}, yields the isomorphisms
\begin{equation}
\begin{gathered}
 H^\bullet_{\hat \mu_1}(\Omega^\bullet(M,\sL))\ :=\ \bigoplus_{k\in\RZ}H^k_{\hat \mu_1}(\Omega^\bullet(M,\sL))~,\\
 H^k_{\hat \mu_1}(\Omega^\bullet(M,\sL))\ \cong\  \bigoplus_{\substack{i+j=k\\ 0\leq i\leq d\\-n+1\leq j\leq 0}} H^i_{\rm dR}(M)\otimes H^j_{\mu_1}(\sL)~,
 \end{gathered}
\end{equation}
where $H^i_{\rm dR}(M)$ denotes the $i$-th de Rham cohomology group.

To construct the $L_\infty$-structure on $H^\bullet_{\hat{\mu}_1}(\Omega^\bullet(M,\sL)$, we note that another consequence of the cochain complex underlying the $L_\infty$-algebra~\eqref{eq:HCSLInfinity} being the tensor product of cochain complexes of vector spaces is that it splits in the sense of~\eqref{eq:splitComplex}; see also Appendix~\ref{app:Hodge}. Hence, we have
\begin{equation}
 \myxymatrix{ \ar@(dl,ul)[]^{h_{\hat{\mu}_1}}&\kern-1cm \Omega^\bullet(M,\sL)~ \ar@<+2pt>@{->>}[rr]^{\kern-1cm p_{\hat{\mu}_1}}  & & ~~H^\bullet_{\hat{\mu}_1}(\Omega^\bullet(M,\sL)) \ar@<+2pt>@{^(->}[ll]^{\kern-1cm e_{\hat{\mu}_1}}},
\end{equation}
where $p_{\hat{\mu}_1}\circ e_{\hat{\mu}_1}=\id$ and $h_{\hat{\mu}_1}:\Omega^\bullet(M,\sL)\to\Omega^\bullet(M,\sL)$ with $\hat{\mu}_1=\hat{\mu}_1\circ h_{\hat{\mu}_1}\circ \hat{\mu}_1$ a contracting homotopy of $P_{\hat{\mu}_1}:=e_{\hat{\mu}_1}\circ p_{\hat{\mu}_1}$. To construct $h_{\hat{\mu}_1}$ explicitly, we 
assume that we have already found\footnote{See Appendix~\ref{app:Hodge} for an explicit example.} a contracting homotopy $h_{\mu_1}:\sL\to\sL$ of $P_{\mu_1}:=e_{\mu_1}\circ p_{\mu_1}$ and construct a contracting homotopy $h_\dd:\Omega^\bullet(M)\to\Omega^\bullet(M)$ of $P_\dd:=e_\dd\circ p_\dd$. In order to write down the latter, we fix a Riemannian metric on $M$ and let $\dd^\dagger$ be the adjoint of $\dd$ with respect to the standard inner product $\langle\alpha_1,\alpha_2\rangle=\int_M\alpha_1\wedge{\star\alpha_2}$ for $\alpha_{1,2}\in\Omega^k(M)$ with $\star$ the Hodge operator for the chosen metric.  Using the Green operator
\begin{equation}\label{eq:GreenLaplace}
 G|_{\im(\dd)\oplus\im(\dd^\dagger)}\ :=\ \Delta^{-1}\eand G|_{\ker(\Delta)}\ :=\ 0 \ewith \Delta\ :=\ \dd\dd^\dagger+\dd^\dagger\dd~,
\end{equation}
we define
\begin{equation}
 h_\dd\ :=\ \dd^\dagger G~.
\end{equation}
Together with the identities
\begin{equation}\label{eq:GreenLaplaceIdentities}
 G\dd\ =\ \dd G~,~~~ G\dd^\dagger\ =\ \dd^\dagger G~,\eand {\star G}\ =\ {G\star}~,
\end{equation}
it is then easily seen that $\dd=\dd\circ h_\dd \circ\dd$. We thus obtain
\begin{equation}
 1\ =\ P_\dd +h_\dd\circ\dd+\dd\circ h_\dd\ =\ P_\dd +\Delta G\quad\Longrightarrow\quad P_\dd\ =\ 1 - \Delta G~,
\end{equation}
which is the projector onto the harmonic forms $\Omega_{\rm h}^\bullet(M)$ under the Hodge decomposition
\begin{equation}
 \Omega^k(M)\ \cong\ \Omega^k_{\rm h}(M)\oplus\dd\Omega^{k-1}(M)\oplus\dd^\dagger\Omega^{k+1}(M)~,
\end{equation} 
cf.~Appendix~\ref{app:Hodge}. Postcomposing the projector with the Hodge isomorphism 
$\Omega_{\rm h}^\bullet(M)\cong H^\bullet_{\rm dR}(M)$, we obtain a projector on de Rham cohomology. We now combine the homotopies $h_\dd$ and $h_{\mu_1}$ as
\begin{equation}\label{eq:ContractingHomotopyHCS}
 h_{\hat{\mu}_1}\ :=\ \tfrac12(h_\dd\otimes 1+1\otimes h_{\mu_1}+P_{\dd}\otimes h_{\mu_1}+h_\dd\otimes P_{\mu_1})~.
\end{equation}
Using $ \hat{\mu}_1=\dd \otimes 1+1\otimes \mu_1$, it now follows that $\hat{\mu}_1=\hat{\mu}_1\circ h_{\hat{\mu}_1}\circ\hat{\mu}_1$, as desired.

Using the contracting homotopy~\eqref{eq:ContractingHomotopyHCS}, it is now easy to adapt the formulas~\eqref{eq:minimalQuasiIsomorphism} for the quasi-isomorphism between $\Omega^\bullet(M,\sL)$ and $H_{\hat{\mu}_1}^\bullet(\Omega^\bullet(M,\sL))$. For the sake of clarity, we shall only display the formulas in homogeneous degree~1. We obtain 
\begin{equation}
\begin{aligned}
\phi_1(a') \ &=\  e_{\hat{\mu}_1}(a')~,\\
 \phi_2(a',a') \ &=\  - h_{\hat{\mu}_1}\Big(\hat{\mu}_2\big(e_{\hat{\mu}_1}(a'),e_{\hat{\mu}_1}(a')\big)\Big)\,,\\
\phi_3(a',a',a')\ &=\   -3h_{\hat{\mu}_1}\Big(\hat{\mu}_2 \big( \phi_2(a', a'),e_{\hat{\mu}_1}(a') \big)  \Big) -  h_{\hat{\mu}_1} \Big( \hat{\mu}_3 \big(e_{\hat{\mu}_1}(a'), e_{\hat{\mu}_1}(a' ),e_{\hat{\mu}_1}(a') \big) \Big)\,,  \\
  \phi'_4(a',a',a',a')\ &= \ - 3h_{\hat{\mu}_1} \Big( \hat{\mu}_2  \big(  \phi_2  (a',a'), \phi_2  (a',a') \big) \Big) 
 - 4 h_{\hat{\mu}_1} \Big( \hat{\mu}_2  \big(  e_{\hat{\mu}_1} (a'), \phi_3  (a', a',a') \big) \Big)\,-\\
 &\kern1cm - 6 h_{\hat{\mu}_1} \Big(\hat{\mu}_3  \big( e_{\hat{\mu}_1}(a'), e_{\hat{\mu}_1}(a'),  
\phi_2 \left( a' , a' \right)  \big) \Big)\,-  \\
&\kern1cm -  h_{\hat{\mu}_1} \Big( \hat{\mu}_4  \big( e_{\hat{\mu}_1}(a'), e_{\hat{\mu}_1}(a'),  
   e_{\hat{\mu}_1}(a'), e_{\hat{\mu}_1}(a') \big) \Big)\,,
\\
  &~\,\vdots
 \end{aligned}
\end{equation}
Hence, the higher products $\hat \mu'_i$ on $H_{\hat{\mu}_1}^\bullet(\Omega^\bullet(M,\sL))$ defined in~\eqref{eq:minimalHigherProducts} are then given for degree~1 elements by
\begin{equation}
\begin{aligned}
\hat{\mu}'_1(a')\ &= \ 0~,\\
 \hat{\mu}'_2(a',a')\ &= \ p_{\hat{\mu}_1}\Big(\hat{\mu}_2\big(e_{\hat{\mu}_1}(a'),e_{\hat{\mu}_1}(a')\big)\Big)\,,\\
 \hat{\mu}'_3(a',a',a')\ &= \ 3p_{\hat{\mu}_1} \Big(\hat{\mu}_2 \big( \phi_2(a', a'),e_{\hat{\mu}_1}(a') \big)  \Big) +
 p_{\hat{\mu}_1} \Big(\hat\mu_3 \big( e_{\hat{\mu}_1}(a'), e_{\hat{\mu}_1}(a'), e_{\hat{\mu}_1}(a')\big) \Big)\,,\\
  \hat{\mu}'_4(a',a',a',a')\ &= \ 3 p_{\hat{\mu}_1} \Big(\hat\mu_2  \big(  \phi_2  (a',a'), \phi_2  (a',a') \big) \Big)  + 4 p_{\hat{\mu}_1} \Big( \hat\mu_2 \big(  e_{\hat{\mu}_1} (a'), \phi_3  (a', a',a') \big) \Big)\,+ \\
 & \kern1cm +  6 p_{\hat{\mu}_1} \Big( \hat\mu_3  \big( e_{\hat{\mu}_1}(a'), e_{\hat{\mu}_1}(a'),  
\phi_2 \left( a' , a' \right)  \big) \Big)\,+  \\
&\kern1cm +   p_{\hat{\mu}_1} \Big( \hat\mu_4  \big( e_{\hat{\mu}_1}(a'), e_{\hat{\mu}_1}(a'),  
   e_{\hat{\mu}_1}(a'), e_{\hat{\mu}_1}(a') \big) \Big)\,,\\
  &~\,\vdots
 \end{aligned}
\end{equation}

We conclude that the resulting classical field theory equivalent to higher Chern--Simons theory has a much simpler or reduced space of fields, but this is compensated by the interactions becoming much more involved. We shall see this phenomenon in more examples in the following.

\paragraph{Supersymmetric extension: topological setting.} 
Let us briefly consider the supersymmetric extension of the above classical higher Chern--Simons theory along the lines discussed in Section~\ref{ssec:HMCAction}. That is, we introduce the additional fields
\begin{equation}
\begin{gathered}
\phi_1\ =\ Y+\phi\ \in\ \Omega_1^\bullet(M,\sL)\ \cong\ \Omega^1(M,\sL_{0})\oplus\Omega^2(M,\sL_{-1})~,\\
\phi_2\ =\ \big[D-\tfrac12\mu_1(\phi)\big]-Z\ \in\ \Omega_2^\bullet(M,\sL)\ \cong\ \Omega^2(M,\sL_{0})\oplus\Omega^3(M,\sL_{-1})~,\\
\psi_1\ =\ \lambda_1+\chi_2\ \in \Pi\Omega_1^\bullet(M,\sL)\ \cong\ \Pi\Omega^1(M,\sL_{0})\oplus\Pi\Omega^2(M,\sL_{-1})~,\\
\psi_2\ =\ \lambda_2+\chi_3\ \in\ \Pi\Omega_2^\bullet(M,\sL)\ \cong\ \Pi\Omega^2(M,\sL_{0})\oplus\Pi\Omega^3(M,\sL_{-1})~,
\end{gathered}
\end{equation}
where, as before, $\Pi$ is the Gra{\ss}mann-parity changing functor. The action~\eqref{eq:4dHCSAction} is then extended to the appropriate specialisation of the action~\eqref{eq:fermionicMCAction},
\begin{equation}
\begin{aligned}
  S_{\rm STMC}\ &:=\ \int_M\Big\{\langle B,F-\tfrac12\mu_1(B)\rangle_\sL+\tfrac{1}{4!}\langle\mu_3(A,A,A),A\rangle_\sL\,+\\
   &\kern2cm+\langle\chi_2,\lambda_2\rangle_\sL+\langle\chi_3,\lambda_1\rangle_\sL+\langle \phi,D-\tfrac12\mu_1(\phi)\rangle_\sL-\langle Z,Y\rangle_\sL\Big\}~.
\end{aligned}
\end{equation}
The fermionic transformations~\eqref{eq:fermionicMCAction} read as
\begin{subequations}
\begin{equation}
\begin{gathered}
Q\lambda_1\ =\ 0~,~~~~
Q\lambda_2\ =\ \CF+D~,\\
QA\ =\ \lambda_1~,~~~~
QY\ =\ \lambda_1~,\\
QD\ =\ -\nabla\lambda_1+\mu_1(\chi_2)
\end{gathered}
\end{equation}
and
\begin{equation}
\begin{gathered}
Q\chi_2\ =\ 0~,\\
Q\chi_3\ =\ H-\nabla\phi-\tfrac12\mu_2(Y,B-\phi)+\tfrac14\mu_3(Y,A,A)-\tfrac{1}{3!}\mu_3(Y,Y,A)+\tfrac{1}{4!}\mu_3(Y,Y,Y)+Z~,\\
QB\ =\ \chi_2~,~~~
Q\phi\ =\ \chi_2~,\\
QZ\ =\ -\tfrac12\mu_2(\lambda_1,B-\phi)+\tfrac14\mu_3(\lambda_1,A,A)-\tfrac{1}{3!}\mu_3(\lambda_1,Y,A)+\tfrac{1}{4!}\mu_3(\lambda_1,Y,Y)~.
\end{gathered}
\end{equation}
\end{subequations}
where the curvatures $\CF$ and $H$ were defined in~\eqref{eq:4DCurvatures}. As 
shown in Section~\ref{ssec:HMCAction}, we have $Q^2=0$ off-shell.

\paragraph{Supersymmetric extension: physical setting.}
The above supersymmetric extension is similar to a topological twist of supersymmetric higher Chern--Simons theory for the $d=4$, $\CN=2$ tensor multiplet coupled to the $\CN=2$ vector multiplet. The Abelian part of the action was already presented in~\cite{deWit:1980lyi,deWit:2006gn} and with our framework, we extend the action to the non-Abelian case on $M=\FR^4$, as we shall now briefly discuss.

For convenience, we switch to spinor notation and replace the coordinates $x^\mu$ by the coordinates $x^{\alpha\ald}$ with $\alpha,\ald=1,2$. This is possible due to the factorisation of the tangent bundle $TM\otimes\FC\cong S_+\otimes S_-$ into the tensor product of the chiral and anti-chiral spin bundles $S_+$ and $S_-$. We also use R-symmetry indices $i,j=1,2$. The $\CN=2$ vector multiplet then consists of a one-form gauge potential, two pairs of Weyl spinors and 5~real auxiliary fields,
\begin{equation}
 \{A_{\alpha\dot\alpha},\lambda^i_\alpha,\hat\lambda_{i\dot\alpha}, D^{ij}=D^{ji},Y,\hat Y\}~,
\end{equation}
which all take values in $\sL_0$. The $\CN=2$ tensor multiplet consists of a two-form gauge potential, two pairs of Weyl spinors and 5~real auxiliary fields,
\begin{equation}
 \{B_{\alpha\beta}=B_{\beta\alpha},B_{\dot\alpha\dot\beta}=B_{\dot\beta\dot\alpha},\chi_{i\alpha},\hat\chi^i_{\dot\alpha}, \phi^{ij}=\phi^{ji},Z,\hat Z\}~,
\end{equation}
which all take values in $\sL_{-1}$. Here, $B_{\alpha\beta}$ and $B_{\ald\bed}$ encode the self-dual and antiself-dual parts of the two-form $B$. The curvatures~\eqref{eq:4DCurvatures} become
\begin{equation}
\begin{gathered}
\CF_{\alpha\beta}\ =\ F_{\alpha\beta}-\mu_1(B_{\alpha\beta})~,\quad\CF_{\dot\alpha\dot\beta}\ =\ F_{\dot\alpha\dot\beta}-\mu_1(B_{\dot\alpha\dot\beta})~,\\
H_{\alpha\dot\beta}\ =\ \varepsilon^{\dot\gamma\dot\delta}\nabla_{\alpha\dot\gamma} B_{\dot\beta\dot\delta}-\varepsilon^{\gamma\delta}\nabla_{\gamma\dot\beta}B_{\alpha\delta}-\tfrac{1}{3!}[\mu_3(A,A,A)]_{\alpha\dot\beta}~.
\end{gathered}
\end{equation}

The action functional
\begin{equation}\label{eq:4dSUSYAction}
\begin{aligned}
S_{\rm SPMC}\ &:=\ \int \dd^4 x\,\Big\{\,\di\Big[\langle B^{\alpha\beta},F_{\alpha\beta}-\tfrac12\mu_1(B_{\alpha\beta})\rangle_\sfL-\langle B^{\dot\alpha\dot\beta},F_{\dot\alpha\dot\beta}-\tfrac12\mu_1(B_{\dot\alpha\dot\beta})\rangle_\sfL\,+\\
&\kern2cm +\tfrac{1}{2\cdot 4!}\langle [\mu_3(A,A,A)]^{\alpha\dot\beta},A_{\alpha\dot\beta}\rangle_\sfL\Big]+\langle\chi^{i\alpha},\lambda_{i\alpha}\rangle_\sfL+\langle\hat\chi^{i\dot\alpha},\hat\lambda_{i\dot\alpha}\rangle_\sfL-\\
  &\kern2cm-\langle Z,Y\rangle_\sfL-\langle\hat Z,\hat Y\rangle_\sfL-\langle\phi^{ij},D_{ij}-\tfrac12 \mu_1(\phi_{ij})\rangle_\sfL\,\Big\}
 \end{aligned}
\end{equation}
is then invariant under the supersymmetry transformations
\begin{subequations}\label{eq:N2SUSY}
\begin{equation}
\begin{gathered}
  \delta_{\rm SUSY}A_{\alpha\dot\beta} \ :=\ \epsilon^i_\alpha\hat\lambda_{i\dot\beta}-\hat\epsilon_{i\dot\beta}\lambda^i_\alpha~,\\
  \delta_{\rm SUSY}\lambda^i_\alpha\ :=\ \di \epsilon^{i\beta}\CF_{\alpha\beta}-\tfrac12\epsilon_{j\alpha}[D^{ij}-\mu_1(\phi^{ij})]+\hat\epsilon^{i\dot\beta} \nabla_{\alpha\dot\beta}Y+\epsilon^i_\alpha\mu_2(Y,\hat Y)~,\\
  \delta_{\rm SUSY}\hat\lambda_{i\dot\alpha}\ :=\ \di \hat\epsilon_i^{\dot\beta}\CF_{\dot\alpha\dot\beta}-\tfrac12\hat\epsilon^j_{\dot\alpha}[D_{ij}-\mu_1(\phi_{ij})]-\epsilon_i^\beta \nabla_{\beta\dot\alpha}\hat Y+\hat\epsilon_{i\dot\alpha}\mu_2(Y,\hat Y)~,\\  
  \delta_{\rm SUSY}D^{ij}\ :=\ \epsilon^{(i\alpha}\nabla_{\alpha\dot\beta}\hat\lambda^{j)\dot\beta}+\hat\epsilon^{(i\dot\alpha}\nabla_{\beta\dot\alpha}\lambda^{j)\beta}+\epsilon^{(i\alpha}\mu_2(Y,\lambda^{j)}_\alpha)+\hat\epsilon^{(i\dot\alpha}\mu_2(\hat Y,\hat\lambda^{j)}_{\dot\alpha})~,\\
  \delta_{\rm SUSY} Y\ :=\ \epsilon^{i\alpha}\lambda_{i\alpha}~,\quad
   \delta_{\rm SUSY}\hat Y\ :=\ \hat\epsilon^{i\dot\alpha}\hat\lambda_{i\dot\alpha}~,
  \end{gathered}
\end{equation}
and
\begin{equation}
\begin{gathered}
  \delta_{\rm SUSY}B_{\alpha\beta}\ :=\ -\epsilon^i_{(\alpha}\chi_{i\beta)}~,\quad
   \delta_{\rm SUSY}B_{\dot\alpha\dot\beta}\ :=\ -\hat\epsilon_{i(\dot\alpha}\hat\chi^i_{\dot\beta)}~,\\
   \delta_{\rm SUSY}\chi_{i\alpha}\ :=\ [\di H_{\alpha\dot\beta}\varepsilon_{ij}-\nabla_{\alpha\dot\beta}\phi_{ij}]\hat\epsilon^{j\dot\beta}+\epsilon_{i\alpha} Z+\epsilon^j_\alpha\mu_2(Y,\phi_{ij})~,\\
   \delta_{\rm SUSY}\hat\chi^i_{\dot\alpha}\ :=\ [\di H_{\beta\dot\alpha}\varepsilon^{ij}+\nabla_{\beta\dot\alpha}\phi^{ij}]\epsilon_j^{\beta}+\hat\epsilon^i_{\dot\alpha} \hat Z-\hat\epsilon_{j\dot\alpha}\mu_2(\hat Y,\phi^{ij})~,\\
   \delta_{\rm SUSY}\phi^{ij}\ :=\ \epsilon^{(i\alpha}\chi^{j)}_\alpha-\hat\epsilon^{(i\dot\alpha}\hat\chi^{j)}_{\dot\alpha}~,\\
    \delta_{\rm SUSY}Z\ :=\ -\hat\epsilon^{i\dot\beta}\nabla_{\alpha\dot\beta}\chi_i^\alpha+\hat\epsilon_i^{\dot\alpha}\mu_2(\hat Y, \hat\chi^i_{\dot\alpha})~,\quad
   \delta_{\rm SUSY}\hat Z\ :=\ -\epsilon_i^\beta\nabla_{\beta\dot\alpha}\hat\chi^{i\dot\alpha}+\epsilon^{i\alpha}\mu_2(Y,\chi_{i\alpha})~.
  \end{gathered}
\end{equation}
\end{subequations}
As in ordinary Chern--Simons theory, all fields except for the gauge potentials appear merely algebraically and are therefore auxiliary and can be integrated out.

\paragraph{Remark on fake curvatures.} We already commented on the general role of fake curvatures in Section~\ref{ssec:HMCE}. Let us briefly consider fake curvatures in the case of higher Chern--Simons theories.

For general $d\geq 3$, the MC potential $a\in\Omega_1^\bullet(M,\sL)$ decomposes as $a=A_1+A_2+\cdots+A_{d-2}$ with $A_k\in\Omega^k(M,\sL_{-k+1})$ while the curvature $f$ has the decomposition $f=F_2+F_3+\cdots+F_{d-1}$ with $F_k\in\Omega^k(M,\sL_{-k+2})$. The curvatures $F_k$ for $k=2,\ldots,d-2$ are known as the {\it $k$-form fake curvatures}.

The gauge parameters of level~$k$ gauge transformations, $c_{-k}$, also decompose into forms of varying degrees and $c_{-k}\in\Omega^0(M,\sL_{-k})\oplus\cdots\oplus\Omega^{d-k-3}(M,\sL_{-d+2})$. We see that the formula~\eqref{eq:GaugeTrafoCurvature} for the gauge transformation of the curvature form $F_{d-1}$ contains a covariant term of the form $\mu_2(F_{d-1},\alpha)$, where $\alpha\in \Omega^0(M,\sL_0)$ is a component of $c_0$ and all other terms are proportional to lower curvatures, $F_j$ with $j<d-1$.

We also note that the successive action of two gauge transformations~\eqref{eq:ClosureGaugeTrafo} does not contain the highest form component of the curvature, $F_{d-1}$: we have $k\leq -1$, and matching the $L_\infty$-degrees makes an appearance of $F_{d-1}$ impossible. 

Altogether, requiring $F_2=\cdots=F_{d-2}=0$ renders the gauge transformations of the highest curvature $F_{d-1}$ covariant and allows to close general gauge transformations. The first point is particularly important for writing down interesting equations of motions, coupling the higher gauge potentials to matter fields.

\paragraph{Comments on (higher) holomorphic Chern--Simons theories.} Finally, let us \linebreak briefly comment on higher holomorphic Chern--Simons theory on a complex manifold $M$. Here, we consider a gauge $L_\infty$-algebra $\sL$ and tensor it with the Dolbeault complex $(\Omega^{0,\bullet}(M),\dparb)$, resulting in homotopy Maurer--Cartan theory for the $L_\infty$-algebra $\Omega^{0,\bullet}(M,\sL)$. Via an extension of the \v Cech--Dolbeault correspondence~\cite{Ivanova:2000af} which is discussed for the Abelian case in~\cite{Saemann:2011nb}, classical solutions to the higher holomorphic Chern--Simons equations of motion correspond to topologically trivial higher holomorphic principal bundles. Over twistor spaces, the latter can be mapped to solutions of various field equations via a Penrose--Ward transform.

Particularly interesting in this context is the complex six-dimensional twistor space considered in~\cite{Saemann:2011nb,Mason:2011nw,Mason:2012va}, as well as the complex three-dimensional twistor space introduced in~\cite{Saemann:2011nb}. Higher holomorphic bundles over these spaces yield solutions to the self-duality equation $H=\star H$ for a three-form curvature $H=\dd B$ in six dimensions as well as the self-dual string equation in four dimensions. The former is part of the Abelian (2,0)-theory, a six-dimensional superconformal field theory whose non-Abelian extension is currently the subject of extensive study. The latter describes Abelian BPS configurations in the (2,0)-theory.

The advantage of the twistor description is now that both supersymmetric and non-Abelian extension of these equations are found by switching to supertwistor spaces and non-Abelian higher principal bundles, see the discussion in the papers~\cite{Saemann:2012uq,Saemann:2013pca,Jurco:2014mva,Jurco:2016qwv}. 

Even in the case of ordinary field theories, higher holomorphic Chern--Simons theories can be of use. The twistor space of $\CN=3$ supersymmetric Yang--Mills theory is a five-dimensional complex manifold and therefore the relevant holomorphic bundles cannot be described in terms of an ordinary holomorphic Chern--Simons action functional. Various auxiliary constructions were tried to remedy the situation (see e.g.~\cite{Mason:2005kn}), but there is a more natural way out. Consider higher holomorphic Chern--Simons theory for $\sL:=\sL_{-2}\oplus \sL_{-1}\oplus \sL_0$ a Lie 3-algebra as done in~\cite{Saemann:2017vuy}. The higher extension of gauge potentials in this theory is merely auxiliary, and the $L_\infty$-algebra of the higher holomorphic Chern--Simons theory reduces to that of ordinary holomorphic Chern--Simons theory. The extension, however, gave us a natural action functional.

\subsection{Yang--Mills theory: second-order formulation}

After having discussed Chern--Simons theories and their higher generalisations, let us now turn to the other prominent gauge theory: Yang--Mills theory. In the following, let $M$ be a smooth compact Riemannian manifold without boundary and real dimension $d$. In addition, let $\sG$ be a Lie group with metric Lie algebra $(\frg,[-,-],\langle-,-\rangle_\frg)$. We define $\sL$ to be $\Omega^\bullet(M,\frg):=\Omega^\bullet(M)\otimes\frg$, the differential graded Lie algebra of $\frg$-valued differential forms on $M$. Furthermore, let  $\dd$ be the exterior derivative and $\dd^\dagger$ its adjoint with respect to the standard inner product $\langle\alpha_1,\alpha_2\rangle=\int_M\alpha_1\wedge{\star\alpha_2}$ for $\alpha_{1,2}\in\Omega^k(M)$ with $\star$ the Hodge operator for the given metric. 

\paragraph{Batalin--Vilkovisky complex.} The field content of plain Yang--Mills theory in second-order formulation consists of a gauge potential $A\in \Omega^1(M,\frg)[0]$ with curvature $F=\dd A+\tfrac12 [A,A]$. We add a ghost $c\in \Omega^0(M,\frg)[1]$ and thus complete the BRST complex, the differential graded algebra description of the gauge algebroid, as explained in Section~\ref{sec:GeneralBVFormalism}. Recall that $[k]$ for $k\in\RZ$ indicates the ghost degree of the object. To complete the BV complex, we also add the antifield $A^+\in\Omega^3(M,\frg)[-1]$ as well as the antifield of the ghost field, $c^+\in \Omega^4(M,\frg)[-2]$. On this space of fields $\frF_{\rm YM_2BV}$, we have the canonical symplectic form 
\begin{equation}\label{eq:SymplecticFormYM2}
  \omega_{\rm YM_2BV}\ :=\  \int_M\left\{\langle\delta A,\delta A^+\rangle_\frg-\langle\delta c,\delta c^+\rangle_\frg\right\}~,
\end{equation}
as introduced in~\eqref{eq:BVSymplecticStructureGeneral}.

The BV action for Yang--Mills theory is derived to be 
\begin{equation}\label{eq:BVActionYM2}
 S_{\rm YM_2BV}\ :=\ \int_M \Big\{\tfrac12 \langle F,{\star F}\rangle_\frg-\langle A^+,\nabla c\rangle_\frg+\tfrac12\langle c^+,[c,c]\rangle_\frg\Big\}~,
\end{equation}
cf.~e.g.~\cite{Batalin:1981jr}, and it is a straightforward exercise to show that $S_{\rm YM_2BV}$ satisfies the classical master equation $\{S_{\rm YM_2BV},S_{\rm YM_2BV}\}_{\rm YM_2BV}=0$, where $\{-,-\}_{\rm YM_2BV}$ is the Poisson bracket induced by~\eqref{eq:SymplecticFormYM2}. Hence, we may define the homological vector field
\begin{equation}\label{eq:HomologicalVectorFieldYM2}
 Q_{\rm YM_2BV}\ :=\ \{S_{\rm YM_2BV},-\}_{\rm YM_2BV}\ewith Q_{\rm YM_2BV}^2\ =\ 0~,
\end{equation}
whose action on a functional $F\in \CCC^\infty(\frF_{\rm YM_2BV})$ reads as
\begin{equation}
\begin{aligned}
 Q_{\rm YM_2BV} F\ &=\ 
 \int_M^*\left\{-\left\langle\frac{\delta S_{\rm BV}}{\delta A^+},\frac{\delta F}{\delta A}\right\rangle_{\frg^*}-
 \left\langle\frac{\delta S_{\rm BV}}{\delta A},\frac{\delta F}{\delta A^+}\right\rangle_{\frg^*}-\right.\\
&\kern2cm\left.- \left\langle\frac{\delta S_{\rm BV}}{\delta c^+},\frac{\delta F}{\delta c}\right\rangle_{\frg^*}-
 \left\langle\frac{\delta S_{\rm BV}}{\delta c},\frac{\delta F}{\delta c^+}\right\rangle_{\frg^*}\right\}~.
\end{aligned}
\end{equation}
On the contracted coordinate functions on $\frF_{\rm YM_2BV}$, we have
\begin{equation}\label{eq:BVTransformationsYM2}
\begin{aligned}
 Q_{\rm YM_2BV}c \ &=\ -\tfrac12[c,c]~,\\
 Q_{\rm YM_2BV} A  \ &=\ \nabla c\ =\ \dd c+[A,c]~,\\
 Q_{\rm YM_2BV}A^+  \ &=\ -{\nabla{\star F}}-[c,A^+]\\
 \ &=\ -\dd {\star \dd} A+\tfrac12 \dd {\star [A,A]}+[A,\star \dd A]+\tfrac12 [A,\star[A,A]]-[c,A^+]~,\\
 Q_{\rm YM_2BV}c^+ \ &=\ {\nabla}{A^+}-[c,c^+]\\
 \ &=\ \dd A^++[A,A^+]-[c,c^+]~.
 \end{aligned}
\end{equation}

\paragraph{\mathversion{bold}$L_\infty$-algebra structure.} The differential graded algebra $(\CCC^\infty(\frF_{\rm YM_2BV}),Q_{\rm YM_2BV})$ is dual to an $L_\infty$-algebra structure on the graded vector space\footnote{In principle, one may apply the natural isomorphisms $\Omega^{d-i}(M,\frg)\cong \Omega^i(M,\frg)$ to be able to identify the second $\mu_1$ with the Hodge Laplacian. This, however, is somewhat unnatural from the BV perspective, as it will modify the canonical symplectic structure~\eqref{eq:SymplecticFormYM2} on $\frF_{\rm YM_2BV}$. It would also make our computations below less straightforward.}
\begin{subequations}
\begin{equation}\label{eq:YMComplex}
 \underbrace{\Omega^0(M,\frg)}_{=:\,\sL_0}\ \xrightarrow{~\mu_1\,:=\, \dd~}\  \underbrace{\Omega^1(M,\frg)}_{=:\,\sL_1}\ \xrightarrow{~\mu_1\,:=\, \dd\star\dd~}\  \underbrace{\Omega^{d-1}(M,\frg)}_{=:\,\sL_2}\ \xrightarrow{~\mu_1\,:=\, \dd~}\  \underbrace{\Omega^d(M,\frg)}_{=:\,\sL_3}~.
\end{equation}
We call this complex the {\it second-order Yang--Mills complex}. Note that the higher products $\mu_i$ are read off the action~\eqref{eq:BVTransformationsYM2} of the homological vector field on the fields using formula~\eqref{eq:Qxi_L_infty}: $\mu_1$ is given by the linear part of the right-hand side of~\eqref{eq:BVTransformationsYM2}, $\mu_2$ by the quadratic part, etc. The coordinate functions on $\sL_0$, $\sL_1$, $\sL_2$, and $\sL_3$ are, respectively, $c$, $A$, $A^+$, and $c^+$. 

In particular, we have the higher products
\begin{equation}\label{eq:ZeitlinProducts}
\begin{gathered}
\mu_1(c_1)\ :=\ \dd c_1~,~~~
\mu_1(A_1)\ :=\ \dd{\star\dd A_1}~,~~~
\mu_1(A^+_1)\ :=\ \dd A^+_1~,\\
\mu_2(c_1,c_2)\ :=\ [c_1,c_2]~,~~~ 
\mu_2(c_1,A_1)\ :=\ [c_1,A_1]~,\\ 
\mu_2(c_1,A^+_2)\ :=\ [c_1,A^+_2]~,~~~
\mu_2(c_1,c^+_2)\ :=\ [c_1,c^+_2]~,\\ 
\mu_2(A_1,A^+_2)\ :=\ [A_1,A^+_2]~,\\ 
\mu_2(A_1,A_2)\ :=\ \dd{\star[A_1,A_2]}+[A_1,{\star\dd A_2}]+[A_2,{\star\dd A_1}]~,\\
 \mu_3(A_1,A_2,A_3)\ :=\ [A_1,\star[A_2,A_3]]+[A_2,\star[A_3,A_1]]+[A_3,\star[A_1,A_2]]
\end{gathered}
\end{equation}
\end{subequations}
for elements $c_i\in \sL_0$, $A_i\in \sL_1$, $A^+_i\in \sL_2$, and $c^+_i\in \sL_3$. We shall denote this $L_\infty$-algebra by $\sL_{\rm YM_2}$. Note that as expected from the formalism, $\sL_{\rm YM_2}$ is cyclic with cyclic structure induced by the components of the symplectic form~\eqref{eq:SymplecticFormYM2}.

We stress again that the $L_\infty$-algebra $\sL_{\rm YM_2}$ encodes all classical information about Yang--Mills theory: it contains the field content, the gauge symmetries, the equations of motions as well as the Noether identities. 

We note that in the case of classical gauge theories with Abelian gauge group, for which $\sL$ is concentrated in degrees $0,\dots,3$, the underlying complex has been studied under the name of {\em detour complex}~\cite{Gover:2006ha,Gover:2006fa}. The non-abelian detour complex differs from the complex underlying $\sL$ by terms  covariantising the $\mu_1$. This relation was explained in~\cite{Zeitlin:2008cc}, where also the homotopy algebra structures underlying the complex~\eqref{eq:YMComplex} were studied.

The $L_\infty$-algebra $\sL_{\rm YM_2}$ was first given in~\cite{Movshev:2003ib,Movshev:2004aw} in its dual formulation as a differential graded algebra. The same $L_\infty$-algebra was then rederived from string field theory considerations and further discussed in~\cite{Zeitlin:2007vv,Zeitlin:2007yf}.

\paragraph{Homotopy Maurer--Cartan action.} Let us now plug $\sL_{\rm YM_2}$ into the homotopy MC action~\eqref{eq:MCAction}. From our discussion in Section~\ref{sec:HMCT}, we expect that we recover the Yang--Mills action. We have with $a=A$
\begin{equation}
\begin{aligned}
 \tfrac12\langle a,\mu_1(a)\rangle_{\sL_{\rm YM_2}}\ &=\ \tfrac12\int_M \langle \dd A, {\star\dd A}\rangle_\frg~,\\
  \tfrac{1}{3!}\langle A,\mu_2(a,a)\rangle_{\sL_{\rm YM_2}}\ &=\ \tfrac12\int_M \langle \dd A, {\star[A,A]}\rangle_\frg\\
  & \ =\ \tfrac14\int_M\big\{ \langle[A,A],{\star\dd A}\rangle_\frg+\langle \dd A, {\star[A,A]}\rangle_\frg\big\}~,\\
    \tfrac{1}{4!}\langle a,\mu_3(a,a,a)\rangle_{\sL_{\rm YM_2}}\ &=\ \tfrac18\int_M \langle [A,A], {\star[A,A]}\rangle_\frg~,
\end{aligned}
\end{equation}
so that indeed
\begin{equation}
 S_{\rm MC}\ =\ \sum_{i=1}^3\frac{1}{(i+1)!}\langle a,\mu_i(a,\ldots,a)\rangle_{\sL_{\rm YM_2}}\ =\ \tfrac12\int_M\langle F,{\star F}\rangle_\frg~.
\end{equation}
In addition, the Yang--Mills equation translates into the flatness condition
\begin{equation}
 \dd{\star F}+[A,{\star F}]\ =\ 0~~~\ \rightarrow\ ~~~\sum_{i=1}^3\frac{1}{i!}\mu_i(a,\ldots,a)\ =\ 0~.
\end{equation}

We can also reproduce the BV action~\eqref{eq:BVActionYM2} from the BV action~\eqref{eq:BVaction}, see also~\eqref{eq:Example2TermAction}, using $\sfa=c_0+a+a^++c_0^+=c+A+A^++c^+$. Indeed, we find
\begin{equation}
\begin{aligned}
 S_{\rm BV}\ &=\ \sum_{i=1}^3\frac{1}{(i+1)!}\langle a,\mu_i(a,\ldots,a)\rangle_{\sL_{\rm YM_2}}-\langle c_0,\mu_1(a^+)\rangle_{\sL_{\rm YM_2}}+\langle a,\mu_2(a^+,c_0)\rangle_{\sL_{\rm YM_2}}\\
 &=\ S_{\rm YM_2BV}~.
\end{aligned}
\end{equation}

In summary, we have obtained a reformulation of Yang--Mills theory as a homotopy MC theory, which is closely related to Chern--Simons theory. 

\paragraph{Minimal model.} Above we observed that the structures of classical Yang--Mills theory are fully captured by the $L_\infty$-algebra $\sL_{\rm YM_2}$. As explained earlier, the appropriate isomorphisms, namely the quasi-isomorphisms, are supposed to lead to theories which are physically equivalent at the classical level. A particularly interesting quasi-isomorphic $L_\infty$-algebra is certainly a minimal model of $\sL_{\rm YM_2}$ in which all possible equivalences have been divided out.

To construct this model explicitly, we shall make use of the Hodge decomposition
\begin{subequations}
\begin{equation}
 \Omega^k(M,\frg)\ \cong\ \Omega^k_{\rm h}(M,\frg)\oplus\dd\Omega^{k-1}(M,\frg)\oplus\dd^\dagger\Omega^{k+1}(M,\frg)~,
\end{equation} 
together with the projectors 
\begin{equation}\label{eq:HodgeMaps}
\begin{gathered}
 P_{\rm h}\,:\,\Omega^k(M,\frg)\ \to\ \Omega^k_{\rm h}(M,\frg)~,~~~P_{\rm e}\,:\,\Omega^k(M,\frg)\ \to\ \dd\Omega^{k-1}(M,\frg)~,\\
 P_{\rm c}\,:\,\Omega^k(M,\frg)\ \to\ \dd^\dagger\Omega^{k+1}(M,\frg)~,
 \end{gathered}
 \end{equation}
 \end{subequations}
 which extract the harmonic, exact, and coexact parts, respectively. The cohomology complex 
\begin{equation}
 \sL'_{\rm YM_2}\ :=\ H^\bullet_{\mu_1}(\sL_{\rm YM_2})
\end{equation} 
of the second-order Yang--Mills complex~\eqref{eq:YMComplex} is then given by
\begin{equation}\label{eq:YMCohomologyComplexYM2}
\underbrace{\Omega^0_{\rm h}(M,\frg)}_{=:\,\sL'_0}\ \xrightarrow{~\mu'_1\,:=\, 0~}\  \underbrace{\Omega^1_{\rm h}(M,\frg)}_{=:\,\sL'_1}\ \xrightarrow{~\mu'_1\,:=\, 0~}\  \underbrace{\Omega^{d-1}_{\rm h}(M,\frg)}_{=:\,\sL'_2}\ \xrightarrow{~\mu'_1\,:=\,0~}\  \underbrace{\Omega^d_{\rm h}(M,\frg)}_{=:\,\sL'_3}~,
\end{equation}
as we shall now argue. First, $\sL'_0$ and $\sL'_3$ are obvious; note also that $\Omega^0_{\rm h}(M,\frg)\cong\frg$. On $\Omega^d(M,\frg)$, we have $\ker\big(\dd{\star \dd}\big)\cong \ker(\dd^\dagger\dd)\cong (P_{\rm h}+P_{\rm e})\Omega^d(M,\frg)$ and so, $H^1_{\mu_1}(\sL_{\rm YM_2})\cong H^1_{\rm dR}(M,\frg)\cong \Omega^1_{\rm h}(M,\frg)$ using the Hodge theorem. Moreover, 
\begin{equation}
 \begin{aligned}
  \ker(\dd)\ &\cong\ \Omega^{d-1}_{\rm h}(M,\frg)\oplus\dd\Omega^{d-2}(M,\frg)\\
   &\cong\ \Omega^{d-1}_{\rm h}(M,\frg)\oplus\dd\dd^\dagger\Omega^{d-1}(M,\frg)\\
  \ &\cong\ \Omega^1_{\rm h}(M,\frg)\oplus\im(-\dd{\star \dd})~,
 \end{aligned}
\end{equation}
where we used the Hodge decomposition and Hodge duality, and therefore $H^2_{\mu_1}(\sL_{\rm YM_2})\cong\Omega^1_{\rm h}(M,\frg)$.

As discussed in Section~\ref{ssec:quasiisomorphism}, the complex $\sL'_{\rm YM_2}$ admits an $L_\infty$-structure. To construct the higher products $\mu'_i$ for $i>1$ on $\sL'_{\rm YM_2}$, we first note that the second-order Yang--Mills complex~\eqref{eq:YMComplex} is split in the sense of~\eqref{eq:splitComplex}. To see this, set 
\begin{equation}
 \dd_0\ :=\ \dd~,~~~\dd_1\ :=\ \dd{\star\dd}~, \eand \dd_2\ :=\ \dd~.
 \end{equation}
Next, we need to find a contracting homotopy $h_k:\sL_k\to\sL_{k-1}$, that is, $\dd_k=\dd_k\circ h_{k+1}\circ\dd_k$ for $k =0,1,2$. Using the Green operator $G$ defined in~\eqref{eq:GreenLaplace}, we can put
\begin{equation}\label{eq:ContractingHomotopyYM2}
 h_1\ :=\ \dd^\dagger G~,~~~h_2\ :=\ (-1)^{d+1}G\star~,\eand h_3\ :=\ \dd^\dagger G~.
\end{equation}
This is seen using the identities~\eqref{eq:GreenLaplaceIdentities}. Furthermore, setting $(\dd_{-1},h_0):=(0,0)$ and $(\dd_3,h_4):=(0,0)$, we have the projectors $P_k$ defined by
\begin{equation}
 1\ =\ P_k +h_{k+1}\circ\dd_k+\dd_{k-1}\circ h_{k}~,
\end{equation}
which are the compositions of the projections $p_k:\sL_k\twoheadrightarrow H^k_{\mu_1}(\sL_{\rm YM_2})$ and the injections $e_k:H^k_{\mu_1}(\sL_{\rm YM_2})\hookrightarrow\sL_k$. That is, the $P_k$ form a projector $P_{\rm h}:\Omega^\bullet(M,\frg)\to\Omega^\bullet_{\rm h}(M,\frg)$ onto the harmonic forms using the Hodge decomposition. Using the contracting homotopy~\eqref{eq:ContractingHomotopyYM2}, it is now easy to adapt the formulas~\eqref{eq:minimalQuasiIsomorphism} for the quasi-isomorphism between $\sL_{\rm YM_2}$ and $\sL'_{\rm YM_2}=H_{\mu_1}^\bullet(\sL_{\rm YM_2})$. For the sake of clarity, we shall again only display the formulas in homogeneous degree~1. We obtain
\begin{equation}
\begin{aligned}
\phi_1(a')\ &= \ e(a')~,\\
 \phi_2(a',a')\ &= \ -G\dd^\dagger[e(a'),e(a')]~,\\
  &~\,\vdots
 \end{aligned}
\end{equation}
Hence, the higher products on $\sL'_{\rm YM_2}=H_{\mu_1}^\bullet(\sL_{\rm YM_2})$ defined in~\eqref{eq:minimalHigherProducts} are then given for degree~1 elements by
\begin{equation}
\begin{aligned}
\mu'_1(a')\ &= \ 0~,\\
 \mu'_2(a',a')\ &= \ 0~,\\
 \mu'_3(a',a',a')\ &= \  3 p\Big(\big[e(a'),{\star P_{\rm h}[e(a'),e(a')]}\big]\Big)\,,\\
  &~\,\vdots
 \end{aligned}
\end{equation}
where we have used the projectors~\eqref{eq:HodgeMaps}.

Altogether, we note that the simplification of the configuration space in the quasi-isomorphism induced a much more complicated structure in the interaction terms. We plan to study this structure in more detail in future work.

\paragraph{Supersymmetric extension.} Let us briefly comment on the supersymmetric extension. We consider the example of $\CN=1$ supersymmetric Yang--Mills theory on a ten-dimensional compact Riemannian spin manifold $M$. The lower-dimensional cases simply follow by dimensional reduction, see e.g.~\cite{Brink:1976bc}. The spin bundle decomposes into the bundles of chiral and anti-chiral spinors $S_\pm$ and we have $S_\pm\cong S^*_\mp$. Let $\sigma_\pm:TM\otimes\FC\to \odot^2 S_\pm$ and consider the complex 
\begin{equation}
\begin{aligned}
 &\underbrace{\Omega^0(M,\frg)}_{=:\,\sL_0}\ \xrightarrow{~\mu_1:=\dd~}\ \underbrace{\Omega^1(M,\frg)\oplus\Gamma(M,\Pi S_+\otimes\frg)}_{=:\,\sL_1}\\
  &\kern2.5cm\xrightarrow{~\mu_1:=\dd\star\dd+\slashed{\CD}~}\  \underbrace{\Omega^{d-1}(M,\frg)\oplus \Gamma(M,\Pi S_-\otimes\frg)}_{=:\,\sL_2}\ \xrightarrow{~\mu_1:=\dd~}\ \underbrace{\Omega^d(M,\frg)}_{=:\,\sL_3}~,
 \end{aligned}
\end{equation}
where $\slashed{\CD}$ is the Dirac operator on $M$ and for any one-form $\omega\in\Omega^1(M)$ we set $\slashed{\omega}:=\sigma_-\intprod\omega$ with the non-vanishing higher products
\begin{equation}\label{eq:YM2Products}
\begin{gathered}
\mu_1(c_1)\ :=\ \dd c_1~,~~~
\mu_1(A_1+\psi_1)\ :=\ \dd{\star\dd A_1}+\slashed{\CD}\psi_1~,~~~
\mu_1(A^+_1)\ :=\ \dd A^+_1~,\\
\mu_2(c_1,c_2)\ :=\ [c_1,c_2]~,~~~ 
\mu_2(c_1,A_1+\psi_1)\ :=\ [c_1,A_1+\psi_1]~,\\ 
\mu_2(c_1,A^+_1+\psi_1^+)\ :=\ [c_1,A^+_1+\psi_1^+]~,~~~
\mu_2(c_1,c^+_2)\ :=\ [c_1,c^+_2]~,\\ 
\mu_2(A_1+\psi_1,A^+_2+\psi_2^+)\ :=\ [A_1,A^+_2]+[\psi_1,\psi^+_2]~,\\ 
\mu_2(A_1+\psi_1,A_2+\psi_2)\ :=\ \dd{\star[A_1,A_2]}+[A_1,{\star\dd A_2}]+[A_2,{\star\dd A_1}]\,+\\
\kern5cm+\,\psi_1(\sigma_-\intprod(\star 1))\psi_2+[\slashed{A}_1,\psi_2]-[\slashed{A}_2,\psi_1]~,\\
 \mu_3(A_1,A_2,A_3)\ :=\ [A_1,\star[A_2,A_3]]+[A_2,\star[A_3,A_1]]+[A_3,\star[A_1,A_2]]~.
 \end{gathered}
\end{equation}
Here, $c_i\in\sL_0$, $\{A_i+\psi_i\}\in\sL_1$, $\{A^+_i+\psi^+_i\}\in\sL_2$, and $c^+_i\in\sL_3$ for $i=1,2,3$. Following the same discussion as in the previous sections, one can check that the MC action~\eqref{eq:MCAction} with $a=A+\psi$ becomes
\begin{equation}
 S_{\rm MC}\ =\ \tfrac12\int_M\big\{\langle F,{\star F}\rangle_\frg+\langle\psi,{\star\slashed{\nabla}\psi}\rangle_\frg\big\}~,
\end{equation}
where $\slashed{\nabla}$ is the covariant Dirac operator, involving the gauge connection one-form $A$. We note that extensions of the Yang--Mills $L_\infty$-algebra $\sL_{\rm YM_2}$ by scalars and Dirac spinor fields coupling to the gauge field were already given in~\cite{Zeitlin:2007fp}.

\subsection{Yang--Mills theory: first-order formulation}

It is known~\cite{Okubo:1979gt} that Yang--Mills theory in four dimensions admits an alternative formulation which only makes use of first-order rather than second-order differential operators and has only cubic interactions. Let $M$ now be a smooth compact four-dimensional Riemannian manifold without boundary and let $\sG$ be a Lie group with Lie algebra $\frg$ with metric $\langle-,-\rangle_\frg$. The metric on $M$ induces the decomposition of differential $2$-forms 
\begin{equation}
\Omega^2(M,\frg)\ \cong\ \Omega^2_+(M,\frg)\oplus\Omega^2_-(M,\frg)
\end{equation}
into self-dual and anti-self-dual parts. Letting $P_\pm:=\frac12(1\pm\star)$ be the corresponding projectors where $\star$ is, as before, the Hodge operator associated with the given metric, we may write $\Omega^2_\pm(M,\frg)=P_\pm\Omega^2(M,\frg)$. 

\paragraph{Batalin--Vilkovisky complex.} The field content of Yang--Mills theory in first-order formulation consists of a gauge potential $A\in \Omega^1(M,\frg)[0]$ with usual curvature $F=\dd A+\tfrac12 [A,A]$ and a Lie-algebra valued, self-dual two-form $B_+\in \Omega^2_+(M,\frg)[0]$, subject to the gauge transformations
\begin{equation}
 \delta_{c}A\ :=\ \nabla c~,~~~\delta_{c}B_+\ :=\ -[c,B_+]~,\eand
 \delta_c F\ =\ -[c,F]
\end{equation}
for $c\in \Omega^0(M,\frg)$. The action reads as
\begin{equation}\label{eq:actionYM1}
 S_{\rm YM_1}\ :=\ \int_M \Big\{\langle F, B_+\rangle_\frg+\tfrac\varepsilon2\langle B_+, B_+\rangle_\frg\Big\}~,
\end{equation}
where the parameter $\eps$ is a positive real number. Note that the Yang--Mills equation $\nabla{\star F}=0$ is equivalent to  $\nabla F_\pm=0$ due to the Bianchi identity $\nabla F=0$. Hence, the equation of motions following from~\eqref{eq:actionYM1}, $B_+=-\frac{1}{\eps}F_+$ and $\nabla B_+=0$, imply the Yang--Mills equations so that~\eqref{eq:actionYM1} is classically equivalent to Yang--Mills theory.

The action~\eqref{eq:actionYM1} is extended to a BV action by adding ghosts $c\in \Omega^0(M,\frg)[1]$ as well as the antifields $A^+\in \Omega^3(M,\frg)[-1]$, $B^+_+\in\Omega^2_+(M,\frg)[-1]$, and $c^+\in\Omega^4(M,\frg)[-2]$. The canonical symplectic form on the space of BV fields $\frF_{\rm BV}$ is
\begin{equation}\label{eq:SymplecticFormYM1}
  \omega_{\rm YM_1BV}\ :=\  \int_M\left\{\langle\delta A,\delta A^+\rangle_\frg+\langle\delta B_+,\delta B^+_+\rangle_\frg-\langle\delta c,\delta c^+\rangle_\frg\right\}~,
\end{equation}
and the BV action reads as 
\begin{equation}\label{eq:BVActionYM1}
\begin{aligned}
 S_{\rm YM_1BV}\ &:=\ \int_M \Big\{\langle F, B_+\rangle_\frg+\tfrac\varepsilon2\langle B_+, B_+\rangle_\frg\,-\\
 &\kern2cm-\langle A^+,{\nabla c}\rangle_\frg-\langle B^+_+,[B_+,c]\rangle_\frg+\tfrac12\langle c^+,{[c,c]}\rangle_\frg\Big\}~.
 \end{aligned}
\end{equation}
Hence, we may define
\begin{equation}\label{eq:HomologicalVectorFieldYM1}
 Q_{\rm YM_1BV}\ :=\ \{S_{\rm YM_1BV},-\}_{\rm YM_1BV}\ewith Q_{\rm YM_1BV}^2\ =\ 0~,
\end{equation}
where $\{-,-\}_{\rm YM_1BV}$ is the Poisson bracket induced by~\eqref{eq:SymplecticFormYM1}. We then obtain
\begin{equation}\label{eq:BVTransformationsYM1}
\begin{aligned}
 Q_{\rm YM_1BV}c\ &=\ -\tfrac12[c,c]~,\\
 Q_{\rm YM_1BV} (B_++A) \ &=\ -[c,B_+]+\nabla c\ =\ \dd c+[B_++A,c]~,\\
 Q_{\rm YM_1BV}(B^+_++A^+) \ &=\ -(F_++\varepsilon B_++[c,B^+_+])-(\nabla B_++[c,A^+])~,\\ 
 &=\ -\varepsilon B_+-P_+\dd A-\dd B_+\,-\\
 &\kern1cm-\tfrac12P_+[A,A]-[A,B_+]-[c,B^+_++A^+]~,\\
 Q_{\rm YM_1BV}c^+\ &=\ \nabla A^++[B_+,B_+^+]-[c,c^+]\\
 \ &=\ \dd A^++[A, A^+]+[B_+,B_+^+]-[c,c^+]~.
 \end{aligned}
\end{equation}

\paragraph{\mathversion{bold}$L_\infty$-algebra structure.} The $L_\infty$-algebra $\sL_{\rm YM_1}$ has now the underlying graded vector space
\begin{subequations}
\begin{equation}\label{eq:YMComplexFirst}
\begin{aligned}
 &\underbrace{\Omega^0(M,\frg)}_{=:\,\sL_0}\ \xrightarrow{~\mu_1\,:=\,\dd~}\  \underbrace{\Omega^2_+(M,\frg)\oplus\Omega^1(M,\frg)}_{=:\,\sL_1}\ \xrightarrow{~\mu_1\,:=\,(\varepsilon+\dd)+P_+\dd~}\\
  &\kern5cm \underbrace{\Omega^2_+(M,\frg)\oplus\Omega^3(M,\frg)}_{=:\,\sL_2}\ \xrightarrow{~\mu_1\,:=\,0+\dd~}\  \underbrace{\Omega^4(M,\frg)}_{=:\,\sL_3}~,
  \end{aligned}
\end{equation}
which we call the {\it first-order Yang--Mills complex}. The non-vanishing higher products are read off the action of the homological vector field $Q_{\rm YM_1BV}$ on $\frF_{\rm BV}=\sL_{\rm YM_1}[1]$ as given in~\eqref{eq:BVTransformationsYM1}:
\begin{equation}\label{eq:HigherProductsYM1}
\begin{gathered}
\mu_1(c_1)\ :=\ \dd c_1~,~~~
\mu_1(B_{+1}+A_1)\ :=\ (\eps B_{+1}+P_+\dd A_1)+\dd B_{+1}~,~~~
\mu_1(A^+_1)\ :=\ \dd A^+_1~,\\
\mu_2(c_1,c_2)\ :=\ [c_1,c_2]~,~~~ 
\mu_2(c_1,B_{+1}+A_1)\ :=\ [c_1,B_{+1}]+[c,A_1]~,\\ 
\mu_2(c_1,B_{+1}^++A^+_1)\ :=\ [c_1,B_{+1}^+]+[c,A^+_1]~,~~~
\mu_2(c_1,c^+_2)\ :=\ [c_1,c^+_2]~,\\ 
\mu_2(B_{+1}+A_1,B_{+2}+A_2)\ :=\ P_+[A_1,A_2]+[A_1,B_{+2}]+[A_2,B_{+1}]~,\\
\mu_2(B_{+1}+A_1,B_{+2}^++A^+_2)\ :=\ [A_1,A_2^+]+[B_1,B_{+2}^+]~.\\ 
\end{gathered}
\end{equation}
\end{subequations}
Here, $c_i\in\sL_0$, $(B_{+i}+A_i)\in\sL_1$, $(B_{+i}^++A_i^+)\in\sL_2$, and $c_i^+\in\sL_3$ for $i=1,2$. An inner product on $\sL_{\rm YM_1}$ is induced by the symplectic form~\eqref{eq:SymplecticFormYM1} and reads as
\begin{equation}\label{eq:InnerProductYM1}
 \langle \alpha_1\otimes t_1,\alpha_2\otimes t_2\rangle_{\sL_{\rm YM_1}}\ :=\ \int_M\alpha_1\wedge{\alpha_2}~\langle t_1,t_2\rangle_\frg~.
\end{equation}
We note that the complex underlying this $L_\infty$-algebra was already discussed in~\cite{Costello:2007ei}.

\paragraph{Homotopy Maurer--Cartan action.} Again, let us briefly check that the homotopy MC action~\eqref{eq:MCAction} for the $L_\infty$-algebra $\sL_{\rm YM_1}$ reproduces the classical action~\eqref{eq:actionYM1} and, after extension to shifted copies, the BV action~\eqref{eq:BVActionYM1}. For degree~$1$ elements $a\in\sL_1$ we have $a=B_++A\in\Omega^2_+(M,\frg)[0]\oplus\Omega^1(M,\frg)[0]$ and so
\begin{equation}
\begin{aligned}
 \tfrac12\langle a,\mu_1(a)\rangle_{\sL_{\rm YM_1}}\ &=\ \int_M\Big\{\langle \dd A, B_+\rangle_\frg+ \tfrac\varepsilon2\langle B_+, B_+\rangle_\frg\Big\}~,\\
  \tfrac{1}{3!}\langle a,\mu_2(a,a)\rangle_{\sL_{\rm YM_1}}\ &=\ \tfrac12\int_M \langle [A,A], B_+\rangle_\frg~.
\end{aligned}
\end{equation}
Consequently, the homotopy MC action~\eqref{eq:MCAction} becomes~\eqref{eq:actionYM1}.

Furthermore, the BV action~\eqref{eq:BVActionYM1} inducing the transformations~\eqref{eq:BVTransformationsYM1} is obtained from the BV action~\eqref{eq:BVaction}, see also~\eqref{eq:Example2TermAction}, using $\sfa=c_0+a+a^++c_0^+=c+B_++A+B^+_++A^++c^+$.

\paragraph{\mathversion{bold}Minimal model.} As in the second-order formalism, it is a rather straightforward exercise to compute a minimal model of the $L_\infty$-algebra $\sL_{\rm YM_1}$. We start from the cohomology complex 
\begin{equation}
 \sL'_{\rm YM_1}\ :=\ H^\bullet_{\mu_1}(\sL_{\rm YM_1})
\end{equation} 
of the first-order Yang--Mills complex~\eqref{eq:YMComplexFirst} using the Hodge decomposition and the Hodge theorem. We obtain the complex
\begin{equation}
\underbrace{\Omega^0_{\rm h}(M,\frg)}_{=:\,\sL'_0}\ \xrightarrow{~\mu'_1\,:=\, 0~}\  \underbrace{\Omega^1_{\rm h}(M,\frg)}_{=:\,\sL'_1}\ \xrightarrow{~\mu'_1\,:=\, 0~}\  \underbrace{\Omega^3_{\rm h}(M,\frg)}_{=:\,\sL'_2}\ \xrightarrow{~\mu'_1\,:=\,0~}\  \underbrace{\Omega^4_{\rm h}(M,\frg)}_{=:\,\sL'_3}~,
\end{equation}
which is the same as the complex~\eqref{eq:YMCohomologyComplexYM2}. Indeed, 
\begin{equation}
 H^0_{\mu_1}(\sL_{\rm YM_1})\ \cong\ \Omega^0_{\rm h}(M,\frg)\eand H^3_{\mu_1}(\sL_{\rm YM_1})\ \cong\ \Omega^4_{\rm h}(M,\frg)
\end{equation}
follow trivially. Furthermore,
\begin{subequations}
\begin{equation}
\ker(P_+\dd+(\varepsilon+\dd))\ \cong\ \ker(\dd^\dagger\dd|_{\Omega^1(M,\frg)})
\end{equation}
so that 
\begin{equation}
 H^1_{\mu_1}(\sL_{\rm YM_1})\ \cong\  \ker(\dd^\dagger\dd|_{\Omega^1(M,\frg)})/\im(\dd)\ \ \cong\  \Omega^1_{\rm h}(M,\frg)
\end{equation}
\end{subequations}
as was already shown in the previous section. Since
\begin{equation}\label{eq:decompo2Forms}
1|_{\Omega_\pm^2(M,\frg)}\ =\ (P_{\rm h}+2P_\pm\circ P_{\rm e})|_{\Omega_\pm^2(M,\frg)}\ =\ (P_{\rm h}+2P_\pm\circ P_{\rm c})|_{\Omega_\pm^2(M,\frg)}~,
\end{equation}
where $P_{\rm h}$, $P_{\rm e}$, and $P_{\rm c}$ were introduced in~\eqref{eq:HodgeMaps}, we obtain
\begin{subequations}
\begin{equation}
\begin{gathered}
\ker(\dd)\ \cong\ \Omega^3_{\rm h}(M,\frg)\oplus\dd\Omega^2(M,\frg)\oplus\Omega_+^2(M,\frg)~,\\
\im(P_+\dd+(\varepsilon+\dd))\ \cong\ \Omega_+^2(M,\frg)\oplus\dd\Omega^2(M,\frg)~.
\end{gathered}
\end{equation}
Hence,
\begin{equation}
H^2_{\mu_1}(\sL_{\rm YM_1})\ \cong\ \Omega^3_{\rm h}(M,\frg)~.
\end{equation}
\end{subequations}

To complete the quasi-isomorphism, let us again construct a contracting homotopy $h_k:\sL_k\rightarrow \sL_{k-1}$. We set
\begin{equation}
\dd_0\ :=\ \begin{pmatrix} 0 \\ \dd \end{pmatrix},~~~
\dd_1\ :=\  \begin{pmatrix} \varepsilon & P_+ \dd\\  \dd & 0 \end{pmatrix},\eand
\dd_2\ :=\ \big(0,\dd\big)~.
\end{equation}
Then, we wish to find $h_k$ such that $\dd_k=\dd_k\circ h_{k+1}\circ\dd_k$ for $k=0,1,2$. Using the Green operator~\eqref{eq:GreenLaplace} and~\eqref{eq:decompo2Forms},
we obtain
\begin{equation}\label{eq:ContractingHomotopyYM1}
 h_1\ =\ \big(0,\dd^\dagger G\big)~,~~~
 h_2\ =\ \begin{pmatrix} \frac{1}{\varepsilon} P_{\rm h} & 2P_+\dd^\dagger G\\ 2\dd^\dagger G & 2\varepsilon \dd^\dagger G\dd G\star\end{pmatrix},\eand 
 h_3\ =\ \begin{pmatrix} 0\\ \dd^\dagger G\end{pmatrix}.
\end{equation}
Furthermore, setting $(\dd_{-1},h_0):=(0,0)$ and $(\dd_3,h_4):=(0,0)$, we have the projectors $P_k$ defined by
\begin{equation}
 1\ =\ P_k +h_{k+1}\circ\dd_k+\dd_{k-1}\circ h_{k}
\end{equation}
projecting $\sL_k$ onto $H^k_{\mu_1}(\sL_{\rm YM_1})$. That is, the $P_k$ yield the projector $P_{\rm h}:\Omega^\bullet(M,\frg)\to\Omega^\bullet_{\rm h}(M,\frg)$. Using the contracting homotopy~\eqref{eq:ContractingHomotopyYM1}, we now adapt the formulas~\eqref{eq:minimalQuasiIsomorphism} for the quasi-isomorphism between $\sL_{\rm YM_1}$ and $\sL'_{\rm YM_1}=H_{\mu_1}^\bullet(\sL_{\rm YM_1})$. As before, for the sake of clarity, we shall only display the formulas in homogeneous degree~1. We obtain 
\begin{equation}
\begin{aligned}
\phi_1(a') \ &= \ e(a')~,\\
  \phi_2(a',a') \ &=\  - \big(\tfrac{1}{\eps}P_{\mathrm{h}}+ 2\mathrm{d}^\dagger G \big)P_+[e(a'),e(a')]~,\\
  &~\,\vdots
 \end{aligned}
\end{equation}
Hence, the higher products on $\sL'_{\rm YM_1}=H_{\mu_1}^\bullet(\sL_{\rm YM_1})$ defined in~\eqref{eq:minimalHigherProducts} are then given for degree~1 elements by
\begin{equation}
\begin{aligned}
\mu'_1(a') \ &=\ 0~,\\
 \mu'_2(a',a')\ &=\ 0~,\\
 \mu'_3(a',a',a')\ &=-\tfrac{3}{\eps} p\Big(\big[[e(a'),P_{\mathrm{h}}P_+ [e(a'),e(a')]\big]\Big)\,,\\
  &~\,\vdots
 \end{aligned}
\end{equation}

\paragraph{Integrating out fields.}
Before showing that both formulations of Yang--Mills theory are $L_\infty$-quasi-isomorphic, we demonstrate, as a warm up, that both formulations are equivalent by `integrating out fields'. We note that a similar computation is found in~\cite{Costello:2011aa}. Recall the action 
\begin{equation}
 S_{\rm YM_1}\ =\ \int_M \Big\{\langle F, B_+\rangle_\frg+\tfrac\varepsilon2\langle B_+, B_+\rangle_\frg\Big\}
\end{equation}
of Yang--Mills theory in the first-order formulation. It is a straightforward exercise to integrate out $B_+$ as it only appears algebraically. We obtain
 \begin{equation}
  S_{\rm YM_1,\,eff}\ =\ -\tfrac{1}{2\eps} \int_M \langle  F_+,  F_+\rangle_\frg\ =\ -\tfrac{1}{4\eps} \int_M\langle  F, {\star  F}\rangle_\frg-\tfrac{1}{4\eps}\int_M \langle  F,  F\rangle_\frg~,
\end{equation}
that is, we find the Yang--Mills action in the second-order formulation plus a topological term. Hence, the two formulations of Yang--Mills theory are equivalent at the level of their equations of motion.

Next, let us recall the BV action~\eqref{eq:BVActionYM1} 
\begin{equation}\label{eq:BVActionYM1Recall}
 \begin{aligned}
 S_{\rm YM_1BV}\ &=\ \int_M \Big\{\langle F, B_+\rangle_\frg+\tfrac\varepsilon2\langle B_+, B_+\rangle_\frg\,-\\
 &\kern2cm-\langle A^+,{\nabla c}\rangle_\frg-\langle B^+_+,[B_+,c]\rangle_\frg+\tfrac12\langle c^+,{[c,c]}\rangle_\frg\Big\}~,
 \end{aligned}
\end{equation}
of Yang--Mills theory in the first-order formulation. Since the ghosts and all the anti-fields are present, integrating out $B_+$ and $B_+^+$ is not as straightforward as above even though they appear only algebraically. To show that the two formulations of Yang--Mills theory are also equivalent in the BV formalism, we first consider the symplectomorphism given by the Hamiltonian\footnote{Costello~\cite{Costello:2011aa} uses $H:=\tfrac{1}{\eps}\int_M \langle F, B_+^+\rangle_\frg$ instead.}
\begin{equation}
 H\ :=\ \tfrac{1}{2\eps}\int_M \langle c, [B_+^+,B_+^+]\rangle_\frg
\end{equation}
for the symplectic form~\eqref{eq:SymplecticFormYM1}. Concretely,
\begin{equation}\label{eq:YM1Symplecto}
\begin{gathered}
 A\ \mapsto\ A+\{H,A\}_{\rm YM_1BV}\ =\ A~,\\
 B_+\  \mapsto\ \ B_++\{H,B_+\}_{\rm YM_1BV}\ =\ B_+-\tfrac{1}{\eps} [c,B_+^+]~,\\
 c\ \mapsto\  c+\{H,c\}_{\rm YM_1BV}\ =\ c~,\\
 A^+\ \mapsto   A^++\{H,A^+\}_{\rm YM_1BV}\ =\ A^+~,\\
 B_+^+\ \mapsto\  B_+^++\{H,B_+^+\}_{\rm YM_1BV}\ =\ B_+^+~,\\
 c^+\ \mapsto\  c^++\{H,c^+\}_{\rm YM_1BV}\ =\ c^++\tfrac{1}{2\eps}[B_+^+,B_+^+]~,
\end{gathered}
\end{equation}
where $ \{-,-\}_{\rm YM_1BV}$ is the Poisson structure induced by~\eqref{eq:SymplecticFormYM1}. Furthermore, it is easy to see that this symplectomorphism preserves the Darboux path integral measure. Upon performing the transformation~\eqref{eq:YM1Symplecto}, the BV action~\eqref{eq:BVActionYM1Recall} becomes
\begin{equation}
\begin{aligned}
  \tilde S_{\rm YM_1BV}\ :=&\ \ S_{\rm YM_1BV} +  Q_{\rm YM_1BV} H\\
 =&\ \int_M \Big\{-\tfrac{1}{2\eps}\langle  F_+,  B_+\rangle_\frg+\tfrac\varepsilon2\langle  B_+,  B_+\rangle_\frg\,-\\
 &\kern2cm-\,\tfrac1\eps\langle  F_+,[c,B_+^+]\rangle_\frg-\langle  A^+,{ \nabla  c}\rangle_\frg+\tfrac12\langle  c^+,{[ c, c]}\rangle_\frg\Big\}~.
 \end{aligned}
\end{equation}
Now we can straightforwardly integrate out $ B_+$ and $ B_+^+$. Indeed, we obtain 
\begin{equation}
  \tilde S_{\rm YM_1BV,\,eff}\ =\ \int_M \Big\{-\tfrac{1}{4\eps}\langle  F, {\star  F}\rangle_\frg-\langle  A^+,{ \nabla  c}\rangle_\frg+\tfrac12\langle  c^+,[ c, c]\rangle_\frg\Big\}
 -\tfrac{1}{4\eps}\int_M \langle  F,  F\rangle_\frg~,
\end{equation}
that is, we find~\eqref{eq:BVActionYM2}  in the second-order formulation plus a topological term. 
 
\paragraph{\mathversion{bold}$L_\infty$-quasi-isomorphism between the formulations of Yang--Mills theory.} 
Since the $L_\infty$-algebras $\sL_{\rm YM_1}$ and $\sL_{\rm YM_2}$ describe equivalent classical field theories, they should be quasi-isomorphic according to our general discussion in Section~\ref{ssec:Classical_FT_and_L_infty}. Let us now show that verifying classical equivalence by giving a quasi-isomorphism is very concise. A first derivation of the quasi-isomorphism was also given in~\cite{Rocek:2017xsj}.

Clearly, we expect a quasi-isomorphism that is based on an $L_\infty$-morphism which is an injective $L_\infty$-morphism $\sL_{\rm YM_2}\embd \sL_{\rm YM_1}$. In the dga-picture, this corresponds to a surjection\linebreak $\Phi:\CCC^\infty(\frF_{\rm YM_1BV})\rightarrow\CCC^\infty(\frF_{\rm YM_2BV})$. For this surjection to be a quasi-isomorphism, we have to verify that 
\begin{equation}\label{eq:quasi_iso_YM}
 \begin{aligned}
  Q_{\rm YM_2BV}\circ \Phi\ =\ \Phi\circ Q_{\rm YM_1BV}~.
 \end{aligned}
\end{equation}

Trying to construct such a $\Phi$, one is led to the surjection defined on the coordinate functions as
\begin{equation}
\begin{gathered}
 \Phi(c)\ :=\ c~,~~~\Phi(B_+)\ :=\ -\tfrac{1}{\eps}F_+~,~~~\Phi(A)\ :=\ A~,\\
\Phi(B_+^+)\ :=\ 0~,~~~\Phi(A^+)\ =\ A^+~,~~~\Phi(c^+)\ :=\ c^+~.
\end{gathered}
\end{equation}
The left-hand side of~\eqref{eq:quasi_iso_YM} reads as 
\begin{equation}
 \begin{aligned}
  &Q_{\rm YM_2BV}\Phi(c+B_++A+B^+_++A^++c^+)\ =\\
  \ &=\ Q_{\rm YM_2BV}(c+A-\tfrac{1}{\eps}F_++A^++c^+)\\
  \ &=\ -\tfrac12[c,c]+\nabla c-\tfrac{1}{\eps}[F_+,c]+\nabla{\star F}+[c,A^+]-\nabla A^+-[c,c^+]~,
 \end{aligned}
\end{equation}
while the right-hand side of~\eqref{eq:quasi_iso_YM} evaluates to
\begin{equation}
 \begin{aligned}
  &\Phi(Q_{\rm YM_1BV}(c+B_++A+B^+_++A^++c^+))\ =\\
  \ &=\ \Phi\big(-\tfrac12[c,c]-[c,B_+]+\nabla c-(F_++\eps B_++[c,B_+^+])\,-\\
  &\kern2cm-\nabla B_+-[c,A^+]+\nabla A^++[B_+,B_+^+]-[c,c^+]\big)\\
  \ &=\ -\tfrac12[c,c]+\nabla c-\tfrac{1}{\eps}[F_+,c]+\nabla(1+\star) F-[c,A^+]+\nabla A^+-[c,c^+]\\
  \ &=\ -\tfrac12[c,c]+\nabla c-\tfrac{1}{\eps}[F_+,c]+\nabla{\star F}-[c,A^+]+\nabla A^+-[c,c^+]~.
 \end{aligned}
\end{equation}
Since both results agree, $\Phi$ defines indeed a morphism of $L_\infty$-algebras. Moreover, this isomorphism is surjective, and because we know that the cohomologies of $\sL_{\rm YM_1}$ and $\sL_{\rm YM_2}$ agree, the $L_\infty$-morphism induces an isomorphism on cohomology. One trivially notes that the symplectic from on $\sL_{\rm YM_2}$ is the pullback of that on $\sL_{\rm YM_1}$ along $\Phi$. Altogether, $\Phi$ defines a cyclic quasi-isomorphism, as expected. 

This short computation shows the power of going back and forth between the bracket formulation of $L_\infty$-algebras and the dga-picture. The direct construction of a quasi-isomorphism in the bracket formulation would have been somewhat lengthier, as was showing the equivalence of the BV actions by integrating out fields in the previous paragraph.

Finally, we note that this quasi-isomorphism also extends to $\CN=4$ supersymmetric Yang--Mills theory using the results about the $\CN=1$ supersymmetric Yang--Mills theory in ten dimensions presented at the end of the previous section.

\subsection{Bagger--Lambert--Gustavsson model}

Due to its relevance to the description of M-theory, which is ultimately the source of much motivation of higher structures, let us also quickly review the Bagger--Lambert--Gustavsson (BLG) model; other Chern--Simons matter theories lead to analogous results. A sub-$L_\infty$-algebra of the $L_\infty$-algebra structure of this model was identified previously in~\cite{IuliuLazaroiu:2009wz}.

\paragraph{Review of the Bagger--Lambert--Gustavsson model.}
Let $M=\FR^{1,2}$. It is convenient to describe the gauge structure of the BLG model using the metric 3-Lie algebra $A_4$. This 3-Lie algebra is a vector space $A_4\cong \FR^4$ with basis $\sigma_a$ and 3-algebra relation and metric structure
\begin{equation}
 [\sigma_a,\sigma_b,\sigma_c]\ =\ {\eps_{abc}}^d\sigma_d\eand \langle \sigma_a,\sigma_b\rangle_{A_4}\ =\ \delta_{ab}~.
\end{equation}
This 3-Lie algebra comes with an associated Lie algebra $\frg_{A_4}\cong \asu(2)\oplus \asu(2)$ of inner derivations acting on $A_4$ and the metric $\langle-,-\rangle_{A_4}$ on $A_4$ induces a metric $\langle-,-\rangle_{\frg_{A_4}}$ of split signature on $\frg_{A_4}$.

The matter fields of the BLG model consist of eight scalars $X^I\in \Omega^0(M,A_4)\otimes \FR^8$ on $M$ with $I,J,\ldots=1,\ldots, 8$ and a Gra{\ss}mann-odd Majorana spinor $\Psi\in \Gamma(M,\Pi S\otimes A_4)$ on $M$ in $\FR^{1,10}$, reduced to 3~dimensions, both taking values in $A_4$. In addition, we have a gauge potential $A\in\Omega^1(M,\frg_{A_4})$ taking values in the Lie algebra $\frg_{A_4}$ associated with $A_4$. Let us decompose the gamma matrices $\Gamma_M$ for $\sSO(1,10)$ as $\Gamma_M\to(\Gamma_\mu,\Gamma_I)$ with $\mu,\nu,\ldots=0,1,2$ and we shall write $\Gamma_{IJK\cdots}$ for the corresponding normalised totally antisymmetric products. The action of the BLG model reads as\cite{Bagger:2007jr,Gustavsson:2007vu}
\begin{equation}\label{eq:BLGaction}
\begin{aligned}
 S_{\rm BLG}\ &:=\ \int_M \Big\{\tfrac12\langle A,\dd A\rangle_{\frg_{A_4}}+\tfrac{1}{3!}\langle A,[A,A]\rangle_{\frg_{A_4}}+\tfrac12\langle X^I,\nabla{\star\nabla X^I}\rangle_{A_4}+\tfrac{\di}{2}\langle \bar\Psi,{\star\slashed{\nabla} \Psi}\rangle_{A_4}\,+\\
&\kern2cm+\tfrac{\di}{4}\langle \bar\Psi,{\star\Gamma_{IJ}}[X^I,X^J,\Psi]\rangle_{A_4}-\tfrac{1}{2\cdot 3!}\langle [X^I,X^J,X^K],{\star[X^I,X^J,X^K]}\rangle_{A_4} \Big\}
 \end{aligned}
\end{equation}
with equations of motion
\begin{equation}\label{eq:BLGeom}
\begin{aligned}
{\nabla{\star\nabla X^I}}+\star\tfrac12[X^J,X^K,[X^I,X^J,X^K]]\ &=\ 0~,\\
\slashed{\nabla}\!_A\Psi+\tfrac12\Gamma_{IJ}[X^I,X^J,\Psi]\ &=\ 0~,\\
\underbrace{\dd A+\tfrac12[A,A]}_{=:\,F}+{\star (X^I\wedge\nabla X^I +\tfrac{\di}{2} \bar\Psi\wedge\Gamma\Psi)}\ &=\ 0~,
\end{aligned}
\end{equation}
where, in local coordinates $x^\mu$, we define $\Gamma:=\dd x^\mu\Gamma_\mu$.

\paragraph{Batalin--Vilkovisky action.}
The action~\eqref{eq:BLGaction} is extended to the corresponding BV action 
\begin{equation}
 S_{\rm BLGBV}\ :=\ S_{\rm BLG}+ S_{\rm gh}~,
\end{equation}
containing the same ghosts and antifields as Chern--Simons theory, $c\in\Omega^0(M,\frg)[1]$, $A^+\in\Omega^2(M,\frg)[-1]$ and $c^+\in\Omega^3(M,\frg)[-2]$, as well as the additional two antifields\linebreak $X^{I+}\in \Omega^3(M,A_4)\otimes \FR^8[-1]$ and $\Psi^+\in \Gamma(M,\Pi S\otimes A_4)[-1]$. Explicitly, 
\begin{equation}
\begin{aligned}
 S_{\rm gh}\ &:=\ \int_M \Big\{-\langle c,\dd A^+\rangle_\frg-\langle c,[A,A^+]\rangle_\frg+\tfrac12 \langle c^+,[c,c]\rangle_\frg\,+\\
 &\kern2cm+\langle X^{I+},c\acton X^{I}\rangle_{A_4}+\star\langle \Psi^+,c\acton \Psi\rangle_{A_4}\Big\}~.
\end{aligned}
\end{equation}
As always, the BV action $S_{\rm BLGBV}$ is the Hamiltonian function for the homological vector field with respect to the canonical symplectic form. The latter encodes the $L_\infty$-algebra structure of the BLG model, and we directly jump to its description.

\paragraph{\mathversion{bold}$L_\infty$-structure.}
The action~\eqref{eq:BLGaction} and the equations~\eqref{eq:BLGeom} can be re-written in $L_\infty$-language. In particular, we consider the complex
\begin{subequations}\label{eq:BGLLinf}
\begin{equation}
\begin{aligned}
 &\underbrace{\Omega^0(M,\frg_{A_4})}_{=:\,\sL_0}\ \xrightarrow{~\mu_1:=\dd~}\ \underbrace{\Omega^1(M,\frg_{A_4})\oplus \Omega^0(M, {A_4})\otimes\FR^8\oplus \Gamma(M,\Pi S\otimes {A_4})}_{=:\,\sL_1}\\
  &\kern0.5cm\xrightarrow{~\mu_1:=\dd+{\dd}{\star\dd}+\slashed{\CD}~}\  \underbrace{\Omega^2(M,\frg_{A_4})\oplus \Omega^3(M, {A_4})\otimes\FR^8\oplus \Gamma(M,\Pi S\otimes {A_4})}_{=:\,\sL_2}\ \\
  &\kern1.5cm\xrightarrow{~\mu_1:=\dd~}\ \underbrace{\Omega^3(M,\frg_{A_4})}_{=:\,\sL_3}~,
 \end{aligned}
\end{equation}
which we call the {\em Bagger--Lambert--Gustavsson complex}, together with the non-vanishing higher products
\begin{equation}
\begin{gathered}
\mu_1(c_1)\ =\ \dd c_1~,~~~\mu_1(A_1+X_1+\Psi_1)\ =\ \dd A_1+{\star\dd}{\star\dd} X_1+\slashed{\CD}\Psi_1~,\\
\mu_1(A_1^++X_1^++\Psi_1^+)\ =\ \dd A_1^+~,\\
\mu_2(c_1,c_2)\ =\ [c_1,c_2]~,~~~\mu_2(c_1,c_2^+)\ =\ [c_1,c^+_2]~,\\ 
\mu_2(c_1,A_1+X_1+\Psi_1)\ =\ [c_1,A_1]+c_1\acton(X_1+\Psi_1)~,\\ 
\mu_2(c_1,A^+_1+X_1^{+}+\Psi^+_1)\ =\ [c_1,A^+_1]+c_1\acton(X_1^{+}+\Psi^+_1)~,\\ 
\mu_2(A_1+X_1+\Psi_1,A_2+X_2+\Psi_2)\ =\hspace{7cm} \\
 \kern-5cm\ =\ [A_1,A_2]+\big\{\slashed{A}_1\acton \Psi_2+\star\dd{\star}(A_1\acton X^I_2)\,+\\
  \hspace{3.3cm}+{\star(A_1\acton\star\dd X^I_2)}+{\star}(X^I_1\wedge\dd X^I_2+\tfrac\di2\bar\Psi_1^I\wedge\Gamma\Psi_2^I)+ (1\leftrightarrow 2) \big\}~,\\
  \mu_2(A_1+X_1+\Psi_1,A_2^++X_2^++\Psi_2^+)\ =\ [A_1,A^+_2]+\frd(X^I_1,X_2^{+I})+\frd(\Psi_1,\Psi_2^+)~,\\
\end{gathered}
\end{equation}
\begin{equation}
\begin{gathered}
  \mu_3(A_1+X_1+\Psi_1,\ldots,A_3+X_3+\Psi_3)\ =\kern6.5cm \\
  \kern1cm\ =\ {\star A_1}\acton({\star A_2}\acton X_3)+\tfrac12\Gamma_{IJ}[X^I_1,X^J_2,\Psi_3]+\star X_1^I\wedge A_2\acton X_3^I+\mbox{cyclic}~,\\
    \mu_5(A_1+X_1+\Psi_1,\ldots,A_5+X_5+\Psi_5)\ =\ \tfrac12[X_1^J,X^K_2,[X_3,X^J_4,X^K_5]]+\mbox{cyclic}~,
\end{gathered}
\end{equation}
\end{subequations}
where $c_i\in\sL_0$, $A_i+X_i+\Psi_i\in\sL_1$, $A^+_i+X^{+}_i+\Psi^+_i\in\sL_2$, and $c^+_i\in\sL_3$ for $i=1,\ldots,5$. In addition, the flavour indices on the $X_i$ and $X_i^+$ are contracted with some basis $\lambda_I$, e.g.~$X_i=X_i^I\lambda_I$ and $\frd:A_4\times A_4\rightarrow \frg$ maps two elements in $A_4$ to the corresponding inner derivation $\frd(\tau_a,\tau_b):=[\tau_a,\tau_b,-]$.

It is rather easy to see that the MC equation~\eqref{eq:MCEquation} translates into~\eqref{eq:BLGeom}. We can endow the above $L_\infty$-algebra with the cyclic inner product
\begin{equation}
\begin{aligned}
\langle\ell_1,\ell_2\rangle_\sL\ &:=\ \int_M\Big\{-\langle c_1,c_2^+\rangle_\frg-\langle c_1^+,c_2\rangle_\frg+\langle A_1,A^+_2\rangle_\frg+\langle A^+_1,A_2\rangle_\frg\,+\\
&\kern2cm+\langle X_1,X^+_2\rangle_{A_4}+\langle X^+_1,X_2\rangle_{A_4}+\star\langle \Psi_1,\Psi^+_2\rangle_{A_4}+\star\langle \Psi^+_1,\Psi_2\rangle_{A_4}\Big\}~,
\end{aligned}
\end{equation}
where $\ell_i=c_i+A_i+X_i+\Psi_i+A_i^++X^+_i+\Psi^+_i+c_i^+$ in the notation used above and spinor indices are contracted with the $\sSpin(1,2)$-invariant metric on $\Pi S$. With this inner product, the MC action~\eqref{eq:MCAction} for the  $L_\infty$-algebra~\eqref{eq:BGLLinf} becomes the BLG action~\eqref{eq:BLGaction}, as expected.

\subsection{Alexandrov--Kontsevich--Schwarz--Zaboronsky construction}

In the cases above, we started from a classical field theory and constructed a corresponding BV action. There are, however, more modern approaches that construct classical field theories directly in their BV form. The most important of these is the Alexandrov--Kontsevich--Schwarz--Zaboronsky (AKSZ) formalism~\cite{Alexandrov:1995kv}. We briefly summarise this construction in the following and derive Chern--Simons theory as an example. For more details and examples, see also~\cite{Cattaneo:2001ys,Roytenberg:2006qz,Kotov:2007nr,Kotov:2010wr} and in particular~\cite{Fiorenza:2011jr} for a modern perspective.\footnote{At the time of the submission of this paper the article~\cite{Hyungrok:2018aa} was posted on the arXiv which deals with the AKSZ construction of higher Chern--Simons theories. See also~\cite{Zucchini:2011aa} for an earlier account.} Examples of AKSZ descriptions of non-topological gauge field theories are found, e.g., in~\cite{Barnich:2010sw,Grigoriev:2012xg}.

\paragraph{Alexandrov--Kontsevich--Schwarz--Zaboronsky data.} We start from an N$Q$-\linebreak manifold $(\Sigma,Q_\Sigma)$, which is endowed with a non-degenerate measure $\mu$ of degree~$-n-1$, $n\in \NN^*$ which is $Q_\Sigma$-invariant. The canonical example here is $\Sigma=T[1]\Sigma_0$ for a compact oriented $(n+1)$-dimensional manifold $\Sigma_0$ without boundary and with $Q_\Sigma$ the de Rham differential. The manifold $\Sigma$ is called the {\em source} and $\Sigma_0$ often corresponds to the world volume of the described objects (point particles, strings, etc.)

We also define a {\em target} (though of as an extended form of the target space), which is a symplectic N$Q$-manifold $(M,Q_M,\omega_M)$ of degree~$n$. As explained in Section~\ref{ssec:L_infty_algebras_and_algebroids}, $Q_M$ is Hamiltonian with Hamiltonian function $\Theta$ satisfying $\{\Theta,\Theta\}=0$, where $\{-,-\}$ is the Poisson bracket induced by $\omega_M$.

The space of fields $\frF_{\rm BV}$ is now the space of maps from $\Sigma$ to $M$. This is, in fact, a very general construction. For example, the mechanics of point particles can be described by maps from their worldline $\FR$ into spacetime. It also leads to vast generalisations of gauge theories after appropriate refinement, see~\cite{Sati:2008eg}. In particular, in the case $\Sigma=T[1]\Sigma_0$ and $M=\frg[1]$ with $Q_M$ the Chevalley--Eilenberg differential of the Lie algebra $\frg$, we obtain the kinematical data of Chern--Simons theory as morphisms of degree~0. Let $\xi^\alpha$ be again the coordinate functions on $\frg[1]$ and let $\dd x^\mu$ be the generators of $\Omega^\bullet(\Sigma_0)$ over $\CCC^\infty(\Sigma_0)$. Then a dga-morphism $a:\CCC^\infty(\frg[1])\rightarrow \Omega^\bullet(\Sigma_0)$ maps
\begin{equation}
 a:\xi^\alpha\ \mapsto\ \dd x^\mu A_\mu^\alpha(x)\ =:\ A^\alpha~,
\end{equation}
such that $a\circ Q_M=Q_\Sigma\circ a$ or, equivalently,
\begin{equation}
 (a\circ Q_M) \xi^\alpha\ =\ a(-\tfrac12 f_{\beta\gamma}{}^\alpha \xi^\beta \xi^\gamma)\ =\ -\tfrac12 f_{\beta\gamma}{}^\alpha A^\beta\wedge A^\gamma \ =\ \dd A^\alpha\ =\ (Q_\Sigma\circ a) \xi^\alpha~.
\end{equation}
We thus obtain a gauge potential $A\in \Omega^1(\Sigma_0,\frg)$ whose curvature $F:=\dd A+\tfrac12[A,A]$ vanishes.

\paragraph{Batalin--Vilkovisky structure.} Note that $\frF_{\rm BV}$ is naturally graded, and the degree will be the ghost number of the fields. In addition, the structures on $\Sigma$ and $M$ endow $\frF_{\rm BV}$ with a homological vector field and a symplectic form. To obtain the symplectic form, note that there is the evaluation map ${\rm ev}:\frF_{\rm BV}\times \Sigma \rightarrow M$, acting as ${\rm ev}(\phi,x):=\phi(x)$.

We can pull back any differential form $\alpha\in \Omega^\bullet(M)$ along ${\rm ev}$ to $\frF_{\rm BV}\times \Sigma$ and subsequently integrate over $\Sigma$, leading to the map
\begin{equation}\label{eq:AKSZ:def_p}
 p(\alpha)\ :=\ \int_\Sigma \mu~{\rm ev}^*\alpha~.
\end{equation}
We can use this map to define the symplectic form 
\begin{equation}
 \omega_{\rm BV}\ =\ p(\omega_M)
\end{equation}
inducing the BV bracket. From the degrees of $\omega_M$ and $\mu$ it is clear that $\omega_{\rm BV}$ is of degree~$-1$. Also, $\omega_{\rm BV}$ is non-degenerate if $\mu$ is non-degenerate.

To construct the homological vector field $Q_{\rm BV}$, note that diffeomorphisms on both $\Sigma$ and $M$ induce an action on $\frF_{\rm BV}$, by pre-composition or post-composition, respectively. Therefore, the two homological vector fields $Q_\Sigma$ and $Q_M$ induce vector fields $\hat Q_\Sigma$ and $\hat Q_M$ on $\frF_{\rm BV}$ and we can choose any linear combination of these to form $Q_{\rm BV}$. The compatibility between $Q_{\rm BV}$ and $\omega_{\rm BV}$ is readily checked, cf.~\cite{Cattaneo:2001ys,Roytenberg:2006qz}. In particular, one can show that the map $p$ defined in~\eqref{eq:AKSZ:def_p} is a symplectomorphism. Thus, 
\begin{equation}
 \{\Theta,\Theta\}\ =\ 0\ \Longleftrightarrow\ \{p(\Theta),p(\Theta)\}_{\rm BV}\ =\ 0~,
\end{equation}
and therefore $p(\Theta)$ is the Hamiltonian of a homological vector field $\hat Q_M$. The contribution of $\hat Q_M$ to the Hamiltonian of $Q_{\rm BV}$ is thus a multiple of $p(\Theta)$.

On the other hand, $\mu$ is invariant under $Q_\Sigma$ and so is ${\rm ev}$ under the simultaneous action on $\frF_{\rm BV}$ and $\Sigma$, which leads to 
\begin{equation}
 \CL_{\hat Q_\Sigma}p\ =\ \hat Q_\Sigma\intprod\dd p+\dd ({\hat Q_\Sigma}\intprod p)\ =\ 0~.
\end{equation}
If the symplectic form $\omega_M$ is exact, $\omega_M=\dd \vartheta$, then the Hamiltonian of $\hat Q_{\Sigma}$ is therefore ${\hat Q_\Sigma}\intprod p(\vartheta)$. Moreover, if $\vartheta= \vartheta_\alpha(\xi) \dd \xi^\alpha$ in some coordinates $\xi^\alpha$ on $M$, then  
\begin{equation}
 {\hat Q_\Sigma}\intprod p(\vartheta)\ =\ \int_{\Sigma} \vartheta_\alpha(\phi)\delta \phi^\alpha~,
\end{equation}
where $\phi^\alpha$ is the coordinate corresponding to $\xi^\alpha$ on $\frF_{\rm BV}$ under the map $\phi:\Sigma\rightarrow M$.

Altogether, the Hamiltonian of $Q_{\rm BV}$, which is a linear combination of ${\hat Q_\Sigma}\intprod p(\vartheta)$ and $p(\Theta)$, is the classical BV action. For a more precise argument regarding to which linear combinations are preferable, see~\cite{Kotov:2007nr,Fiorenza:2011jr}.

\paragraph{Example: Chern--Simons theory.} As a simple example, consider the case $\Sigma=T[1]\Sigma_0$ with $Q_\Sigma=\dd$ for a compact oriented three-dimensional manifold $\Sigma_0$. As target, choose $M=\frg[1]$ where $\frg$ is a metric Lie algebra with coordinates $\xi^\alpha$ and metric $\langle \tau_\alpha,\tau_\beta\rangle_\frg=\omega_{\alpha\beta}$ inducing a symplectic form $\omega_\frg=\tfrac12\omega_{\alpha\beta}\dd\xi^\alpha\wedge \dd \xi^\beta$ of degree~2. The Hamiltonian $\Theta$ of $Q_{\frg[1]}$ is $\Theta=\tfrac{1}{3!}f_{\alpha\beta\gamma}\xi^\alpha\xi^\beta\xi^\gamma$ with $f_{\alpha\beta\gamma}:=f_{\alpha\beta}{}^\delta\omega_{\delta\gamma}$ and corresponds to the 3-cocycle $\langle -,[-,-]\rangle_\frg$. 

The maps from $\Sigma$ to $\frg[1]$ form the space of $\frg$-valued forms on $\Sigma_0$, and we have $\frF_{\rm BV}=\Omega^\bullet(\Sigma_0,\frg)$. The symplectic form reads as
\begin{equation}
 \omega_{\rm BV}\ :=\ \int_{\Sigma_0} \tfrac12 \omega_{\alpha\beta} \delta \phi^\alpha \wedge \delta \phi^\beta\ =\ \int_{\Sigma_0} \langle \delta \phi, \delta \phi\rangle_\frg
\end{equation}
for $\phi\in \Omega^\bullet(\Sigma_0,\frg)$. Note that $\omega_{\frg}$ is exact (see Section~\ref{ssec:QManifolds}) with symplectic potential $\vartheta=\tfrac12 \xi^\alpha\omega_{\alpha\beta}\dd \xi^\beta$, and the two contributions to $S_{\rm BV}$ from degree~0 maps are
\begin{equation}
 {\hat Q_\Sigma}\intprod p(\vartheta)\ =\ \tfrac12 \int_{\Sigma_0} \langle \phi,\dd \phi\rangle_\frg\eand p(\Theta)\ =\ \tfrac{1}{3!}\int_{\Sigma_0} \langle \phi,[\phi,\phi]\rangle_\frg~,
\end{equation}
where $\phi\in \Omega^\bullet(\Sigma_0,\frg)$. To see this, note that the pullback along the evaluation map yields
\begin{equation}
 {\rm ev}^* \xi^\alpha\ =\ (x,\phi^\alpha (x))
\end{equation}
and 
\begin{equation}
 {\hat Q_\Sigma}\intprod p(\vartheta)\ =\ {\hat Q_\Sigma}\intprod  \int_{\Sigma_0} \tfrac12\omega_{\alpha\beta}\phi^\alpha {\rm D} \phi^\beta\ =\ \tfrac12 \int_{\Sigma_0} ~\phi^\alpha\omega_{\alpha\beta}\dd \phi^\beta~,
\end{equation}
where ${\rm D}$ is the de Rham differential on $\frF_{\rm BV}\times \Sigma$.

If we decompose $\phi$ into forms of homogeneous degree, $\phi=c+A+A^++c^+$, and linearly combine both of the above contributions, we obtain the classical BV action of Chern--Simons theory,
\begin{equation}
\begin{aligned}
 S_{\rm BV}\ &=\ \int_{\Sigma_0}\Big\{\tfrac12\langle A,\dd A\rangle_\frg+\tfrac{1}{3!}\langle A,[A,A]\rangle_\frg\,-\\
 &\kern2cm-\langle c,\dd A^+\rangle_\frg+\langle A,[A^+,c]\rangle_\frg+\tfrac12\langle c^+,[c,c]\rangle_\frg\Big\}~.
 \end{aligned}
\end{equation}

\acknowledgements 

We would like to thank Alexandros Arvanitakis, James Grant, Jan Gutowski, Alexander Schenkel, Paul Skerritt, and Alessandro Torrielli for useful discussions. We are also grateful to all the participants of the \href{http://www.maths.dur.ac.uk/lms/109/index.html}{\it EPSRC/LMS Durham Symposium on Higher Structures in M-Theory} for fruitful conversations. We would like to thank Grigorios Giotopoulos, Tommaso Macrelli, Dominik Rist and Jim Stasheff for useful comments on first versions of this paper.  B.J.~was supported by the GA\v CR Grant 18-07776S. L.R.~is partially supported by the EPSRC grant EP/N509772. C.S.~was supported in part by the STFC Consolidated Grant ST/L000334/1 {\it Particle Theory at the Higgs Centre}. M.W.~was supported in part by the STFC Consolidated Grant ST/L000490/1 {\it Fundamental Implications of Fields, Strings, and Gravity}.

\datamanagement

No additional research data beyond the data presented and cited in this work are needed to validate the research findings in this work.

\appendices

\subsection{\texorpdfstring{$L_\infty$}{L-infty}-algebras and their morphisms from coalgebras}\label{app:L-infty-morphisms}

Below we explain in detail the connection between $L_\infty$-algebras and codifferential coalgebras. We also derive the structure equation for morphisms of $L_\infty$-algebras from morphisms of codifferential coalgebras. The relevant original reference for this material is~\cite{Lada:1994mn}, helpful may also be the detailed discussions in~\cite{Schuhmacher:0405485,Fregier:2013dda,Khudaverdyan:2015wba}.

\paragraph{Preliminaries.} Given a real graded vector space $V$, we define the following associative algebras:
\begin{equation}
 \begin{aligned}
    \mbox{tensor algebra}&:\,&\bigotimes\nolimits^\bullet \sV\ &:=\ \FR\oplus \sV \oplus (\sV\otimes \sV)\oplus~\cdots\ =\ \bigoplus_{k\geq0}\bigotimes\nolimits^k \sV~,\\
 \mbox{symmetric tensor algebra}&:\,&\bigodot\nolimits^\bullet \sV\ &:=\ \FR\oplus \sV \oplus (\sV\odot \sV) \oplus ~\cdots\ =\ \bigoplus_{k\geq0}\bigodot\nolimits^k \sV~,\\
  \mbox{antisymmetric tensor algebra}&:\,&\bigwedge\nolimits^\bullet \sV\ &:=\ \FR\oplus \sV \oplus (\sV\wedge \sV) \oplus~\cdots\ =\ \bigoplus_{k\geq0}\bigwedge\nolimits^k \sV~,\\
   \mbox{reduced tensor algebra}&:\,&\bigotimes\nolimits^\bullet_0 \sV\ &:=\ \sV \oplus (\sV\otimes \sV) \oplus~ \cdots\ =\ \bigoplus_{k\geq1}\bigotimes\nolimits^k \sV~~,\\
  \mbox{reduced sym.~tensor algebra}&:\,&\bigodot\nolimits^\bullet_0 \sV\ &:=\ \sV \oplus (\sV\odot \sV) \oplus~ \cdots\ =\ \bigoplus_{k\geq1}\bigodot\nolimits^k \sV~,\\
  \mbox{reduced antisym.~tensor algebra}&:\,&\bigwedge\nolimits^\bullet_0 \sV\ &:=\ \sV \oplus (\sV\wedge \sV) \oplus~\cdots\ =\ \bigoplus_{k\geq1}\bigwedge\nolimits^k \sV~,\\
 \end{aligned}
\end{equation}
cf.~Section~\ref{ssec:DGAs}. Here, $\odot$ and $\wedge$ denote the graded symmetric  and antisymmetric tensor products, with weight one, e.g.
\begin{equation}
 v_1\odot v_2\ :=\ v_1\otimes v_2+(-1)^{|v_1|\,|v_2|}v_2\otimes v_1~.
\end{equation}
These tensor products yield embeddings of $\bigodot^\bullet \sV$ and $\bigwedge^\bullet \sV$ into $\bigotimes^\bullet \sV$ as well as $\bigodot^\bullet_0 \sV$ and $\bigwedge^\bullet_0 \sV$ into $\bigotimes^\bullet_0 \sV$. We also have projectors from the reduced tensor algebra $\bigotimes^\bullet_0 V$ to both reduced symmetric and antisymmetric algebras:
\begin{equation}\label{eq:projectors_on_odot}
\begin{aligned}
 \pr_{\odot}(v_1\otimes \dots \otimes v_i)\ &:=\ \sum_{\sigma \in S_i} \eps(\sigma;v_1,\dots,v_i) v_{\sigma(1)}\odot \dots \odot v_{\sigma(i)}~,\\
 \pr_{\wedge}(v_1\otimes \dots \otimes v_i)\ &:=\ \sum_{\sigma \in S_i} \chi(\sigma;v_1,\dots,v_i) v_{\sigma(1)}\wedge \ldots \wedge v_{\sigma(i)}~,
\end{aligned}
\end{equation}
where $\eps(\sigma;v_1,\dots,v_i)$ and $\chi(\sigma;v_1,\dots,v_i)$ are the symmetric and antisymmetric Koszul signs of a permutation $\sigma$. Explicitly, we have 
\begin{equation}
 v_1\odot\dots\odot v_i\ =\ \eps(\sigma;v_1,\dots,v_i) v_{\sigma(1)}\odot \dots \odot v_{\sigma(i)}
\end{equation}
and 
\begin{equation}
 \ell_1\wedge\ldots\wedge \ell_i\ =\ \chi(\sigma;\ell_1,\dots,\ell_i) \ell_{\sigma(1)}\wedge \ldots \wedge \ell_{\sigma(i)}
\end{equation}
for $\ell_1,\ldots,\ell_i\in \sV$.  Using the shift isomorphism $s^\bullet$ defined in~\eqref{eq:isomorphism_asym_sym}, we obtain the identity
\begin{equation}\label{eq:relation_chi_eps}
 \chi(\sigma;\ell_1,\dots,\ell_i)\ =\ (-1)^{\sum_{j=1}^{i-1}(i-j)(|\ell_j|+|\ell_{\sigma(j)}|)} \eps(\sigma;s \ell_1,\dots,s \ell_i)~,
\end{equation}
which we shall use later. 

\paragraph{Reduced symmetric coalgebra.} Consider now the reduced algebras $\bigwedge^\bullet_0 \sV$ and $\bigodot_0^\bullet \sV$ as introduced in~\eqref{eq:reduced_algebras}. Together with the reduced comultiplication,
\begin{equation}
\begin{gathered}
 \Delta_0\,:\, \bigodot\nolimits^\bullet_0 \sV\ \rightarrow\  \bigodot\nolimits^\bullet_0 \sV\otimes  \bigodot\nolimits^\bullet_0 \sV~,\\
v_1\odot \dots \odot v_i\ \mapsto\ \sum_{j+k=i}\sum_{\sigma\in{\rm Sh}(j;k)} \eps(\sigma;v_1,\ldots,v_i)(v_{\sigma(1)}\odot \dots \odot v_{\sigma(j)})\otimes (v_{\sigma(j+1)}\odot \dots \odot v_{\sigma(i)})~, 
\end{gathered}
\end{equation}
$\bigodot^\bullet_0 \sV$ becomes a cocommutative coalgebra. 

Given functions $f_1,f_2: \bigodot^\bullet_0 V\rightarrow \bigodot^\bullet_0 V$, we define the symmetrised tensor product
\begin{equation}\label{eq:odot_functions}
 f_1\odot f_2\ :=\  m_\odot\circ (f_1\otimes f_2)\circ \Delta_0~,
\end{equation}
where $m_\odot(v_1\otimes v_2):=v_1\odot v_2$ for $v_{1,2}\in V$. Explicitly,
\begin{equation}
 \begin{aligned}
  &(f_1\odot f_2)(v_1\odot \dots \odot v_i)\ =\\
  \ &=\ \sum_{j+k=i}\sum_{\sigma\in{\rm Sh}(j;k)} \eps(\sigma;v_1,\ldots,v_i)f_1(v_{\sigma(1)}\odot \dots \odot v_{\sigma(j)})\odot f_2 (v_{\sigma(j+1)}\odot \dots \odot v_{\sigma(i)})~.
 \end{aligned}
\end{equation}
Formula~\eqref{eq:odot_functions} generalises to expressions $f_1\odot\dots\odot f_i$ using unshuffles with $i$ ordered subsets.

\paragraph{Codifferentials.} A {\em codifferential} on a coalgebra $(\bigodot_0^\bullet \sV,\Delta_0)$ is a linear map $D:\bigodot_0^\bullet \sV\rightarrow \bigodot_0^\bullet \sV$ of degree~1, which is a nilquadratic coderivation,
\begin{equation}
 \Delta_0 D\ =\ (D\otimes 1)\Delta_0+(1 \otimes D) \Delta_0\eand D^2\ =\ 0~.
\end{equation}
The first equation is the coalgebra analogue of the Leibniz rule. The second equation is equivalent to 
\begin{equation}\label{eq:D2_in_components}
 \sum_{i=1}^\infty \sum_{j+k=i} D_{k+1}(D_j\odot \id^{\odot k})\ =\ 0
\end{equation}
for all $i\geq 1$, where $D_k:\bigodot_0^k\sV\rightarrow \sV$ is the restriction of the codifferential,
\begin{equation}
 D_k\ :=\ \pr_{\sV}\circ D\circ \pr_{\odot^k_0\sV}~.
\end{equation}
Note that we can pull back $D_k$ along the projection $\pr_{\odot}$ defined in~\eqref{eq:projectors_on_odot} to a map $D_k:\bigotimes_0^k\sV\rightarrow \sV$. The condition $D^2=0$ then simply translates to 
\begin{equation}\label{eq:post_comp_shuffles}
 \sum_{j+k=i} D_{k+1}(D_j\otimes \id^{\otimes k})\circ \sum_{\sigma\in {\rm Sh}(j;i)} \eps(\sigma,-) \sigma(-)\ =\ 0
\end{equation}
for every $i>0$, where the sum is taken over all $(j;i)$-unshuffles, $\eps(\sigma,-)$ is the Koszul sign of the unshuffle and $\sigma(-)$ is the application of the unshuffle to elements of $\bigotimes^i_0 \sV$.

\paragraph{\mathversion{bold}$L_\infty$-algebras from codifferentials.} The restricted codifferentials on the coalgebra\linebreak $(\bigodot_0^\bullet \sV,\Delta_0)$ now induce a set of totally antisymmetric, multilinear products $\mu_i$ on $\bigwedge_0^\bullet \sL$ with $\sV=\sL[1]$. We define
\begin{equation}\label{eq:def_mu_i_2}
 \mu_i\ :=\ (-1)^{\frac12 i(i-1)+1}s^{-1}\circ D_i\circ s^{\otimes i}
\end{equation}
with $s^{\otimes i}$ as in~\eqref{eq:isomorphism_asym_sym}. We can now insert
\begin{equation}
 \id^{\otimes i}\ =\ (-1)^{\frac12i(i-1)} s^{\otimes i}\circ (s^{-1})^{\otimes i} 
\end{equation}
into the conditions~\eqref{eq:D2_in_components} equivalent to $D^2=0$ and concatenate with $s^{-1}$ and $s^{\otimes i}$ to obtain
\begin{equation}
\begin{aligned}
 0\ &=\ \sum_{j+k=i} s^{-1}\circ D_{k+1}\circ (D_j\otimes \id^{\otimes k})\circ \sum_{\sigma\in {\rm Sh}(j;i)} \eps(\sigma,-) \sigma(-)\circ s^{\otimes i}\\
 \ &=\ \sum_{j+k=i} (-1)^{\frac12k(k+1)+\frac12i(i-1)}s^{-1}\circ D_{k+1}\circ s^{\otimes {k+1}} \\
 &\hspace{1cm}\circ(s^{-1})^{\otimes (k+1)}\circ (D_j\otimes \id^{\otimes k})\circ s^{\otimes i}\circ (s^{-1})^{\otimes i}\circ\sum_{\sigma\in {\rm Sh}(j;i)} \eps(\sigma,-) \sigma(-)\circ s^{\otimes i}\\
 \ &=\ \sum_{j+k=i} (-1)^{\frac12k(k+1)+k(j+1)+\frac12i(i-1)}s^{-1}\circ D_{k+1}\circ s^{\otimes {k+1}} \\
 &\hspace{1cm}\circ \big((s^{-1}\circ D_j\circ s^{\otimes j})\otimes (s^{-1})^{\otimes k}\big)\circ (\id^{\otimes j}\otimes s^{\otimes k})\circ (s^{-1})^{\otimes i}\circ\sum_{\sigma\in {\rm Sh}(j;i)} \eps(\sigma,-) \sigma(-)\circ s^{\otimes i}\\
 \ &=\ \sum_{j+k=i} (-1)^{\frac12i(i-1)+\frac12j(j-1)+k(j+1)+\frac12k(k-1)}\mu_{k+1}\circ \\
 &\hspace{1cm}(\mu_j\otimes \id^{\otimes k})\circ (s^{-1})^{\otimes i}\circ \sum_{\sigma\in {\rm Sh}(j;i)} \eps(\sigma,-) \sigma(-)\circ s^{\otimes i}\\
 \ &=\ \sum_{j+k=i} (-1)^{\frac12i(i-1)+k}\mu_{k+1}\circ (\mu_j\otimes \id^{\otimes k})\circ \sum_{\sigma\in {\rm Sh}(j;i)} \chi(\sigma,-) \sigma(-)~,
\end{aligned}
\end{equation}
where we used the identity
\begin{equation}
 (s^{-1})^{\otimes i}\circ\sum_{\sigma\in {\rm Sh}(j;i)} \eps(\sigma,-) \sigma(-)\circ s^{\otimes i}\ =\ (-1)^{\frac12i(i-1)}\sum_{\sigma\in {\rm Sh}(j;i)} \chi(\sigma,-) \sigma(-)~.
\end{equation}
Evaluating the above result on $\ell_1\otimes \dots \otimes \ell_n\in\bigotimes^n\sL$, we obtain the higher homotopy Jacobi identity,
\begin{equation}
 \sum_{j+k=i} \sum_{\sigma\in \Sh(j;i)} (-1)^{k}\chi(\sigma;\ell_1,\dots, \ell_i) \mu_{k+1}(\mu_j(\ell_{\sigma(1)},\dots,v_{\sigma(j)}),\ell_{\sigma(j+1)},\dots,\ell_{\sigma(i)})\ =\ 0~.
\end{equation}

\paragraph{Higher products and the differential graded algebra picture.} Let us briefly link the higher products $\mu_i$ and the differential graded algebra picture of an $L_\infty$-algebra. We start with a graded vector space $\sL$ with basis $\tau_\alpha$ and corresponding coordinate functions $\xi^\alpha$ on $\sL[1]$ with respect to $s\tau_\alpha$. The choice of basis defines structure constants $m^\alpha_{\beta_1\cdots\beta_k}$ via the equation
\begin{equation}\label{eq:def_struct_consts}
 Q\xi^\alpha\ =\ \sum_{k\geq 1} \frac{(-1)^{|\alpha|}}{k!}m^\alpha_{\beta_1\cdots\beta_k}\xi^{\beta_1}\cdots \xi^{\beta_k}~,
\end{equation}
where for each $k$, $|\beta_1|+\dots+|\beta_k|=|\alpha|+1$. The coordinate functions $\xi^\alpha$ are now maps $\sL[1]\rightarrow \FR$, satisfying $\xi^\alpha(X)=\xi^\alpha(X^\beta s\tau_\beta)=X^\alpha$, where $X^\alpha\in \FR$ are the coordinates of the vector $X\in \sL[1]$ with respect to the basis $s\tau_\alpha$. Correspondingly, $\xi^{\beta_1}\cdots \xi^{\beta_k}$ is a function $\bigodot^k_0 \sL[1]\rightarrow \FR$, with
\begin{equation}
 (\xi^{\beta_1}\cdots \xi^{\beta_k})(s\tau_{\gamma_1}\odot \cdots \odot s\tau_{\gamma_k})\ =\ (-1)^{\sum_{i=2}^k\sum_{j=1}^{i-1}|\beta_i||\beta_j|}\delta^{\beta_1}_{(\gamma_1}\cdots \delta^{\beta_k}_{\gamma_k)}~,
\end{equation}
where the symmetrisation of the indices $\gamma_1,\dots,\gamma_k$ is evidently graded. Note that by precomposition with the projection $\pr_{\bigodot^k_0 \sL[1]}$, $\xi^{\beta_1}\cdots \xi^{\beta_k}$ becomes a map $\bigodot^\bullet_0\sL[1]\rightarrow \FR$.

We now contract both sides of~\eqref{eq:def_struct_consts} by $\tau_\alpha$ from the left and apply the result to $(s\tau_{\gamma_1}\odot \dots \odot s\tau_{\gamma_k})$ to obtain
\begin{equation}
\begin{aligned}
 \tau_\alpha (Q\xi^\alpha)(s\tau_{\gamma_1}\odot \dots \odot s\tau_{\gamma_k})\ &=\ \sum_{k\geq 1} \frac{(-1)^{|\alpha|}}{k!}\tau_\alpha m^\alpha_{\beta_1\cdots\beta_k}(\xi^{\beta_1}\cdots \xi^{\beta_k})(s\tau_{\gamma_1}\odot \cdots \odot s\tau_{\gamma_k})~,\\
 \tau_\alpha\xi^\alpha D(s\tau_{\gamma_1}\odot \dots \odot s\tau_{\gamma_k})\ &=\
 (-1)^{\sum_{i=2}^k\sum_{j=1}^{i-1}|\gamma_i||\gamma_j|}\tau_\alpha m^\alpha_{\gamma_1\cdots\gamma_k}~,\\
 s^{-1}(D(s\tau_{\gamma_1}\odot \dots \odot s\tau_{\gamma_k}))\ &=\
 (-1)^{\sum_{i=2}^k\sum_{j=1}^{i-1}|\gamma_i||\gamma_j|}\tau_\alpha m^\alpha_{\gamma_1\cdots\gamma_k}~. 
\end{aligned}
\end{equation}
Using~\eqref{eq:def_mu_i_2}, we now compute
\begin{equation}\label{eq:trans_Q_mu}
\begin{aligned}
 \mu_i(\tau_{\alpha_1},\dots,\tau_{\alpha_i})\ &=\ ((-1)^{\frac12i(i-1)}s^{-1}\circ D_i\circ s^{\odot i})(\tau_{\alpha_1},\dots,\tau_{\alpha_k})\\
 \ &=\ ((-1)^{\frac12i(i-1)+\sum_{j=1}^i(i-j)|\tau_{\alpha_j}|})(s^{-1}(D_i(s\tau_{\alpha_1},\dots,s\tau_{\alpha_i}))\\
 \ &=\ (-1)^{\frac12i(i-1)+\sum_{j=2}^i\sum_{k=1}^{j-1}(|\tau_{\alpha_j}|+1)(|\tau_{\alpha_k}|+1)+\sum_{j=1}^i(i-j)|\tau_{\alpha_j}|}~\tau_\beta m^{\beta}_{\alpha_1\cdots \alpha_i}\\
 \ &=\ (-1)^{\sum_{j=2}^i\sum_{k=1}^{j-1}|\tau_{\alpha_j}||\tau_{\alpha_k}|+\sum_{j=1}^i(j+1)|\tau_{\alpha_j}|}~\tau_\beta m^{\beta}_{\alpha_1\cdots \alpha_i}~.
\end{aligned} 
\end{equation}
Similarly, we have
\begin{equation}\label{eq:comp_contracted_ccf}
 \begin{aligned}
 \tau_\alpha (Q\xi^\alpha)\ &=\ \sum_{i\geq 1} \frac{(-1)^{|\alpha|}}{i!}\tau_\alpha m^\alpha_{\beta_1\cdots\beta_i}(\xi^{\beta_1}\cdots \xi^{\beta_i})~,\\
 (Q\xi)\ &=\ -\sum_{i\geq 1} \frac{(-1)^{\sum_{j=2}^i\sum_{k=1}^{j-1}|\tau_{\beta_j}||\tau_{\beta_k}|+\sum_{j=1}^i(j+1)|\tau_{\beta_j}|}}{i!}
 \mu_i(\tau_{\beta_1},\dots,\tau_{\beta_i})(\xi^{\beta_1}\cdots \xi^{\beta_i})\\
 \ &=\ -\sum_{i\geq 1} \frac{(-1)^{\sum_{j=2}^i\sum_{k=1}^{j-1}|\tau_{\beta_j}|(2|\tau_{\beta_k}|+1)+\sum_{j=1}^i(j+1)|\tau_{\beta_j}|}}{i!}
 \hat \mu_i(\xi,\dots,\xi)\\
 \ &=\ -\sum_{i\geq 1} \frac{1}{i!} \hat \mu_i(\xi,\dots,\xi)~,
 \end{aligned}
\end{equation}
where we used the higher products $\hat \mu_i$ on $\sL_\CCC$ as defined in~\eqref{eq:GhostProd}. 

\paragraph{\mathversion{bold}$L_\infty$-morphisms from coalgebra morphisms.}
Recall that a {\em morphism of coalgebras} from $(\bigodot_0^\bullet \sV ,\Delta_0)$ to $(\bigodot_0^\bullet \sV',\Delta'_0)$ is a map $\Phi:\bigodot_0^\bullet \sV \rightarrow \bigodot_0^\bullet \sV' $ of degree~0 which satisfies
\begin{equation}
 \Delta'_0\circ \Phi\ =\ (\Phi\otimes \Phi)\circ \Delta_0~.
\end{equation}
If $\bigodot_0^\bullet \sV $ and $\bigodot_0^\bullet \sV' $ are both endowed with a codifferential $D$ and $D'$, respectively, we also demand that
\begin{equation}\label{eq:CoMoCoDi}
 \Phi \circ D\ =\  D'\circ \Phi~.
\end{equation}

Consider now two codifferential coalgebras $(\bigodot_0^\bullet \sV ,\Delta_0,D)$ and $(\bigodot_0^\bullet \sV' ,\Delta'_0,D')$ corresponding to two $L_\infty$-algebras. We can restrict a morphism 
\begin{equation}
 \Phi:(\mbox{$\bigodot$}_0^\bullet \sV ,\Delta_0,D)\ \rightarrow\ (\mbox{$\bigodot$}_0^\bullet \sV' ,\Delta'_0,D')
\end{equation}
to the maps
\begin{equation}
 \Phi^j\ :=\ \pr_{\bigodot^j\sV' }\circ \Phi~.
\end{equation}
Note that $\Phi$ is uniquely reconstructed from $\Phi^1$ since $\Phi$ is a morphism of coalgebras, cf.~e.g.~\cite[Prop.~1.2]{Schuhmacher:0405485}. In particular, using the further decomposition
\begin{equation}
 \Phi^j_i\ :=\ \pr_{\bigodot^j\sV' }\circ \Phi|_{\bigodot^i\sV }~,
\end{equation}
we have explicitly
\begin{equation}\label{eq:phi_action}
 \Phi^j_i\ =\ \sum_{k_1+\dots+k_j=i}~\frac{1}{k_1!\cdots k_j!}~\frac{1}{j!}~(\Phi_{k_1}^1\odot\dots\odot \Phi_{k_j}^1)~.
\end{equation}
Note that the maps $\Phi_{k_1}^1\odot\dots\odot \Phi_{k_j}^1$ act on the totally graded symmetrised elements of $\bigodot^i\sV $ and the result is then projected onto $\bigodot^j_0\sV'$. The degrees of all the maps $\Phi$, $\Phi_i$, $\Phi_i^j$ and $\Phi|_{\odot^i\sV }$ are zero, since $\Phi$ is of degree zero, and all other maps originate from restriction and concatenation with projectors.

Condition~\eqref{eq:CoMoCoDi} decomposed into the maps $\Phi^1_i$ and $D_i$ reads as
\begin{multline}
 \sum_{j+k=i}~\Phi^1_{k+1} \left(D_j\otimes \id^{\otimes k}\right)\
 =\ \sum_{j=1}^i ~\sum_{k_1+\dots+k_j=i}~D'_j\circ~\frac{1}{k_1!\cdots k_j!}~\frac{1}{j!}~(\Phi_{k_1}^1\odot\dots\odot \Phi_{k_j}^1)~.
\end{multline}
We multiply this equation by factors of $s$ and $s^{-1}$, restrict it and rewrite both sides, using the shifted morphisms\footnote{The degree of the map $\phi_i$ follows from its definition together with $|\Phi^1_i|=0$ and $|s|=1$.}
\begin{equation}
 \phi_i \ :=\  (-1)^{\frac12 i(i-1)} s^{-1}\circ \Phi_i^1\circ s^{\otimes i}\ewith|\phi_i|\ =\ i-1~,
\end{equation}
and the postcomposition with unshuffles, cf.~\eqref{eq:post_comp_shuffles}, as
\begin{equation}
 \begin{aligned}
    &\sum_{j+k=i}~(-1)^{\frac12 k(k+1)+\frac12 i(i-1)}s^{-1}\circ \Phi^1_{k+1}\circ s^{\otimes (k+1)}\circ (s^{-1})^{\otimes (k+1)}\circ \left(D_j\otimes \id^{\otimes k}\right)\circ s^{\otimes i}\,\circ\\
    &\hspace{1cm}\circ (s^{-1})^{\otimes i}\circ \sum_{\sigma\in {\rm Sh}(j;i)} \eps(\sigma,-) \sigma(-)\circ s^{\otimes i}\\
    &~~\ =\ \sum_{j+k=i}~(-1)^{\frac12 k(k-1)+k(j+1)+\frac12 j(j-1)}\phi_{k+1}\circ \left(\mu_j\otimes \id^{\otimes k}\right)\circ\sum_{\sigma\in {\rm Sh}(j;i)} \chi(\sigma,-) \sigma(-)\\
 \end{aligned}
\end{equation}
and
\begin{equation}
 \begin{aligned}
  &\sum_{j=1}^i~\sum_{k_1+\dots+k_j=i}~\frac{(-1)^{\frac12i(i-1)+\frac12j(j-1)}}{j!}~s^{-1}\circ D'_j\circ s^{\otimes j}\circ (s^{-1})^{\otimes j}\circ~(\Phi_{k_1}^1\otimes\dots\otimes \Phi_{k_j}^1)\circ s^{\otimes i}\,\circ\\
  &\hspace{1cm}\circ (s^{-1})^{\otimes i}\circ \sum_{\sigma\in {\rm Sh}(j;i)} \eps(\sigma,-) \sigma(-)\circ s^{\otimes i}\\
  &~~\ =\ \sum_{j=1}^i~\sum_{k_1+\dots+k_j=i}~\frac{1}{j!}~\mu'_j\circ (s^{-1})^{\otimes j}\circ~(\Phi_{k_1}^1\otimes\dots\otimes \Phi_{k_j}^1)\circ s^{\otimes i}\,\circ\\
  &\hspace{1cm}\circ \sum_{\sigma\in {\rm Sh}(j;i)} \chi(\sigma,-) \sigma(-)\\
  &~~\ =\ \sum_{j=1}^i~\sum_{k_1+\dots+k_j=i}~\frac{(-1)^{\sum_{m=1}^{j-1}k_m(j-m)}}{j!}~\mu'_j\circ~(s^{-1}\circ\Phi_{k_1}\circ s^{k_1}\otimes\dots\otimes s^{-1}\circ\Phi_{k_j}^1\circ s^{k_j})\,\circ\\
  &\hspace{1cm}\circ \sum_{\sigma\in {\rm Sh}(j;i)} \chi(\sigma,-) \sigma(-)\\
  &~~\ =\ \sum_{j=1}^i~\sum_{k_1+\dots+k_j=i}~\frac{(-1)^{\frac12 (k_1+1)k_1+\dots+\frac12 (k_j+1)k_j+\sum_{m=1}^{j-1}k_m(j-m)}}{j!}~\mu'_j\circ~(\phi_{k_1}\otimes\dots\otimes \phi_{k_j})\,\circ\\
  &\hspace{1cm}\circ \sum_{\sigma\in {\rm Sh}(k_1,\dots,k_{j-1};i)} \chi(\sigma,-) \sigma(-)~.\\
 \end{aligned}
\end{equation}
We obtain
\begin{equation}
 \begin{aligned}
  &\sum_{j+k=i}~(-1)^{\frac12 i(i-1)+k}\phi_{k+1}\circ \left(\mu_j\otimes \id^{\otimes k}\right)\circ\sum_{\sigma\in {\rm Sh}(j;i)} \chi(\sigma,-) \sigma(-)\\
  &~~\ =\ \sum_{j=1}^i~\sum_{k_1+\dots+k_j=i}~\frac{(-1)^{\frac12 (k_1+1)k_1+\dots+\frac12 (k_j+1)k_j+\sum_{m=1}^{j-1}k_m(j-m)}}{j!}~\mu'_j\,\circ\\
  &\hspace{1cm}\circ~(\phi_{k_1}\otimes\dots\otimes \phi_{k_j})\circ \sum_{\sigma\in {\rm Sh}(k_1,\dots,k_{j-1};i)} \chi(\sigma,-) \sigma(-)~.
 \end{aligned}
\end{equation}
Applied to $\ell_1\otimes \dots\otimes \ell_i\in \sL^{\otimes i}$, we find 
\begin{equation}
 \begin{aligned}
  &\sum_{j+k=i}\sum_{\sigma\in {\rm Sh}(j;i)}~(-1)^{k}\chi(\sigma,-)\phi_{k+1}(\mu_j(\ell_{\sigma(1)},\dots,\ell_{\sigma(k)}),\ell_{\sigma(k)+1},\dots,\ell_{\sigma_i})\\
  &~~\ =\ \sum_{j=1}^i~\sum_{k_1+\dots+k_j=i}~\frac{(-1)^{\sum_{1\leq m<n\leq j}k_mk_n+\sum_{m=1}^{j-1}k_m(j-m)+\sum_{m=2}^j(1-k_m)\sum_{k=1}^{k_1+\cdots+k_{m-1}}|\ell_{\sigma(k)}|_\sL}}{j!}\,\times~\\
  &\hspace{1cm}\times\sum_{\sigma\in {\rm Sh}(k_1,\dots,k_{j-1};i)} \chi(\sigma,-) \mu'_j(\phi_{k_1}(\ell_{\sigma(1)},\dots,\ell_{\sigma(k_1)}),\dots,\phi_{k_j}(\ell_{\sigma(i-k_j+1)},\dots,\ell_{\sigma(i)}))~,
 \end{aligned}
\end{equation}
where Koszul signs are inserted to account for the permutation of the $\ell_{\sigma(m)}$ past the $\phi_{k_n}$. We also used 
\begin{equation}
 (-1)^{\frac12 i(i-1)}\ =\ (-1)^{\sum_{m=1}^j \frac12 k_m(k_m-1)+\sum_{1\leq m<n\leq j}k_mk_n}
\end{equation}
for $k_1+\dots+k_j=i$.

\subsection{Cochain complexes and Hodge--Kodaira decomposition}\label{app:Hodge}

An $L_\infty$-algebra $\sL$ has an underlying cochain complex $(\sL,\mu_1)$. Morphisms and quasi-isomor\-phisms of $L_\infty$-algebras are specialisations of cochain maps and quasi-isomorphisms between the underlying cochain complexes. In the following, we briefly review a few relevant notions; a good reference here is~\cite{Weibel:1994aa}.

A comment on our nomenclature: usually, one would work with the homology of chain complexes with a differential of degree~$-1$. To avoid as much confusion as possible, and due to the prominence of the de Rham complex as key example of cochain complexes, we will use the terms {\em cochains} and {\em cohomology} with differentials of degree~$1$.

\paragraph{Cochain complexes.}
A {\em cochain complex $(C,\dd)$} over an associative ring $R$ is a family of $R$-modules $C^k$ for $k\in \RZ$ and morphisms $\dd_k:C^k\rightarrow C^{k+1}$ of degree~1 satisfying $\dd_{k+1}\circ\dd_k=0$. The {\em $k$-cocycles} of the cochain complex are defined as $Z^k:=\ker(\dd_k)\subseteq C^k$ while the {\em $k$-coboundaries} of the cochain complex are defined as $B^k:=\im(\dd_{k-1})\subseteq C^k$. The cohomology of the cochain complex is then $H^k_\dd(C):=Z^k/B^k$.

A {\em morphism} of cochain complexes $\phi:(C,\dd)\rightarrow (C',\dd')$ is a family of maps $\phi_k:C^k\rightarrow C'\,\!^{k}$ of degree~0 such that $\phi_{k+1}\circ\dd_k=\dd'_k\circ\phi_k$. A {\em quasi-isomorphism} of cochain complexes is a morphism of cochain complexes which induces an isomorphism on the cohomologies.  

A {\em cochain homotopy} between two morphisms of cochain complexes $\phi:(C,\dd)\rightarrow (C',\dd')$ and $\psi:(C,\dd)\rightarrow (C',\dd')$ of cochain complexes is a family of morphisms $h_k:C^{k+1}\rightarrow C'\,\!^{k}$ of degree~$-1$ such that $\phi_k-\psi_k=h_{k}\circ\dd_k + \dd'_{k-1}\circ h_k$. If such a cochain homotopy exists, we say that $\phi$ and $\psi$ are homotopic. A cochain homotopy is the correct notion of a {\em 2-morphism} of cochain complexes. A {\em homotopy equivalence} between two cochain complexes $C$ and $C'$ is a pair of morphisms $\phi:C \rightarrow C'$ and $\psi:C'\rightarrow C$ such that $\psi\circ\phi$ and $\phi\circ\psi$ are homotopic to the respective identity maps. In the special case when $\psi\circ\phi=1$ and $\phi\circ\psi$ is homotopic to the identity, we call this a {\it contracting homotopy} of $\phi\circ\psi$.\footnote{This is a particular case of a {\em strong deformation retract}.} 

\paragraph{Split cochain complexes and Hodge--Kodaira decomposition.} 
Suppose $(C,\dd)$ is a cochain complex of vector spaces. Then, we can always choose decompositions 
\begin{subequations}\label{eq:split_decomposition}
\begin{equation}
\begin{gathered}
 C^k/\ker(\dd_k)\ \cong\ \im(\dd_{k})~~\Longrightarrow~~~ C^k/Z^k\ \cong\ B^{k+1}~~~\Longrightarrow~~~ C^k\ \cong\ Z^k\oplus Z_c^k ~,\\
 H^k_\dd(C)\ =\ Z^k/B^k ~~~\Longrightarrow~~~Z^k\ \cong\ B^k\oplus B_c^k~,\\
 Z_c^k\ \cong\ B^{k+1}\eand B_c^k\ \cong\ H^k_\dd(C)~,
 \end{gathered}
\end{equation}
since short exact vector space sequences always split\footnote{Note that the same holds evidently true for tensor products of split cochain complexes.}. We thus have
\begin{equation}
 C^k\ \cong\ H^k_\dd(C)\oplus B^k\oplus B^{k+1}~.
\end{equation}
\end{subequations}

Consequently, we can define maps $h_k:C^k\to C^{k-1}$ of degree~$-1$ by the compositions
\begin{equation}
 h_k\,:\,C^k\ \twoheadrightarrow\ Z^k\ \twoheadrightarrow\ B^k\ \cong\ Z_c^{k-1}\ \hookrightarrow\ C^{k-1}
\end{equation}
of the projections $C^k\twoheadrightarrow Z^k$ and $Z^k\twoheadrightarrow B^k$, the isomorphism $B^k\cong Z_c^{k-1}$, and the inclusion $Z_c^{k-1} \hookrightarrow C^{k-1}$. Since $h_{k+1}(B^{k+1})\cong Z_c^{k}$, these maps satisfy $\dd_k=\dd_k\circ h_{k+1}\circ\dd_k$.

This motivates the following definition. A general cochain complex $(C,\dd)$ is called {\it split} whenever there is a family of morphisms $h_k:C^{k}\rightarrow C\,\!^{k-1}$ of degree~$-1$, called the {\it splitting maps}, such that $\dd_k=\dd_k\circ h_{k+1}\circ\dd_k$. 

In this case, we then automatically have\footnote{Whilst $(h_{k+1}\circ\dd_k)\circ(\dd_{k-1}\circ h_k)=0$ follows trivially since $\dd_k\circ\dd_{k-1}=0$, if $(\dd_{k-1}\circ h_k)\circ(h_{k+1}\circ\dd_k)\neq 0$ one may always re-define $h_k$ and set $\tilde h_k:=h_k-h_k\circ h_{k+1}\circ \dd_k$. Then, $ \dd_k=\dd_k\circ \tilde h_{k+1}\circ \dd_k$ as well, and it is easy to check that $(\dd_{k-1}\circ \tilde h_k)\circ(\tilde h_{k+1}\circ\dd_k)=0$ and also all other identities hold with $\tilde h_k$ instead of $h_k$. Note that we then have the more strict relation $\tilde h_k\circ \tilde h_{k+1}\circ\dd_k=0$.}
\begin{equation}
\begin{gathered}
 (h_{k+1}\circ\dd_k)^2\ =\ h_{k+1}\circ\dd_k~,~~~(\dd_{k-1}\circ h_k)^2\ =\ \dd_{k-1}\circ h_k~,\\
 (h_{k+1}\circ\dd_k)\circ(\dd_{k-1}\circ h_k)\ =\ (\dd_{k-1}\circ h_k)\circ(h_{k+1}\circ\dd_k)\ =\ 0~,\\
  \end{gathered}
\end{equation}
which, in turn, yield the decomposition 
\begin{subequations}
\begin{equation}
  1\ =\ P_k +h_{k+1}\circ\dd_k+\dd_{k-1}\circ h_{k}
\end{equation}
with
\begin{equation}
\begin{gathered}
 P_k^2\ =\ P_k~,~~~P_k\circ\dd_{k-1}\ =\ \dd_k\circ P_k\ =\ 0~,\\
 P_k\circ(\dd_{k-1}\circ h_k)\ =\ (\dd_{k-1}\circ h_k)\circ P_k\ =\ P_k\circ (h_{k+1}\circ\dd_k)\ =\ (h_{k+1}\circ\dd_k)\circ P_k\ =\ 0~.
  \end{gathered}
\end{equation}
\end{subequations}
Consequently,
\begin{equation}
C^k\ \cong\ \underbrace{\im(P_k)}_{\cong\,H^k_\dd(C)}\oplus\underbrace{\im(\dd_{k-1}\circ h_{k})}_{\cong\, B^k}\oplus\underbrace{\im(h_{k+1}\circ\dd_k)}_{=:\, Z^k_c}\ \cong\ H^k_\dd(C)\oplus B^k\oplus Z^k_c~.
\end{equation}
This decomposition is known as the {\it abstract Hodge--Kodaira decomposition}. Furthermore, the projector $P_k$ induces a surjection 
$p_k:C^k\twoheadrightarrow H^k_\dd(C)$ and an injection $e_k:H^k_\dd(C)\hookrightarrow C^k$ with $p_k\circ e_k=1$ by means of $P_k=e_k\circ p_k$. Thus, we obtain the diagram
\begin{equation}
 \myxymatrix{ \ar@(dl,ul)[]^h  C~\ar@<+2pt>@{->>}[rr]^{\kern-15pt p} & & ~~H^\bullet_\dd(C) \ar@<+2pt>@{^(->}[ll]^{\kern-15pt e}}.
\end{equation}

Put differently, a splitting of a cochain complex $(C,\dd)$ is equivalent to having morphisms of cochain complexes $p_k:C^k\twoheadrightarrow H^k_\dd(C)$ and $e_k:H^k_\dd(C)\hookrightarrow C^k$ such that $p_k\circ e_k=1$ and $1-e_k\circ p_k=h_{k+1}\circ\dd_k+\dd_{k-1}\circ h_{k}$. In other words, such a splitting is equivalent to a contracting homotopy $h_k:C^k\to C^{k-1}$ of $P_k=e_k\circ p_k$. Note that $p_k$ is a quasi-isomorphisms of cochain complexes between $C$ and $H^\bullet_\dd(C)$ and so is $e_k$ between  $H^\bullet_\dd(C)$ and $C$. 

\paragraph{\mathversion{bold}Extension to $L_\infty$-algebras.} 
Consider the cochain complex $(\sL,\mu_1)$ underlying an $L_\infty$-algebra $(\sL,\mu_i)$ with a choice of decomposition~\eqref{eq:split_decomposition}
\begin{equation}\label{eq:decomp_L_infty}
 \sL\ \cong\ B\oplus B_c\oplus Z_c\ewith  B_c\ \cong\ H^\bullet_{\mu_1}(\sL)~.
\end{equation}
We can use the inverse of the isomorphism between $B_c$ and $H^\bullet_{\mu_1}(\sL)$ to define a strict $L_\infty$-morphism to $B_c$. Composition with an $L_\infty$-quasi-isomorphism between $\sL$ and $H^\bullet_{\mu_1}(\sL)$ then yields an $L_\infty$-quasi-isomorphism between $\sL$ to $B_c$. Thus, any projection $\sL\twoheadrightarrow B_c$ can be extended to an $L_\infty$-quasi-isomorphism $\sL\rightarrow B_c$. Consequently, we can use the abstract Hodge--Kodaira decomposition to find a minimal model. See Section~\ref{ssec:quasiisomorphism} for explicit formulas.

\paragraph{Example.} Let $\sL$ be an $L_\infty$-algebra together with decompositions $\sL=\ker(\mu_1)\oplus V=\im(\mu_1)\oplus W$. We have
\begin{equation}
 \mu_1\,:\, \ker(\mu_1)\oplus V\ \rightarrow\ \im(\mu_1)\oplus W
\end{equation}
and $\mu_1$ is invertible as a map from $V\rightarrow \im(\mu_1)$ with inverse $\mu_1^{-1}|_{\im(\mu_1)}$. Define $h:\sL\rightarrow \sL$ as the map of degree~$-1$ 
\begin{equation}
 h\ :=\ \mu_1^{-1}|_{\im(\mu_1)}\circ {\rm pr}_{\im(\mu_1)}~,
\end{equation}
where ${\rm pr}_{\im(\mu_1)}:\sL\rightarrow \im(\mu_1)$ is the orthogonal projection. Note that the map $h$ satisfies $\mu_1\circ h \circ \mu_1=\mu_1$, and we can use it as a starting point for the abstract Hodge--Kodaira decomposition. 

\subsection{Lemmata}\label{app:lemmata}

In this appendix, we shall prove various formulas used throughout the main text, sometimes in more than one of the three possible descriptions of $L_\infty$-algebras: higher brackets, differential coalgebra, differential graded algebra. While some of the calculations are slightly involved, they are illuminating in one sense or another.

Throughout this appendix, let $\sL$ be an $L_\infty$-algebra with higher products $\mu_i$. We shall occasionally assume that $\sL$ is cyclic with an inner product $\langle-,-\rangle_\sL$.

\paragraph{Preliminaries.}
Recall that the Cauchy product of two (absolutely convergent) series $\sum_{i\geq 0} a_i$ and $\sum_{i\geq 0} b_i$ is  
\begin{equation}\label{eq:Cauchy2Series}
\sum_{i\geq 0} a_i \sum_{j\geq 0} b_j\ =\ \sum_{i,j\geq 0} a_i b_j\ =\ \sum_{i\geq 0} \sum_{j=0}^ia_j b_{i-j}\ =\ \sum_{i\geq 0} \sum_{j+k=i} a_j b_{k}~.
\end{equation}
Hence,
\begin{equation}\label{eq:Cauchy}
\sum_{i,j\geq 0} \frac{1}{i!j!} a_i b_j\ =\ \sum_{i\geq 0}\frac{1}{i!} \sum_{j+k=i} \binom{i}{j} a_j b_k~.
\end{equation}
Furthermore, note that~\eqref{eq:Cauchy2Series} generalises to
\begin{equation}\label{eq:CauchyjSeries}
 \sum_{k_1\geq 0} a^{(1)}_{k_1}\cdots \sum_{k_j\geq 0} a^{(j)}_{k_j}\ =\ \sum_{i\geq 0}\sum_{k_1+\cdots+k_j=i}a^{(1)}_{k_1}\cdots a^{(j)}_{k_j}
\end{equation}
for the product of $j$  (absolutely convergent) series and so,
\begin{equation}\label{eq:CauchyGeneral}
\begin{aligned}
 &\sum_{k_1\geq 0} \frac{1}{k_1!}a^{(1)}_{k_1}\cdots  \sum_{k_j\geq 0} \frac{1}{k_j!} a^{(j)}_{k_j}\ =\\
 &\kern1cm\ =\ \sum_{i\geq 0}\frac{1}{i!}\sum_{k_1+\cdots+k_j=i} \binom{i-k_1}{k_2}\cdots\binom{i-k_1-\cdots-k_{j-2}}{k_{j-1}}a^{(1)}_{k_1}\cdots a^{(j)}_{k_j}~.
 \end{aligned}
\end{equation}

\paragraph{Tensor product $L_\infty$-algebras}

Let $\sA$ be a dg-algebra. We wish to verify the higher homotopy Jacobi identities~\eqref{eq:homotopyJacobi} of the higher products $\mu'_i$ defined in~\eqref{eq:LinftyExtensionHP} as well as the cyclicity~\eqref{eq:LinftyExtensionIP} of the tensor product $L_\infty$-algebra $\sL_\sA$. To this end, let us use the abbreviation $\sfl_i:=a_i\otimes\ell_i$ for homogeneous $a_i\in\sA$ and $\ell_i\in\sL$ together with $|\sfl_i|_\sA:=|a_i|_\sA$ and $|\sfl_i|_\sL:=|\ell_i|_\sL$. It is sufficient to consider only those elements since the result for general elements follows from linearity.

The higher homotopy identities~\eqref{eq:homotopyJacobi} of the higher products~\eqref{eq:LinftyExtensionHP} are
\begin{equation}\label{eq:homotopyJacobi_cp}
 \sum_{j+k=i}\sum_{\sigma }\chi(\sigma;\sfl_1,\ldots,\sfl_{j+k})(-1)^{k}\mu'_{k+1}(\mu'_j(\sfl_{\sigma(1)},\ldots,\sfl_{\sigma(j)}),\sfl_{\sigma(j+1)},\ldots,\sfl_{\sigma(j+k)})\ =\ 0~.
\end{equation}
We have 
\begin{subequations}
\begin{equation}
 \mu_i(\ell_{\sigma(1)},\dots ,\ell_{\sigma(i)}\ =\ \chi(\sigma;\ell_1,\dots,\ell_i)\mu_i(\ell_1,\dots,\ell_i)
\end{equation}
and
\begin{equation}\label{eq:Lhat_ident2}
\begin{aligned}
 \chi(\sigma;\sfl_1,\dots, \sfl_i)\ &=\ \chi(\sigma; a_1,\dots, a_i)\chi(\sigma;\ell_1,\dots, \ell_n)\,\times\\
 &\hspace{1cm}\times (-1)^{\sum_{j=2}^i|a_{\sigma(j)}|_\sA\sum_{k=1}^{j-1}|\ell_{\sigma(k)}|_\sL+\sum_{j=2}^i|a_{j}|_\sA\sum_{k=1}^{j-1}|\ell_{k}|_\sL}~.
\end{aligned}
\end{equation}
\end{subequations}

To prove~\eqref{eq:homotopyJacobi_cp}, we first focus on the terms containing the differential $\dd$. For $i=1$, we have
\begin{equation}
\begin{aligned}
 \mu'_{1}(\mu'_1(\sfl_1))\ &=\ \mu'_1(\dd a_1 \otimes \ell_1+(-1)^{|a_1|_\sA}a_1\otimes \mu_1(\ell_1))\\
 \ &=\ (-1)^{|a_1|_\sA+1}\dd a_1\otimes \mu_1(\ell_1)+(-1)^{|a_1|_\sA}\dd a_1\otimes \mu_1(\ell_1)\\
 \ &=\ 0~.
\end{aligned}
\end{equation}
For $i>1$, the relevant terms are 
\begin{subequations}
\begin{equation}\label{eq:Lhat_identity_dd}
\begin{aligned}
 &\dd(\mu'_i(\sfl_1,\ldots,\sfl_i))+\sum_{\sigma }\chi(\sigma;\sfl_1,\ldots,\sfl_{i})(-1)^{i-1}\mu'_{i}(\dd(\sfl_{\sigma(1)}),\sfl_{\sigma(2)},\ldots,\sfl_{\sigma(i)})\ =\\
&=\ s_1\,\dd(a_{1}\cdots a_{i})\otimes \mu_i( \ell_{1},\ldots,\ell_{i})\,+\\
 &\hspace{1cm}+\sum_{\sigma }\chi(\sigma;\sfl_1,\ldots,\sfl_{i})\,s_2(\sigma)\,\big[(\dd a_{\sigma(1)})a_{\sigma(2)}\cdots a_{\sigma(i)}\big]\otimes \mu_{i}(\ell_{\sigma(1)}, \ldots,\ell_{\sigma(i)})\\
&=\ s_1\,\sum_{\sigma} \chi(\sigma;a_1,\dots,a_i)\chi(\sigma;\ell_1,\dots, \ell_i)\big[(\dd a_{\sigma(1)})a_{\sigma(2)}\cdots a_{\sigma(i)}\big]\otimes \mu_{i}(\ell_{\sigma(1)}, \ldots,\ell_{\sigma(i)})\,+\\
 &\hspace{1cm}+\sum_{\sigma }\chi(\sigma;\sfl_1,\ldots,\sfl_{i})\,s_2(\sigma;\sfl_1,\ldots,\sfl_{i})\,\big[(\dd a_{\sigma(1)})a_{\sigma(2)}\cdots a_{\sigma(i)}\big]\otimes \mu_{i}(\ell_{\sigma(1)}, \ldots,\ell_{\sigma(i)})
\end{aligned}
\end{equation}
with signs
\begin{equation}
\begin{aligned}
 s_1\ &:=\ (-1)^{i\sum_{j=1}^i|a_{j}|_\sA+\sum_{j=2}^{i} |a_{j}|_\sA\sum_{k=1}^{j-1} |\ell_{k}|_{\sL}}~,\\
 s_2(\sigma;\sfl_1,\ldots,\sfl_{i})\ &:=\ (-1)^{i-1}(-1)^{i(|a_{\sigma(1)}|_\sA+1+\sum_{j=2}^i|a_{\sigma(j)}|_\sA)+\sum_{j=2}^{i} |a_{\sigma(j)}|_\sA\sum_{k=1}^{j-1} |\ell_{\sigma(k)}|_{\sL}}\\
 &\phantom{:}=\ -(-1)^{i\sum_{j=1}^i|a_{\sigma(j)}|_\sA+\sum_{j=2}^{i} |a_{\sigma(j)}|_\sA\sum_{k=1}^{j-1} |\ell_{\sigma(k)}|_{\sL}}~.
\end{aligned}
\end{equation}
\end{subequations}
The expression~\eqref{eq:Lhat_identity_dd} clearly vanishes since
\begin{equation}
 \chi(\sigma;\sfl_1,\ldots,\sfl_{i})\,s_2(\sigma;\sfl_1,\ldots,\sfl_{i})\ =\ -s_1\,\chi(\sigma; a_1,\dots, a_i)\chi(\sigma;\ell_1,\dots, \ell_i)
\end{equation}
due to~\eqref{eq:Lhat_ident2}.

The remaining terms in~\eqref{eq:homotopyJacobi_cp} combine to
\begin{equation}
\begin{aligned}
 &\sum_{j+k=i}\sum_{\sigma }s_3(\sigma;\sfl_1,\ldots,\sfl_{i})\,\times\\
 &\kern1cm\times(a_1\cdots a_{j+k})\otimes\mu_{k+1}(\mu_j(\ell_{\sigma(1)},\ldots,\ell_{\sigma(j)}), \ell_{\sigma(j+1)},\ldots,\ell_{\sigma(j+k)})\ =\ 0
 \end{aligned}
\end{equation}
with the sign $s_3(\sigma;\sfl_1,\ldots,\sfl_{i})$ given by
\begin{equation}
\begin{aligned}
 s_3(\sigma;&\sfl_1,\ldots,\sfl_{i})\ :=\\
 \ &:=\ \chi(\sigma;\sfl_1,\ldots,\sfl_{j+k})(-1)^{k} \chi(\sigma;a_1,\dots a_{j+k})\,\times\\
 &\hspace{1cm}\times (-1)^{j\sum_{m=1}^j|a_{\sigma(m)}|_\sA+\sum_{m=2}^{j} |a_{\sigma(m)}|_\sA\sum_{n=1}^{m-1} |\ell_{\sigma(n)}|_{\sL}}\,\times \\
 &\hspace{1cm}\times (-1)^{(k+1)\sum_{m=1}^{j+k}|a_{m}|_\sA+\sum_{m=1}^{k} |a_{\sigma(j+m)}|_\sA~(j+\sum_{n=1}^{j+m-1}|\ell_{\sigma(n)}|_{\sL})}\\
 \ &=\ (-1)^k\chi(\sigma; \ell_1,\dots, \ell_{j+k})(-1)^{\sum_{m=2}^{j+k}|a_{\sigma(m)}|_\sA\sum_{n=1}^{m-1}|\ell_{\sigma(n)}|_\sL+\sum_{m=2}^{j+k}|a_{m}|_\sA\sum_{n=1}^{m-1}|\ell_{n}|_\sL}\,\times \\ 
 &\hspace{1cm}\times (-1)^{j\sum_{m=1}^j|a_{\sigma(m)}|_\sA+\sum_{m=2}^{j} |a_{\sigma(m)}|_\sA\sum_{n=1}^{m-1} |\ell_{\sigma(n)}|_{\sL}}\,\times \\
 &\hspace{1cm}\times (-1)^{(k+1)\sum_{m=1}^{j+k}|a_{m}|_\sA+\sum_{m=1}^{k} |a_{\sigma(j+m)}|_\sA~(j+\sum_{n=1}^{j+m-1}|\ell_{\sigma(n)}|_{\sL})}\\
 \ &=\ (-1)^k\chi(\sigma; \ell_1,\dots, \ell_{j+k})\underbrace{(-1)^{\sum_{m=2}^{j+k}|a_{m}|_\sA\sum_{n=1}^{m-1}|\ell_{n}|_\sL+(j+k+1)\sum_{m=1}^{j+k}|a_{m}|_\sA}}_{=:\,s_4}~,
\end{aligned}
\end{equation}
where we used again~\eqref{eq:Lhat_ident2}. Note that $s_4$ contributes an overall sign so that~\eqref{eq:homotopyJacobi_cp} reduces to the homotopy Jacobi identity~\eqref{eq:homotopyJacobi} on $\sL$.

Next suppose that $\sL$ and $\sA$ are equipped with inner products $\langle-,-\rangle_\sL$ and $\langle-,-\rangle_\sA$. Let us verify the cyclicity of the inner product $\langle-,-\rangle_{\sL_\sA}$ defined in~\eqref{eq:LinftyExtensionIP}. Using the definitions of the higher products $\mu'_i$ given~\eqref{eq:LinftyExtensionHP} we obtain for $i=1$
\begin{equation}
\begin{aligned}
 &\langle \sfl_1,\mu'_1(\sfl_2)\rangle_{\sL_\sA}\ =\\
  &\kern1cm\ =\ \langle a_1\otimes\ell_1,\dd a_2\otimes\ell_2+(-1)^{|\sfl_2|_\sA}a_2\otimes\mu_1(\ell_2)\rangle_{\sL_\sA}\\
 &\kern1cm\ =\ (-1)^{|\sfl_1|_\sL(|\sfl_2|_\sA+1)}\langle a_1,\dd a_2\rangle_\sA~\langle\ell_1,\ell_2\rangle_\sL+(-1)^{(|\sfl_1|_\sL+1)|\sfl_2|_\sA}\langle a_1,a_2\rangle_\sA~\langle\ell_1,\mu_1(\ell_2)\rangle_\sL\\
 &\kern1cm\ =\ -(-1)^{|\sfl_1|_\sL(|\sfl_2|_\sA+1)+|\sfl_2|_\sA(|\sfl_1|_\sA+1)+|\sfl_1|_\sA+|\sfl_1|_\sL|\sfl_2|_\sL}\langle a_2,\dd a_1\rangle_\sA~\langle\ell_2,\ell_1\rangle_\sL\,-\\
 &\kern2cm-(-1)^{(|\sfl_1|_\sL+1)|\sfl_2|_\sA+|\sfl_1|_\sA|\sfl_2|_\sA+|\sfl_1|_\sL+|\sfl_2|_\sL+|\sfl_1|_\sL|\sfl_2|_\sL}\langle a_2,a_1\rangle_\sA~\langle\ell_2,\mu_1(\ell_1)\rangle_\sL\\
 &\kern1cm\ =\ -(-1)^{|\sfl_1|_\sL(|\sfl_2|_\sA+1)+|\sfl_2|_\sA(|\sfl_1|_\sA+1)+|\sfl_1|_\sA+|\sfl_1|_\sL|\sfl_2|_\sL+|\sfl_2|_\sL(|\sfl_1|_\sA+1)}\langle a_2\otimes\ell_2,\dd a_1\otimes\ell_1\rangle_{\sL_\sA}\,-\\
 &\kern2cm-(-1)^{(|\sfl_1|_\sL+1)|\sfl_2|_\sA+|\sfl_1|_\sA|\sfl_2|_\sA+|\sfl_1|_\sL+|\sfl_2|_\sL+|\sfl_1|_\sL|\sfl_2|_\sL+|\sfl_2|_\sL|\sfl_1|_\sA}\langle a_2\otimes\ell_2,a_1\otimes\mu_1(\ell_1)\rangle_{\sL_\sA}\\
 &\kern1cm\ =\ (-1)^{1+|\sfl_1|_{\sL_\sA}|\sfl_2|_{\sL_\sA}+|\sfl_1|_{\sL_\sA}|\sfl_2|_{\sL_\sA}}\langle \sfl_2,\mu'_1(\sfl_1)\rangle_{\sL_\sA}~,
\end{aligned}
\end{equation}
while for $i\geq2$ we find
\begin{equation}
\begin{aligned}
& \langle \sfl_1,\mu'_i(\sfl_2,\ldots,\sfl_{i+1})\rangle_{\sL_\sA}\ =\\
&\kern1cm\ =\ (-1)^{i\sum_{j=2}^{i+1}|\sfl_j|_\sA+|\sfl_1|_{\sL}\sum_{j=2}^{i+1}|\sfl_j|_\sA+\sum_{j=3}^{i+1}|\sfl_j|_\sA\sum_{k=2}^{j-1}|\sfl_k|_{\sL}}\,\times\\[-3pt]
&\kern2cm\times\langle a_1,a_2\cdots a_{i+1}\rangle_\sA~\langle \ell_1,\mu_i(\ell_2,\ldots,\ell_{i+1})\rangle_{\sL}\\[3pt]
&\kern1cm\ =\ (-1)^{i\sum_{j=2}^{i+1}|\sfl_j|_\sA+|\sfl_1|_{\sL}\sum_{j=2}^{i+1}|\sfl_j|_\sA+\sum_{j=3}^{i+1}|\sfl_j|_\sA\sum_{k=2}^{j-1}|\sfl_k|_{\sL}}\,\times\\
&\kern2cm\times(-1)^{i+i(|\sfl_1|_{\sL}+|\sfl_{i+1}|_{\sL})+|\sfl_{i+1}|_{\sL}\sum_{j=1}^{i}|\sfl_j|_{\sL}+|\sfl_{i+1}|_\sA\sum_{j=1}^{i}|\sfl_j|_\sA}\,\times\\[-3pt]
&\kern2cm\times\langle a_{i+1},a_1\cdots a_i\rangle_\sA~\langle \ell_{i+1},\mu_i(\ell_1,\ldots,\ell_i)\rangle_{\sL}\\[3pt]
&\kern1cm\ =\ (-1)^{i\sum_{j=2}^{i+1}|\sfl_j|_\sA+|\sfl_1|_{\sL}\sum_{j=2}^{i+1}|\sfl_j|_\sA+\sum_{j=3}^{i+1}|\sfl_j|_\sA\sum_{k=2}^{j-1}|\sfl_k|_{\sL}}\,\times\\
&\kern2cm\times(-1)^{i+i(|\sfl_1|_{\sL}+|\sfl_{i+1}|_{\sL})+|\sfl_{i+1}|_{\sL}\sum_{j=1}^{i}|\sfl_j|_{\sL}+|\sfl_{i+1}|_\sA\sum_{j=1}^{i}|\sfl_j|_\sA}\,\times\\
&\kern2cm\times(-1)^{i\sum_{j=1}^{i}|\sfl_j|_\sA+|\sfl_{i+1}|_{\sL}\sum_{j=1}^{i}|\sfl_j|_\sA+\sum_{j=2}^{i}|\sfl_j|_\sA\sum_{k=1}^{j-1}|\sfl_k|_{\sL}}\,\times\\
&\kern2cm\times\langle\sfl_{i+1},\mu'_i(\sfl_1,\ldots,\sfl_{i})\rangle_{\sL_\sA}~,
\end{aligned}
\end{equation}
and therefore
\begin{equation}
\begin{aligned}
&(-1)^{i\sum_{j=2}^{i+1}|\sfl_j|_\sA+|\sfl_1|_{\sL}\sum_{j=2}^{i+1}|\sfl_j|_\sA+\sum_{j=3}^{i+1}|\sfl_j|_\sA\sum_{k=2}^{j-1}|\sfl_k|_{\sL}}\times\\
&\kern1cm\times(-1)^{i+i(|\sfl_1|_{\sL}+|\sfl_{i+1}|_{\sL})+|\sfl_{i+1}|_{\sL}\sum_{j=1}^{i}|\sfl_j|_{\sL}+|\sfl_{i+1}|_\sA\sum_{j=1}^{i}|\sfl_j|_\sA}\times\\
&\kern1cm\times(-1)^{i\sum_{j=1}^{i}|\sfl_j|_\sA+|\sfl_{i+1}|_{\sL}\sum_{j=1}^{i}|\sfl_j|_\sA+\sum_{j=2}^{i}|\sfl_j|_\sA\sum_{k=1}^{j-1}|\sfl_k|_{\sL}}\ =\\
&=\ (-1)^{i+i(|\sfl_1|_{\sL_\sA}+|\sfl_{i+1}|_{\sL_\sA})+|\sfl_{i+1}|_{\sL}\sum_{j=1}^{i}|\sfl_j|_{\sL_\sA}+|\sfl_{i+1}|_\sA\sum_{j=1}^{i}|\sfl_j|_\sA}\,\times\\
 &\kern1cm\times(-1)^{|\sfl_1|_{\sL}\sum_{j=2}^{i+1}|\sfl_j|_\sA+\sum_{j=2}^{i}|\sfl_j|_\sA\sum_{k=1}^{j-1}|\sfl_k|_{\sL}+\sum_{j=3}^{i+1}|\sfl_j|_\sA\sum_{k=2}^{j-1}|\sfl_k|_{\sL}}\\
 &=\ (-1)^{i+i(|\sfl_1|_{\sL_\sA}+|\sfl_{i+1}|_{\sL_\sA})+|\sfl_{i+1}|_{\sL}\sum_{j=1}^{i}|\sfl_j|_{\sL_\sA}+|\sfl_{i+1}|_\sA\sum_{j=1}^{i}|\sfl_j|_\sA}\,\times\\
 &\kern1cm\times(-1)^{|\sfl_1|_{\sL}(|\sfl_2|_\sA+\cdots+|\sfl_i|_\sA+|\sfl_{i+1}|_\sA)}\,\times\\
 &\kern1cm\times(-1)^{|\sfl_1|_{\sL}|\sfl_2|_\sA+(|\sfl_1|_{\sL}+|\sfl_2|_{\sL})|\sfl_3|_\sA+\cdots+(|\sfl_1|_{\sL}+\cdots+|\sfl_{i-1}|_{\sL})|\sfl_i|_\sA}\,\times\\
 &\kern1cm\times(-1)^{|\sfl_2|_{\sL}|\sfl_3|_\sA+(|\sfl_2|_{\sL}+|\sfl_3|_{\sL})|\sfl_4|_\sA+\cdots+(|\sfl_2|_{\sL}+\cdots+|\sfl_{i-1}|_{\sL})|\sfl_{i}|_\sA+(|\sfl_2|_{\sL}+\cdots+|\sfl_{i}|_{\sL})|\sfl_{i+1}|_\sA}\\
  &=\ (-1)^{i+i(|\sfl_1|_{\sL_\sA}+|\sfl_{i+1}|_{\sL_\sA})+|\sfl_{i+1}|_{\sL}\sum_{j=1}^{i}|\sfl_j|_{\sL_\sA}+|\sfl_{i+1}|_\sA\sum_{j=1}^{i}|\sfl_j|_\sA}\,\times\\
  &\kern1cm\times(-1)^{|\sfl_{i+1}|_\sA\sum_{j=1}^{i}|\sfl_j|_{\sL}+2\sum_{j=1}^{i-1}|\sfl_j|_{\sL}\sum_{k=j+1}^{i}|\sfl_k|_\sA}\\
   &=\ (-1)^{i+i(|\sfl_1|_{\sL_\sA}+|\sfl_{i+1}|_{\sL_\sA})+|\sfl_{i+1}|_{\sL_\sA}\sum_{j=1}^{i}|\sfl_j|_{\sL_\sA}}~.
\end{aligned}
\end{equation}
Altogether, we obtain the cyclicity
\begin{equation}
\begin{aligned}
&\langle \sfl_1,\mu'_i(\sfl_2,\ldots,\sfl_{i+1})\rangle_{\sL_\sA}\ =\\
&\kern1cm\ =\ (-1)^{ i+i(|\sfl_1|_{\sL_\sA}+|\sfl_{i+1}|_{\sL_\sA})+|\sfl_{i+1}|_{\sL_\sA}\sum_{j=1}^{i}|\sfl_j|_{\sL_\sA}}\langle\sfl_{i+1},\mu'_i(\sfl_1,\ldots,\sfl_{i})\rangle_{\sL_\sA}
\end{aligned}
\end{equation}
for $i\in\NN$, as claimed.

\paragraph{Bianchi identity.}
Let us verify the Bianchi identity~\eqref{eq:BianchiIdentity} for the curvature $f$ defined in~\eqref{eq:Curvature}. To this end, recall the homotopy Jacobi identity~\eqref{eq:homotopyJacobi} for a gauge potential $a\in\sL_1$,
\begin{equation}
 \sum_{j+k=i} (-1)^j \binom{i}{j} \mu_{j+1}(\mu_k(a,\ldots,a),a,\ldots,a)\ =\ 0~.
\end{equation}
Making use of~\eqref{eq:Cauchy}, we rewrite
\begin{equation}
\begin{aligned}
\sum_{i\geq0}\frac{(-1)^i}{i!}\mu_{i+1}(f,a,\ldots,a)\ &=\ \sum_{i\geq0}\frac{(-1)^i}{i!j!}\mu_{i+1}(\mu_j(a,\ldots,a),a,\ldots,a)\\
&=\ \sum_{i\geq0} \frac{1}{i!}  \sum_{j+k=i} (-1)^j \binom{i}{j} \mu_{j+1}(\mu_k(a,\ldots,a),a,\ldots,a)\ =\ 0~.
\end{aligned}
\end{equation}

Let us prove the same statement using the $Q$-manifold morphism language, and in particular formula~\eqref{eq:Qxi_L_infty},
\begin{equation}
 Q \xi\ =\ -\sum_{i\geq 1} \frac{1}{i!} \hat\mu_i(\xi,\dots,\xi)
\end{equation}
for $\xi=\xi^\alpha\otimes \tau_\alpha\in (\sL[1])^*\otimes \sL$. We can evaluate this function on 
\begin{equation}
\begin{aligned}
 \de^{sa}\ &:=\ 1+\de^{sa}_0\\
  &\phantom{:}=\ 1+sa+sa\otimes sa+sa\otimes sa\otimes sa+\cdots\\
 &\phantom{:}=\ 1+sa+\tfrac12 sa\odot sa+\tfrac{1}{3!}sa\odot sa \odot sa+\cdots~,
\end{aligned}
\end{equation}
for $a\in\sL_1$. Since $sa$ is even, no Koszul signs appear and we obtain
\begin{equation}
 (Q\xi)(\de^{sa}_0)\ =\ -\sum_{i\geq 1}\frac{1}{i!}\hat\mu_i(a,\ldots,a)\ =\ -f~.
\end{equation}
This equation will prove very helpful. It also shows why the definition of $f$ is so natural.

A first advantage of this alternative perspective is a trivial derivation of the Bianchi identity. We compute
\begin{equation}
\begin{aligned}
 0\ &=\ (-Q^2\xi)(\de^{sa}_0) \\
 &=\ (Q f(\xi)(\de^{sa}_0) \\
  &=\ \left(Q\sum_{i\geq 1}\frac{1}{i!}\hat\mu_i(\xi,\ldots,\xi)\right)(\de^{sa}_0)\\
  &=\ \left(\sum_{i\geq 0}\frac{(-1)^{2-(i+1)}}{i!}\hat\mu_{i+1}(Q\xi,\xi,\ldots,\xi)\right)(\de^{sa}_0)\\
   &=\ \left(\sum_{i\geq 0}\frac{(-1)^i}{i!}\hat\mu_{i+1}(f(\xi),\xi,\ldots,\xi)\right)(\de^{sa}_0),\\
   &=\ \sum_{i\geq 0}\frac{(-1)^i}{i!}\hat\mu_{i+1}(f,a,\ldots,a)~.
\end{aligned}
\end{equation}

Let us also give the coalgebra picture description, which shall be useful later on. Here, we have
\begin{equation}\label{eq:coalgebra_trick}
 D(\de^{sa}_0)\ =\ \sum_{i\geq 1} \tfrac{1}{i!} D_i(sa,\dots,sa)\odot \de^{sa}\ =\ sf\odot \de^{sa}~,
\end{equation}
because
\begin{equation}
\begin{aligned}
 sf\ &=\ s\sum_{i\geq 1}\tfrac{1}{i!}\mu_i(a,\dots,a)\\
 &=\ s\sum_{i\geq 1}\tfrac{(-1)^{\frac12 i(i-1)}}{i!}s^{-1}\circ D_i\circ s^{\otimes i}(a,\dots,a)\\
 &=\ \sum_{i\geq 1}\tfrac{1}{i!}D_i(sa,\dots,sa)~.
\end{aligned}
\end{equation}
The Bianchi identity follows here from $D^2$ as follows:
\begin{equation}
\begin{aligned}
 0\ &=\ (s^{-1}\circ {\rm pr}_{\bigodot^1 \sL[1]}\circ D\circ D)(\de^{sa}_0)\\
 &=\ (s^{-1}\circ {\rm pr}_{\odot^1 \sL[1]}\circ D)(sf\odot \de^{sa})\\
 &=\ \sum_{i\geq 1}\frac{(-1)^{i}}{i!}\mu_{i+1}(f,a,\dots,a)~.
\end{aligned}
\end{equation}

\paragraph{Commutator of gauge transformations.}
We wish to prove~\eqref{eq:CommutatorGT}. Using~\eqref{eq:Cauchy}, we obtain 
\begin{equation}\notag
\begin{aligned}
\delta_{c_0}\delta_{c'_0}a\ &=\ \sum_{i,j\geq0}\frac{1}{i!j!}\mu_{i+2}(\mu_{j+1}(a,\ldots,a,c'_0),a,\ldots,a,c_0)\\
&=\ \sum_{i\geq0}\frac{1}{i!}\sum_{j+k=i}\binom{i}{j}\mu_{j+2}(\mu_{k+1}(a,\ldots,a,c'_0),a,\ldots,a,c_0)\\
&=\ \sum_{i\geq2}\frac{1}{(i-2)!}\sum_{j+k=i}\binom{i-2}{j-1}\mu_{j+1}(\mu_{k}(a,\ldots,a,c'_0),a,\ldots,a,c_0)\\
&=\ \sum_{i\geq2}\frac{1}{(i-2)!}\sum_{j+k=i}\binom{i-2}{k-1}\mu_{j+1}(\mu_{k}(a,\ldots,a,c'_0),a,\ldots,a,c_0)\\
&=\ \delta_{c'_0}\delta_{c_0}a\,+\\
&\kern1cm+\sum_{i\geq2}\frac{1}{(i-2)!}\sum_{j+k=i}(-1)^{j}\binom{i-2}{j-2}\mu_{j+1}(\mu_{k}(a,\ldots,a),a,\ldots,a,c_0,c'_0)\,+\\
&\kern1cm+\sum_{i\geq2}\frac{1}{(i-2)!}\sum_{j+k=i}(-1)^{j} \binom{i-2}{k-2}\mu_{j+1}(\mu_{k}(a,\ldots,a,c_0,c'_0),a,\ldots,a)\\
\end{aligned}
\end{equation}
\pagebreak
\begin{equation}
\begin{aligned}
&=\ \delta_{c'_0}\delta_{c_0}a\,+\\
&\kern1cm+\sum_{i,j\geq0}\frac{1}{i!j!}(-1)^{i}\mu_{i+3}(\mu_{j}(a,\ldots,a),a,\ldots,a,c_0,c'_0)\,+\\
&\kern1cm+\sum_{i,j\geq0}\frac{1}{i!j!}\mu_{i+1}(a,\ldots,a,\mu_{j+2}(a,\ldots,a,c_0,c'_0))~,
\end{aligned}
\end{equation}
where in the fifth step we have used the homotopy Jacobi identity~\eqref{eq:homotopyJacobi} for the elements $(a,\ldots,a,c_0,c'_0)$,
\begin{equation}
\begin{aligned}
&\sum_{j+k=i}(-1)^{j}\binom{i-2}{j-2}\mu_{j+1}(\mu_{k}(a,\ldots,a),a,\ldots,a,c_0,c'_0)\,+\\ 
&\kern2cm-\sum_{j+k=i}\binom{i-2}{k-1}\mu_{j+1}(\mu_{k}(a,\ldots,a,c_0),a,\ldots,a,c'_0)\,+\\
&\kern2cm+\sum_{j+k=i}\binom{i-2}{k-1}\mu_{j+1}(\mu_{k}(a,\ldots,a,c'_0),a,\ldots,a,c_0)\,+\\
&\kern2cm+\sum_{j+k=i}(-1)^{j} \binom{i-2}{k-2}\mu_{j+1}(\mu_{k}(a,\ldots,a,c_0,c'_0),a,\ldots,a)\ =\ 0~.
\end{aligned}
\end{equation}
Hence, using the expression~\eqref{eq:Curvature} of the curvature, we find~\eqref{eq:CommutatorGT}:
\begin{subequations}
\begin{equation}
  [\delta_{c_0},\delta_{c'_0}]a\ =\ \delta_{c''_0}a+\sum_{i\geq0}\frac{1}{i!}(-1)^{i}\mu_{i+3}(f,a,\ldots,a,c_0,c'_0)
\end{equation}
with
\begin{equation}
  c''_0\ :=\ \sum_{i\geq0}\frac{1}{i!}\mu_{i+2}(a,\ldots,a,c_0,c'_0)~.
\end{equation}
\end{subequations}

A proof via the $Q$-manifold evaluation map $\xi$ is possible, but rather technical and not very enlightening.

\paragraph{Gauge transformation of the curvature.} The gauge transformation of the curvature is derived using the Bianchi identity~\eqref{eq:BianchiIdentity} on $\Omega^\bullet(I,\sL)$,
\begin{equation}
\begin{aligned}
 0\ &=\ \sum_{i\geq0}\frac{(-1)^i}{i!}\hat \mu_{i+1}(\sff(t),\sfa(t),\ldots,\sfa(t))\\
 \ &=\  \sum_{i\geq0}\frac{(-1)^i}{i!}\mu_{i+1}(f(t),a(t),\ldots,a(t))\,+\\
 &\kern2cm+\dd t\otimes\left\{\der{t} f(t)-\sum_{i\geq0}\frac{(-1)^i}{i!} \mu_{i+2}(f(t),a(t),\ldots,a(t), c(t))\right\}\\
 \ &=\ \dd t\otimes\left\{\der{t} f(t)-\sum_{i\geq0}\frac{(-1)^i}{i!} \mu_{i+2}(f(t),a(t),\ldots,a(t), c(t))\right\}~,
\end{aligned}
\end{equation}
from which we read off the gauge transformation of the curvature,
\begin{equation}
 \delta_{c_0} f\ :=\ \left.\der{t}\right|_{t=0}  f(t)\ =\ \sum_{i\geq 0}\frac{(-1)^i}{i!}\mu_{i+2}(f,a,\ldots,a,c_0)~.
\end{equation}

Alternatively, we can perform the direct computation using brackets. Upon making use of~\eqref{eq:Cauchy}, we find
\begin{equation}
\begin{aligned}
\delta_{c_0} f\ &=\ \sum_{i\geq 0}\frac{1}{i!}\mu_{i+1}(\delta_{c_0}a,a,\ldots,a)\\
&=\ \sum_{i,j\geq 0}\frac{1}{i!j!}\mu_{i+1}(\mu_{j+1}(a,\ldots,a,c_0),a,\ldots,a)\\
&=\ \sum_{i\geq 0}\frac{1}{i!}\sum_{j+k=i}\binom{i}{j}\mu_{j+1}(\mu_{k+1}(a,\ldots,a,c_0),a,\ldots,a)\\
&=\ \sum_{i\geq 1}\frac{1}{(i-1)!}\sum_{j+k=i}\binom{i-1}{j}\mu_{j+1}(\mu_{k}(a,\ldots,a,c_0),a,\ldots,a)\\   
&=\ \sum_{i\geq 1}\frac{1}{(i-1)!}\sum_{j+k=i}\binom{i-1}{k-1}\mu_{j+1}(\mu_{k}(a,\ldots,a,c_0),a,\ldots,a)\\
&=\ \sum_{i\geq 1}\frac{1}{(i-1)!}\sum_{j+k=i}(-1)^{j-1}\binom{i-1}{j-1}\mu_{j+1}(\mu_{k}(a,\ldots,a),a,\ldots,a,c_0)\\
&=\ \sum_{i\geq 0}\frac{1}{i!}\sum_{j=0}^i(-1)^{i-j}\binom{i}{i-j}\mu_{i-j+2}(\mu_{j}(a,\ldots,a),a,\ldots,a,c_0)\\
&=\ \sum_{i,j\geq 0}\frac{(-1)^i}{i!j!}\mu_{i+2}(\mu_j(a,\ldots,a),a,\ldots,a,c_0)\\
&=\ \sum_{i\geq 0}\frac{(-1)^i}{i!}\mu_{i+2}(f,a,\ldots,a,c_0)~,
\end{aligned}
\end{equation}
where in the sixth step have used the homotopy Jacobi identity~\eqref{eq:homotopyJacobi} for $(a,\ldots,a,c_0)$,
\begin{equation}
\begin{aligned}
&\sum_{j+k=i}(-1)^{j}\binom{i-1}{j-1}\mu_{j+1}(\mu_{k}(a,\ldots,a),a,\ldots,a,c_0)\,+\\ 
&\kern2cm+\sum_{j+k=i}\binom{i-1}{k-1}\mu_{j+1}(\mu_{k}(a,\ldots,a,c_0),a,\ldots,a)\ =\ 0~.
\end{aligned}
\end{equation}
Altogether we recover~\eqref{eq:GaugeTrafoCurvature}.

\paragraph{(Higher) gauge-of-gauge transformations}
Let us verify~\eqref{eq:ClosureGaugeTrafo}. Firstly, using~\eqref{eq:Cauchy}, we find
\begin{equation}
\begin{aligned}
  \delta_{c_{-1}}(\delta_{c_0} a)\ &=\ \sum_{i\geq0} \frac{1}{i!}\mu_{i+1}(a,\ldots,a,\delta_{c_{-1}}c_0) \\
  &=\  \sum_{i\geq0} \frac{(-1)^i}{i!}\mu_{i+1}(\delta_{c_{-1}}c_0, a,\ldots,a)\\
  &=\   \sum_{i,j\geq0} \frac{(-1)^i}{i!j!}\mu_{i+1}(\mu_{j+1}(a,\ldots,a,c_{-1}), a,\ldots,a)\\
  &=\ \sum_{i\geq 0}\frac{1}{i!}\sum_{j=0}^i(-1)^j\binom{i}{j}\mu_{j+1}(\mu_{i-j+1}(a,\ldots,a,c_{-1}), a,\ldots,a)\\
   &=\ \sum_{i\geq 1}\frac{1}{(i-1)!}\sum_{j+k=i}(-1)^j\binom{i-1}{k-1}\mu_{j+1}(\mu_{k}(a,\ldots,a,c_{-1}), a,\ldots,a)\\
    &=\ \sum_{i\geq 1}\frac{1}{(i-1)!}\sum_{j+k=i}(-1)^{j-1}\binom{i-1}{j-1}\mu_{j+1}(\mu_{k}(a,\ldots,a), a,\ldots,a,c_{-1})\\
    &=\ \sum_{i\geq 0}\frac{1}{i!}\sum_{j=0}^i(-1)^{i-j}\binom{i}{i-j}\mu_{i-j+2}(\mu_{j}(a,\ldots,a),a,\ldots,a,c_{-1})\\
&=\ \sum_{i,j\geq 0}\frac{(-1)^i}{i!j!}\mu_{i+2}(\mu_j(a,\ldots,a),a,\ldots,a,c_{-1})\\
&=\ \sum_{i\geq 0}\frac{(-1)^i}{i!}\mu_{i+2}(f,a,\ldots,a,c_{-1})~,   
  \end{aligned}
\end{equation}
where we have used the homotopy Jacobi identity~\eqref{eq:homotopyJacobi} for $(a,\ldots,a,c_{-1})$ in the sixth step,
\begin{equation}
\begin{aligned}
&\sum_{j+k=i}(-1)^{j}\binom{i-1}{j-1}\mu_{j+1}(\mu_{k}(a,\ldots,a),a,\ldots,a,c_{-1})\,+\\ 
&\kern2cm+\sum_{j+k=i}(-1)^{j}\binom{i-1}{k-1}\mu_{j+1}(\mu_{k}(a,\ldots,a,c_{-1}),a,\ldots,a)\ =\ 0~.
\end{aligned}
\end{equation}
This establishes the first part of~\eqref{eq:ClosureGaugeTrafo}.

As for the second part, the gauge transformation~\eqref{eq:GaugeTrafo} of $a$ (odd degree) and the gauge-of-gauge transformation of $c_{0}$ (even degree) make it clear how to extend this to $c_{-k}$ for all $k\in\NN$. Indeed, a straightforward calculation shows that 
\begin{equation}
  \delta_{c_{-k-2}} (\delta_{c_{-k-1}}c_{-k})\ =\ \sum_{i\geq0} \frac{(-1)^i}{i!}\mu_{i+2}(f, a,\ldots,a,c_{-k-2})~.
\end{equation}

\paragraph{Covariant derivative.}
Next, we verify~\eqref{eq:GTNabla} and~\eqref{eq:NableSquared}. For~\eqref{eq:GTNabla}, we make use of the definitions~\eqref{eq:GaugeTrafo},~\eqref{eq:GaugeTrafoMatter}, and~\eqref{eq:CovariantDerivative} together with~\eqref{eq:Cauchy} to obtain
\begin{equation}
\begin{aligned}
 \delta_{c_0}(\nabla\phi)\ &=\ \sum_{i\geq0}\frac{1}{i!}\Big[\mu_{i+2}(\delta_{c_0}a,a,\ldots,a,\phi)+(-1)^{i(|\phi|_\sL+1)}\mu_{i+1}(\delta_{c_0}\phi,a,\ldots,a)\Big]\\
  &=\ \sum_{i,j\geq0}\frac{1}{i!j!}\Big[\mu_{i+2}(\mu_{j+1}(a,\ldots,a,c_0),a,\ldots,a,\phi)\,-\\
 &\kern1cm-(-1)^{i(|\phi|_\sL+1)}\mu_{i+1}(\mu_{j+2}(a,\ldots,a,c_0,\phi),a,\ldots,a)\Big]\\
   &=\ \sum_{i\geq0}\frac{1}{i!}\sum_{j+k=i}\binom{i}{j}\Big[\mu_{j+2}(\mu_{k+1}(a,\ldots,a,c_0),a,\ldots,a,\phi)\,-\\
 &\kern1cm-(-1)^{j(|\phi|_\sL+1)}\mu_{j+1}(\mu_{k+2}(a,\ldots,a,c_0,\phi),a,\ldots,a)\Big]\\
    &=\ \sum_{i\geq2}\frac{1}{(i-2)!}\sum_{j+k=i}\binom{i-2}{j-1}\mu_{j+1}(\mu_{k}(a,\ldots,a,c_0),a,\ldots,a,\phi)\,-\\
 &\kern1cm-\sum_{i\geq2}\frac{1}{(i-2)!}\sum_{j+k=i}(-1)^{j(|\phi|_\sL+1)}\binom{i-2}{j}\mu_{j+1}(\mu_{k}(a,\ldots,a,c_0,\phi),a,\ldots,a)\\
    &=\ \sum_{i\geq2}\frac{1}{(i-2)!}\sum_{j+k=i}\binom{i-2}{k-1}\mu_{j+1}(\mu_{k}(a,\ldots,a,c_0),a,\ldots,a,\phi)\,-\\
 &\kern1cm-\sum_{i\geq2}\frac{1}{(i-2)!}\sum_{j+k=i}(-1)^{j(|\phi|_\sL+1)}\binom{i-2}{k-2}\mu_{j+1}(\mu_{k}(a,\ldots,a,c_0,\phi),a,\ldots,a)\\
 &=\  \sum_{i\geq2}\frac{1}{(i-2)!}\sum_{j+k=i}(-1)^{j}\binom{i-2}{j-2}\mu_{j+1}(\mu_{k}(a,\ldots,a),a,\ldots,a,c_0,\phi)\,+\\
 &\kern1cm+ \sum_{i\geq2}\frac{1}{(i-2)!}\sum_{j+k=i}(-1)^{(j+1)|\phi|_\sL}\binom{i-2}{k-1}\mu_{j+1}(\mu_{k}(a,\ldots,a,\phi),a,\ldots,a,c_0)\\
 &=\  \sum_{i\geq2}\frac{1}{(i-2)!}\sum_{j+k=i}(-1)^{j}\binom{i-2}{j-2}\mu_{j+1}(\mu_{k}(a,\ldots,a),a,\ldots,a,c_0,\phi)\,+\\
 &\kern1cm+ \sum_{i\geq2}\frac{1}{(i-2)!}\sum_{j+k=i}(-1)^{(j+1)|\phi|_\sL}\binom{i-2}{j-1}\mu_{j+1}(\mu_{k}(a,\ldots,a,\phi),a,\ldots,a,c_0)\\
  &=\  \sum_{i\geq0}\frac{1}{i!}\sum_{j=0}^i (-1)^{j}\binom{i}{j}\Big[\mu_{j+3}(\mu_{i-j}(a,\ldots,a), a,\ldots,a,c_0,\phi)\,+\\
   &\kern1cm+(-1)^{(j+1)|\phi|_\sL}\mu_{j+2}(\mu_{i-j+1}(a,\ldots,a,\phi),a,\ldots,a,c_0)\Big]\\
   &=\ \sum_{i,j\geq0}\frac{1}{i!j!}\Big[(-1)^i\mu_{i+3}(\mu_{j}(a,\ldots,a), a,\ldots,a,c_0,\phi)\,+\\
   &\kern1cm+(-1)^{i|\phi|_\sL}\mu_{i+2}(\mu_{j+1}(a,\ldots,a,\phi),a,\ldots,a,c_0)\Big]~,\\
    &=\ \sum_{i,j\geq0}\frac{1}{i!j!}\Big[(-1)^i\mu_{i+3}(\mu_{j}(a,\ldots,a), a,\ldots,a,c_0,\phi)\,-\\
   &\kern1cm-\mu_{i+2}(a,\ldots,a,c_0,\mu_{j+1}(a,\ldots,a,\phi))\Big]~,
 \end{aligned}
\end{equation}
where in the sixth step we have used the homotopy Jacobi identity~\eqref{eq:homotopyJacobi} for the elements $(a,\ldots,a,c_0,\phi)$,
\begin{equation}
\begin{aligned}
&\sum_{j+k=i}(-1)^{j}\binom{i-2}{j-2}\mu_{j+1}(\mu_{k}(a,\ldots,a),a,\ldots,a,c_0,\phi)\,-\\ 
&\kern2cm-\sum_{j+k=i}\binom{i-2}{k-1}\mu_{j+1}(\mu_{k}(a,\ldots,a,c_0),a,\ldots,a,\phi)\,+\\
&\kern2cm+\sum_{j+k=i}(-1)^{(j+1)|\phi|_\sL}\binom{i-2}{k-1}\mu_{j+1}(\mu_{k}(a,\ldots,a,\phi),a,\ldots,a,c_0)\,+\\
&\kern2cm+\sum_{j+k=i}(-1)^{j(|\phi|_\sL+1)} \binom{i-2}{k-2}\mu_{j+1}(\mu_{k}(a,\ldots,a,c_0,\phi),a,\ldots,a)\ =\ 0~.
\end{aligned}
\end{equation}
Hence, using~\eqref{eq:Curvature} and~\eqref{eq:CovariantDerivative}, we obtain
\begin{equation}
 \delta_{c_0}(\nabla\phi)\ =\  -\sum_{i\geq0}\frac{1}{i!}\mu_{i+2}(a,\ldots,a,c_0,\nabla\phi)+\sum_{i\geq0}\frac{(-1)^i}{i!}\mu_{i+3}(f,a,\ldots,a,c_0,\phi)~,
\end{equation}
as required.

To verify~\eqref{eq:NableSquared}, consider
\begin{equation}
\begin{aligned}
\nabla^2\phi\ &=\ \sum_{i\geq 0}\frac{1}{i!}\mu_{i+1}(a,\ldots,a,\nabla\phi)\\
&=\ \sum_{i,j\geq 0}\frac{(-1)^{i|\phi|_\sL}}{i!j!}\mu_{i+1}(\mu_{j+1}(a,\ldots,a,\phi),a,\ldots,a)\\
&=\ \sum_{i\geq 0}\frac{1}{i!}\sum_{j+k=i}(-1)^{j|\phi|_\sL}\binom{i}{j}\mu_{j+1}(\mu_{k+1}(a,\ldots,a,\phi),a,\ldots,a)\\
&=\ \sum_{i\geq 1}\frac{1}{(i-1)!}\sum_{j+k=i}(-1)^{j|\phi|_\sL}\binom{i-1}{j}\mu_{j+1}(\mu_{k}(a,\ldots,a,\phi),a,\ldots,a)\\   
&=\ \sum_{i\geq 1}\frac{1}{(i-1)!}\sum_{j+k=i}(-1)^{j|\phi|_\sL}\binom{i-1}{k-1}\mu_{j+1}(\mu_{k}(a,\ldots,a,\phi),a,\ldots,a)\\
&=\ \sum_{i\geq 1}\frac{1}{(i-1)!}\sum_{j+k=i}(-1)^{j-1}\binom{i-1}{j-1}\mu_{j+1}(\mu_{k}(a,\ldots,a),a,\ldots,a,\phi)\\
&=\ \sum_{i\geq 0}\frac{1}{i!}\sum_{j=0}^i(-1)^{i-j}\binom{i}{i-j}\mu_{i-j+2}(\mu_{j}(a,\ldots,a),a,\ldots,a,\phi)\\
&=\ \sum_{i,j\geq 0}\frac{(-1)^i}{i!j!}\mu_{i+2}(\mu_j(a,\ldots,a),a,\ldots,a,\phi)\\
&=\ \sum_{i\geq 0}\frac{(-1)^i}{i!}\mu_{i+2}(f,a,\ldots,a,\phi)~,
\end{aligned}
\end{equation}
where we have used~\eqref{eq:Cauchy} in the third step and the homotopy Jacobi identity~\eqref{eq:homotopyJacobi} for $(a,\ldots,a,\phi)$,
\begin{equation}
\begin{aligned}
&\sum_{j+k=i}(-1)^{j}\binom{i-1}{j-1}\mu_{j+1}(\mu_{k}(a,\ldots,a),a,\ldots,a,\phi)\,+\\ 
&\kern2cm+\sum_{j+k=i}(-1)^{j|\phi|_\sL}\binom{i-1}{k-1}\mu_{j+1}(\mu_{k}(a,\ldots,a,\phi),a,\ldots,a)\ =\ 0~,
\end{aligned}
\end{equation}
in the sixth step. Altogether, we arrive at~\eqref{eq:NableSquared},
\begin{equation}
\nabla^2\phi\ =\ \sum_{i\geq 0}\frac{(-1)^i}{i!}\mu_{i+2}(f,a,\ldots,a,\phi)~.
\end{equation}

\paragraph{\mathversion{bold}$L_\infty$-morphisms and Maurer--Cartan elements in the coalgebra picture.} In the following, we explain formulas~\eqref{eq:morphism_on_a},~\eqref{eq:morphism_on_f}, and~\eqref{eq:morphism_on_omega} in detail. Recall that a morphism $\phi:\sL\rightarrow \sL'$ corresponds to a morphism of coalgebras $\Phi:\bigodot^\bullet_0\sL[1]\rightarrow \bigodot^\bullet_0\sL'[1]$ and satisfies $D\circ \Phi=\Phi\circ D$. In the dual, dga-picture, we have a morphism $\Phi^*:\CCC^\infty(\sL'[1])\rightarrow \CCC^\infty(\sL[1])$ satisfying $\Phi^*\circ Q=Q'\circ \Phi^*$. 

From equation~\eqref{eq:phi_action}, it follows that
\begin{subequations}
\begin{equation}
\begin{aligned}
\Phi(\de^{sa}_0)\ &=\ \Phi(sa+\tfrac12 sa\odot sa+\tfrac{1}{3!}sa\odot sa \odot sa+\cdots)\\
 &=\ \Phi^1_1(sa)+\tfrac12 \Phi^1_2(sa\odot sa)+\tfrac12 \Phi^1_1(sa)\odot \Phi^1_1(sa)+\tfrac{1}{3!}\Phi^1_3(sa\odot sa\odot sa)\,+\\
 &\kern1cm+\tfrac12 \Phi^1_2(sa\odot sa)\odot \Phi^1_1(sa)+\tfrac{1}{3!}\Phi^1_1(sa)\odot \Phi^1_1(sa)\odot \Phi^1_1(sa)+\cdots\\
 &=\ \de^{sa'}_0~,
\end{aligned}
\end{equation}
where
\begin{equation}
 a'\ :=\ \sum_{i\geq 1} \frac{1}{i!}\phi_i(a,\dots,a)~.
\end{equation}
\end{subequations}

We can then use equation~\eqref{eq:coalgebra_trick} to compute the curvature of $a'$ as
\begin{equation}
\begin{aligned}
 f'\ &=\ (s^{-1}\circ {\rm pr}_{\bigodot^1 \sL[1]}\circ D')(\de^{sa'}_0)\\\
  &=\ (s^{-1}\circ {\rm pr}_{\bigodot^1 \sL[1]}\circ D'\circ \Phi)(\de^{sa}_0)\\
 &=\ (s^{-1}\circ {\rm pr}_{\bigodot^1 \sL[1]}\circ \Phi\circ D)(\de^{sa}_0)\\
 &=\ (s^{-1}\circ {\rm pr}_{\bigodot^1 \sL[1]}\circ \Phi)(sf\odot\de^{sa})\\
 &=\ s^{-1}(\Phi_1^1(sf)+\tfrac12 \Phi_2^1(sf\odot sa)+\dots)\\
 &=\ \sum_{i\geq 0}\frac{(-1)^i}{i!}\phi_{i+1}(f,a,\ldots,a)~.
\end{aligned}
\end{equation}

Furthermore, using equation~\eqref{eq:coalgebra_trick}, we can write gauge transformations as follows in the coalgebra picture,
\begin{equation}\label{eq:gauge_trafos}
 \delta_{c_0} a\ =\ (s^{-1}\circ {\rm pr}_{\bigodot^1\sL[1]}\circ D)(sc_0\odot \dd e^{sa})~.
\end{equation}
This allows us to compare different gauge orbits,
\begin{equation}
\begin{aligned}
 \delta_{c'_0} a'\ &=\ s^{-1}\circ {\rm pr}_{\bigodot^1\sL[1]} D'(sc'_0\odot \de^{sa'})\\
 &=\ s^{-1}\circ {\rm pr}_{\bigodot^1\sL[1]}\circ D'\circ \Phi(sc_0\odot \de^{sa})\\
 &=\ s^{-1}\circ {\rm pr}_{\bigodot^1\sL[1]}\circ \Phi \circ D (sc_0\odot \de^{sa})\\
 &=\ s^{-1}\circ {\rm pr}_{\bigodot^1\sL[1]}\circ \Phi(s\delta_{c_0} a\odot \de^{sa}+sc_0\odot sf\odot \de^{sa})~,
\end{aligned}
\end{equation}
and we conclude that
\begin{equation}
 c_0'\ =\ \sum_{i\geq 0} \frac{1}{i!}\phi_{i+1}(a,\dots,a,c_0)~.
\end{equation}

\paragraph{\mathversion{bold}$L_\infty$-morphisms and Maurer--Cartan elements using brackets.}

Consider the definition~\eqref{eq:L_infty_morphism} of a general  $L_\infty$-morphism $(\sL,\mu_i)\to (\sL',\mu'_i)$ evaluated at $(\ell_1,\ldots,\ell_i)=(a,\ldots,a)$ for $a\in\sL_1$. Then, the left-hand-side of~\eqref{eq:L_infty_morphism}  becomes
\begin{subequations}
\begin{equation}\label{eq:LHSQuasiIsoCurvature}
\begin{aligned}
 &\sum_{j+k=i}(-1)^k\binom{i}{k}\phi_{k+1}(\mu_j(a,\ldots,a),a,\ldots,a)\ =\\
  & \ =\ i! \sum_{k_1+k_2=i}\frac{(-1)^{k_1}}{k_1!}\phi_{k_1+1}\left(\frac{1}{k_2!}\mu_{k_2}(a,\ldots,a),a,\ldots,a\right)~,
  \end{aligned}
\end{equation}
while the right-hand-side reads as
\begin{equation}\label{eq:RHSQuasiIsoCurvature}
\begin{aligned}
 &\sum_{j=1}^i\frac{1}{j!} \sum_{k_1+\cdots+k_j=i}\binom{i}{k_1}\binom{i-k_1}{k_2}\cdots\binom{i-k_1-\cdots-k_{j-2}}{k_{j-1}}\,\times\\
 &\kern1cm\times\mu'_j(\phi_{k_1}(a,\ldots,a),\ldots,\phi_{k_j}(a,\ldots,a))\\
 \ &=\ i! \sum_{j=1}^i\frac{1}{j!} \sum_{k_1+\cdots+k_j=i}\mu'_j\left(\frac{1}{k_1!}\phi_{k_1}(a,\ldots,a),\ldots,\frac{1}{k_j!}\phi_{k_j}(a,\ldots,a)\right).
\end{aligned}
\end{equation}
\end{subequations}
Hence, upon equating~\eqref{eq:LHSQuasiIsoCurvature} and~\eqref{eq:RHSQuasiIsoCurvature}, we obtain
\begin{equation}\label{eq:LHS=RHSQuasiIsoCurvature}
\begin{aligned}
 \sum_{k_1+k_2=i}&\frac{(-1)^{k_1}}{k_1!}\phi_{k_1+1}\left(\frac{1}{k_2!}\mu_{k_2}(a,\ldots,a),a,\ldots,a\right)\ =\\ 
 & \ =\ \sum_{j=1}^i\frac{1}{j!} \sum_{k_1+\cdots+k_j=i}\mu'_j\left(\frac{1}{k_1!}\phi_{k_1}(a,\ldots,a),\ldots,\frac{1}{k_j!}\phi_{k_j}(a,\ldots,a)\right).
 \end{aligned}
\end{equation}
Thus, setting
\begin{equation}\label{eq:DefLIsoA}
 a'\ :=\ \sum_{i\geq 1}\frac{1}{i!}\phi_i(a,\ldots,a)
\end{equation}
and using the Cauchy product formula~\eqref{eq:CauchyjSeries}, we obtain from~\eqref{eq:LHS=RHSQuasiIsoCurvature} the relation
\begin{subequations}
\begin{equation}\label{eq:CurvatureUnderLIso}
 \sum_{i\geq 0}\frac{(-1)^i}{i!}\phi_{i+1}(f,a,\ldots,a)\ =\ f'~,
\end{equation}
where
\begin{equation}
 f\ :=\ \sum_{i\geq1}\frac{1}{i!}\mu_i(a,\ldots,a)\eand  f'\ :=\ \sum_{i\geq1}\frac{1}{i!}\mu'_i(a',\ldots,a')~.
\end{equation}
\end{subequations}
are the corresponding curvatures. Thus, we conclude that under $L_\infty$-morphisms, MC elements are mapped to MC elements.

Recall the formula~\eqref{eq:GaugeTrafo} for gauge transformations,
\begin{equation}\label{eq:appGTLM}
 a\ \ \mapsto\  a+\delta_{c_0}a\ewith \delta_{c_0}a\ =\ \sum_{i\geq 0}\frac{1}{i!}\mu_{i+1}(a,\ldots,a,c_0) \efor c_0\ \in\ \sL_0~.
\end{equation}
We wish to study~\eqref{eq:DefLIsoA} under such transformations. 

Generally, we have
\begin{equation}\label{eq:GeneralDeformationALM}
\begin{aligned}
 \sum_{i\geq 1}\frac{1}{i!}\phi_{i}(a+\delta_{c_0}a,\ldots,a+\delta_{c_0}a)\ &=\  \underbrace{\sum_{i\geq 1}\frac{1}{i!}\phi_{i}(a,\ldots,a)}_{=\,a'}+\underbrace{\sum_{i\geq 0}\frac{1}{i!}\phi_{i+1}(\delta_{c_0}a,a,\ldots,a)}_{=:\,\Delta a'}\\
 \ &=\ a'+\Delta a'~.
 \end{aligned}
\end{equation}
To compute $\Delta a'$, we again consider equation~\eqref{eq:L_infty_morphism} for a general  $L_\infty$-morphism, and this time we evaluate it at $(\ell_1,\ldots,\ell_i)=(a,\ldots,a,c_0)$ for $c_0\in\sL_0$ and $a\in\sL_1$. The left-hand-side of that equation becomes
\begin{subequations}
\begin{equation}\label{eq:LHSQuasiIsoGT}
\begin{aligned}
&\sum_{j+k=i}\left[(-1)^k\binom{i-1}{k-1}\phi_{k+1}(\mu_{j}(a,\ldots,a),a,\ldots,a,c_0)\,+\right.\\ 
&\kern2cm+\left.\binom{i-1}{j-1}\phi_{k+1}(\mu_{j}(a,\ldots,a,c_0),a,\ldots,a)\right]\ =\\
&=\ (i-1)!\sum_{k_1+k_2=i}\left[\frac{(-1)^{k_1}}{(k_1-1)!}\phi_{k_1+1}\left(\frac{1}{k_2!}\mu_{k_2}(a,\ldots,a),a,\ldots,a,c_0\right)\right.\!+\\ 
&\kern2cm+\left. \frac{1}{k_1!}\phi_{k_1+1}\left(\frac{1}{(k_2-1)!}\mu_{k_2}(a,\ldots,a,c_0),a,\ldots,a\right)\right]~,
\end{aligned}
\end{equation}
while the right-hand-side reads as
\begin{equation}\label{eq:RHSQuasiIsoGT}
\begin{aligned}
 &\sum_{j=1}^i\frac{1}{(j-1)!} \sum_{k_1+\cdots+k_j=i}\binom{i-1}{k_1}\binom{i-1-k_1}{k_2}\cdots\binom{i-1-k_1-\cdots-k_{j-2}}{k_{j-1}}\,\times\\
 &\kern1cm\times\mu'_j(\phi_{k_1}(a,\ldots,a),\ldots,\phi_{k_j}(a,\ldots,a,c_0))\ =\\
 \ &=\ (i-1)! \sum_{j=1}^i\frac{1}{(j-1)!} \sum_{k_1+\cdots+k_j=i}\,\times\\
 &\kern1cm\times\mu'_j\left(\frac{1}{k_1!}\phi_{k_1}(a,\ldots,a),\ldots,\frac{1}{k_{j-1}!}\phi_{k_{j-1}}(a,\ldots,a),\frac{1}{(k_j-1)!}\phi_{k_j}(a,\ldots,a,c_0)\right).
\end{aligned}
\end{equation}
\end{subequations}
Hence, upon equating~\eqref{eq:LHSQuasiIsoGT} and~\eqref{eq:RHSQuasiIsoGT}, we obtain
\begin{equation}\label{eq:LHS=RHSQuasiIsoGT}
\begin{aligned}
 &\sum_{k_1+k_2=i}\left[\frac{(-1)^{k_1}}{(k_1-1)!}\phi_{k_1+1}\left(\frac{1}{k_2!}\mu_{k_2}(a,\ldots,a),a,\ldots,a,c_0\right)\right.\!+\\ 
&\kern1cm+\left. \frac{1}{k_1!}\phi_{k_1+1}\left(\frac{1}{(k_2-1)!}\mu_{k_2}(a,\ldots,a,c_0),a,\ldots,a\right)\right]\ =\\
 \ &=\  \sum_{j=1}^i\frac{1}{(j-1)!} \sum_{k_1+\cdots+k_j=i}\,\times\\
 &\kern1cm\times\mu'_j\left(\frac{1}{k_1!}\phi_{k_1}(a,\ldots,a),\ldots,\frac{1}{k_{j-1}!}\phi_{k_{j-1}}(a,\ldots,a),\frac{1}{(k_j-1)!}\phi_{k_j}(a,\ldots,a,c_0)\right).\\
 \end{aligned}
\end{equation}
Next, we set
\begin{equation}
 c'_0\ :=\ \sum_{i\geq 0}\frac{1}{i!}\phi_{i+1}(a,\ldots,a,c_0)
\end{equation}
and use~\eqref{eq:DefLIsoA} and~\eqref{eq:appGTLM}, the definition
\begin{equation}
 \delta_{c'_0}a'\ :=\ \sum_{i\geq 0}\frac{1}{i!}\mu'_{i+1}(a',\ldots,a',c'_0)~,
\end{equation}
and the Cauchy product formula~\eqref{eq:CauchyjSeries} to obtain
\begin{equation}
 -\sum_{i\geq 0}\frac{(-1)^i}{i!}\phi_{i+2}(f,a,\ldots,a,c_0)+ \sum_{i\geq 0}\frac{1}{i!}\phi_{i+1}(\delta_{c_0}a,a,\ldots,a)\ =\ \delta_{c'_0}a'
\end{equation}
from~\eqref{eq:LHS=RHSQuasiIsoGT}. Upon comparing this with~\eqref{eq:GeneralDeformationALM}, we find
\begin{equation}
 \Delta a'\ =\ \delta_{c'_0}a'+\sum_{i\geq 0}\frac{(-1)^i}{i!}\phi_{i+2}(f,a,\ldots,a,c_0)~.
\end{equation}
Consequently, for MC elements this reduces to
\begin{equation}
 \Delta a'\ =\ \delta_{c'_0}a'\quad\Longrightarrow\quad \sum_{i\geq 1} \frac{1}{i!}\phi_{i}(a+\delta_{c_0}a,\ldots,a+\delta_{c_0}a)\ =\ a'+\delta_{c'_0}a'~,
\end{equation}
and, combining this with~\eqref{eq:CurvatureUnderLIso}, we realise that gauge equivalent MC configurations are mapped to gauge equivalent MC configurations under $L_\infty$-morphisms.

\paragraph{A curvature identity.}
Let $\sL$ be equipped with an inner product $\langle-,-\rangle_\sL$ and let $f$ be the curvature as defined in~\eqref{eq:Curvature}. We wish to prove that
\begin{equation}\label{eq:CurvatureIdentity}
 \langle f,f\rangle_\sL\ =\ 0~.
\end{equation}

Firstly,
\begin{equation}
\begin{aligned}
\langle f,f\rangle_\sL\ &=\ \sum_{i,j\geq 0}\frac{1}{i!j!}\langle \mu_i(a,\ldots,a), \mu_j(a,\ldots,a)\rangle_\sL\\
&=\ \sum_{i\geq 0}\frac{1}{i!}\underbrace{\sum_{j+k=i}\binom{i}{j}\langle \mu_j(a,\ldots,a), \mu_k(a,\ldots,a)\rangle_\sL}_{=:\,F_i}~,
\end{aligned}
\end{equation}
and therefore
\begin{equation}\label{eq:SomeEq}
\begin{aligned}
 F_i\ &=\ \sum_{j=0}^i\binom{i}{j}\langle \mu_j(a,\ldots,a), \mu_{i-j}(a,\ldots,a)\rangle_\sL\\
 \ &=\ -\sum_{j=0}^i(-1)^j\binom{i}{j}\langle a, \mu_j(\mu_{i-j}(a,\ldots,a),a,\ldots,a)\rangle_\sL\\
 \ &=\ -\sum_{j=1}^{i-1}(-1)^j\binom{i}{j}\langle a, \mu_j(\mu_{i-j}(a,\ldots,a),a,\ldots,a)\rangle_\sL\\
  \ &=\ \sum_{j=0}^{i-2}(-1)^j\binom{i}{j+1}\langle a, \mu_{j+1}(\mu_{i-j-1}(a,\ldots,a),a,\ldots,a)\rangle_\sL~.
\end{aligned}
\end{equation}
Hence,
\begin{equation}
\begin{aligned}
 F_{i+1}\ &=\ \sum_{j=0}^{i-1}(-1)^j\binom{i+1}{j+1}\langle a, \mu_{j+1}(\mu_{i-j}(a,\ldots,a),a,\ldots,a)\rangle_\sL\\
 \ &=\ \sum_{j=0}^{i}(-1)^j\binom{i+1}{j+1}\langle a, \mu_{j+1}(\mu_{i-j}(a,\ldots,a),a,\ldots,a)\rangle_\sL\\
 \ &=\ \sum_{j=0}^{i}(-1)^j\left[\binom{i}{j}+\binom{i}{j+1}\right]\langle a, \mu_{j+1}(\mu_{i-j}(a,\ldots,a),a,\ldots,a)\rangle_\sL\\
  \ &=\ \sum_{j=0}^{i}(-1)^j\binom{i}{j+1}\langle a, \mu_{j+1}(\mu_{i-j}(a,\ldots,a),a,\ldots,a)\rangle_\sL~,
\end{aligned}
\end{equation}
where in the last step we have use the Bianchi identity~\eqref{eq:BianchiIdentity}. Therefore,
\begin{subequations}
\begin{equation}\label{eq:FiA}
\begin{aligned}
F_{i+1}\ &=\ \sum_{j=0}^{i-1}\binom{i}{j+1}\langle \mu_{j+1}(a,\ldots,a),\mu_{i-j}(a,\ldots,a)\rangle_\sL\\
&=\   \sum_{j=1}^{i}\binom{i}{j}\langle \mu_{j}(a,\ldots,a),\mu_{i+1-j}(a,\ldots,a)\rangle_\sL~.
\end{aligned}
\end{equation}
However, from the first line of~\eqref{eq:SomeEq}, we also have
\begin{equation}\label{eq:FiB}
 F_{i+1}\ =\ \sum_{j=1}^{i}\binom{i+1}{j}\langle \mu_j(a,\ldots,a), \mu_{i+1-j}(a,\ldots,a)\rangle_\sL~.
\end{equation}
\end{subequations}
Furthermore, for any $A_{ij}=A_{ji}$ we have the identity
\begin{equation}
\begin{aligned}
 &\binom{i+1}{1}A_{1i}+\binom{i+1}{2}A_{2i-1}+\cdots +\binom{i+1}{i-1}A_{i-12}+\binom{i+1}{i}A_{i1}\ =\\
 &\kern2cm\ =\ 2  \left[\binom{i}{1}A_{1i}+\binom{i}{2}A_{2i-1}+\cdots +\binom{i}{i-1}A_{i-12}+\binom{i}{i}A_{i1}\right].
 \end{aligned}
\end{equation}
Hence, using the symmetry of the inner product we take  $A_{ij}:=\langle \mu_i(a,\ldots,a), \mu_{j}(a,\ldots,a)\rangle_\sL$, which implies that the sum in~\eqref{eq:FiB} is twice the sum in~\eqref{eq:FiA}, that is,
\begin{equation}
\begin{aligned}
 &\sum_{j=1}^{i}\binom{i+1}{j}\langle \mu_j(a,\ldots,a), \mu_{i+1-j}(a,\ldots,a)\rangle_\sL\ =\\
 &\kern2cm\ =\ 2  \sum_{j=1}^{i}\binom{i}{j}\langle \mu_{j}(a,\ldots,a),\mu_{i+1-j}(a,\ldots,a)\rangle_\sL~.
\end{aligned}
\end{equation}
Consequently, we must have $F_{i+1}=0$ and so  $\langle f,f\rangle_\sL=0$, as claimed.

\paragraph{Becchi--Rouet--Stora--Tyutin transformations.} Let us verify~\eqref{eq:BRSTSquared}. Firstly, we have 
\begin{equation}
 \begin{aligned}
  Q_{\rm BRST} a\ &=\ \sum_{i\geq0} \frac{1}{i!}\mu_{i+1}(a,\ldots,a,c_0)~,\\
  Q_{\rm BRST} c_0\ &=\  -\sum_{i\geq0} \frac{1}{i!}\Big[\mu_{i+1}(a,\ldots,a,c_{-1})+\frac{1}{2!} \mu_{i+2}(a,\ldots,a,c_0,c_0)\Big]
  \end{aligned}
\end{equation}
and so
\begin{equation*}
 \begin{aligned}
  Q^2_{\rm BRST} a\ &=\ \sum_{i\geq0} \frac{1}{i!}\big[\mu_{i+2}(Q_{\rm BRST} a,a\ldots,a,c_0)+(-1)^i\mu_{i+1}(Q_{\rm BRST} c_0,a,\ldots,a)\big]\\
  \ &=\  \sum_{i\geq0} \frac{1}{i!j!}\Big[\mu_{i+2}(\mu_{j+1}(a,\ldots,a,c_0),a\ldots,a,c_0)\,+\\
  &\kern1cm -(-1)^i\mu_{i+1}(\mu_{j+1}(a,\ldots,a,c_{-1}), a,\ldots,a)\,-\\
  &\kern1cm-\frac{(-1)^i}{2!}\mu_{i+1}(\mu_{j+2}(a,\ldots,a,c_0,c_0),a,\ldots,a)\Big]\\
  \ &=\ \sum_{i\geq0}\frac{1}{i!}\sum_{j+k=i}\binom{i}{j}\Big[\mu_{j+2}(\mu_{k+1}(a,\ldots,a,c_0), a,\ldots,a,c_0)\,+\\
   &\kern1cm -(-1)^j\mu_{j+1}(\mu_{k+1}(a,\ldots,a,c_{-1}), a,\ldots,a)\,-\\
  &\kern1cm-\frac{(-1)^j}{2!}\mu_{j+1}(\mu_{k+2}(a,\ldots,a,c_0,c_0),a,\ldots,a)\Big]\\
  \ &=\ \sum_{i\geq2}\frac{1}{(i-2)!}\sum_{j+k=i}\binom{i-2}{j-1}\mu_{j+1}(\mu_{k}(a,\ldots,a,c_0), a,\ldots,a,c_0)\,+\\
  &\kern1cm - \sum_{i\geq1}\frac{1}{(i-1)!}\sum_{j+k=i}(-1)^j\binom{i-1}{j}\mu_{j+1}(\mu_k(a,\ldots,a,c_{-1}), a,\ldots,a)\,-\\
  &\kern1cm-\sum_{i\geq2}\frac{1}{(i-2)!}\sum_{j+k=i}\frac{(-1)^j}{2!}\binom{i-2}{j}\mu_{j+1}(\mu_{k}(a,\ldots,a,c_0,c_0),a,\ldots,a)\\
\end{aligned}
\end{equation*}
\begin{equation}
 \begin{aligned}
   \ &=\ \sum_{i\geq2}\frac{1}{(i-2)!}\sum_{j+k=i}\binom{i-2}{k-1}\mu_{j+1}(\mu_{k}(a,\ldots,a,c_0), a,\ldots,a,c_0)\,+\\
  &\kern1cm - \sum_{i\geq1}\frac{1}{(i-1)!}\sum_{j+k=i}(-1)^j\binom{i-1}{k-1}\mu_{j+1}(\mu_k(a,\ldots,a,c_{-1}), a,\ldots,a)\,-\\
  &\kern1cm-\sum_{i\geq2}\frac{1}{(i-2)!}\sum_{j+k=i}\frac{(-1)^j}{2!}\binom{i-2}{k-2}\mu_{j+1}(\mu_{k}(a,\ldots,a,c_0,c_0),a,\ldots,a)\\
  \ &=\  -\sum_{i\geq1}\frac{1}{(i-1)!}\sum_{j+k=i}(-1)^{j-1}\binom{i-1}{j-1}\mu_{j+1}(\mu_k(a,\ldots,a), a,\ldots,a,c_{-1})\,+\\
  &\kern1cm +\sum_{i\geq2}\frac{1}{(i-2)!}\sum_{j+k=i}\frac{(-1)^{j}}{2!}\binom{i-2}{j-2}\mu_{j+1}(\mu_{k}(a,\ldots,a),a,\ldots,a,c_0,c_0)\\
   \ &=\  \sum_{i\geq0}\frac{1}{i!}\sum_{j=0}^i (-1)^{j}\binom{i}{j}\Big[-\mu_{j+2}(\mu_{i-j}(a,\ldots,a), a,\ldots,a,c_{-1})\,+\\
   &\kern1cm+\frac{1}{2!}\mu_{j+3}(\mu_{i-j}(a,\ldots,a),a,\ldots,a,c_0,c_0)\Big]\\
   \ &=\ \sum_{i,j\geq0}\frac{(-1)^i}{i!j!}\Big[-\mu_{i+2}(\mu_{j}(a,\ldots,a), a,\ldots,a,c_{-1})\,+\\
   &\kern1cm+\frac{1}{2!}\mu_{i+3}(\mu_{j}(a,\ldots,a),a,\ldots,a,c_0,c_0)\Big]~,
     \end{aligned}
\end{equation}
where we used the homotopy Jacobi identity~\eqref{eq:homotopyJacobi} for $(a,\ldots,a,c_0,c_0)$ in the sixth step,
\begin{equation}
\begin{aligned}
&\frac{1}{2!}\sum_{j+k=i}(-1)^{j}\binom{i-2}{j-2}\mu_{j+1}(\mu_{k}(a,\ldots,a),a,\ldots,a,c_0,c_0)\,-\\ 
&\kern2cm-\sum_{j+k=i}\binom{i-2}{k-1}\mu_{j+1}(\mu_{k}(a,\ldots,a,c_0),a,\ldots,a,c_0)\,+\\
&\kern2cm+\frac{1}{2!}\sum_{j+k=i}(-1)^{j} \binom{i-2}{k-2}\mu_{j+1}(\mu_{k}(a,\ldots,a,c_0,c_0),a,\ldots,a)\ =\ 0~,
\end{aligned}
\end{equation}
and for $(a,\ldots,a,c_{-1})$,
\begin{equation}
\begin{aligned}
&\sum_{j+k=i}(-1)^{j}\binom{i-1}{j-1}\mu_{j+1}(\mu_{k}(a,\ldots,a),a,\ldots,a,c_{-1})\,+\\ 
&\kern2cm+\sum_{j+k=i}(-1)^{j} \binom{i-1}{k-1}\mu_{j+1}(\mu_{k}(a,\ldots,a,c_{-1}),a,\ldots,a)\ =\ 0~,
\end{aligned}
\end{equation}
respectively. Altogether, using the curvature $f$ defined in~\eqref{eq:MCEquation}, we arrive at
\begin{equation}
Q^2_{\rm BRST}a\ =\ \sum_{i\geq0}\frac{(-1)^i}{i!}\Big[-\mu_{i+2}(f, a,\ldots,a,c_{-1})+\frac{1}{2!}\mu_{i+3}(f,a,\ldots,a,c_0,c_0)\Big]~.
\end{equation}

\bibliographystyle{latexeu}

\bibliography{littleone}

\end{document}